\pdfoutput=1

\documentclass[oneside,11pt]{CUEDthesisPSnPDF}

\usepackage{amsmath}
\usepackage{graphicx}
\usepackage{subfigure}
\usepackage{amssymb}
\usepackage{url}
\usepackage{appendix}

\newcommand{\myvec}[1]{\vec{#1}\,}

\newcommand{\beq}{\begin{equation}}
\newcommand{\eeq}{\end{equation}}
\newcommand{\bea}{\begin{eqnarray}}
\newcommand{\eea}{\end{eqnarray}}

\newcommand{\mathsym}[1]{{}}

\def\BeginSingle{\def\baselinestretch{1.0}\large\normalsize}

\begin{document}

\thispagestyle{empty}
  \let\footnotesize\small
  \let\footnoterule\relax
  \BeginSingle
  \vspace*{-40mm}
  \bigskip
  \bigskip
  \centerline{\includegraphics[height=30mm]{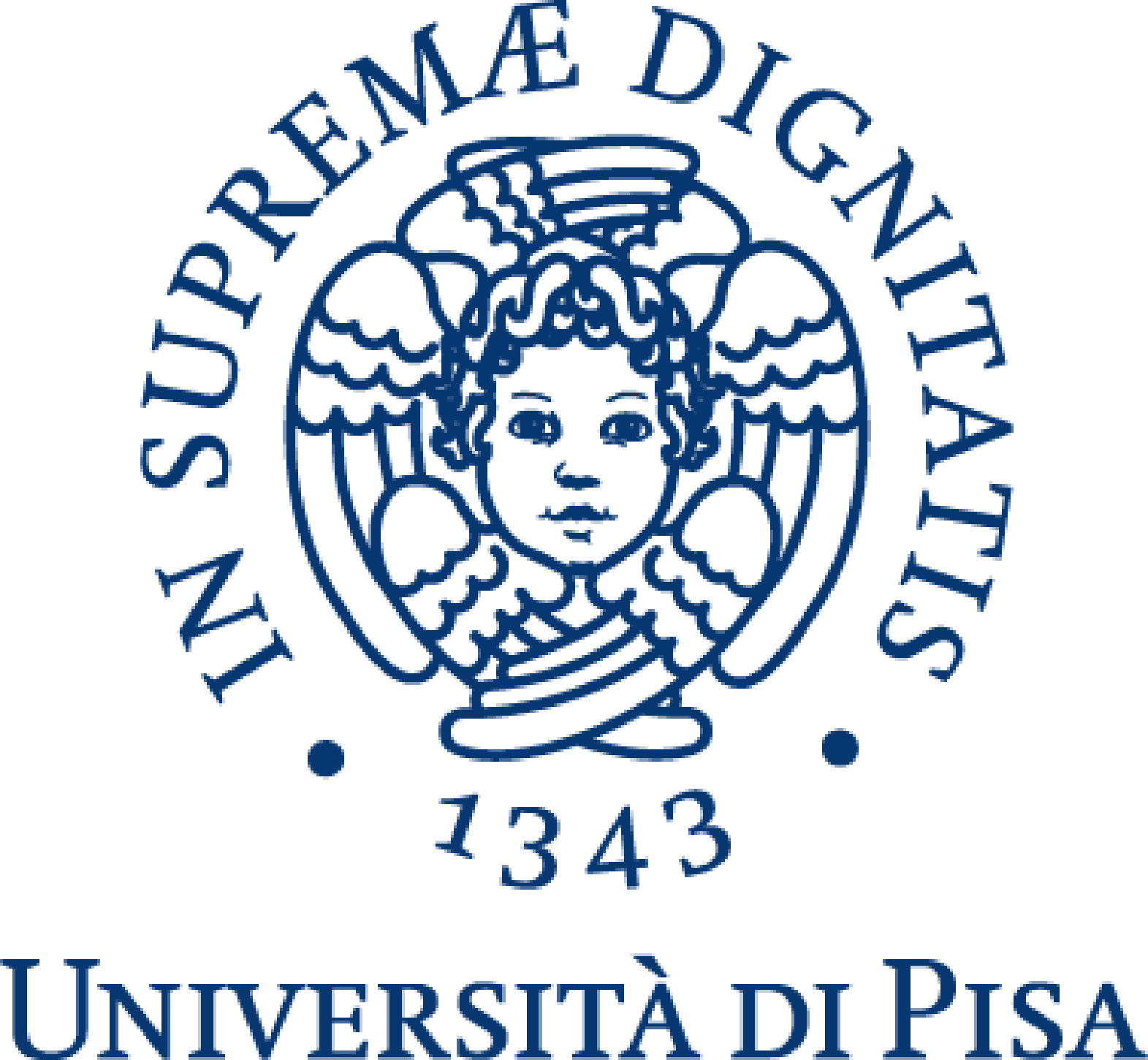}}
  \begin{center}
\vspace{10mm}
 \textsc{\Large GRADUATE COURSE IN PHYSICS}\medskip\\
    \textsc{\Large University of Pisa}\bigskip\\
\rule{50mm}{0.01mm}\medskip\\
       The School of Graduate Studies in Basic Sciences "GALILEO
       GALILEI"
\vspace{20mm}

    \vfill
    {\huge \bf Researches on Non-standard Optics for Advanced Gravitational Waves Interferometers.  \par}
    \bigskip\bigskip
    PhD thesis ($XVIII^\circ$ cycle ) \par\bigskip

      \href{mailto:agresti@df.unipi.it}{{\large Juri Agresti}}
 \par

  \end{center}\par
  \vfill
\vspace{20mm}
      Supervisor:
\vspace{2mm}
  \begin{quote}
    {\large Dr. R. DeSalvo}\newline
  \end{quote}\par

\vspace{10mm}
      Internal Supervisor:
  \vspace{2mm}
  \begin{quote}
    {\large Prof. F. Fidecaro}\newline
  \end{quote}\par

  \vfill
  \vspace{5mm}
  \begin{center}
    \rule{40mm}{0.01mm}\\
      {  2007}\par\bigskip \bigskip
LIGO-P080010-00-R

  \end{center}

% turn of those nasty overfull and underfull hboxes
\hbadness=10000
\hfuzz=50pt

% Put all the style files you want in the directory StyleFiles and usepackage like this:
%\usepackage{StyleFiles/watermark}

 \renewcommand\baselinestretch{1.2}
\baselineskip=18pt plus1pt

% A page with the abstract on including title and author etc may be
% required to be handed in separately. If this is not so, then comment
% the below 3 lines (between '\begin{abstractseparte}' and
% 'end{abstractseparate}'), normally like a declaration ... needs some more
% work, mind as environment abstracts creates a new page!
% \begin{abstractseparate}
%   \input{Abstract/abstract}
% \end{abstractseparate}

% Using the watermark package which is in StyleFiles/
% and to remove DRAFT COPY ONLY appearing on the top of all pages comment out below line
%\watermark{DRAFT COPY ONLY}

%set the number of sectioning levels that get number and appear in the contents
\setcounter{secnumdepth}{3}
\setcounter{tocdepth}{3}

\frontmatter
 % Thesis Dedictation ---------------------------------------------------

\begin{dedication} %this creates the heading for the dedication page

``I' ve seen things you people wouldn' t believe. Attack ships on
fire off the shoulder of Orion. I watched C-beams ... glitter in
the dark near Tannhauser Gate. All those ... moments will be lost
... in time, like tears ... in rain. Time ... to die.''

Roy Batty, \textit{Blade Runner}, 1982.

\end{dedication}

% ----------------------------------------------------------------------

%%% Local Variables:
%%% mode: latex
%%% TeX-master: "../thesis"
%%% End:

% Thesis Acknowledgements ------------------------------------------------

%\begin{acknowledgementslong} %uncommenting this line, gives a different acknowledgements heading
\begin{acknowledgements}    %this creates the heading for the acknowlegments

First of all, I would like to thank my advisor, Riccardo DeSalvo,
for his support, encouragement and for giving me so much freedom
to explore and discover new areas of physics. I must say I am very
pleased with the opportunity of being his student; for the
challenging and rich experience and intellectual growth it
represented, and for the energy he transmitted to me when moving
throughout such a long way.

Many thanks to my co-workers, from whom I received invaluable
guidance and knowledge:  Marco Giacinto Tarallo,  Erika
D'Ambrosio, Phil Willems, John Miller, Virginio Sannibale,
Innocenzo Pinto and Vincenzo Galdi. In particular, I single out
one of them, who have been not only a collaborator but also
genuine friend. With Marco I enjoyed  discussing  research topics
but also Borbonian, WWE and QUEFEF problems and had the best
racquetball matches ever.

During my stay in Pasadena I had the opportunity of sharing my way
with people from the almost all over the world  (even if most of
them from Italy) and this was a real treasure. My housemates,
Kalin, Snejana, Ganesh, Detchev family, my friends Artan, Alberto,
Ciro e Simona, Misha e Cinzia, Monica, Paola, Enrico,  and all the
other ``Caltech little Italy'' friends; each of them in their own
particular and unforgettable way made these years more than a nice
memory.

Thanks to family and friends for their presence and positive
spirit that helped me to further my PhD adventure.

I express my deepest gratitude and all my love to Chiara, who has
been bringing so much more into my life than I could ever dream. I
thank her for the patience, trust and support she has given during
these hard times, while I have been working over a thousand miles
away.

\vspace{2 cm}

The author gratefully acknowledge the support of the United States
National Science Foundation for the construction and operation of
the LIGO Laboratory and the Science and Technology Facilities
Council of the United Kingdom, the Max-Planck-Society, and the
State of Niedersachsen/Germany for support of the construction and
operation of the GEO600 detector. The author also gratefully
acknowledge the support of the research by these agencies and by
the Australian Research Council, the Council of Scientific and
Industrial Research of India, the Istituto Nazionale di Fisica
Nucleare of Italy, the Spanish Ministerio de Educacion y Ciencia,
the Conselleria d'Economia Hisenda i Innovacio of the Govern de
les Illes Balears, the Scottish Funding Council, the Scottish
Universities Physics Alliance, The National Aeronautics and Space
Administration, the Carnegie Trust, the Leverhulme Trust, the
David and Lucile Packard Foundation, the Research Corporation, and
the Alfred P. Sloan Foundation.

\end{acknowledgements}
%\end{acknowledgmentslong}

% ------------------------------------------------------------------------

%%% Local Variables:
%%% mode: latex
%%% TeX-master: "../thesis"
%%% End:

\tableofcontents
\listoffigures
%\printglossary  %% Print the nomenclature
%\addcontentsline{toc}{chapter}{Nomenclature}

\mainmatter

%%% Thesis Introduction --------------------------------------------------
\chapter{Introduction}
\ifpdf
    \graphicspath{{IntroductionFigs/PNG/}{IntroductionFigs/PDF/}{IntroductionFigs/}}
\else
    \graphicspath{{IntroductionFigs/EPS/}{IntroductionFigs/}}
\fi

Gravitational waves\footnote{In this thesis we decided to present
only the subjects which are strictly related to the research
topics of our work. The reader is addressed to
references~\cite{Carroll, Gravitation, ph237} for an introduction
to the subject or an in-depth treatment of gravitational waves
physics.}, one of the more discussed predictions of Einstein's
General Theory of Relativity may be detected within the next few
years. General Relativity predicts that massive accelerating and
rotating astrophysical objects emit gravitational waves which
propagate through space with the speed of light. Gravitational
waves are most simply thought of as ripples in the space-time
fabric, their effect being to change the separation of adjacent
masses on earth or in space; this tidal effect is the basis of all
present detectors. Sources such as interacting black holes,
coalescing compact binary systems, supernova explosions and
pulsars are all possible candidates for detection; observing
signals from them will significantly boost our understanding of
the Universe.

 The first gravitational wave
detectors were based on the effect of these tidal forces on the
fundamental resonant mode of aluminum bars at room temperature.
Initial instruments were constructed by Joseph Weber~\cite{Weber}.
Following the lack of confirmed detection of signals, aluminum bar
systems operated at and below the temperature of liquid helium
were developed and work in this area is still
underway~\cite{Allegro,Auriga,Exp-Naut}. However the most
promising design of gravitational wave detectors, offering the
possibility of very high sensitivities over a wide range of
frequency, uses widely separated test masses freely suspended as
pendulums on earth~\cite{Saulson} or in a drag free craft in
space; laser interferometry provides a means of sensing the motion
of the masses produced as they interact with a gravitational wave.
Ground based detectors of this type have the sensitivity to
observe sources whose radiation is emitted at frequencies above a
few Hz, and space borne detectors will be developed for
implementation at lower frequencies. Already gravitational wave
detectors of long baseline are operating in a number of places
around the world; in the USA (LIGO project~\cite{LIGO} led by a
Caltech/ MIT consortium) , in Italy (VIRGO project~\cite{Virgo}, a
joint Italian/French venture), in Germany (GEO $600$
project~\cite{Geo}, a UK/German collaboration) and in Japan (TAMA
300 project ~\cite{Tama}) . LISA ~\cite{LISA}, a space-based
detector proposed by a collaboration of European and US research
groups, is one of the most challenging large scale experiment of
the next future.  This detector array should have the capability
of detecting gravitational wave signals from many astrophysical
events in the Universe, providing unique information on testing
aspects of General Relativity and opening up a new field of
astronomy.

 Gravitational wave strengths are
characterized by the gravitational wave amplitude $h$, given by $h
= 2\Delta L/ L $, where $\Delta L$ is the change in separation of
two masses a distance $L$ apart. Unlike electromagnetism,
gravitational radiation field is quadrupole in nature and this
shows up in the pattern of the interaction of the waves with
matter. The problem for the experimental physicist is that the
predicted magnitudes of the amplitudes or strains in space in the
vicinity of the earth caused by gravitational waves even from the
most violent astrophysical events are extremely small, of the
order of $10^{-21}$ or lower (for a review of the gravitational
wave sources and expected signal strength see~\cite{Cut-Thorne}).
Indeed current theoretical models on the event rate and strength
of such events suggest that in order to detect a few events per
year, from coalescing neutron star binary systems for example, an
amplitude sensitivity close to $10^{-22}$ over timescales as short
as a millisecond is required. If the Fourier transform of a likely
signal is considered, it is found that the energy of the signal is
distributed over a frequency range or bandwidth which is
approximately equal to $1/$timescale. Thus detector noise levels
must have an amplitude spectral density lower than $\simeq
10^{-23}(Hz)^{-1/2}$ over the frequency range of the signal. The
weakness of the signal means that limiting noise sources like the
thermal motion of molecules in the detector (thermal noise),
seismic or other mechanical disturbances, and noise associated
with the detector readout, whether electronic or optical, must be
reduced to a very low level. For signals above $\approx 10 Hz$
ground based experiments are possible, but for lower frequencies
where local fluctuating gravitational gradients and seismic noise
on earth become a problem, it is best to consider developing
detectors for operation underground~\cite{AspenRic} or in
space~\cite{LISA}.

Gravitational wave detector using laser interferometry, offers the
possibility of very high sensitivities over a wide range of
frequency. It uses test masses which are widely separated and
freely suspended as pendulums to isolate against seismic noise;
laser interferometry provides a means of sensing the motion of
these masses produced as they interact with a gravitational wave.

\begin{figure}[htb]
\begin{center}
\includegraphics[width=0.9\textwidth]{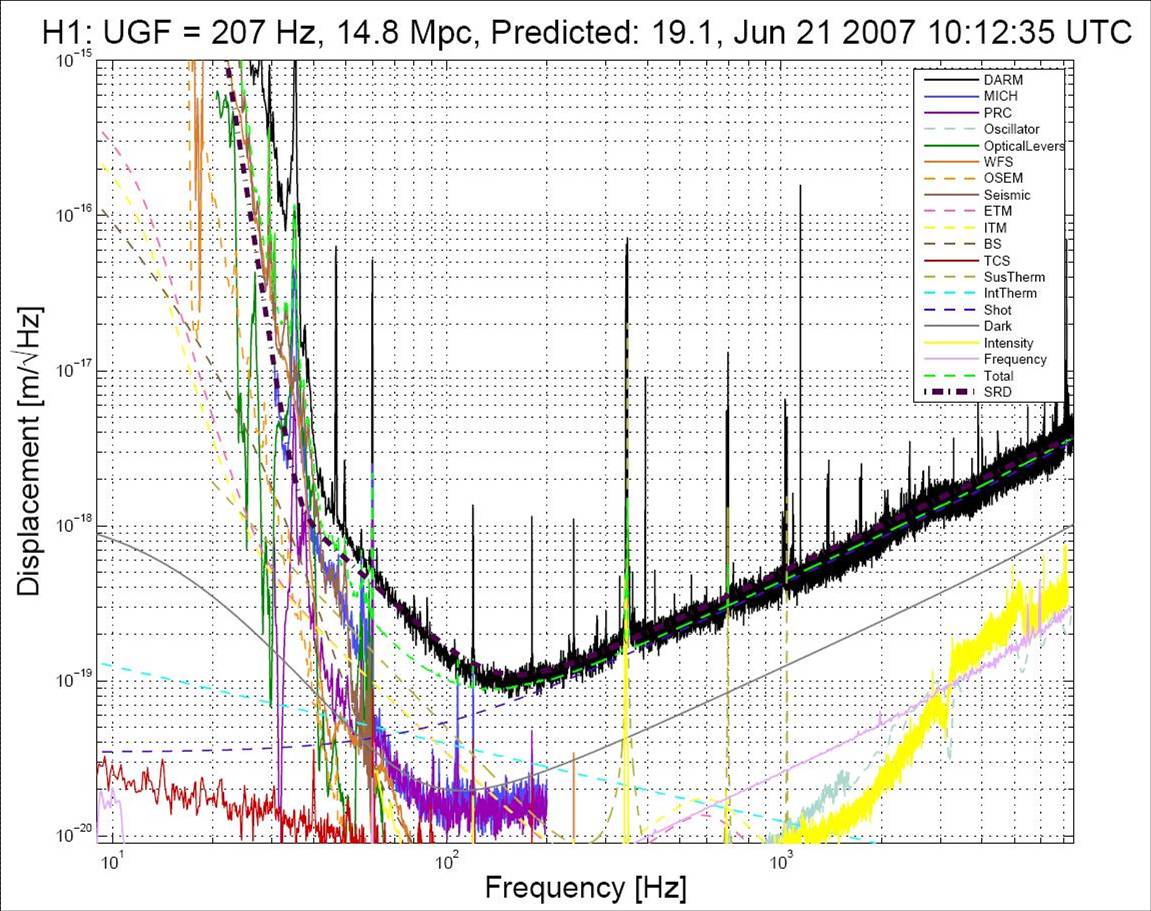}
\end{center}
\caption{Sensitivity of the LIGO detector $H1$ during the science
run $S5$.} \label{fig:LIGOnoise}
\end{figure}

This technique is based on the Michelson interferometer and is
particularly suited to the detection of gravitational waves as
they have a quadrupole nature. Waves propagating perpendicular to
the plane of the interferometer will result in one arm of the
interferometer being increased in length while the other arm is
decreased and vice versa. The induced change in the length of the
interferometer arms results in a small change in the intensity of
the light observed at the interferometer output.  The sensitivity
of an interferometric gravitational wave detector is limited by
noise from various sources. Taking this frequency dependent noise
floor into account, the American LIGO  detectors have  a
sensitivity (shown in Fig.~\ref{fig:LIGOnoise}), which would allow
a reasonable probability for detecting gravitational wave sources.

In order to observe a full range of sources and to initiate
gravitational wave astronomy, a sensitivity or noise performance
approximately ten times better in the mid-frequency range and
several orders of magnitude better at $10$ Hz, is desired. Such a
performance is planned for a future LIGO upgrade, Advanced
LIGO~\cite{AdLIGO}.

In general, for ground based detectors the most important
limitations to sensitivity result from the effects of seismic and
other ground-borne mechanical noise, thermal noise associated with
the test masses and their suspensions and shot noise in the
photocurrent from the photodiode which detects the interference
pattern.

\begin{figure}[htb]
\begin{center}
\includegraphics[width=0.8\textwidth]{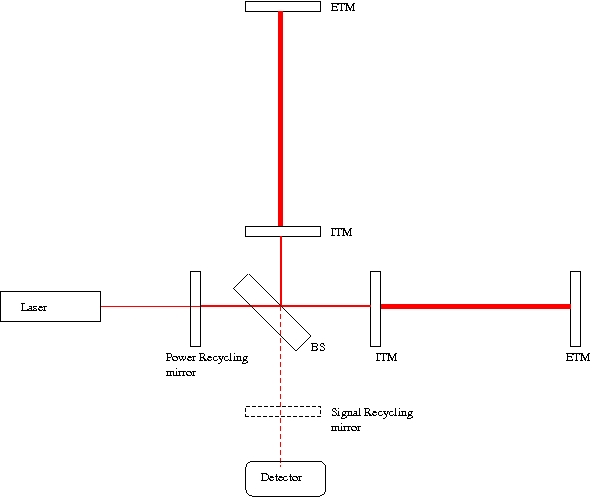}
\end{center}
\caption{Optical scheme of the LIGO GW interferometers.}
\label{fig:LIGOsch}
\end{figure}

Most modern designs implement improved versions of a simple
Michelson interferometer (see Fig.~\ref{fig:LIGOsch}). A simple
Michelson interferometer has an antenna response function which
has a maximum sensitivity for $\tau_{rt}=\tau_{gw}/2$,where
$\tau_{rt}= 2L/c$ is the round-trip travel time for photons
leaving and returning to the beamsplitter and $\tau_{gw}$ is the
period of the gravitational wave signal ($L$ the length of each
arm). A simple calculation from this expression shows that for
frequency between $10 Hz$ and $1 kHz$, the optimal antenna length
is of order $10^5 m$ to $10^7 m$. This is much larger than would
be feasible for an earth based detector.

 \begin{figure}[htb]
\begin{center}
\includegraphics[width=\textwidth]{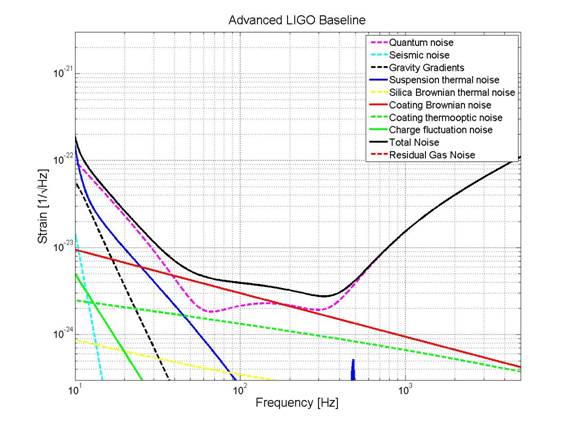}
\end{center}
\caption{Advanced LIGO sensitivity design.} \label{fig:AdLIGOB}
\end{figure}

 The situation can be helped greatly if a multi-pass arrangement
is used in the arms of the interferometer as this multiplies up
the apparent movement by the number of bounces the light makes in
the arms. The multiple beams can either be separate as in an
optical delay line, like the GEO $600$ configuration, or may lie
on top of each other as in a Fabry-Perot resonant cavity, used in
all others interferometric gravitational waves detectors. A
Fabry-Perot cavity consists of a partially transmitting input
mirror and a high reflective rear mirror. If the length of the
Fabry-Perot is adjusted to a multiple of the laser wavelength the
cavity becomes resonant. The light power inside the cavity builds
up and simulates the effect of sending the light forth and back
multiple times. However, in this case the number of bounces is not
a fixed quantity, but rather an averaged effective value.
Optimally, the light should be stored for a time comparable to the
characteristic timescale of the signal. Thus if signals of
characteristic timescale $1\,ms$ are to be searched for, the
number of bounces should be approximately $50$ for an arm length
of $4 km$.

If the mirrors have low optical losses and if the rear mirror is a
high reflector, most of the power incident to a Fabry-Perot arm
cavity will be reflected back to the beam splitter. Ideally, the
anti-symmetric port of the Michelson interferometer is set on a
dark fringe to minimize shot noise~\cite{Edelstein}. Then a
differential length change induced by a gravitational wave will
leave through the anti-symmetric port with the highest possible
signal-to-noise ratio. This in turn means that most of the
injected light will leave the interferometer through the symmetric
port and be lost. By placing an additional partially transmitting
mirror at input one can form yet another cavity, the power
recycling cavity, and recycle most of the otherwise lost light.
The interferometer response is then enhanced by the power
recycling gain (the shot noise is reduced by the additional power
build-up in the power recycling cavity) and this configuration is
called Power-recycled Michelson with Fabry-Perot arm cavities. By
adding a partially transmitting mirror to the anti-symmetric
output port the gravitational wave signal can be made
resonant~\cite{Strain}. This makes it possible to shape the
interferometer response, so that its sensitivity is improved in a
narrow frequency band around the signal resonance. In general,
this means that the sensitivity outside the resonant frequency
band will be worse. This is not a problem at lower frequencies
where the interferometer is usually limited by seismic noise. If
both power and signal recycling are implemented the configuration
is called dual recycled.

\section{Topics discussed in this thesis}

Thermal noise of the test masses is expected to be a limiting
factor in Advanced GW interferometers. As an example,
Fig.~\ref{fig:AdLIGOB} shows the expected sensitivity curve for
Advanced LIGO interferometer with the main noise contribution
enlightened.  Many research groups work on $R\&D$ activities
finalized to improve the thermal noise performance of next
generation detectors. Some research lines deal with cryogenic
temperature, other with improved or new materials, other with
optical beam shaping and optimization of the mirror geometry
and/or coating. Non Gaussian beams have been proposed years
ago~\cite{reducing,erika} to reduce a particular type of thermal
noise (substrate thermoelastic).

In this thesis we provide a quantitative  analysis of the impact
of  non-Gaussian beams on different kinds of thermal
noises\footnote{In particular we  show that the thermal noise of
the mirror's dielectric coating is greatly reduced by using flat
profile beams. This resolved a question raised ~\cite{Tourn-Elba}
on the effectiveness of this type of beam on coating noise}. We
show that the mesa beam implementation could boost the Advanced
LIGO sensitivity considerably: even with a rough estimation
(without re-optimizing the detector for the introduction of mesa
beams), the  binary neutron star inspiral range increases from
$175$ Mpc to $225$ Mpc.

We illustrate the importance of uniform sampling of the mirror
surface to reduce thermal noise and the limitation brought by the
use of excited modes with nodes on the mirror surface.

 We developed the theory of mesa beam, in view of a
future implementation in advanced GW interferometers of the mesa
beam idea,  focusing on the analytical derivation of the
quantities (beam width, divergence, $M^2$ factor, etc...), which
are chosen as ``ISO standard''~\cite{ISO} reference parameters for
the characterization of an optical beam.

 We also analytically
proved a new duality relation between optical cavities with
non-spherical mirrors. This derivation provides a unique mapping
between the eigenvalues and eigenvectors of two cavities  whose
mirrors shapes are related by a simple relation. This duality
allows the direct application of beam property calculations
performed in a case to geometries of the other configuration.

 The interest of
the GW community in this new beam technology led us to the
construction and testing of a prototype mesa beam Fabry-Perot
cavity with mexican-hat mirror. Part of the work of this thesis
was devoted to the development of new simulation programs of
optical systems\footnote{Available at
\url{http://www.ligo.caltech.edu/~jagresti/}}. These programs
provided the theoretical expected behavior of our experiment, in
particular cavity's modes structure and misalignments sensitivity
to be confronted with the experimental results. We developed new
simulation packages to analyze the performance of our cavity
prototype with real imperfect mirrors, using the measured mirrors
maps. The model developed can include uniform and non-uniform
scattering and absorption losses, as well as the effects of mirror
heating A particular attention has been devoted to keep theses
simulation programs very easy to use and very easy to
change/upgade\footnote{ Most of the time we used
\textit{Mathematica}$^{\circledR}$ for its capability of
analytical manipulation and numerical analysis or
\textit{Matlab}$^{\circledR}$ for its ability in manipulating
large matrices.} by anyone involved in optics
research\footnote{The great simplicity and versatility of these
programs turned to be fundamental properties for the usage by
other $R\&D$ research groups.}. The good agreement between theory
and experiment validated the mathematical tools here developed
thus allowing safe extrapolation to the larger optical systems
needed in GW observatories.

 We also explored another complementary way of reducing the mirror
thermal noise, beside the beam shaping, that is the multi-layered
coating thickness optimization. We  show it to be effective in
reducing the coating noise and explore the possible implications
for GW interferometers in terms of sensitivity.

 During this
analysis we developed an independent model for the coating
effective elastic parameters, which is based on the well
understood subject of homogenization theory.

%%% ----------------------------------------------------------------------

%%% Local Variables:
%%% mode: latex
%%% TeX-master: "../thesis"
%%% End:

\chapter{Paraxial beam and optical cavities}
\ifpdf
    \graphicspath{{Chapter1Figs/PNG/}{Chapter1Figs/PDF/}{Chapter1Figs/}}
\else
    \graphicspath{{Chapter1Figs/EPS/}{Chapter1Figs/}}
\fi

\section{Introduction}

 This chapter is a brief description of the techniques used in this thesis to
 study paraxial laser beam propagation and optical resonators. It
 gives an overview of the main concepts related to standard
 spherical mirrors optics, and some less known concepts related to
 generic paraxial beam and resonators which is one of the main
 subject of this work.
The reader familiar with these techniques can jump to
Chapter~\ref{Ch2}.

\section{Paraxial laser beams}
Radiation from lasers is different from conventional optical light
because it is very close to be monochromatic. Although each laser
has its own fine spectral distribution and noise properties, the
electric and magnetic fields from lasers are considered to have
minimal phase and amplitude variations in the first-order
approximation. Like microwaves from a maser, electromagnetic
radiation with a precise phase and amplitude is described most
accurately by Maxwell's wave equations. For devices with
structures that have dimensions very much larger than the
wavelength, e.g. in a multimode fiber or in an optical system
consisting of lenses, prisms or mirrors, the rigorous analysis of
Maxwell's vector wave equations becomes very complex and tedious:
there are too many modes in such a large space. It is difficult to
solve Maxwell's vector wave equations for such cases, even with
large computers. Even if we find the solution, it would contain
fine features (such as the fringe fields near the lens) which are
often of little or no significance to practical applications. In
these cases we look for a simple analysis which can give us just
the main features (i.e. the amplitude and phase) of the dominant
component of the electromagnetic field in directions close to the
direction of propagation and at distances reasonably far away from
the aperture. When one deals with laser radiation fields which
have slow transverse variations and which interact with devices
that have overall dimensions much larger than the optical
wavelength $\lambda$, the fields can often be approximated as
transverse electric and magnetic (TEM) waves. In TEM waves both
the dominant electric field and the dominant magnetic field
polarization lie approximately in the plane perpendicular to the
direction of propagation. For such waves, we usually need only to
solve the scalar wave equations to obtain the amplitude and the
phase of the dominant electric field along its local polarization
direction. The dominant magnetic field can be calculated directly
from the dominant electric field. Under these assumption, the
Maxwell equations for the beam propagation in free space (or
homogeneous and isotropic medium), can be replaced by the scalar
wave equation (Helmholtz equation)

\begin{equation}\label{Helmholtz}
    [\nabla^2+k^2]E(x,y,z)=0
\end{equation}

where $E(x,y,z)$ is the phasor amplitude of a field component that
is sinusoidal in time, $\nabla^2$ is the Laplacian operator and
$k$ is the laser wavenumber. Then, since laser beams are usually
sufficiently collimated, we can describe their diffraction
properties using the paraxial wave approximation. If the field is
expected to propagate mainly in the $z$ direction, with a slow
variation of amplitude and phase along the transverse direction,
it is convenient to write the field in the following way

\begin{equation}\label{envelope}
    E(x,y,z)=u(x,y,z)e^{- i k z}
\end{equation}

where $u$ is the complex scalar wave amplitude, called envelope
function, which describes the transverse profile of the beam.
Inserting this into the (Helmholtz) equation ~\eqref{Helmholtz} we
find the reduced equation

\begin{equation}\label{reduced}
    \Big(\frac{\partial^2}{\partial x^2}+\frac{\partial^2}{\partial
    y^2}+\frac{\partial^2}{\partial z^2}- 2 i k \frac{\partial}{\partial
    z}\Big)u=0
\end{equation}

If the $z$ dependence of the envelope function is slow compared to
the optical wavelength and to the transverse variations of the
field, we can drop the second order partial derivative in $z$ in
~\eqref{reduced} and obtain the \textit{paraxial wave equation}

\begin{equation}\label{paraxial}
\frac{\partial^2 u}{\partial x^2}+\frac{\partial^2 u}{\partial
    y^2}- 2 i k \frac{\partial u}{\partial z}=0
\end{equation}

This equation can be written for a generic set of transverse
coordinates as

\begin{equation}\label{paraxialGC}
    \nabla^2_t u(\textbf{s},z)- 2 i k \frac{\partial u(\textbf{s},z)}{\partial
    z}=0
\end{equation}

where $\textbf{s}$ refers to the specific coordinate system,
orthogonal to the $z$ direction, and $ \nabla^2_t$ means the
Laplacian operator in these coordinates.

\section{Integral approach}

Another equally valid and effective way of analyzing paraxial wave
propagation is to employ the Huygens-Fresnel principle in the
paraxial approximation. Consider an aperture at a plane $z_0$
illuminated with a light field distribution $E_0(x_0,y_0,z_0)$
within the aperture. Then for a point lying somewhere after the
aperture, say at $P$ with coordinates $(x,y,z)$, the net field is
given by adding together spherical waves emitted from each point
$P_0$ in the aperture. Each spherical wavelet takes on the
strength and phase of the field at the point where it originates.
Mathematically, this summation takes the form

\begin{align}\label{Huyg-Fres}
     E(x,y,z) &=\frac{i}{ \lambda}\int_{S_0} E_0(x_0,y_0,z_0)\frac{e^{-\imath k
    \rho}}{\rho}\cos\theta \, dS_0 \\
    \rho &= \sqrt{(z-z_0)^2+(x-x_0)^2+(y-y_0)^2} \notag
\end{align}

where $\theta$ is the angle under which the aperture element
centered at $P_0$ is seen from the observation point $P$ and
$\rho$ is the distance between these two points. This equation can
be rigorously derived from the Rayleigh-Sommerfeld scalar
diffraction theory as shown in ~\cite{goodman} and rely on the
approximation $\rho \gg \lambda$. The paraxial and the Fresnel
approximations to diffraction theory consist in approximating the
obliquity factor $\cos \theta$ by unity and, once expanded the
distance $\rho$ in a power series in the form

\begin{equation}\label{roexp}
    \rho = z-z_0 +\frac{(x-x_0)^2+(y-y_0)^2}{2(z-z_0)}+\ldots
\end{equation}

replacing the denominator of \eqref{Huyg-Fres} by simply $z-z_0$
and retaining the quadratic terms in the exponent of the phase
shift factor $e^{-i k \rho}$. With these approximations we obtain
the paraxial approximation of the Huygens-Fresnel integral
\eqref{Huyg-Fres}

\begin{equation}\label{parax-int}
     E(x,y,z) =\frac{i e^{ -\imath k (z-z_0)}}{\lambda (z-z_0)}\int E_0(x_0,y_0,z_0)
     e^{ -\imath k \frac{(x-x_0)^2+(y-y_0)^2}{2(z-z_0)} } \, dx_0
     dy_0
\end{equation}

The integral is extended to the entire plane $z_0$ with the
assumption that the input field distribution $E_0(x_0,y_0,z_0)$ is
a paraxial optical beam which transverse profile going to zero
outside the region $S_0$. The integral \eqref{parax-int} provides
a way to propagate an arbitrary optical wavefront from an input
plane $z_0$ to any later plane $z$. In the following we will
assume $z_0=0$. It is useful to note that this integral has
exactly the form of a convolution product in the $x,y$ coordinates
between the input field $E_0(x_0,y_0,z_0=0)$ and the paraxial
diffraction kernel $h(x,y,z)$

\begin{equation}\label{h}
    h(x,y,z)=\frac{i e^{ -\imath k z}}{\lambda z}
    e^{-\imath k \frac{x^2+y^2}{2z} }
\end{equation}

Using the $2D$ Fourier transform the integral equation
\eqref{parax-int} can be transformed into a simple algebraic
product. Consider a complex function (e.g. an optical wave
amplitude) $f(x,y)$ of two real variables, of integrable square
modulus. Its two dimensional Fourier transform is defined by

\begin{equation}\label{FTdef}
    \widetilde{f}(p,q)=\int_{\mathbf{R}^2}dx\, dy\, e^{- i(p x + q
    y)}f(x,y)
\end{equation}
and the reciprocal transform by

\begin{equation}\label{invFTdef}
   f(x,y) =\frac{1}{4 \pi^2}\int_{\mathbf{R}^2}dp\, dq \,e^{i(p x + q
    y)}\widetilde{f}(p,q)
\end{equation}

The Fourier transform of the propagator $h$ can be easily computed

\begin{equation}\label{FTprop}
    \widetilde{h}(p,q,z)=e^{-\imath k z}e^{i z \frac{p^2+q^2}{2 k}}
\end{equation}

and the propagation equation for the transformed field become

\begin{equation}\label{propFT}
\widetilde{E}(p,q,z)=\widetilde{h}(p,q,z) \widetilde{E_0}(p,q,0)
\end{equation}

This equation can be retrieved starting from the Helmholtz
equation in the paraxial approximation. Taking the Fourier
transform of the equation \eqref{paraxial} with respect to~$x,y$
we obtain the following equation for the Fourier transformed
envelope function

\begin{equation}\label{FTparax}
    [2 i k \partial_z +(p^2+q^2)]\widetilde{u}(p,q,z)=0
\end{equation}
the solution of which is of the form

\begin{equation}\label{FT}
    \widetilde{u}(p,q,z)=\widetilde{u}(p,q,0)e^{i z \frac{p^2+q^2}{2 k}}
\end{equation}

in which we recover the propagator \eqref{FTprop} once we
reintroduce the phase factor $e^{-\imath k z}$ for the complete
field.

\subsection{Paraxial plane waves decomposition}

The scalar wave equation \eqref{Helmholtz} gives us a formal
method for propagating an optical wave forward in space. The main
concept in Fourier optics is that we use a Fourier transform to
break up an arbitrary field into plane-wave components.  Then we
propagate the plane waves components, and the linearity of the
propagation equations allows the reconstruction of the propagated
field. Let's see how this works. Recall the scalar plane wave
solution to the wave equation:

\begin{equation}\label{planewave}
    E(\mathbf{r})=E_0 e^{-i \mathbf{k} \cdot \mathbf{r}}=E_0
    e^{ -i (k_x x + k_y y + k_z z)}
\end{equation}

where the wave vector $\mathbf{k} = (k_x, k_y, k_z)$ indicates the
propagation direction of the wave. Let's let the optical axis lie
along the z-axis. Then the wave vector $\mathbf{k}$ has a
direction determined by its angles with respect to the z-axis,

\begin{equation}\label{thetaxy}
    \theta_x=\arccos\left(\frac{k_x}{k}\right) \quad \mbox{and} \quad \theta_y=\arcsin\left(\frac{k_y}{k}\right)
\end{equation}

where $k_x$ and $k_y$ are the transverse spatial frequencies.

That is, in the $z = 0$ plane, the electric field has harmonic
spatial dependence in the x- and y-directions:

\begin{equation}\label{planez}
    E(x, y, z = 0) = E_0 e^{- i (k_x x + k_y y )}
\end{equation}

 This is the central point
of Fourier optics: harmonic phase variation in the (x, y) plane
(say, at z = 0) corresponds to a plane wave in a particular
direction determined by \eqref{thetaxy}. That is, the frequency of
the harmonic variation determines the direction of propagation. A
general field profile $E(x, y)$ at $z = 0$ can thus be written as
a superposition of plane waves via the Fourier transform:

\begin{equation}\label{FTdecomposition}
    E(x,y)=\frac{1}{4 \pi^2}\int_{\mathbf{R}^2}dk_x\, dk_y \,e^{i(k_x x +
    k_y
    y)}\widetilde{E}(k_x,k_y)
\end{equation}
where the spatial frequency distribution is given by

\begin{equation}\label{FTdecomposition2}
    \widetilde{E}(k_x,k_y)=\int_{\mathbf{R}^2}dx\, dy \,e^{- i(k_x x +
    k_y
    y)}E(x,y)
\end{equation}

 Having transformed $E(x, y)$ into the spatial-frequency domain,
 we can propagate this spatial-frequency distribution forward to
 any other plane $z$ by multiplying it by the phase shift factor
$e^{-i k_z z}$, where $k_z=\sqrt{k^2-k_x^2-k_y^2}$

\begin{equation}\label{FTdecoprop}
    \widetilde{E}(k_x,k_y,z)= \widetilde{E}(k_x,k_y) e^{-i k_z z}
\end{equation}

However in the paraxial approximation, since $k_{x,y}\ll k$ , the
longitudinal component $k_z$ can be written in the form

\begin{equation}\label{kparax}
k_z=\sqrt{k^2-k_x^2-k_y^2}\approx k
\Big[1-\frac{1}{2}\Big(\frac{k_x^2+k_y^2}{k^2}\Big) \Big]
\end{equation}

Thus, the free-space transfer function becomes

\begin{equation}\label{paraxplane}
e^{-i k_z z}\approx e^{ -i k z} e^{i z \frac{k_x^2+k_y^2}{2 k}}
\end{equation}

where the first factor is an overall phase factor corresponding to
plane-wave propagation along the optical axis, and the second
factor generates the evolution of the spatial profile. Combining
Eqs.~\eqref{propFT} and \eqref{paraxplane} we obtain the
expression for the propagation of the spatial-frequency components
of the paraxial beam which is equal to Eq.~\eqref{propFT},
obtained by applying Fourier theorems to the Huygens-Fresnel
integral.

\section{Gaussian beam}\label{sec:GB}

The Gaussian beam is the simplest model of a directed beam that
satisfies the Helmholtz equation in the paraxial approximation
\eqref{paraxial}. It also turns out that the outputs of spherical
mirror resonators and lasers are often Gaussian beams
(approximatively), and therefore the theory of Gaussian beams is
widely used in gravitational waves interferometric detectors. In
most applications the standard Gaussian beam is expressed as

\begin{equation}\label{GB}
    E_G(\mathbf{r})= E_0 \frac{w_0}{w(z)}
    \exp \left [-\frac{r^2}{w^2(z)}\right]
    \exp \left [-i k z +i \arctan \left (\frac{z}{z_0}\right )\right]
    \exp \left [- i k \frac{r^2}{2 R(z)}\right ]
\end{equation}

In this expression, we have used polar coordinates $(r, z)$ with
$r =\sqrt{x^2+y^2}$; $E_0$ is  an overall field-amplitude
constant; $z_0$ is a constant called the Rayleigh length (or
``Rayleigh range'');

\begin{equation}\label{waist}
    w_0=\sqrt{\frac{\lambda z_0}{\pi}}
\end{equation}

is the beam waist parameter or minimum beam radius

\begin{equation}\label{wz}
    w(z)=w_0 \sqrt{1+ \left( \frac{z}{z_0} \right )^2}
\end{equation}

is the beam radius as a function of $z$, and

\begin{equation}\label{Rz}
    R(z)= z \left [1+\left( \frac{z_0}{z} \right)^2 \right]
\end{equation}
is the radius of curvature of the wave front. The intensity is the
square modulus of Eq.~\eqref{GB}

\begin{equation}\label{IntGB}
    I(r,z)= |E_0|^2 \left (\frac{w_0}{w(z)} \right)^2 \exp \left [-\frac{2 r^2}{w^2(z)}\right]
\end{equation}

Clearly, the intensity falls off in the radial direction like a
Gaussian function, hence the name \textit{Gaussian beam}. Also,
note that the beam spot size  $w(z)$ correspond to the distance
from the optic axis to the point of $1/e^2$ attenuation of the
beam intensity  ($1/e$ in electric field amplitude).

The beam radius $w(z)$ traces out a hyperbolic curve in $z$. Near
the focus at $z = 0$, the spot size achieves its minimum value
$w_0$. At large distances from the focus, the hyperbola approaches
its asymptotes, given by $(w_0/z_0)z$. The Rayleigh length $z_0$
marks the crossover between these two regimes; thus,

\begin{align*}
    & w(z)\sim w_0 \qquad \mbox{for}\quad |z|\ll z_0 \\
    & w(z)\sim \left (\frac{w_0}{z_0} \right ) z
    \qquad \mbox{for}\quad |z|\gg z_0
\end{align*}

for the regions near and far away from the focus, respectively.
From this we see that in the far field, the beam propagates in the
form of a cone of half angle $\theta_0$, called beam divergence,
given by

\begin{equation}\label{diverGB}
    \theta_0\approx \tan \theta_0=\frac{w_0}{z_0}=\frac{\lambda}{\pi w_0}
\end{equation}
where we have used the small angle approximation for the paraxial
Gaussian beam. The region in which a gaussian beam can be
considered collimated is roughly $2z_0$ around the focal plane
$z=0$. Recall that the longitudinal phase factor has the form

\begin{equation}\label{long-phase}
 \exp \left [-i k z + i \arctan \left (\frac{z}{z_0}\right )\right]
\end{equation}

The first term in the phase is simply the phase of a plane wave
$ikz$ propagating in the same direction and with the same optical
frequency as the Gaussian beam. The second term is called the Gouy
phase shift and represents a small departure from planarity. The
longitudinal phase is dominated by the $ikz$ term, but the Gouy
term is important as well. It represents a phase retardation
compared to the plane wave. Because of the \textit{arctan} form,
the retardation amounts to a total of $\pi$ in phase over all $z$.
Gouy effects are generic to focusing-beam-type solutions to the
wave equation. As we will see, the Gouy phase is important in
computing the resonant frequencies of optical resonators.

Examining the expression of the radius of curvature of the
wavefront \eqref{Rz} we can see that at the waist $z=0$, the
wavefront is flat, and that in the far field region $(|z|\gg z_0)$
the radius increases as $R(z)\approx z$, i.e., the gaussian beam
becomes essentially like a spherical wave centered at the beam
waist. The minimum radius of curvature occurs for the wavefront at
a distance from the waist given by $z=z_0$, with a radius value
$R=2 z_0$. In a cavity, the boundary conditions imposed by the
cavity mirrors require that the curvature of the spherical mirrors
and the curvature of the wave fronts match. This allows the wave
to map back on itself. This is one reason why Gaussian beams can
exist in resonators as we will see later.

 Despite the complex form
of the Gaussian beam, relatively little information is needed to
completely specify it. For example, if we know where $z = 0$ is
located, the value of $w_0$, and the optical wavelength $\lambda$
all the other parameters are uniquely fixed. Alternately, it is
sufficient to know $w_0$ and $R(z)$ at some distance $z$, or it is
sufficient to know $w(z)$ and $R(z)$ at some distance $z$. It is
useful the definition of a complex parameter for the gaussian
beam, called $q$-parameter

\begin{equation}\label{q-par-GB}
    q(z)=z+i z_0
\end{equation}

Using the definition \eqref{Rz} and \eqref{wz} it is easy to
calculate $1/q$ in the more useful form

\begin{equation}\label{invq-par-GB}
    \frac{1}{q(z)}=\frac{1}{R(z)} - i \frac{\lambda}{\pi w^2(z)}
\end{equation}

The propagation of a gaussian beam in free space can be simply
computed using the q-parameters; the propagation law between two
planes along the optical axis $z$ for $q(z)$ is established as

\begin{equation}\label{prop-q-free}
    q(z_2)=q(z_1)+z_2-z_1
\end{equation}

This result is a particular case of the propagation low of a
gaussian beam through various optical structures modelled with the
so-called ``ABCD'' formalism of geometric optics which we will
describe briefly in the next section.

\section{$ABCD$ transformation }

Matrix optics has been well established a long time ago. Within
the paraxial approach, it provides a modular transformation
describing the effect of an optical system as the cascaded
operation of its components. Then each simple optical system is
given by its matrix representation. Before presenting the results
of the application of the matrix optics to the Gaussian beam
transformation, we need to analyze the basis of this approach
(e.g., see Chapter $15$ of Ref.~\cite{siegman}). In paraxial
optics, the light is presented as ray trajectories that are
described, at a given meridional plane, by its height and its
angle with respect to the optical axis of the system. These two
parameters can be arranged as a column vector. The simplest
mathematical object relating two vectors (besides a multiplication
by a scalar quantity) is a matrix. In this case, the matrix is a
$2\times 2$ matrix that is usually called the $ABCD$ matrix
because its elements are labelled as $A, B, C,$ and $D$. If the
input plane and the output plane are in the same optical medium,
then the determinant of the $ABCD$ matrix is unity, $AD-BC=1$, and
the matrix is called unimodular. In the absence of loss, the
matrix elements are real, they are complex otherwise.

 The relation can be written as:

\begin{equation}\label{ABCD}
    \left(%
\begin{array}{c}
  x_2 \\
  x'_2 \\
\end{array}%
\right)=\left(%
\begin{array}{cc}
  A & B \\
  C & D \\
\end{array}%
\right) \left(%
\begin{array}{c}
  x_1 \\
  x'_1 \\
\end{array}%
\right)
\end{equation}

where the column vector with subindex $1$ stands for the input
ray, and the subindex $2$ stands for the output ray. An
interesting result of this previous equation is obtained when a
new magnitude is defined as the ratio between height and angle.

\begin{figure}[!htbp]
  \begin{center}
    \leavevmode
      \includegraphics[height=2in]{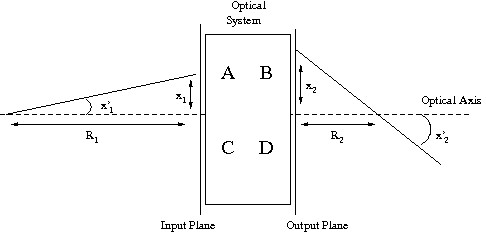}
    \caption{Ray transformation trough an $ABCD$ system}
    \label{FigABCD}
  \end{center}
\end{figure}

From Fig.\ref{FigABCD}, this parameter coincides with the distance
between the ray-optical axis intersection and the position of
reference for the description of the ray. This distance is
interpreted as the radius of curvature of a wavefront departing
from that intersection point and arriving to the plane of interest
where the column vector is described. When this radius of
curvature is obtained by using the matrix relations, the following
result is found:

\begin{equation}\label{ABCD-R}
    R_2=\frac{A R_1 + B}{C R_1 + D}
\end{equation}

 This expression is known as the $ABCD$ law for the radius of
curvature. It relates the input and output radii of curvature of a
spherical wavefront passing through an optical system described by
its $ABCD$ matrix. In recent years, a very useful generalized form
of paraxial beam propagation has been developed, in which a
generalized Huygens type of integral, describes paraxial wave
propagation trough cascade sequences of optical elements, in terms
of $ABCD$ matrices. Consider an optical system
(Fig.~\ref{FigABCD}) described by an $ABCD$ matrix with real
elements and illuminated by a beam with a transverse amplitude
distribution $E_1(x_1,y_1)$ at the input plane. The amplitude
distribution at the output plane can be written as

\begin{equation}\label{Huyg-ABCD}
E_2(x_2,y_2)= \frac{i e^{-i k L}}{\lambda B}\int E_1(x_1,y_1)e^{-i
\frac{k}{2 B}\left[A(x_1^2+y_1^2)-2(x_1 x_2 + y_1 y_2)+
D(x_2^2+y_2^2) \right]}dx_1 dy_1
\end{equation}

or for an optical system with cylindrical symmetry

\begin{equation}\label{Huyg-ABCD-Cyl}
E_2(r_2,\phi_2)= \frac{i e^{-i k L}}{\lambda B}\int
E_1(r_1,\phi_1)e^{-i \frac{k}{2 B}\left[A r_1^2-2r_1
r_2\cos(\phi_1-\phi_2) +  D r_2^2 \right]} r_1 dr_1 d\phi_1
\end{equation}

If we now consider the transformation of a gaussian beam through a
paraxial system described by an $ABCD$ matrix, it is possible to
prove (ref..) that the complex $q$-parameter introduced in
\eqref{invq-par-GB} transforms according to the simple relation

\begin{equation}\label{ABCD-q}
    q_2=\frac{A q_1 + B}{C q_1 + D}
\end{equation}

where $q_1$ and $q_2$ are the $q$-parameters before and after the
optical system, respectively. The equation \eqref{ABCD-q} has the
same form of the equation \eqref{ABCD-R} for the transformation of
the radius of curvature of a spherical wave passing through the
optical elements; the q-parameter is also called complex radius of
curvature for a gaussian beam. The results of the application of
the $ABCD$ law can be written in terms of the real radius of
curvature $R$ and the Gaussian width $w$ by properly taking the
real and imaginary parts of the resulting complex radius of
curvature. When a Gaussian beam propagates along an $ABCD$ optical
system, its complex radius of curvature changes according to the
$ABCD$ law. The new parameters of the beam are obtained from the
value of the new complex radius of curvature. However, there
exists an invariant parameter that remains the same throughout
$ABCD$ optical systems. This invariant parameter is the product of
the minimum beam width $(w_0)$ with the divergence of the beam

\begin{equation}\label{inv-GB}
    w_0 \theta_0 =\frac{\lambda}{\pi}
\end{equation}

Using this relation, we can conclude that a good collimation (very
low value of the divergence) will be obtained when the beam is
wide. On the contrary, a high focused beam will be obtained by
allowing a large divergence angle.

\section{Higher order modes}\label{HOM}

In section \ref{sec:GB}, only one solution of \eqref{paraxialGC},
was discussed, i.e., a light beam with the property that its
intensity profile in every beam cross section is given by the same
function, namely, a Gaussian. The width of this Gaussian
distribution changes as the beam propagates along its axis. There
are other solution of \eqref{paraxialGC} with similar properties,
and they are discussed in this section. These solutions form a
complete and orthogonal set of function and are called the
\textit{modes of propagation}. Every arbitrary distribution of
monochromatic light can be expanded in terms of these modes. This
procedure is largely used in the modal analysis of the light
circulating in the gravitational waves interferometers.

\textit{$\left.a \right)$ Modes in cartesian coordinates}

\begin{figure}[!htbp]
  \begin{center}
    \leavevmode
      \includegraphics[width=0.6\textwidth]{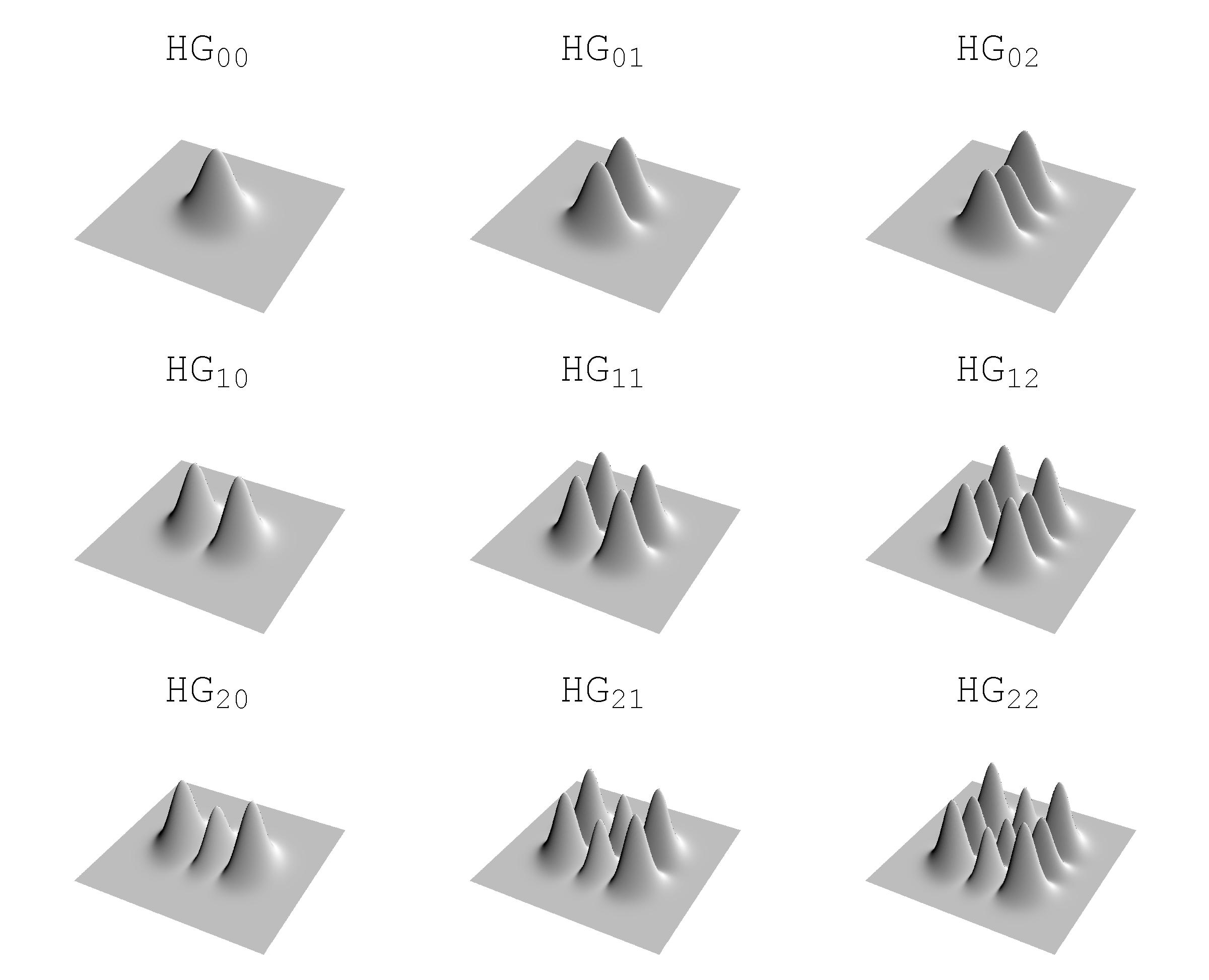}
    \caption{Hermite Gauss modes}
    \label{FigHG}
  \end{center}
\end{figure}

Using cartesian coordinates, a more general solution of the
paraxial wave equation is given by the Hermite-Gaussian (HG) beam
which can be written as

\begin{align}\label{eq:HG}
    HG_{m n}(x,y,z)=& \sqrt{\frac{2}{\pi 2^{m+n}\, m! n!\, w^2(z)}}\; H_m \left
    (\frac{\sqrt{2}x}{w(z)}\right )H_n \left
    (\frac{\sqrt{2}y}{w(z)}\right )\exp \left
    [-\frac{x^2+y^2}{w^2(z)}\right] \nonumber \\
    &\exp \left [-i k z + i(m+n+1) \arctan \left (\frac{z}{z_0}\right )\right]
    \exp \left [-i k \frac{x^2+y^2}{2 R(z)}\right ]
\end{align}

where the functions $H_m(X)$ are the Hermite polynomials of order
$m$ and the parameters $w(z),R(z)$ and $z_0$ are the same as for
the lowest-order gaussian mode as given in Sec. \ref{sec:GB}. In
general, the intensity pattern of the $HG_{m,n}$ mode,
Fig.~\ref{FigHG}, has $m$ dark bands across the $x$-direction and
$n$ dark bands across the $y$-direction, corresponding to the
zeros of the Hermite polynomials. Alternately, the intensity
pattern has a grid of $m + 1$ bright spots in the x-direction and
$n+ 1$ in the y-direction.

\textit{$\left.b \right)$ Modes in cylindrical coordinates}

\begin{figure}[!htbp]
  \begin{center}
    \leavevmode
      \includegraphics[width=0.6\textwidth]{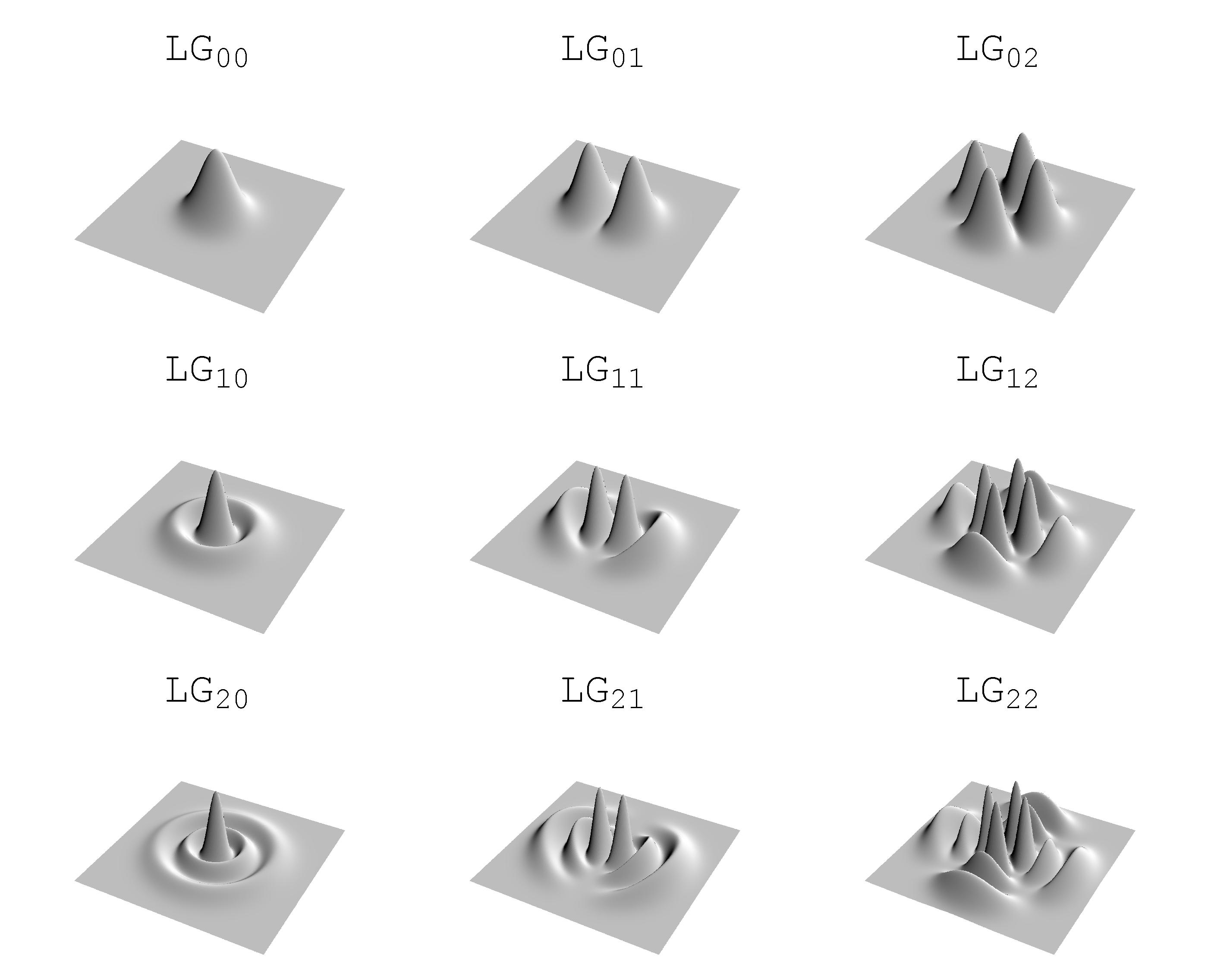}
    \caption{Laguerre Gauss modes}
    \label{FigLG}
  \end{center}
\end{figure}

An alternative but equally valid family of solution to the
paraxial wave equation \eqref{paraxialGC} can be written in
cylindrical coordinates $(r,\phi,z)$ and are called Laguerre
Gaussian (LG) beams

\begin{align}\label{eq:LG}
LG_{p m}(r,\phi,z)=& \sqrt{\frac{4 p!}{\pi (1+\delta_{m 0})
(m+p)!}} \; \left
    (\frac{\sqrt{2}r}{w(z)}\right )^m L_p^m \left
    (\frac{2 r^2}{w^2(z)}\right )\frac{\exp \left
    [-\frac{r^2}{w^2(z)}\right]}{w(z)} \nonumber \\
    &\exp \left [-i k z + i(2p+m+1) \arctan \left (\frac{z}{z_0}\right
    )\right]\cos(m \phi)
    \exp \left [-i k \frac{r^2}{2 R(z)}\right ]
\end{align}
where the integer $p\geq 0$ is the radial index and the integer
$m$ is the azimuthal mode index (see Fig.~\ref{FigLG} for the
Intensity distribution); the $L_p^m$ are the generalized Laguerre
polynomials and all other quantities $R(z), w(z), z_0$ are exactly
the same as in the Hermite-gaussian case.

\section{Generic paraxial beam and ABCD
transformation}\label{sec:GenParBeam}

The propagation and transformation of Gaussian, Hermite-Gaussian
and Laguerre-Gaussian light beams through paraxial systems are
governed by the $ABCD$ law \eqref{ABCD-q}. However, since most of
the work in this thesis deals with non-Gaussian (non-HG and
non-LG) beams, it is useful to introduce a formalism for the
analysis of a non-Gaussian beam by using an extension of the
complex beam $q$-parameter. We are going to introduce the most
characteristic parameters for a generic paraxial beam, defined in
terms of the moments of the intensity distribution and its Fourier
transform. These parameters have been introduced
in~\cite{SiegmanR,BelangerABCD,PorrasABCD}. In the case of
orthogonal astigmatic beams, the field is separable in two
orthogonal one-dimensional components. If the amplitude
distribution in one of these transversal directions is denoted by
$\psi(x)$, then the width $w(\psi)$ of the beam is given by

\begin{equation}\label{gen-width}
    w(\psi)= 2 \left ( \frac{\int_{-\infty}^\infty |\psi(x)|^2 x^2
    dx}{I(\psi)}- x^2(\psi) \right)^\frac{1}{2}
\end{equation}

where
\begin{equation}\label{int}
    I(\psi)=\int_{-\infty}^\infty |\psi(x)|^2
    dx
\end{equation}
is the total intensity in the transverse direction and

\begin{equation}\label{mean-pos}
    x(\psi)= \frac{1}{I(\psi)}\int_{-\infty}^\infty |\psi(x)|^2 x
    dx
\end{equation}

is the mean value of the transversal position of the beam. When
$x(\psi)$ is zero, we will say that the beam is on axis. It is
easy to check that in the case of a Gaussian distribution, the
width is the Gaussian width defined in the previous sections.

As we saw in the definition of the divergence for Gaussian beams,
the divergence is related to the spreading of the beam along its
propagation. This concept is described analytically by the Fourier
transform of the amplitude distribution, i.e., also named as the
angular spectrum. In this section the Fourier transform of the
amplitude distribution is defined in a slightly different way with
respect to \eqref{FTdef}; introducing a $2 \pi$ factor in the
exponent we have

\begin{equation}\label{FT-1D}
    \widetilde{\psi}(\xi)=\int_{-\infty}^\infty \psi(x) e^{- i 2 \pi \xi x}
    dx
\end{equation}

where $\xi$ is the transverse spatial frequency that is related to
the angle by means of the wavelength. The angular width that is
taken as the divergence of the beam is related to the mean-square
deviation of the Fourier transform of the amplitude
\begin{equation}\label{div-gen}
    \theta_0(\widetilde{\psi})= 2 \lambda \left (\frac{\int_{-\infty}^\infty |\widetilde{\psi}|^2 \xi^2 d\xi
    }{I(\widetilde{\psi})}- \xi^2(\widetilde{\psi}) \right )^\frac{1}{2}
\end{equation}
where due to the Parseval's theorem $I(\widetilde{\psi})=I(\psi)$
and

\begin{equation}\label{mean-freq}
    \xi(\widetilde{\psi})=\frac{1}{I(\widetilde{\psi})}\int_{-\infty}^\infty |\widetilde{\psi}|^2 \xi d\xi
\end{equation}
is the mean value of the transversal spatial frequency which is
related to the angle between the optical axis and the direction of
the propagation of the beam by $\alpha= - \lambda
\xi(\widetilde{\psi})$.  The mean value of the position,
$x(\psi)$, and the slope of the beam, $\alpha$ transform according
to the geometrical optics rules \eqref{ABCD} when the beam passes
trough the $ABCD$ system.

Another parameter defined in analogy with the Gaussian beam case
is the radius of curvature. For totally coherent laser beams, it
is also possible to define an effective or generalized radius of
curvature for arbitrary amplitude distributions. This radius of
curvature is the radius of the spherical wavefront that best fits
the actual wavefront of the beam. This fitting is made by
weighting the departure from the spherical wavefront with the
irradiance distribution. The analytical expression for this radius
of curvature can be written as follows:

\begin{equation}\label{gen-curv}
    \frac{1}{R(\psi)}= \frac{i \lambda}{\pi I(\psi)
    w^2(\psi)}\int_{-\infty}^\infty \left (\frac{\partial \psi(x)}{\partial
    x}\psi^*(x) - \psi(x)\frac{\partial \psi^*(x)}{\partial
    x}\right )[x - x(\psi)] dx
\end{equation}

By using the previous definitions, it is possible to introduce a
generalized complex radius of curvature as follows:

\begin{equation}\label{q-gen}
    \frac{1}{q(\psi)}=\frac{1}{R(\psi)}-i \sqrt{\frac{\theta_0^2(\widetilde{\psi})}{w^2(\psi)}-\frac{1}{R^2(\psi)}}
\end{equation}
Now the transformation of the complex radius of curvature can be
carried out by applying the ABCD law, as it can be proved by using
the generalized Huygens integral \eqref{Huyg-ABCD} or
\eqref{Huyg-ABCD-Cyl}, which relates the output beam to the input
one and the ABCD elements. It is important to note that there are
three parameters involved in the calculation of the generalized
complex radius of curvature: $w^2(\psi),
\theta_0^2(\widetilde{\psi})$ and $R(\psi)$. The application of
the ABCD law provides two equations: one for the real part, and
one for the imaginary. Therefore we will need another relation
involving these three parameters to solve the problem of the
transformation of those beams by ABCD optical systems. This third
relation is given by the invariant parameter $M^2$  called
\textit{beam propagation factor}\footnote{Sometimes it is also
called beam quality factor. }.

For the Gaussian beam case, we have found a parameter that remains
invariant through $ABCD$ optical systems. Now in the case of
totally coherent non-Gaussian beams, we can define a new parameter
that will have the same properties. It will be constant along the
propagation through $ABCD$ optical systems. Its definition in
terms of the previous characterizing parameters is:

\begin{equation}\label{M2-gen}
    M^2 = \frac{\pi}{\lambda}w(\psi)\sqrt{\theta_0^2(\widetilde{\psi})-\frac{w^2(\psi)}{R^2(\psi)}}
\end{equation}
This invariance, along with the results obtained from the $ABCD$
law applied to the generalized complex radius of curvature, allows
to calculate the three resulting parameters for an $ABCD$
transformation. The value of the $M^2$ parameter has an
interesting meaning. It is related to the divergence that would be
obtained if the beam having an amplitude distribution $\psi$ is
collimated at the plane of interest. The collimation should be
considered as having an effective, or generalized, radius of
curvature equal to infinity. From the definition of $R(\psi)$,
this is an averaged collimation. The divergence of this collimated
beam is the minimum obtainable for such a beam having a
generalized width of $w(\psi)$.  The propagation factor $M^2$ can
be directly related to the product of the minimum width (waist
size) of the beam (defined as variance in the $x$ coordinate)
times the divergence (defined as variance in the $\xi$
coordinate). The value of $M^2$ for a Gaussian beam is 1, which
follows directly by \eqref{inv-GB}. It is not possible to find a
lower value of the $M^2$ for actual, realizable beams. This
property, along with its definition in terms of the variance in
$x$ and $\xi$, resembles very well an uncertainty principle. Using
these definitions, the transformation of the beam width,
divergence and radius of curvature of a generic paraxial beam by a
real $ABCD$ system from an input plane to an output plane (for
simplicity in the following formulas we will assume that the beam
is on axis and that the slope is zero at the input plane) can be
written as

\begin{align}
     w^2(\psi_2)& = w^2(\psi_1)\left [A+\frac{B}{R(\psi_1)}\right]^2+
    B^2 \frac{M^4 \lambda^2}{\pi^2 w^2(\psi_1)} \\
     \theta_0^2(\widetilde{\psi}_2)&= w^2(\psi_1)\left [C+\frac{D}{R(\psi_1)}\right]^2+
    D^2 \frac{M^4 \lambda^2}{\pi^2 w^2(\psi_1)} \\
     \frac{1}{R(\psi_2)}&= \frac{w^2(\psi_1)}{w^2(\psi_2)}\left [A+\frac{B}{R(\psi_1)}\right]\left
    [C+\frac{D}{R(\psi_1)}\right]+ B D \frac{M^4 \lambda^2}{\pi^2 w^2(\psi_1) w^2(\psi_2)}
\end{align}

In particular, if $z=0$ is the plane of the smallest width, the
free propagation of the beam $(A=D=1,C=0$ and $B=z )$ produces
these equations

\begin{align}
    w^2(\psi_z)&=w^2(\psi_0) \left[1+ z^2
    \frac{\theta_0^2(\widetilde{\psi}_0)}{w^2(\psi_0)}\right ] = w^2(\psi_0) \left[1+ \left(\frac{z}{z_R}\right)^2\right
    ]\\
    R(\psi_z)&= z \left[1+
    \frac{w^2(\psi_0)}{z^2 \quad \theta_0^2(\widetilde{\psi}_0)}\right
    ]= z \left[1+ \left(\frac{z_R}{z}\right)^2\right
    ]
\end{align}

where we have introduced a generalized Rayleigh distance $z_R=
\frac{\pi w^2(\psi_0)}{M^2 \lambda}$. Since in the next chapter we
will deal with cylindrical symmetric laser beam, it is useful to
write the definitions of width, divergence, and radius of
curvature in polar coordinates. If the beam is centered and
aligned with the optical axis we have

\begin{align}\label{WRD-cyl}
    w(\psi)&= 2 \sqrt{\frac{\pi}{I(\psi)}\int_0^\infty |\psi(r)|^2 r^3
    dr}\\
    \frac{1}{R(\psi)}&=\frac{i \lambda}{I(\psi)
    w^2(\psi)}\int_{0}^\infty \left (\frac{\partial \psi(r)}{\partial
    r}\psi^*(r) - \psi(r)\frac{\partial \psi^*(r)}{\partial
    r}\right ) r^2  dr \\
    \theta_0(\widetilde{\psi})&= 2 \lambda \sqrt{\frac{\pi}{I(\widetilde{\psi})}\int_0^\infty |\widetilde{\psi}(\rho)|^2 \rho^3  d\rho}
\end{align}

where $I(\psi)=I(\widetilde{\psi})$ is the integrated intensity in
the transversal plane and $\rho$ is the radial polar coordinate in
the spatial frequency plane of the 2D Fourier transform of the
field. It can be proved, using the integral transformation
\eqref{Huyg-ABCD-Cyl}, that the three transformation formulas are
identical to the orthogonal case, and therefore the conservation
$(M^2)$ and the ABCD law remain valid for cylindrical symmetric
beams. Once we have introduced this moment approach for the
propagation of a general paraxial beam, it is useful to
characterize the pointing stability, ~\cite{Morin}, of a laser
beam with a misalignment factor $|\eta_m|^2$, which provides a
global comparison between the misaligned beams and is expressed in
the form

\begin{equation}\label{misa-gen}
    |\eta_m|^2 = \left |\frac{\int_{-\infty}^\infty \psi(x) [\psi(x-\delta)e^{-i k \alpha x}]^* dx}{\int_{-\infty}^\infty |\psi(x)|^2
    dx}\right|^2
\end{equation}

for the one dimensional case, where $\eta_m$ is called the
misalignment superposition integral, $\psi(x)$ is the field of the
perfectly aligned beam and its misaligned copy is described by
$\psi(x-\delta)e^{-i k \alpha x}$ where $\delta$ and $\alpha$
represent the transverse and angular shifts, respectively, of the
misaligned beam in a given plane. The level of misalignment can be
quantified by the number $|\eta_m|^2$ , which takes a maximum
unitary value for a perfectly aligned beam and goes to zero as the
misalignment is increased. This number has all the desired
features. It takes into account both transverse and angular shifts
as well as the width and the divergence of the beam. In essence,
this number quantifies by how much the misaligned beam differs in
phase and intensity from the perfectly aligned beam in any plane,
since it is invariant under propagation through an ideal optical
system. For small shifts equation \eqref{misa-gen} reduces to

\begin{equation}\label{misa-small}
|\eta_m|^2\approx 1- (M^2)^2 \left (
\frac{\alpha^2}{\theta_0^2}+\frac{\delta^2}{w_0^2}\right)
\end{equation}

where $M^2, w_0$ and $\theta_0$ are the beam propagation factor,
waist width and far-field divergence angle of the beam, in
accordance with the second-order moments definition
\eqref{M2-gen}, \eqref{gen-width} and \eqref{div-gen}
respectively. In the derivation of \eqref{misa-small} it was
assumed that the transverse shift was measured at the waist plane.
The variation range of the misalignment factor is $0\leq
|\eta_m|^2 \leq1 $. The larger value of the parameters multiplying
the transverse and angular shifts, means that the beam is more
sensitive to the misalignment. It is important to stress that the
formula \eqref{misa-small} is a useful criterion to characterize
the misalignment sensitivity of different beam geometry and this
aspect will be investigated in Sec.~\ref{sec:MBmis}.

\section{Optical cavities with spherical mirrors}
An optical cavity \footnote{Optical resonators are often called
cavities. This term has been taken over from microwave technology
(masers), where resonators really look like closed cavities, while
optical resonators normally have a rather ``open'' kind of setup
which doesn't really look like a cavity.} or optical resonator is
an arrangement of mirrors that forms a standing wave cavity
resonator for light waves. Light confined in a resonator will
reflect multiple times from the mirrors, and due to the effects of
interference, only certain patterns and frequencies of radiation
will be sustained by the resonator, with the others being
suppressed by destructive interference. In general, radiation
patterns which are reproduced on every round-trip of the light
through the resonator are the most stable, and these are known as
the modes of the resonator. Cavity modes are self-consistent field
distributions of light, more precisely, electric field
distributions which are self-reproducing (apart from a possible
loss of power) in each round trip. The most common types of
optical cavities consist of two facing plane (flat) or spherical
mirrors. Now we have seen in Sec.~\ref{sec:GB} that the wavefronts
of a Gaussian beam (and its higher order partners HG and LG beams)
are paraboloidal surfaces with radii of curvature given by
\eqref{Rz}. Suppose that we fit a pair of curved mirrors to this
beam at any two points along the beam in such a way that the radii
of curvature of the mirrors are exactly matched to the wavefront
radii of the gaussian beam at those two points. If the transverse
size of the mirrors is substantially larger than the gaussian spot
size of the beam, each of these mirrors will essentially reflect
the gaussian beam exactly back on itself, with exactly reversed
wavefront curvature and direction. These two mirrors thus form an
optical resonator which can support both the lowest-order gaussian
mode, and (at slightly different wavelengths) the higher-order HG
or LG  modes as resonant modes of the cavity. Given two curved
mirrors with radii of curvature $R_1$ and $R_2$, separated by a
distance $L$, the calculation of the beam parameters follows
directly from imposing that the wavefront curvature must match the
mirror curvature at each mirror position. It is useful to
introduce the resonator $g$ parameters, $g_1$ and $g_2$, which are
a  standard notation in the field of optical resonators

\begin{equation}\label{gpar-def}
    g_1=1-\frac{L}{R_1}\quad \mbox{and}\quad g_2=1-\frac{L}{R_2}
\end{equation}

In terms of these parameters we can find that the self-reproducing
gaussian beam has a waist size $w_0$ given by

\begin{equation}\label{w0-gpar}
    w_0^2=\frac{L \lambda}{\pi}\sqrt{\frac{g_1 g_2 (1-g_1 g_2)}{(g_1+g_2-2 g_1
    g_2)^2}},
\end{equation}

and the spot sizes $w_1$ and $w_2$ at the mirrors surfaces
\begin{equation}\label{w12-gpar}
    w_1^2=\frac{L \lambda}{\pi} \sqrt{\frac{g_2}{g_1 (1 - g_1
    g_2)}}\quad \mbox{and} \quad w_2^2=\frac{L \lambda}{\pi} \sqrt{\frac{g_1}{g_2 (1 - g_1
    g_2)}}
\end{equation}
It is obvious from these expressions that real and finite
solutions for the gaussian beam spot sizes can exist only if the
$g_1,g_2$ parameters are confined to a stability range defined by
\begin{equation}\label{gpar-stab}
    0\leq g_1 g_2 \leq 1
\end{equation}
This condition is called stability range because this also exactly
describes the condition required for two mirrors with radii $R_1$
and $R_2$ and spacing $L$ to form a stable periodic focusing
system for optical rays. The stability of an optical cavity can be
analyzed from the ray-optics point of view. Light rays that bounce
back and forth between the spherical mirrors of a laser resonator
experience a periodic focusing action. The effect on the rays is
the same as in a periodic sequence of lenses or, for more general
type of resonator, of $ABCD$ optical elements. It can be easily
proved that the rays passing through these equivalent periodic
systems, are periodically refocused if the elements of the $ABCD$
matrix obeys the inequality

\begin{equation}\label{ABCD-stab}
    -1\leq \frac{A+D}{2} \leq 1
\end{equation}

For a laser resonator with spherical mirrors the stability
condition \eqref{ABCD-stab} reduces to \eqref{gpar-stab}. The
simplest resonator configuration to analyze are the symmetric
resonators (baseline for Advanced LIGO) which have equal mirror
curvature $R$ and hence $g$ parameters $g_1=g_2=g=1-L/R$. The
waist of the gaussian resonant mode is then obviously in the
center of the resonator, with waist and end mirror spot sizes
given by

\begin{equation}\label{simm-GB}
    w_0^2=\frac{L \lambda}{\pi}\sqrt{\frac{1+g}{4 (1-g)}}\quad
    \mbox{and} \quad w_1^2=w_2^2=\frac{L \lambda}{\pi}\sqrt{\frac{1}{1-g^2}}
\end{equation}

Another very often used configuration is the half-symmetric
resonator (i.e. Virgo long arms cavities) in which one mirror is
planar $R_1=\infty$ so that $g_1=1$, and the other curved. The
waist in this situation will be located on mirror number $1$, with
spot sizes given by

\begin{equation}\label{halfsimm-GB}
    w_0^2=w_1^2\frac{L \lambda}{\pi}\sqrt{\frac{g_2}{1-g_2}}\quad
    \mbox{and} \quad w_2^2=\frac{L \lambda}{\pi}\sqrt{\frac{1}{g_2(1-g_2)}}
\end{equation}

The central point in the stability diagram, $g_1=g_2=0$,
corresponds to the symmetric confocal\footnote{This is referred to
as confocal resonator because the focal points of the two mirrors
coincide with each other at the center of the resonator.}
resonator where the radii of curvature of the mirrors are exactly
equal to the separation length $L$. The beam waist and spot sizes
at the end mirrors are given by
\begin{equation}\label{confocal-GB}
    w_0^2=\frac{L \lambda}{2 \pi}\quad
    \mbox{and} \quad w_2^2=w_1^2=\frac{L \lambda}{\pi}
\end{equation}
The confocal resonator has overall the smallest average spot
diameter along its length of any stable resonator, although other
resonator may have smaller waist size at one plane within the
resonator. Moreover the confocal resonator is also highly
insensitive to misalignment of either mirror.

We have seen that if a gaussian beam is found resonant for a given
(infinite) spherical resonator, then all its Hermite-gaussian (or
Laguerre-gaussian) functions can be resonant modes for that
cavity. However for each of the transverse mode patterns, there
are only certain optical frequencies for which the optical phase
is self-consistently reproduced after each round trip (i.e., the
round-trip phase shift is an integer multiple of $2\pi$). These
are called the mode frequencies or resonance frequencies and are
equidistantly spaced. The cavity modes are labelled by three
indices in which the first refers to the axial mode number and the
other two to the transverse mode indices. Using the expression for
the total phase shift of a travelling gaussian beam, and imposing
the boundary condition on the wavefronts at the mirror positions,
we have that the resonance frequencies of the
axial-plus-transverse modes in the cavity are given by

\begin{align}
    \nu_{q m n}&= \left ( q + (m+n+1)\frac{\arccos \pm \sqrt{g_1
    g_2}}{\pi} \right) \frac{c}{2 L} \quad \mbox{for HG} \\
 \nu_{q p m}&= \left ( q + (2p+m+1)\frac{\arccos \pm \sqrt{g_1
    g_2}}{\pi} \right) \frac{c}{2 L} \quad \mbox{for LG}
\end{align}
where the $+$ sign applies in the upper right quadrant $(g_1, g_2
> 0)$ of the stability diagram and the $-$ sign applies in the lower
left quadrant. The frequency spacing, $\frac{c}{2 L}$, of the
cavity axial modes is called free spectral range (FSR). The
confocal resonator represent a situation where all the
even-symmetry transverse modes of the cavity are exactly
degenerate at the axial mode frequencies of the laser, and all the
odd-symmetry modes are exactly degenerate at the half-FSR
positions midway between the axial mode locations. This is the
reason way, despite the optimum performance in term of
misalignment stability, this configuration is not employed in GW
interferometric detector, which must be able to operate on the
single fundamental gaussian mode.

\section{Optical cavities: General Theory}

In the preceding section aperture diffraction effects due to the
finite size of the mirrors were neglected. There, it was mentioned
that resonators used in Fabry Perot optical cavities usually take
the form of an open structure consisting of a pair of mirrors
facing each other. Such a structure with finite mirror apertures
is intrinsically lossy and, unless energy is supplied to it
continuously, the electromagnetic field in it will decay. In this
case a mode of the resonator is a slowly decaying field
configuration whose relative distribution does not change with
time~\cite{Fox-Li}. The problem of the open resonator is a
difficult one and a rigorous solution is yet to be found. However,
if certain simplifying assumptions are made, the problem becomes
tractable and physically meaningful results can be obtained. The
simplifying assumptions involve essentially the quasi-optic nature
of the problem; specifically, they are 1) that the dimensions of
the resonator are large compared to the wavelength and 2) that the
field in the resonator is substantially transverse electromagnetic
(TEM).

 So long as those assumptions are valid, the
Fresnel-Kirchhoff formulation of Huygens principle can be invoked
to transform  the paraxial Helmholtz equation \eqref{paraxial}
with the appropriate boundary conditions for the cavity,
 into the familiar linear homogeneous Fredholm integral equation
for defining the eigenmodes of an optical resonator. At a
reference plane along the optical axis we have~\cite{siegman}

\begin{equation}\label{gen-eigen}
    \gamma_n u_{n}(x,y)=\int K(x,y;x',y') u_n(x',y') dx' dy'
\end{equation}

 where $K(x,y;x',y')$ is the Huygens kernel that
propagates the complex optical wave amplitude $u(x,y)$ through one
complete round trip around the resonator. The kernel
$K(x,y;x',y')$ in the Fresnel approximation can be given in terms
of the ABCD matrix elements using equations \eqref{Huyg-ABCD} and
\eqref{Huyg-ABCD-Cyl} for cartesian and cylindrical coordinates
respectively.

 The eigenfunctions
$u_n(x,y)$ that satisfy this integral equation form the
lowest-order and higher-order transverse modes of the resonator.
Each eigenfunction $u_n$ for Eq. \eqref{gen-eigen} has its
corresponding complex eigenvalue $\gamma_n$. The fractional power
loss per round trip for the n-th eigenmode is given by $1 -
|\gamma_n|^2$ and the phase shift accumulated during a round-trip
is given by $\arg(\gamma_n)$.

Questions have been raised as to whether the lossy open resonator
can be viewed as having resonant modes in either a physical or
mathematical sense. The kernel of the Huygens integral above is
symmetric  but not unitary due to the boundary conditions over the
finite mirror surface. Therefore, the different eigenfunctions
$u_n$ are not power orthogonal in the usual sense. The
orthogonality relations over the cross section are of the form
$\int u_n u_m dx dy \propto \delta_{nm}$ rather than $\int u_n^*
u_m dx dy \propto \delta_{nm}$. Hence, the total power $\int |u|^2
dx dy$ cannot be expressed as the sum of the powers carried by the
individual modes. But, the eigenfunctions $u_n$, although perhaps
not normal modes in the usual sense, are still self-reproducing
mode distributions in the sense that each such mode when excited
will retain the same transverse form, attenuated only in over-all
amplitude, after any number of transits around the resonator.
Further, it seems clear both theoretically and experimentally
that, as long as power is provided to the resonator to compensate
for  the losses, at least the lowest-loss mode, and in principle
any of the higher-loss modes, can maintain a steady-state
oscillation. In this sense it would seem that the functions $u_n$
are clearly meaningful transverse modes for an open optical
resonator. For optical resonators having rectangular mirrors
(i.e., mirrors having different radii of curvature in the $x-y$
plane) the eigenvalue equation can be separated into two separate
equations, each corresponding to the eigenvalue equation for a
cylindrical strip mirror in the $x$ or $y$ variable. For mirrors
of circular symmetry in the $x-y$ plane the equation can be
converted into cylindrical $r, \phi $ coordinates and then solved
separately for modes of different azimuthal symmetry characterized
by different $exp(\pm i m \phi)$ variations. As a representative
example we will recall the eigenvalue equation for optical
cavities with spherical mirror which is of large use in GW
community. In the following, and in  future Chapters, we may refer
to field amplitude at the mirror surface and not at a plane
orthogonal to the optical axis\footnote{In the paraxial
approximation the transition from the field referred to the mirror
surface to the field referred to a plane tangent to the mirror
surface is given (as explained on pg.\pageref{Refl}) by a phase
transformation $u(x,y)=v(x,y) e^{i k \frac{x^2+y^2}{2 R} }$, where
$R$ is the radius of curvature of the mirror and $v$ and $u$ are
the field at the mirror surface and at the tangent plane
respectively. }. For the case of circular mirrors, the equation
\eqref{gen-eigen} is reduced to the one-dimensional form by using
cylindrical coordinates\footnote{The angular integrations can be
computed analytically and led to the Bessel functions of the first
kind and $m$th order.}

 \begin{align}\label{eigen-g1g2}
    \gamma_{p m} R_{p m}(r_1) &= - \left(\frac{k}{L}\right)^2 i^{2m }e^{-2 i k
    L}\int_0^{a_1} r_1'  {\cal K}_m(r_1,r_1')R_{p m}(r_1') dr_1'  \\
    {\cal K}_m(r_1,r_1')&= \int_0^{a_2} r_2 dr_2 J_m(\frac{k r_1
    r_2}{L}) J_m(\frac{k r_2
    r_1'}{L})\exp \left[-\frac{i k}{2 L} \left(g_1 r_1^2 + g_1
    r_1'^2 + 2 g_2 r_2^2 \right) \right] \nonumber
\end{align}

where $g_1$ and $g_2$ are the cavity's g-parameters defined in
\eqref{gpar-def}.

This equation can be analytically solved only in the very special
case $g_1=g_2=0$, (a confocal resonator). In general must be
solved by numerical analysis, as we will see in
Sec.~\ref{sec:FEM}. In the case of infinite mirrors the
propagation between mirror surfaces is mathematically equivalent
to a Fractional Fourier Transform, and the eigenfunctions of this
linear operator are exactly the Hermite Gauss or Laguerre Gauss
functions introduced in Sec.~\ref{HOM}. The Fractional Fourier
Transform has recently attracted much attention in optics studies
as a tool to analyze paraxial optical systems
imaging~\cite{FracFT}.

\section{Optical response of a Fabry-Perot cavity}

\begin{figure}[htb]
\begin{center}
\includegraphics[width=0.8\textwidth]{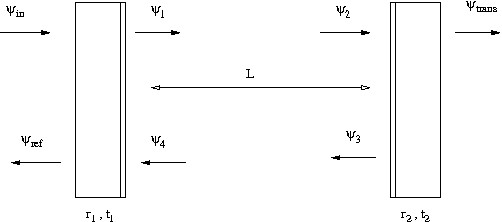}
\end{center}
\caption{Electromagnetic fields in a Fabry-Perot resonator.}
\label{fig:FP}
\end{figure}

For completeness we recall the basics principles of the
Fabry-Perot optical cavities ~\cite{Malik}(they constitute the
fundamental configuration for the present and future GW
interferometers and the focus of the $R\&D$ activity related to
non Gaussian optics that will be analyzed in the following
chapter). For simplicity we  assume that the system is perfectly
mode matched, that means the transverse field propagation law
doesn't change inside and outside the cavity. Let consider the
input field as a generic $\psi_{in}$ and $r_i$, $t_i$ and ${\cal
L}$ are the field reflectivity, transmittivity and losses for each
mirror [see Fig.\ref{fig:FP}]. The loss coefficient comprehends
the absorption in the coating and any other source of optical
power loss for the considered mode ($r^2+t^2=1-{\cal L}$).

 The equations for the steady-state fields
are:

\begin{equation}\label{eq:FPfields}
\left\{
\begin{array}{ll}
    \psi_1  & =\hbox{$ t_1 \psi_{in}- r_1 \psi_4$;} \\
    \psi_{trans} & =\hbox{$t_2 \psi_1 e^{-i \phi}$;} \\
    \psi_3 & =\hbox{$-r_2 \psi_2$;} \\
    \psi_{ref} &=\hbox{$- r_1 \psi_{in}+ t_1 \psi_4$.} \\
\end{array}
\right.
\end{equation}

where $\phi$ is the phase gained (or loss) by the field in transit
from mirror $1$ to mirror $2$ (it includes the axial phase shift $
k L$ and the Guoy phase shift). The resulting field are

\begin{align}\label{eq:FPfields1}
\psi_1  & = \frac{t_1 e^{i \phi}}{1-r_1 r_2 e^{ i 2 \phi}}
\psi_{in}; \\\label{eq:FPfieldtrans}
    \psi_{trans} & =\frac{t_1 t_2 e^{i \phi}}{1-r_1 r_2 e^{ i 2 \phi}} \psi_{in}; \\\label{eq:FPfieldsrefl}
    \psi_{ref} &=\left(-r_1 +\frac{t_1^2 r_2 e^{i \phi}}{1-r_1 r_2 e^{ i 2 \phi}} \right).
\end{align}

The circulating power, $P_{circ}$, stored in the cavity is given
by

\begin{align}
    P_{circ}= &|\psi_1|^2= \frac{t_1^2}{(1-r_1 r_2)^2 + 4 r_1 r_2
    \sin^2(\phi)} \, P_{in} \\
    = & \frac{t_1^2}{(1-r_1 r_2)^2}\frac{P_{in}}{1+\frac{4 r_1 r_2}{(1- r_1
    r_2)^2}\sin^2(\phi)}= g^2 P_{in} {\cal A}(\phi)
\end{align}

where we have introduced the Fabry-Perot gain $g$ and the Airy
function ${\cal A}(\phi)$. The maximum stored  power corresponds
to the peak of the Airy function, i.e. when the detuning phase
$\phi = n \pi$, where $n$ is an integer. The detuning phase can be
changed to match the resonance condition both by varying the
length of the cavity and/or the frequency of the input light. The
half-width $\delta \phi$ of the resonances is determined by the
finesse, {\cal F}, of the cavity

\begin{equation}\label{finesse}
    \delta \phi = \frac{\pi}{2 {\cal F} }, \quad \mbox{with}\quad {\cal
    F}= \frac{\pi \sqrt{r_1 r_2}}{1-r_1 r_2}
\end{equation}
The finesse is also related to the effective number of round-trips
of the light inside the FP cavity by ${\cal F}\approx \pi N_{eff}$
and to the storage time of the cavity $\tau_{sto}$ by
$\tau_{sto}\approx \frac{\cal F} {\pi} \frac{2 L}{c}$.

% ------------------------------------------------------------------------

%%% Local Variables:
%%% mode: latex
%%% TeX-master: "../thesis"
%%% End:

\chapter{Analytic and Numeric investigation of Optical
Cavities}\label{Ch2}

 \ifpdf
    \graphicspath{{Chapter2Figs/PNG/}{Chapter2Figs/PDF/}{Chapter2Figs/}}
\else
    \graphicspath{{Chapter2Figs/EPS/}{Chapter2Figs/}}
\fi

\section{Duality Relation}

\subsection{Introduction}

We discuss  here a duality relation between deviations from
sphericity (or planarity) of mirrors in FP cavities. This duality
proved a key point for our theoretical and experimental studies of
cavities with non gaussian beam profiles ~\cite{Duality}. In his
recent work on a tilt instability for advanced LIGO
interferometers~\cite{Savov-Vyat}, P.~Savov discovered numerically
a unique duality relation between the eigenspectra of paraxial
optical cavities with non-spherical mirrors: he found  a
one-to-one mapping between eigenstates and eigenvalues of cavities
deviating from flat mirrors by $h(r)$ and cavities deviating from
concentric mirrors by $-h(r)$, where $h$ need not be a  small
perturbation. In the following section, we analytically prove and
generalize this result. In this work, we prove this remarkable
correspondence analytically, for an even broader category of
cavities: those whose mirror shapes remain invariant under the
parity operation, identified as spatial reflection in the two
dimensional $\myvec r$-space (which is also equivalent to a
$180^{\circ}$ rotation around the cavity axis). Eigenmodes of such
cavities can be put into eigenstates of parity, and we show that
all corresponding eigenmodes of dual cavities have the same
intensity profiles at the mirrors, with their eigenvalues
satisfying
\begin{equation}
\label{eq:intro:dual}
 \gamma_{\rm c}^k=(-1)^{p_k+1} e^{-2ikL} (\gamma_{\rm f}^k)^*\,,
\end{equation}
where $(-1)^{p_k}$ is the parity of the $k$th eigenmode;
subscripts c and f denote nearly concentric and nearly flat
mirrors, respectively. We then illustrate its application to
interferometric gravitational-wave detectors; in particular, we
employ it to confirm the numerical results of Savov and Vyatchanin
for the impact of optical-pressure torques on LIGO's Fabry-Perot
arm cavities (i.e.\ the tilt instability), when the mirrors are
designed to support beams with rather flat intensity profiles over
the mirror surfaces. This unique mapping might be very useful in
future studies of alternative optical designs for LIGO
interferometers, when an important feature is the intensity
distribution on the cavity optics. While such a duality relation
is well-known between cavities with spherical mirrors, i.e., those
with $h(\vec r)\sim \alpha \vec r\,^2$~(for example
see~\cite{Gordon,siegman,KL}), to our best knowledge no such
relations had been established between generic cavities.
 In the following sections we
present the work done in collaboration with E. D'Ambrosio. Another
interesting approach to the duality relation was formulated by P.
Savov and Y.Chen and is presented together with ours in a joint
paper~\cite{Duality}.

\subsection{Analytical proof for mirror-to-mirror propagation}

\subsubsection{Cartesian Coordinates}

\begin{figure}[htbp]
  \centering
  \subfigure[Nearly flat mirrors]{
     \label{fig:flat}
     \includegraphics[width=0.45\textwidth]{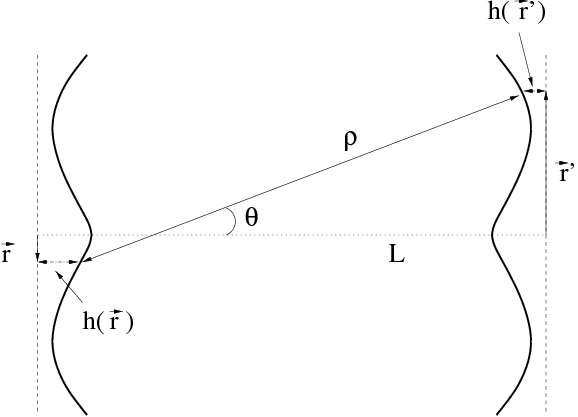}}
  \hspace{0.2in}
  \subfigure[Nearly concentric mirrors]{
    \label{fig:sfer}
      \includegraphics[width=0.45\textwidth]{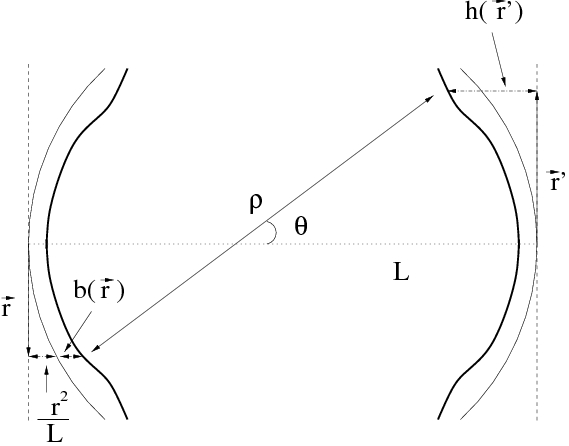}}
 \caption{Geometric constructions for the
 propagators \eqref{eq:aa} and \eqref{eq:aaconc}.}
 \label{fig:prop}
 \end{figure}

In this section we focus on field distributions on mirror
surfaces, and restrict ourselves to cavities with two identical
mirrors facing each other. The extension to the non symmetric
cavity is presented later. We adopt the Fresnel-Kirchhoff
diffraction formula to propagate fields from mirror surface to
mirror surface (see e.g.~\cite{Fox-Li}).  In this formalism, the
field amplitude $v_1(\myvec r')$ on the surface of mirror 1
propagates into
\begin{equation}
v_2(\vec r)=\int\mbox{d}^2\vec r\,'\;{\cal K}(\vec r,\vec
r\,')\,v_1(\vec r\,')
\end{equation}
on mirror 2, via the propagator \beq {\cal K}(\vec r,\myvec r')
=\frac{i k}{4 \pi \rho}(1 + \cos\theta)e^{-i k \rho} \qquad\qquad
k=\frac{2\pi}{\lambda}\,, \eeq from $\myvec r'$ (on mirror 1) to
$\myvec r$ (on mirror 2), where $\rho$ denotes the (3-D) spatial
distance between these two points and $\theta$ stands for the
angle between the cavity axis and the reference straight line, as
is illustrated in Fig.~\ref{fig:flat}. We know that the
Fresnel-Kirchhoff integral eigenequation\begin{equation}
\label{eq:eigenK} \gamma \, v(\vec r)=\int\mbox{d}^2\vec
r\,'\;{\cal K}(\vec r,\vec r\,')\,v(\vec r\,')
\end{equation}
univocally determines the eigenmodes $v$ and eigenvalues $\gamma$
of the cavity.

Applying the paraxial approximation
\begin{equation}
 \label{eq:rho}
\theta \approx 0\,,\quad \rho \approx L +\frac{|\myvec r -\myvec
r'|^2}{2L}-h(\myvec r) - h(\myvec r')\,,
\end{equation}
and we can use
\begin{equation}
{\cal K}^{h}_{\rm f}(\vec r,\vec r\,')=\frac{ik}{2\pi L}e^{-ikL}
e^{ikh(\vec r)}e^{-\frac{ik}{2L}|\vec r-\vec r\,'|^2}e^{ikh(\vec
r\,')}\,. \label{eq:aa}
\end{equation}
in the integral eigenequation.

Here the mirror surfaces deviate by $h(\myvec r)$ from a flat
reference, and the subscript f is used to reflect this convention.
From here on, we will also refer to $\mathcal{K}_{\rm f}^h$ as the
{\it nearly flat propagator}. We now consider two slightly
deformed concentric mirrors (see Fig.~\ref{fig:sfer}) so that the
mirrors height with respect to the flat reference surface is
\begin{equation}
h(\myvec{r})=\vec r\,^2/L\,+b(\myvec{r})\, , \label{eq:conv}
\end{equation}
where the height $b(\myvec{r})$ is the deviation from the
concentric spherical surface (note that concentric spherical
mirrors have their radii of curvature equal to $L/2$, and thus
surface height $r^2/L$). Inserting Eq.~\eqref{eq:conv} into
Eq.~\eqref{eq:rho}, we obtain the propagator for a {\it
nearly-concentric} cavity,
\begin{eqnarray}
{\cal K}^{b}_{\rm c}(\vec r,\vec r\,')&=&\frac{ik}{2\pi L}e^{-ikL}\nonumber \\
&& e^{ikb(\vec r)} e^{+\frac{ik}{2L}|\vec r+\vec
r\,'|^2}e^{ikb(\vec r\,')}\,.\quad \label{eq:aaconc}
\end{eqnarray}
 We use the term {\it nearly concentric} propagator for ${\cal K}^{b}_{\rm c}(\vec r,\vec r\,')$.
Although we use the terms {\it nearly-flat} and {\it
nearly-concentric}, $h$ and $b$ are not required to be small; in
fact, they can represent any deviation from perfectly flat and
concentric spherical mirrors.

Now let us consider mirrors that are then invariant under parity,
i.e., those in which we also have \beq h(\vec r)= h(-\vec
r)\,,\qquad b(\vec r)=b(-\vec r)\,. \eeq so that $\mathcal{K}_{\rm
f,\,c}$ are both invariant under a spatial reflection
\begin{equation}
\left\{\vec r, \vec r\,' \right\}\leftrightarrow\left\{-\vec
r,-\vec r\,' \right\}
\end{equation}
and therefore, we have
\begin{equation}
\label{commute} \mathcal{P}\mathcal{K}=\mathcal{K}\mathcal{P}\,,
\end{equation}
where we have defined
\begin{equation}
\mathcal{P}v(\vec{r})=v(-\vec{r})\,.
\end{equation}
for two dimensional reflection. Equation~\eqref{commute} implies
that all eigenmodes can be put into forms with definite parity. We
derive the following relation between nearly flat and nearly
concentric propagators, as constructed:
\begin{equation}
\label{eq:correspondence} \left[\mathcal{K}_{\rm f}^h(-\vec r,\vec
r\,')\right]^* =-e^{2ikL}\mathcal{K}_{\rm c}^{-h}(\vec r,\vec
r\,')\,,
\end{equation}
that is equivalent to:
\begin{equation}
\label{eq:duality} \mathcal{P} \left[\mathcal{K}_{\rm
f}^{h}\right]^* =-e^{2ikL} \mathcal{K}_{\rm c}^{-h}\,.
\end{equation}
Suppose we have an eigenstate $v_{\rm f}$ of $\mathcal{K}_{\rm
f}^{h}$, i.e., an eigenstate of a cavity with mirror deviating by
$(+h)$ from flat surface, and we compute its eigenvalue
$\gamma_{\rm f}$ and know the parity eigenvalue $(-1)^p$:
\begin{eqnarray}
\label{eq:v1}
\mathcal{K}_{\rm f}^{h}v_{\rm f} &=&\gamma_{\rm f}\, v_{\rm f}\,,\\
\label{eq:v2} \mathcal{P}v_{\rm f}&=&(-1)^p v_{\rm f}\,.
\end{eqnarray}
By applying Eqs.~\eqref{eq:duality}--\eqref{eq:v2}, we derive the
correspondance\begin{equation} \mathcal{K}_{\rm c}^{-h} v_{\rm
f}^*=e^{-2ikL}(-1)^{p+1}\gamma_{\rm f}^*v_{\rm f}^*\,.
\end{equation}
which identifies $v_{\rm c} \equiv v_{\rm f}^*$ as the
corresponding eigenstate of $\mathcal{K}_{\rm c}^{-h}$, that is
eigenstate of the corresponding resonator we denote the {\it
dual}. The eigenvalue is $\gamma_{\rm c}\equiv
e^{-2ikL}(-1)^{p+1}\gamma_{\rm f}^*$. We also induce that the
parity is still $(-1)^p$. The reverse is straightforward and the
result is an established one-to-one correspondence between dual
cavities. We summarize this mapping in Table~\ref{tab:duality:EJ}.
It is obvious to note that the corresponding eigenstates, $v_{\rm
f}$ and $v^*_{\rm f}$,  have the same intensity profiles on the
mirror surfaces; for infinite mirrors, we know $v_{\rm f}(\vec r)$
is real-valued (see Appendix b of ~\cite{Duality}), so it is an
eigenstate of the dual configuration itself.

\begin{table}[t]
\centerline{
\begin{tabular}{r|c|c}
\hline\hline
& Nearly Flat & Nearly Concentric\\
\hline
Kernel & $\mathcal{K}_{\rm f}^{h}$ & $\mathcal{K}_{\rm c}^{-h}$\\
Eigenstate & $v_{\rm f}$ & $v^*_{\rm f}$ \\
Parity & $(-1)^p$ & $(-1)^p$ \\
Half-trip eigenvalue & $\gamma_{\rm f}$ & $e^{-2ikL}(-1)^{p+1}\gamma_{\rm f}^*$ \\
\hline Round-trip eigenvalue &  $\eta_{\rm f}$ &
$e^{-4ikL}\eta_{\rm f}^*$
\\
\hline\hline
\end{tabular}
} \caption{\label{tab:duality:EJ} Correspondence of propagation
kernels, eigenstates, parities, and eigenvalues between dual
configurations.}
 \end{table}

 For cavities with identical mirrors facing each other, the full, round-trip propagator is just the square of the half-trip one. From Eqs.~\eqref{eq:correspondence} and \eqref{commute}, we have
\begin{equation}
\left[\left[\mathcal{K}_{\rm
f}^h\right]^2\right]^*=e^{4ikL}\left[\mathcal{K}_{\rm
c}^{-h}\right]^2
\end{equation}
which means that the same duality correspondence exists between
eigenstates of the full propagator, with their eigenvalues related
by
\begin{equation}
\eta_{\rm c} = e^{-4ikL}\eta_{\rm f}^*\,.
\end{equation}

Note that when $h(\vec r)=r^2/(2L)$ the two dual cavities are
identical to each other. Using the relation that links the
eigenvalues of two dual resonators, we can determine the spectrum
$$\gamma_c=\pm e^{-2ikL}\gamma_{f}^{*}=\gamma_f=e^{-ikL+in\pi/2}$$
where $n\in{\cal N}$. The resulting separation between the
eigenvalues is the Gouy phase
$$e^{i\theta_G}=e^{i\arccos(1-L/R)}\qquad R=L$$
computed for confocal resonators~\cite{siegman,KL}.

\subsubsection{Cylindrically symmetric mirrors}

In most LIGO applications, cavity mirrors still have cylindrical
shapes: $h(\myvec{r})=h(|\myvec{r}|)$. This allows us to decouple
radial and azimuthal degrees of freedom, and simplify the
eigenvalue problem. We shall follow roughly the notation
of~\cite{KL}.

We adopt the cylindrical coordinate system:
\begin{equation}
\myvec{r} = r(\cos\varphi,\sin\varphi)\,.
\end{equation}
Since  $\mathcal{K}$ is now invariant under rotation along the
$z$-axis, all eigenmodes can be put into eigenstates of rotation:
\begin{equation}
v(r,\varphi) = R(r) e^{-i m\varphi}\,,\quad m={\rm integer}\,.
\end{equation}
Inserting this into the eigenequation~\eqref{eq:eigen} and
performing analytically the angular integration we obtain the
radial eigenequation
\begin{equation}
\gamma_{nm} R_{nm}(r) = \int_0^a
K_{\mathrm{f}(m)}^h(r,r')R_{nm}(r')r'dr'\,,
\end{equation}
where for each angular mode number $m$ we have indexed the radial
eigenstates by $n$, and
\begin{equation}
\label{eq:radialK:f}
K_{\mathrm{f}(m)}^h(r,r')=\frac{i^{m+1}k}{L}J_m\left(\frac{kr
r'}{L}\right)e^{ik\left[-L+h(r)+h(r')-\frac{r^2+r'^2}{2L}\right]}\,
\end{equation}
is a symmetric radial propagator, in the {\it nearly-flat}
description.\footnote{Here we have used $\displaystyle
J_n(z)=\frac{1}{2\pi i^n}\int_0^{2\pi}
e^{iz\cos\varphi}e^{in\varphi}d\varphi$, where $J_n(z)$ is the
$n$th order Bessel function of the first kind. }  Since
$K_{\mathrm{f}(m)}^h(r,r')$ is symmetric, we obtain orthogonality
relations between radial eigenfunctions:
\begin{equation}
\int_0^a  R_{n_1m}(r) R_{n_2m}(r) rdr =\delta_{n_1n_2}\,.
\end{equation}

Using Eq.~\eqref{eq:conv} again, for a configuration with $b(r)$
correction from concentric spherical mirrors, we obtain the radial
kernel of the {\it nearly-concentric} description:
\begin{equation}
\label{eq:radialK:c}
K_{\mathrm{c}(m)}^b(r,r')=\frac{i^{m+1}k}{L}J_m\left(\frac{kr
r'}{L}\right)e^{ik\left[-L+b(r)+b(r')+\frac{r^2+r'^2}{2L}\right]}\,.
\end{equation}
Comparing Eqs.~\eqref{eq:radialK:c} and \eqref{eq:radialK:f}, we
obtain:
\begin{equation}
(-1)^{m+1}\left[K_{\mathrm{f}(m)}^h \right]^*  =
e^{2ikL}K_{\mathrm{c}(m)}^{-h}\,.
\end{equation}
This is a radial version of Eq.~\eqref{eq:duality};  here we know
explicitly that all $m$-eigenstates have parity $(-1)^m$.

Following a similar reasoning as done in the previous section,
{\it for each} angular mode number $m$, we can establish a
one-to-one correspondence between radial eigenstates of a
nearly-flat configuration to those of the dual configuration:
\begin{equation}
\left[R_{nm}\right]_{\rm c}=\left[R_{nm}\right]_{\rm f}^*\,.
\end{equation}
The mapping of the eigenvalues are given by
\begin{equation}
\left[\gamma_{nm}\right]_{\rm c}
=(-1)^{m+1}e^{-2ikL}\left[\gamma_{nm}\right]_{\rm f}^*\,.
\end{equation}
Similarly, the round-trip eigenstates have the same
correspondence, their eigenvalues related by
\begin{equation}
\left[\eta_{nm}\right]_{\rm c}
=e^{-4ikL}\left[\eta_{nm}\right]_{\rm f}^*\,.
\end{equation}

\subsubsection{Duality relation for non-identical mirrors}

In this section we will study the duality relation when the
mirrors shapes are not identical, but each still symmetric under a
$180^\circ$ rotation around the cavity axis. Since now the field
distributions of eigenstates over the two mirror surfaces are not
the same, we have to study the eigenvalue problem associated with
the round-trip propagator, instead of the individual
surface-to-surface ones. Nevertheless, we can still use the
propagators~\eqref{eq:aa} and~\eqref{eq:aaconc} to build a system
of integral equations relating field distributions $v_1(\vec r_1)$
and $v_2(\vec r_2)$ over the two mirror surfaces. [Throughout this
section, we use the subscripts $1$ and $2$ to refer to quantities
associated with mirrors $1$ and $2$, respectively.] If the mirrors
deviate from parallel planes by $h_{1,2}(\vec r)$, we have:

 \bea \label{noneq1}
&& \gamma_1 v_{1}(\vec r_{1}) = \int_{S_2} \mbox{d}^2\vec r_2 \;{\cal K}_{12}(\vec r_1 , \vec r_2)\, v_{2}(\vec r_{2})\,, \\
\label{noneq2} &&\gamma_2  v_{2}(\vec r_{2}) = \int_{S_1}
\mbox{d}^2\vec r_1\; {\cal K}_{21}(\vec r_2 , \vec r_1)\,
v_{1}(\vec r_{1})\,, \eea

where $\gamma_{1,2}$ give the attenuation and phase shift
experienced by the optical field in transit from one mirror to the
other and

\bea &&{\cal K}_{12}(\vec r_1 , \vec r_2)=\frac{i k e^{-ikL}
}{2\pi L}
 e^{ikh_1(\vec r_1)-\frac{ik}{2L}|\vec r_1-\vec r_2|^2 +ikh_2(\vec r_2)},\qquad\\
&&{\cal K}_{21}(\vec r_2 , \vec r_1)=\frac{i k e^{-ikL} }{2\pi L}
 e^{ikh_2(\vec r_2)-\frac{ik}{2L}|\vec r_2-\vec r_1|^2+ikh_1(\vec r_1)},\qquad
\eea

are the propagators from mirror 2 to mirror 1, and from mirror 1
to mirror 2, respectively. The equations ~\eqref{noneq1} and
\eqref{noneq2} give the field at each mirror in terms of the
reflected field at the other but they can be combined to form the
round-trip equation which states that the field at each mirror
must reproduce itself after one round-trip. In the following, we
will add a subscript $\rm f$ or $\rm c$ to make a distinction
between quantities related to the nearly-flat or nearly-concentric
case.
 \bea
&& \eta_f \, v_{1f}(\vec r_1) = \int_{S_1 '} \mbox{d}^2\myvec r_1' \;{\cal K}_{1f}^{h_1h_2}(\vec r_1 , \myvec r_1')\, v_{1f}(\myvec r_1') ,\qquad \\
&& \eta_f \, v_{2f}(\vec r_2) =\int_{S_2 '} \mbox{d}^2\myvec r_2'
\;{\cal K}_{2f}^{h_2h_1}(\vec r_2 , \myvec r_2')\, v_{2f}(\myvec
r_2'), \qquad \eea where the common eigenvalue $\eta_f$ is given
by $\gamma_{1f}\gamma_{2f}$ and the round-trip propagators \bea
\label{eq:gen:f} &&{\cal K}_{1f}^{h_1h_2}(\vec r_1 , \myvec r_1')=
\int_{S_2} \mbox{d}^2\vec r_2 \;
 {\cal K}_{12f}(\vec r_1 , \vec r_2)\, {\cal K}_{21f}(\vec r_2 , \vec r_1)\nonumber \\
&&{\cal K}_{2f}^{h_2h_1}(\vec r_2 , \myvec r_2')=
(1\leftrightarrow 2)\cdot {\cal K}_{1f}^{h_1h_2}(\vec r_1 , \myvec
r_1') \eea

In the nearly-concentric configuration, using kernels of the form
~\eqref{eq:aaconc} for the propagation from one mirror to the
other and combining them as done for the nearly-flat
configuration, we obtain the following nearly-concentric
round-trip equation for the field distribution over the mirror $1$
(similar formulas for the mirror $2$ with the substitution
$1\leftrightarrow 2$ throughout the argument).

\bea \label{eq:gen:c}
&& \eta_c \, v_{1c}(\vec r_1) = \int_{S_1 '} \mbox{d}^2\myvec r_1' \;{\cal K}_{1c}^{b_1b_2}(\vec r_1 , \myvec r_1')\, v_{1c}(\myvec r_1') \\
&&{\cal K}_{1c}^{b_1b_2}(\vec r_1 , \myvec r_1')= - \int_{S_2} \mbox{d}^2\vec r_2 \; e^{-2ikL}\Big(\frac{k}{2\pi L}\Big)^2 \cdot\\
&&\cdot \,e^{\frac{ik}{2L}|\vec r_1+\vec r_2|^2+\frac{ik}{2L}|\vec
r_2+\myvec r_1'|^2+ikb_1(\vec r_1)+ikb_1(\myvec r_1')+2ikb_2(\vec
r_2)} \nonumber \eea where $b_{1,2}$ are the mirrors deviations
from concentric surfaces. Using the assumed symmetry properties of
the mirrors, the propagators for the nearly-flat and
nearly-concentric cavity fulfills this relation

\bea \label{eq:gen:dual}
{\cal K}_{1c}^{-h_1-h_2}(\vec r_1 , \myvec r_1')&=& e^{-4ikL}[{\cal K}_{1f}^{h_1h_2}(-\vec r_1 , -\myvec r_1')]^* \nonumber \\
&=&  e^{-4ikL}[{\cal K}_{1f}^{h_1h_2}(\vec r_1 , \myvec r_1')]^*
\eea

Equation~\eqref{eq:gen:dual}, together with Eqs.~\eqref{eq:gen:f}
and \eqref{eq:gen:c}, provides us with a more general duality
relation, for cavities with non-identical mirrors: as long as the
corresponding mirrors of two cavities $A$ and $B$  satisfy \beq
h_{\alpha A}(\vec r)=\frac{\vec r\,^2}{L} - h_{\alpha B}(\vec
r)\,, \quad \alpha=1,2\,, \eeq the eigenstates and eigenvalues of
the two cavities will be related by: \bea v_{\alpha  A} =
v_{\alpha B}^*\,,\quad \eta_A = e^{-4ikL} \eta_B^*\,, \quad
\alpha=1,2\,. \eea

\section{Eigenvalues and FEM analysis}\label{sec:FEM}

In the preceding section we have analyzed an important relation
which is satisfied by optical cavities in terms of eigenvalues and
eigenfunctions of certain integral operator. In real application
to   GW Interferometers it is necessary to evaluate the
eigenvalues and optical modes in situation where the mirror are
not spherical and/or the finite size of the mirror become
important. A few examples will be given below. We developed our
own calculus packages in order to compute the optical modes and
frequency spectrum of a FP cavity with non-spherical mirrors,
which will be described in Sec.~\ref{sec:MHprot}.  This package
found valuable applications also in the analysis of parametric
instability\footnote{Knowledge of the diffraction losses for
higher order modes is important for predicting opto-acoustic
parametric instabilities. These arise due to the resonant
scattering of the cavity fundamental mode $\omega_0$ against test
mass acoustic modes $\omega_t$ into other optical cavity modes
$\omega_1$, which satisfy the condition $\omega_0\sim \omega_t +
\omega_1$. The parametric gain $R_0$ for this process determines
whether the system is stable $(R_0<1)$ or unstable $(R_0\geq 1)$.
The parametric gain depends linearly on the quality factor with
higher order modes, which, on the optical side of the interaction,
depends inversely on the diffraction loss ~\cite{ParInst}. Thus it
is very important to be able to obtain accurate estimates of modal
diffraction losses in the optical cavities.}\cite{Barriga} of
advanced detectors, due to its capability of obtaining, in a very
efficient\footnote{In term of computational time. Other approaches
like FFT propagation could obtain the same results but requiring
up to two orders of magnitude more time than the approach
described here (depending on the finesse of the cavity). } and
precise way, the eigenspectrum of an optical cavity. The equation
which is the object of the next sections is

\begin{equation}
\gamma v(\vec r)=\int_{Mirror} \mbox{d}^2\vec r\,'{\cal K}(\vec
r,\vec r\,')v(\vec r\,') \label{eq:eigen}
\end{equation}
where the 2-D vectors $\vec r$ and $\vec r\,'$ are the coordinates
of the projections of mirror-surface points  on planes orthogonal
to the cavity axis at the mirrors positions, $\mathcal{K}$ is the
round trip (or half round trip for symmetric cavity as Ad-LIGO)
propagator and  $\gamma$ and $v(\vec r)$ are the eigenvalues and
eigenmodes (field distribution over the mirror surface) of the
optical cavity. If the cavity mirrors are not equals, there is one
eigenvalue equation for each field on the mirror surface; for the
first mirror we have

\bea  \gamma \, v_{1}(\vec r_1) &=& \int_{S_1 '} \mbox{d}^2\myvec
r_1' \;{\cal K}_{1}^{h_1h_2}(\vec r_1 , \myvec r_1')\,
v_{1}(\myvec r_1')    \label{eq:eigen2}\\
 {\cal K}_{1}^{h_1 h_2}(\vec r_1 , \myvec r_1')&=& - \int_{S_2} \mbox{d}^2\vec r_2 \; e^{-2ikL}\left(\frac{k}{2\pi
 L}\right)^2
 \, \\
 && \cdot \exp \left\{\frac{-ik}{2L}|\vec r_1+\vec r_2|^2 -\frac{i k}{2L}|\vec
r_2+\myvec r_1'|^2+i k h_1(\vec r_1)+i k h_1(\myvec r_1')+2i k
h_2(\vec r_2)\right\}\nonumber \eea

where $h_1$ and $h_2$ are the heights of the two mirrors respect
with two planes orthogonal to the optical axis and separated by a
distance $L$. The equation for the transverse mode at the second
surface is obtained with the substitution $(1\leftrightarrow 2)$.
These equations can be solved analytically only in very special
cases (i.e. confocal spherical resonators, that is
$g_1=g_2=0$,~\cite{Boyd}). In general some approximated solutions
are found using different approaches. The finite-element method
(FEM) has become a very powerful tool for the approximate solution
of boundary-value problems governing diverse physical phenomena.
We developed some general numerical tools for the FEM analysis of
optical cavities of interest for the GW community working on the
optical set-ups.

\subsection{Cylindrical symmetric cavity}\label{sec:FEMCyl}
We will now solve the problem stated above for a cylindrical
symmetric cavity, which is of interest for the application on the
ideal perfect optical cavities of the GW interferometers both in
their present design with spherical mirrors and in new
configurations recently explored. It is obvious that for standard
spherical optics the height function is given by $h(r)=r^2/(2
R_{oc})$, where $R_{oc}$ is the radius of curvature of the mirror.
 As long
as cavity mirrors posses cylindrical symmetry, we can separate the
radial and azimuthal degrees of freedom, and simplify the search
for eigenmodes. We adopt the cylindrical coordinate system:
$\myvec{r} = r(\cos\varphi,\sin\varphi)$. Since  $\mathcal{K}$ is
invariant under rotation along the cavity axis, all eigenmodes can
be written in the form of
\begin{equation*}
v(r,\varphi) = R(r) e^{-i m\varphi}\,,\quad m={\rm integer}\,.
\end{equation*}

Inserting this into the equation Eq.~\eqref{eq:eigen2}, and
performing analytically the angular integration~\footnote{Here we
have used $\displaystyle J_m(z)=\frac{1}{2\pi i^n}\int_0^{2\pi}
e^{iz\cos\varphi}e^{im\varphi}d\varphi$, in which $J_m(z)$ is the
$m$th order Bessel function of the first kind.}, we have the
reduced radial eigen-equation ($a$ is the mirror radius)
\begin{equation}
\gamma_{pm} R_{1,pm}(r_1) = \int_0^{a_1} K_{1,(m)}^{h_1
h_2}(r_1,r_1')R_{1,pm}(r_1')r_1'dr_1'\,,\label{eq:eigenradial}
\end{equation}

\begin{align}
K_{1,(m)}^{h_1 h_2}(r_1,r_1')&= - i^{2 m} \frac{k^2}{L^2}\, e^{
-2ikL} \int_0^{a_2} dr_2 r_2 J_{m}\!\left(\frac{k r_1
r_2}{L}\right)\,J_{m}\!\left(\frac{k r_2 r_1'}{L}\right) \\
& \cdot \exp\left\{\frac{-ik}{2L}\left( r_1^2+  2 r_2^2 + r_1'^2
\right) +i k h_1( r_1)+i k h_1( r_1')+2i k
h_2(r_2)\right\}\nonumber
\end{align}

where for each angular number $m$ we have indexed the eigenstates
by $p$ which is the radial mode number.

The problem is thus reduced  to a series of one-dimensional
integral equation, one for each $m$. These are homogeneous
Fredholm equations of the second kind and there are well
documented standard techniques for their numerical
solutions~\cite{LD,KA}.

There are two general methods of solving \eqref{eq:eigenradial}.
One is based on iterative techniques applied to a discrete grid
that extract the dominant or several of the lowest-loss modes; the
classical Fox and Li algorithm falls in this category.

 For stable resonators the iterative method usually
converges slowly because even higher-order mode losses are quite
low. The second method, called Nystr\"{o}m method,  requires the
choice of some approximate quadrature rules in order to reduce the
integral equation to a matrix eigenvalue problem of dimension
equal to the number of integration points $N$. The matrix method
has two main advantages: it extracts the lowest N modes and their
eigenvalues at one time, and its accuracy is determined by the
order $N$ of the matrix.
 It is certainly
possible to use low-order quadrature rules like trapezoidal,
mid-point or Simpson's rules, but since we are looking for a quite
accurate solution of the eigenvalue problem (the modulus of the
eigenvalues must be calculated with high precision for a realistic
estimation of the diffraction losses) this would require a large
number of integration points and therefore a lot of computation
time. We used  the Gaussian quadrature rule in order to have a
greater accuracy with fewer points.

Typically, the value of the radial coordinate  is bound by the
radius of the mirror. We denote as $a$ this maximum value. Let
$x_i$ be the abscissas and $w_i$ the weight factors for
Gauss–Legendre quadrature in the range $(0, a)$. This technique
has fine meshing at the edges of the computational window and more
course meshing in the center.

 The
equation \eqref{eq:eigenradial} for a certain $m$ can be
approximated  as

\begin{equation} \gamma R(x_i) = \sum_{j=1}^N K(x_i,x_j)x_j w_j R(x_j)
\end{equation}

Defining $R_i$ the vector $R(x_i)$ and $\widetilde{K}_{i j}$ the
matrix $K(x_i,x_j)x_j w_j$ we have transformed the integral
eigenvalue problem Eq.~\eqref{eq:eigenradial} to a matrix
eigenvalue problem $\gamma R=\widetilde{K} R$ which can be solved
for $\gamma$ and $R$ by any standard mathematical package. The
convergence of the eigenvalues and eigenvectors of the discretized
problem to those of the continuous eigenvalue equation
\eqref{eq:eigenradial} for sufficient large $N$ is guaranteed by
mathematical theorems~\cite{Atkinson}.

Once the eigenvectors are determined, a particular mode may be
evaluated at any point $r$ (not a sample point) by calculating the
sum\footnote{This formula has been demonstrated to be the best
interpolating function between the grid points. This is called
Nystr\"{o}m interpolation formula.}:

\begin{equation}\label{R-reconstruc}
    R_{p m}(r)\simeq \sum_{j}^N K_{m}(r,x_j)x_j w_j R_{p m}(x_j)
\end{equation}

 The complex
eigenvectors $v_{1;pm}(r,\varphi) = R_{1;pm}(r) e^{-i m\varphi}$
correspond to the resonator mode patterns at the first mirror
surface. The fractional power loss per transit of the mode due to
diffraction effects at the mirrors is given by ${\cal{L}}=
1-|\gamma_{pm}|^2$. As is usual in optical resonators, the
eigenvalues can be sorted in decreasing magnitude order. The phase
shift $\phi_{pm}$ is given by the angle of $\gamma_{pm}$ , which
is the phase shift suffered (or enjoyed) by the wave field in a
roundtrip. The resonant condition requires that the total phase
shift  along the axis of the cavity be an entire multiple of $2
\pi$ rad; thus, separating the longitudinal phase
shift\footnote{The minus signs are due to the sign convention in
the propagator.}, $\phi_{pm}=-2kL + \beta_{pm}= -2\pi q$. The
resonance frequencies of the axial-plus-transverse modes in the
cavity is given by

\begin{equation}\label{eq:res-freq}
    f_{pm}= f_0 \left(q + \frac{\beta_{pm}}{2 \pi}\right)
\end{equation}
 where $f_0=\frac{c}{2L}$ is the free spectral range of the cavity, i.e the
 frequency spacing between successive longitudinal resonances.
We implemented the described procedure in a
$Mathematica^\circledR$ program, using its built-in routines for
the solution of eigenvalue problem\footnote{As a technical remark
is useful to point to the fact that when diffraction losses are
very small, i.e. when the absolute value of the eigenvalues is
close to one, it may be necessary to increase the significant
digits in all the computations and one of the advantage in
$Mathematica^\circledR$ is that it can handle approximate real
numbers with any number of digits.}.

\begin{figure}[htb]
\begin{center}
\includegraphics[width=0.8\textwidth]{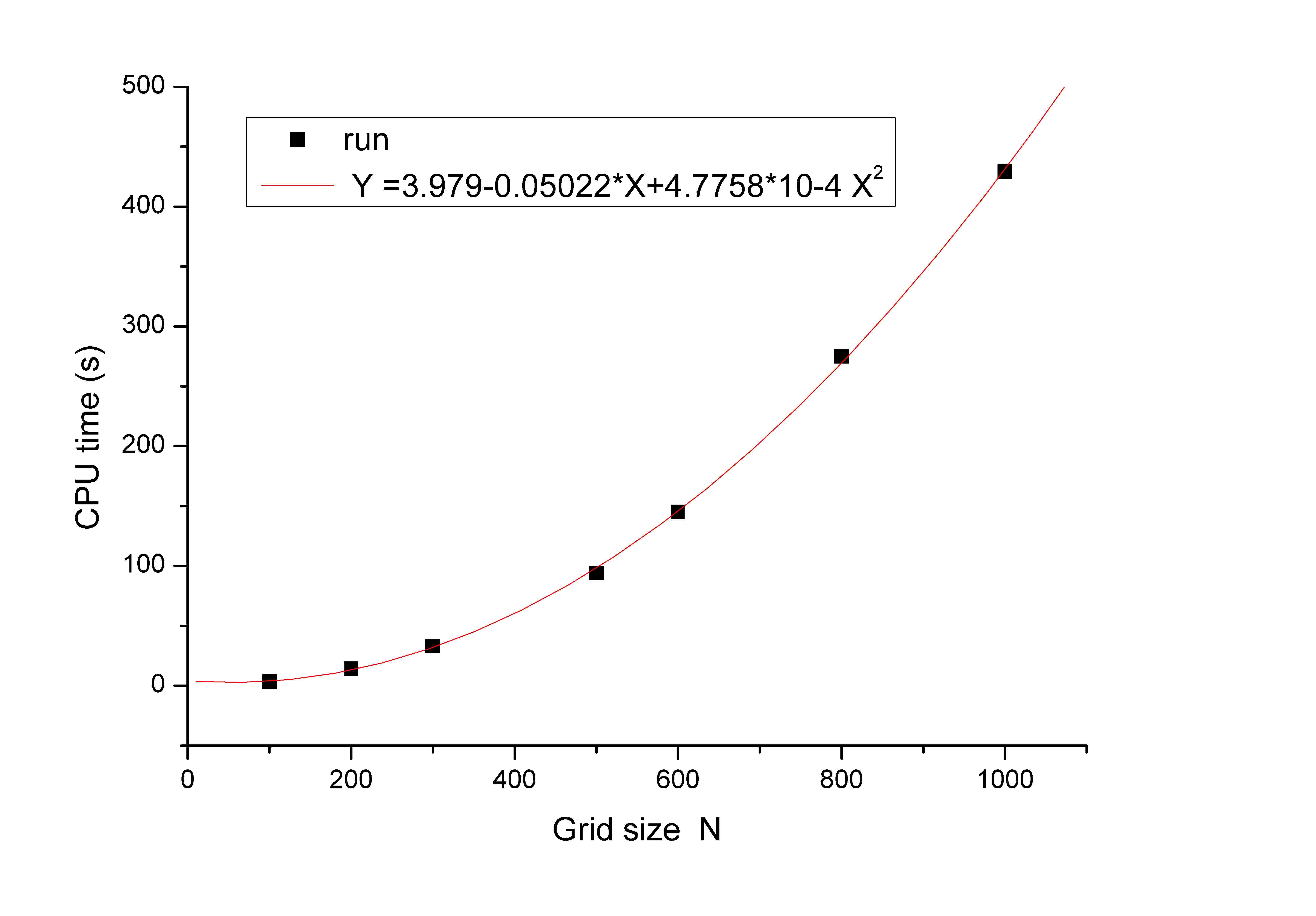}
\end{center}
\caption{CPU time for the computation of $10$ eigenvalues for
Ad-LIGO like cavity. [PC Laptop ,CPU $2$ GHz, RAM $780$ MB]}
\label{fig:Quadrature}
\end{figure}

 In Fig.~\ref{fig:Quadrature}
we show that the CPU time for the computation of ten eigenvalues
as function of the grid size $N$. The advantage of using the
Gaussian quadrature rule is evident if we consider that a $1000$
grid nodes in an equal spaced mid-point rectangular quadrature
rule can achieve an accuracy of $10^{-6}$ in the magnitude of the
eigenvalues; by comparison a Gaussian quadrature rule with $100$
nodes reaches an accuracy of $10^{-13}$ in the eigenvalue
magnitude value.

\subsection{Advantages with respect to iterative method }

The advantages of the eigenvector method are:

\begin{itemize}
    \item The propagation matrix has to be calculated only once,
then the main task is to solve the set of eigenvalues  of the
transit matrix, and each eigenvalue can directly derive one set of
eigenvectors, which just represents one mode distribution on the
mirror, without hundreds of iteration.
    \item Phase shift and
amplitude loss per roundtrip can be easily evaluated from the
eigenvalues.
    \item The multi-modes can be obtained with a single calculation.
    \item There is no need of an initial field distribution,
which has to be carefully given in the Fox–Li or in the Prony
method to get the convergent results.
\end{itemize}

\begin{figure}[htb]
\begin{center}
\includegraphics[width=\textwidth]{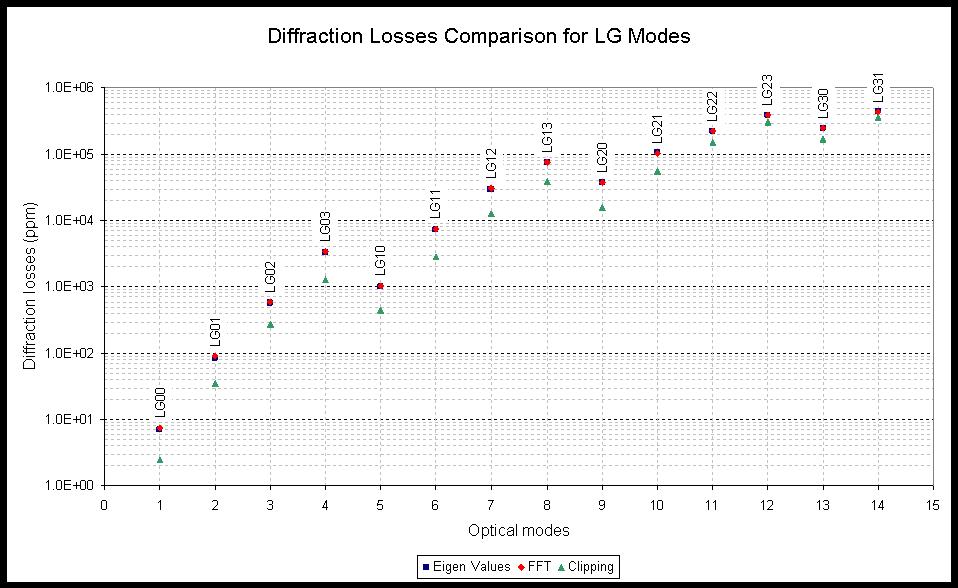}
\end{center}
\caption{Comparison between diffraction losses calculated using
eigenvalue FEM algorithm of this thesis, FFT program and clipping
approximation in the case of Ad-LIGO FP cavities with  mirrors
diameter of $31.4$ cm.[Courtesy of P. Barriga] }
\label{fig:DiffLoss}
\end{figure}

As an example of the capabilities of this simulation tool let us
mention the use in the calculation of realistic diffraction losses
in the Ad-LIGO FP cavities, for the estimation of parametric
instability. In eleven lines of \textit{Mathematica}$^\circledR$
code we are able to compute the diffraction losses of ten higher
order modes in less than four seconds of CPU time using  $100$
grid nodes. Barriga \textit{et al.}~\cite{Barriga} used an
iterative process based on FFT algorithm, which requires one
hour\footnote{Barriga in private communication.} of CPU time for
the computation of each mode diffraction loss. As it is shown in
Fig.~\ref{fig:DiffLoss}, the agreement between the two
calculations is very good with a maximum deviation of around $5
\%$ due to intrinsic limitation of the FFT algorithm for this kind
of analysis.

\subsection{Arbitrary mirror shape}\label{sec:FEM-Gen}

In the preceding paragraph we analyzed the problem
\eqref{eq:eigen2} in the very special case of cylindrically
symmetric mirrors and ideal mirrors with no imperfection. In real
application, we have to analyze broader situations, where the
mirrors deviate from the ideal shape by some amount,i.e. intrinsic
manufacturing defects or deformations due to thermal effects. It
is thus very useful to have an algorithm to solve
Eq.~\eqref{eq:eigen2} without the restriction of
Sec.~\ref{sec:FEMCyl}. In literature there are two well known
technics for approximate solution of this $2D$ problem both based
on iterative procedure; to find the steady-state mode
distribution, an arbitrary initial wave distribution is propagated
repeatedly around the resonator,undergoing changes from transit to
transit and loosing energy by diffraction. After many iterations,
the wave converges to a steady-state stationary mode. In one case
the propagation between mirrors was carried out by expanding the
optical wave in an eigenfunction expansion using as a basis set
the Hermite-Gaussian beam functions characteristic of free-space
optical beam propagation. Because these functions are normal modes
for free-space propagation, this basis set has both conceptual and
computational advantages, as demonstrated in~\cite{SiegmanMODE}.
However, this expansion has also the same practical disadvantage
characteristic of most such normal mode expansions. If one expands
a discrete $N \times N$ point two-dimensional function in an $M
\times M$ series of basis functions, the amount of computation
required goes up in proportion to $(NM)^2$, or in proportion to
$N^4$ if $M = N$. The other approach,~\cite{SiegmanFFT}, is based
on a Fast Fourier Transform algorithm to compute the propagation
between the surfaces and will be described in more details in the
next Section. The FFT method has the advantage of requiring a
smaller computational effort which increase as $N^2 \log_2 N$
respect with the $N^4$ of the other iterative approach in the real
space. Nevertheless if the diffraction losses are small, the
iterative process could converge very slowly and require a long
computational time. For this reason we developed a procedure for
the calculation of the eigenmodes, that reduce the problem
\eqref{eq:eigen2} to a matrix eigenvalue problem which can be
solved by standard numerical routines. The problem is that in a
simple discretized version of \eqref{eq:eigen2} over two $2D$
grids, one for each mirror, we have an equation between $2D$
matrix related by a $4D$ kernel. We use the following trick  to
reduce it to a standard matrix problem. First we choose a
numerical integration rule for the evaluation of integrals in a
compact bi-dimensional domain (mirror surface). We will use, for
simplicity, a mid-point rectangle quadrature formula,\footnote{The
extension to more efficient quadrature formulas in polygonal
domain is quite straightforward.} such that the mirror is sampled
at equally spaced points in the $x,y$ coordinates and assume that
the mirror have equal radii $a$. We first select the dimension of
the $2D$ grid $N^2$ and the spacing between nodes in this grid

\begin{equation}\label{DxDy}
    \Delta_x=\Delta_y =\frac{2 a}{N}
\end{equation}

It is however a waste of memory to keep track of the node points
that fall outside the mirror boundary; we therefore build a
$M\times 2$, with $M<N\times N$ matrix, that contains only the
coordinates of cells inside the mirrors

\begin{equation}\label{nM}
    n_M=\begin{pmatrix}
      x_1 & y_1 \\
      \cdot & \cdot \\
      \cdot & \cdot \\
      x_{M} & y_{M} \\
    \end{pmatrix}
\end{equation}

The equation \eqref{eq:eigen2} is then reduced to the matrix
equation
\begin{equation}\label{eq:autoKM}
    \gamma v_k = \sum_{j=1}^M K_{k j}\; v_j ,\quad k=1\cdots M
\end{equation}

where the matrix $K$ is given by

\begin{align}
    K_{k j}&= - \left(\frac{k}{2 \pi L}\right)^2\, \Delta_x^2\, e^{
-2ikL}\,\sum_{l=1}^M e^{ -\frac{i k }{2 L}\left [ (x_k-x_l)^2
+(y_k-y_l)^2+(x_l-x_j)^2 +(y_l-y_j)^2 \right]} \nonumber \\
& e^{i k \left[h_1(x_k,y_k) + h_1(x_j,y_j) +2 h_2(x_l,y_l)\right]}
\end{align}

This eigenvalue equation can be solved by any standard
mathematical package. We implemented the above described routine
in a $\textit{Matlab}^\circledR$ code because of its efficiency in
dealing with large matrices. With respect to the $1D$ grid of
dimension $N$, the size of the matrix $K$ increases as $N^4$,
requiring a notable amount of memory to store its values and
compute the eigenvectors.

\subsection{Constraints on FEM analysis}

It is interesting to point out some constraint of FEM analysis
applied to Eq.\eqref{eq:eigen2} since it is often an overlooked
aspect in papers related to this technique for optical simulation.
The grid size $N$ is set by a compromise between accuracy of the
calculations and computational effort required. The kernel in the
integral equation is a rapidly oscillating function of the
coordinates and sampling it at the grid nodes introduces
inevitably some errors. A faithful discrete representation of this
function is obtained applying a sort of \textit{sampling theorem}
of signal analysis. The fundamental constraint is that the phase
difference between two adjacent point must be less than $\pi$.
There are two contribution to this phase, one independent on the
mirror profiles and the other directly given by the height of the
mirrors. They give the following constraints on the maximum
spacing allowed between grid nodes

\begin{equation}\label{Drmax}
    \Delta x_{max} < \frac{L \lambda}{2 x_i} \Big|_{max} \, , \quad
    \mbox{and}\quad \Delta h\left[\Delta x\right]\Big|_{max}<\frac{\lambda}{2}
\end{equation}

These constraints set, for a given quadrature rule, the number of
minimum grid points; it is important to point out that the first
constraint is determined by the laser wavelength and the cavity
length, whereas  the second constraint is set by the mirror
geometry and the laser wavelength. Once these minimum requirements
are fulfilled, a further increase in $N$ results in a better
accuracy in the extraction of the eigenvalues and eigenvectors.

\section{FFT}\label{sec:FFT}

A new package based on $Mathematica^\circledR$ was developed to
simulate an optical cavity when realistic mirrors are included. It
can moreover deal with many features including mirror tilts and
shifts, beam mismatch and/or misalignment, diffractive loss from
finite mirror apertures, mirror surface figure and substrate
inhomogeneity profiles, in particular, using deformation phase
maps that are adapted from measurements of real mirror surfaces
and substrates, as well as fluctuations of reflection and
transmission intensity across the mirror profiles.
 Similar programs~\cite{VinetFFT,Bochner} developed by many other research groups, have been used for a variety of
applications by the gravitational wave community, including
numerous design help and performance estimation tasks. We decided
to build our own code because we needed a fast and reliable tool
for the optical simulation of our mesa beam cavity; the time we
would have spent in learning and modifying the existent programs
was employed to write and debug our own code. Incidentally, thanks
to its simplicity and versatility, the programs revealed itself
very useful  for investigations related to Ad-LIGO and Virgo
thermal compensation system.

The required procedure is to expand the propagating wave at any
given plane into a series of plane-waves or spatial- frequency
components, travelling at different angles in $k-$space. The
component plane waves are propagated from one transverse plane to
the next by a simple propagation formula, Eq.~\eqref{FTprop}. The
transformations from coordinate space to $k-$space and back again
are formally equivalent to Fourier transformations,
Eq.~\eqref{FTdef}, and can be very efficiently carried out using
the fast Fourier transform (FFT) algorithm. Fourier transformation
carried out using the fast Fourier transform (FFT) algorithm
provides the opportunity for a significant increase in
computational efficiency. The computational labor required to
expand a two-dimensional $N\times N$ function in a two-dimensional
$N \times N$ Fourier expansion only increases as $N^2 log_2 N$,
which is a much smaller number than $N^4$ for moderate-to-large
$N$ values.
 However,
propagation from plane to plane is not sufficient for our purpose.
In general we need to represent the propagation from the surface
of one mirror to the other, which could be of standard spherical
shape, or even non-spherical as we will see in
Sec.~\ref{sec:MHprot}. One fundamental point for the application
of this algorithm is that the action of a mirror on the optical
field, can be modelled without using a diffraction integral.
Following \cite{VinetFFT} it is convenient to discuss this issue
in the Fourier space. If $\psi(x,y,0)$ is the field in the plane
$z=0$, and $\psi(x,y,z)$ the field \textit{on} the mirror surface,
we have

\begin{equation}\label{FT-surf}
    \widetilde{\psi}(k_x,k_y,z)= e^{-i z \sqrt{k^2-k_x^2-k_y^2}}\, \widetilde{\psi}(k_x,k_y,0)
\end{equation}

The argument of the imaginary exponent can be expanded keeping in
mind that we are dealing with paraxial optics

\begin{equation}\label{expan-parax}
z \sqrt{k^2-k_x^2-k_y^2}\approx k z - \frac{z (k_x^2+k_y^2)}{2 k}
\end{equation}

In order to estimate the order of magnitude of the second term
with respect to the first one let us remind that the divergence of
a beam is defined as the second moment of the Fourier transformed
field, Eq.\eqref{eq:WRD}; Therefore we have

\begin{equation}\label{exp-div}
\frac{z (k_x^2+k_y^2)}{2 k}\sim k z \frac{\theta_0^2}{2 \pi} =
\frac{z M^2}{z_R \pi}
\end{equation}

where we have used the definition of the generalized Rayleigh
range $z_R$ introduced in Sec.~\ref{sec:GenParBeam}. Since the
second term is a factor $\theta_0^2$ smaller than the first one,
we can conclude that for surface height $z\ll z_R$,  with good
accuracy we can write

\begin{equation}\label{FT-surf2}
    \widetilde{\psi}(k_x,k_y,z)= e^{-i k z}\, \widetilde{\psi}(k_x,k_y,0)
\end{equation}

The expression of the field \textit{on} the mirror surface of
equation $z=f(x,y)$ is related to the field at a reference plane
$z=0$ by

\begin{equation}\label{Refl}
    \psi_{surf}(x,y)= \psi_{plane}(x,y)e^{- i k f(x,y)}
\end{equation}

with this approximation is obvious that  the reflection and
transmission operators from a mirror with non planar surfaces are
described by

\begin{equation}\label{T-R}
    R(x,y)= r e^{i g(x,y)} d(x,y) \quad \mbox{and} \quad T(x,y)=t
    e^{i e(x,y)} d(x,y)
\end{equation}

where $t$ and $r$ are the ordinary scalar amplitude transmission
and reflection coefficients and $g(x,y)$ and $e(x,y)$ are the
functions representing the local phase change due to reflection or
transmission trough a mirror (these functions depend on the mirror
surface shape and on the material refraction index).

Therefore we can represent propagation in free space by a phase
factor in the Fourier space according to equation \eqref{FT}, and
the interaction with a mirror by phase factors in the direct
space.

 A simple FP cavity like Fig.~\ref{fig:FP} can be thus modelled using just
 multiplication of elements of matrices and FFT routines. For example the
 intra-cavity field $\psi_1$ obeys the steady-state  equation

 \begin{equation}\label{FFT-trip}
    \psi_1= T_1 \psi_{in} + R_{1,r} \, P\, R_{2,l}\,P \, e^{i\varphi}\, \psi_1\quad
    \mbox{where}\quad P= {\cal F}^{-1} G_L {\cal F} \quad
    \mbox{and}\quad G_L=e^{-\imath k L}e^{i L \frac{k_x^2+k_y^2}{2 k}}
\end{equation}

where ${\cal F}$ indicate the Fourier transform, $P$ is the
propagator down the cavity length $L$, the reflection operators
distinguish between right $(r)$ or left $(l)$ incidence and
$\varphi$ is a uniform phase added to achieve the fine tuning of
the cavity (it plays the same role as a microscopic length
change). Let us assume that we choose a square grid in the real
space of dimension $N \times N$ and the corresponding grid in the
Fourier space.
 Thus each mirror reflection or transmission operation is reduced
to a pixel-by-pixel multiplication  between a mirror operator map
and the e-field slice on the reference plane near to it; the
propagation between two mirrors is given by an FFT, followed by a
element-by-element multiplication between the transformed matrix
and the $G_L$ matrix, and then an inverse FFT. The simulation
program models a static optical cavity, which means that we do not
follow the time evolution of the field inside the cavity but we
iterate the equation \eqref{FFT-trip} until a steady-state is
obtained for the field. The simple field relaxation algorithm
requires a pre-specified threshold of accuracy: the iteration
process ends when the field calculated at the $M^{th}$ iteration
is equal, within the accuracy requested, to the field calculated
at the $M-1^{th}$ iteration. Other more sophisticated relaxation
algorithm are described in ~\cite{Bochner} and will not be adopted
in this work.

\subsection{FFT limitations}

The propagation calculation using FFT procedure require proper
attention to questions of sampling and aliasing. These  problems
have been deeply discussed in~\cite{Bochner} and we will point the
fundamental issues. Let $a$ being the mirror radius and consider a
square grid of dimension $N^2$ and side length ${\cal{W}}$ which
contains the mirror at its center\footnote{For the moment we do
not set a constraint on the relative size ${\cal W}/a$. It will be
a crucial point in the aliasing discussion later.}. Let $\Delta x$
being the spacing, in the $x/y$ directions, between the grid nodes
(${\cal W}=N\,\Delta x$). These parameters set the k-space grid
length $k_{max}$ and spacing $\Delta k$

\begin{equation}\label{Dk-kmax}
    k_{max}=\frac{\pi}{\Delta x}=\frac{\pi N}{\cal W},\qquad \Delta k=\frac{2 \pi}{N \Delta
    x}= \frac{2 \pi}{\cal W}, \qquad k_i= i \,\Delta k , \quad i=
    -\frac{N}{2}+1,\ldots,\frac{N}{2}
\end{equation}

It is well known that sampling a function which has power at
spatial frequency above the Nyquist frequency, $f_N=1/(2 \Delta
x)$, causes aliasing. One possible solution is increasing the
highest spatial frequency in order to push this limit above the
spectral distribution of the function. To quantify the
requirements for the $k$-space gridding, let us remember that
$k_x/k$ gives the propagation direction of one Fourier component
(plane wave) and that the beam divergence, $\Theta_0$ is the
second moment of the Fourier transformed field. In paraxial beam
propagation, the k-spectrum of the beam is therefore limited by
$\gamma \Theta_0$, where $\gamma$ is a constant of order unity.
The maximum propagation angle and the $k$-resolution pose the
following constraints to the grid choice

\begin{equation}\label{k-constr}
    \frac{k_{max}}{k}> \gamma \Theta_0\,\mapsto {\cal W}< \frac{N \lambda}{2 \gamma
    \Theta_0},\quad \frac{\gamma \Theta_0}{\frac{\Delta k}{k}}\ll
    1\, \mapsto {\cal W} \gg \frac{\lambda}{\gamma \Theta_0}
\end{equation}

Remembering that for a generic paraxial beam, $M^2=W_0 \Theta_0
\pi/\lambda$, we see that the second constraint is easily
satisfied, but the first needs some care: it requires a large
number of grid nodes $N\gg 1$ but also constrains the resolution
$\Delta x $ of the grid by $\Delta x \ll \frac{\pi W_0}{\gamma
M^2}$. It is moreover important to consider the aliasing effect
due to large-angle scatter from  mirror deformations: a very
simple way to take care of the power scattered by one fourier cell
into the next replica, is to increase the window size ${\cal W}$
well above the mirror diameter, and apply the so called
''zero-padding'' method. It simply forces to zero the power
outside the mirror surface. We thus have that for most
application, ${\cal W}=4 a$ is a good choice. There is also
another aspect that has to be considered among this program
constraints. Since the actions of the mirror are represented by
pixel-per-pixel phase multiplication, the optical path difference
between two neighboring pixels must be smaller than $\lambda/2$.
This sets some limitations on the mirror shape and tilt we are
able to simulate with a given grid finesse. For our optical cavity
($L=7.32 $m, $N=128, \Delta x= 0.35$ mm, $a=1.3 $cm) we have the
limitation $1.5$ mrad  on tilt we can model.

\section{Mesa beam }\label{sec:MesaBeam}

\subsection{Introduction}

The fundamental Gaussian $TEM_{00}$ has great flexibility, is used
in many applications, and has an easy and attractive mathematical
form. Nevertheless, many cases require  different intensity
profiles: a cylindrical volume with a diameter $d = 3w$ is just a
quarter filled by a gaussian intensity distribution, therefore a
gaussian beam is not appropriate for high power laser
applications. The case of interest here is that of the reduction
of thermal noise of mirrors. Most gravitational wave
interferometric detectors (GWID) measure the variation in phase
between light beams resonating in two perpendicular cavities
caused by the passage of a gravitational wave. Any physical
displacement of the reflective surfaces of the cavity mirrors also
creates phase variations and thus contributes noise to the
measurement. In particular, random displacements of the test
masses' reflective surfaces due to thermodynamical fluctuations is
a major source of fundamental noise in the frequency range of
maximum sensitivity for next generation interferometers
(Fig.~\ref{fig:AdLIGOB}).

Gaussian laser beams are typically used to measure the position of
the mirrors in these new detectors.  Because of the peaked nature
of the gaussian beams, they are not well suited to average over
the localized thermal fluctuations.  It is possible to
significantly reduce the  test mass thermal noise using modified
optics (``graded-phase mirrors \cite{belange1}'') that reshape the
beam from the conventional gaussian intensity profile into a
flat-top beam profile. Similarly, flat topped beams would be of
interest to reduce mirror thermal noise in metrology and in
frequency and distance standards.

 One set of potential
beam profiles that satisfy the request of a flat-top, more uniform
transverse power distributions, is given by the supergaussian
functions, with analytical form $SG(r)=e^{- r^{\gamma}}$. Such a
field pattern can be obtained shaping a gaussian beam with a
diffractive optic, and use a refractive lens to produce a focused
flat top intensity. In the case of gravitational wave
interferometers, the optimized displacement sensitivity also
requires  very small diffraction losses ($\simeq 10$ ppm per
bounce) in their Fabry-Perot arms. This precludes ordinary
refractive techniques to flatten the resonant beam shape in the
cavities. It was shown  that certain optical resonators formed by
two facing graded phase mirrors - i.e. aspherical mirrors - can
have a fundamental mode with supergaussian
features~\cite{belange2}. The mirror profile, in this case, can be
designed to match the phase wavefront for a prescribed intensity
profile, while maintaining small diffraction losses. A different
analytical approach to design flattened beam was proposed by
Gori~\cite{Gori}: the flat top beam can be obtained as a sum of
Laguerre Gauss modes in the cylindrical coordinate system. The
main advantage of such transverse field distribution is that the
field that they produce upon propagation can be evaluated in a
simple way without introducing any approximation except, of
course, that the paraxial regime is assumed to hold.
Tovar~\cite{Tovar} has proposed a new class of beams, called
multi-Gaussian beams, that can reproduce the flatness of a
super-gaussian transverse field. They consists of a small sum  of
finite-width gaussian beams side by side with the same width,
phase curvature and absolute phase. Unlike the flattened Gaussian
beams, each of the multi-Gaussian beam components can be traced
individually without resorting to further series expansion. Hence,
this approach has an analytical form more desirable  to study the
beam diffraction characteristics.

Furthermore, beam propagation must be exactly known in order to
characterize interferometer performances.

Thorne and others proposed and theoretically studied
~\cite{reducing,oshaug,erika} the possibility of using a
particular class of aspherical mirrors, the so-called
``Mexican-hat'' (MH) mirrors, in a Fabry-Perot cavity  to generate
a wider, flat top laser beam, the \textit{mesa beam}. Such a beam
is predicted to significantly reduce all sources of test mass
thermal noise by better averaging over its surface fluctuations
~\cite{reducing, oshaug, erika, Aspen, VinetFB, TNmeeting}. A
complete analysis of this problem will be given in
Sec.~\ref{sec:TNMBG}.

\subsection{Mesa Beam and Mexican Hat mirrors cavity}
The mesa-beam must have an intensity distribution that is nearly
flat across the light beam, and then falls as rapidly as possible
(to minimize diffraction losses) at the beam's edges.  For a given
 length $L$, and laser wavelength $\lambda$ there is a particular Gaussian beam, called
 minimal-Gaussian
beam, whose radius increases by a factor $\sqrt{2}$ between the
beam waist (at the cavity's center plane) and the cavity's
mirrors, which provides a minimally spreading Gaussian beam. The
mesa electromagnetic field,in a quasi-flat cavity, is a
superposition of minimal-Gaussian fields whose axes are parallel
to the cavity axis and lie within a cylinder of radius $b$
centered on the cavity axis. The (non-normalized) field
distribution at the mirror plane is

\begin{equation} U_{FM}(r)= \int_{r'\leq b} \mbox{ d}^2\myvec r' e^{-\frac{(\myvec
r -\myvec r')^2(1-i)}{2w_0^2}} = 2\pi \int_0^b e^{/\frac{(r^2 +
r'^2)(1-i)}{2w_0^2}} I_0\Big[\frac{r r'(1-i)}{w_0^2}\Big] r'
\mbox{ d}r' \label{eq:MBint}
\end{equation}

where $I_0$ is the modified Bessel function of zeroth order and
$w_0=\sqrt{L/k}$ is the waist of the minimal-Gaussian (where  k is
the wavenumber). For a cavity to have the mesa beam as an
eigenmode, the surfaces of the mirrors must coincide the mesa
field's surfaces of constant phase at that distance from the
waist(this is true for infinite mirrors, but also a quite good
approximation for finite mirrors if the diffraction losses are
small). The resulting height distribution as a function of radius
$r$ is given by

\beq h_{MH}(r)= \frac{Arg[U(r)]- Arg[U(0)]}{k} \eeq

\begin{figure}[htbp]
  \centering
  \subfigure[Intensity distributions (a.u.) for a Gaussian beam (\textbf{red}) and a Flat
   Top beam (\textbf{blue}) having the same diff. losses.]{
     \label{fig:subfig:power}
     \includegraphics[width=0.45\textwidth]{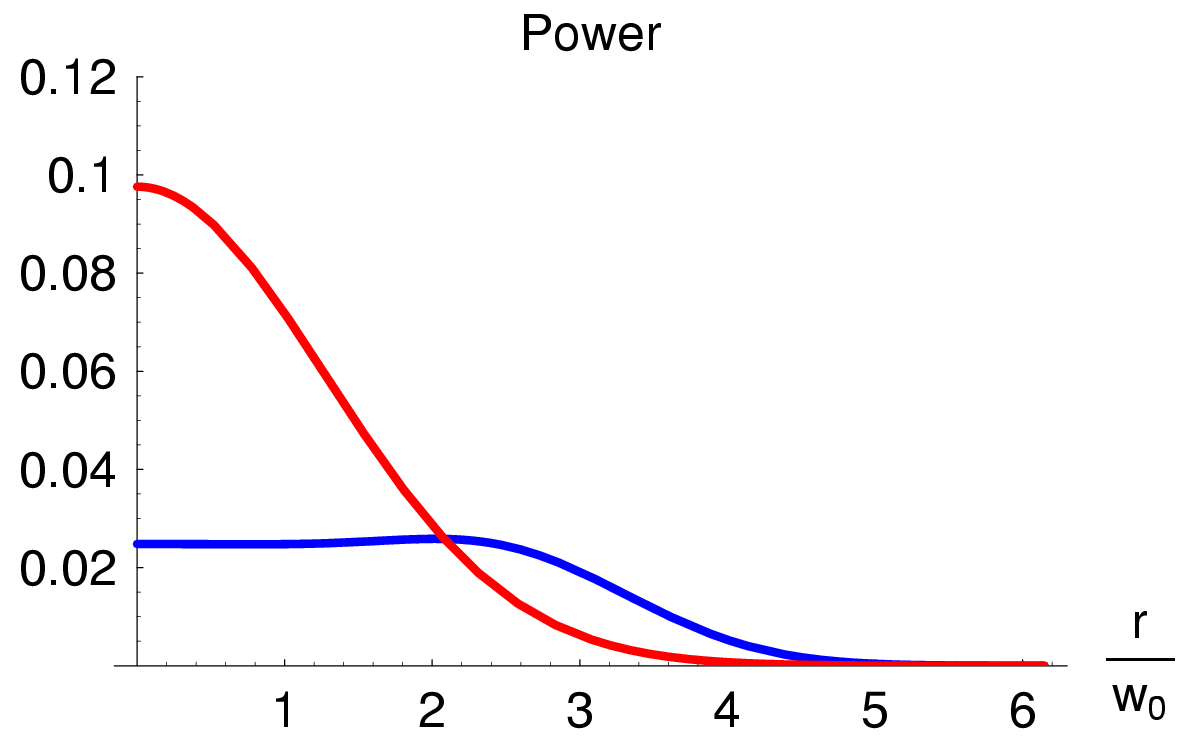}}
     \hspace{0.2in}
\subfigure[Mirror profiles: spherical mirror (\textbf{red})
supporting the Gaussian beam and Mexican Hat mirror
(\textbf{blue})supporting the mesa beam.]{
      \label{fig:subfig:mirr}
      \includegraphics[width=0.45\textwidth]{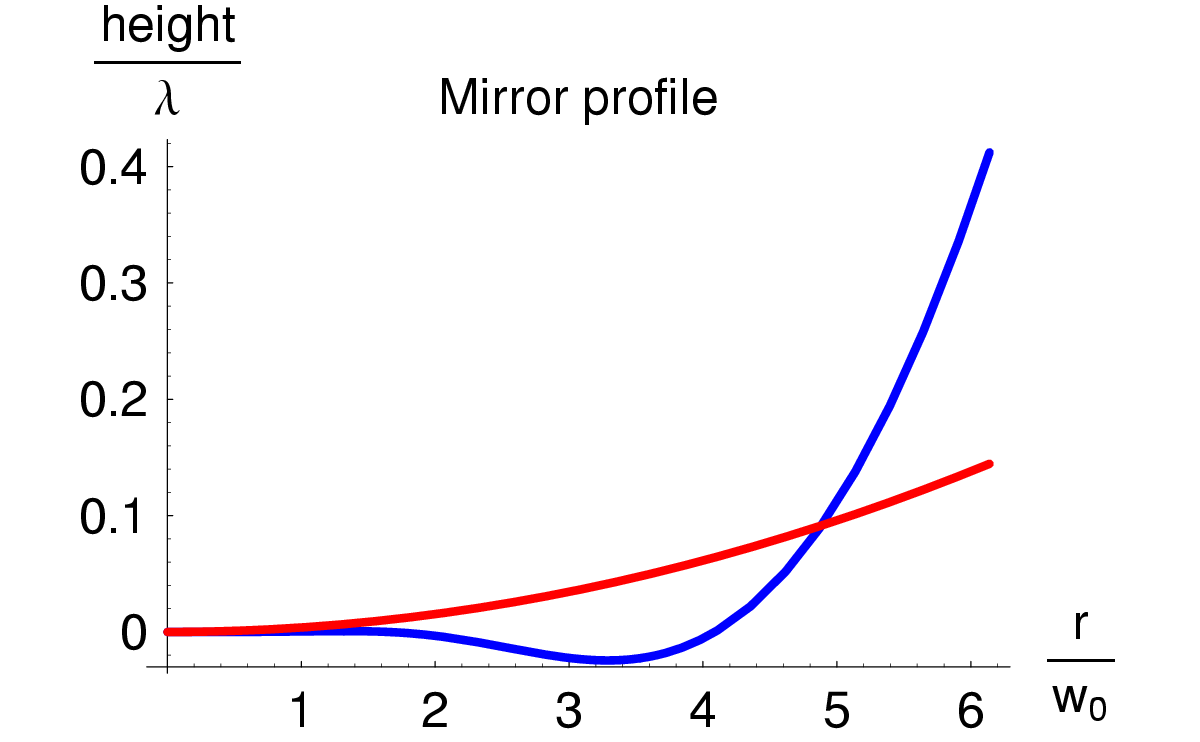}}
 \caption{Comparison between a Gaussian beam and a mesa beam.}
 \label{fig:comparison}
 \end{figure}

The resulting mirror's radial profile is shown in
Fig.~\ref{fig:subfig:mirr} along with the spherical (nearly-flat)
mirror profile which support a Gaussian beam with the same
diffraction losses; notice the shallow bump in the middle and the
flaring outer edges which resemble a Mexican hat (sombrero) and
gave the mirror its name.

 The corresponding optical design has shown strong tilt instability at high power~\cite{Savov-Vyat}than the corresponding Gaussian beam. K.S.Thorne proposed
  a different version of the {\it mesa beam}, that is supported by {\it nearly concentric} and opportunely
   shaped mirrors; this new version provides the same intensity profile at the cavity
   mirrors surface
   (and thus the same thermal noise), but introduces a weaker tilt instability (better than Gaussian beam cavities with
   the corresponding nearly concentric spherical mirrors analyzed by Sigg and Sidles~\cite{sidles}) --- as calculated
    by Savov and Vyatchanin~\cite{Savov-Vyat}  and by us in Sec.~\ref{sec:TiltInst}. A general method to design a family of optical cavities
 has been proposed by Bondarescu and Thorne~\cite{Bond-T}, from nearly flat resonators to nearly concentric
  ones. This kind of mesa beam is constructed by coherently overlapping minimal Gaussian beams
   with  axes in diverging directions sharing a common
   point at the cavity waist. The non-normalized
   field distribution at the waist plane (cavity middle point) is
   given by

   \begin{equation}\label{Uconc}
    U_{CM}^{\rm waist}(r)\!\!=\!\!
\frac{1}{\pi w_0^2}\int_{0}^{p}r_0 \mbox{d}r_0
\int_{0}^{2\pi}\mbox{d}\phi \,
 e^{-\frac{ r^2+2ir_0r\cos(\phi)}
{ w_{0}^{2}}}\,= \, \frac{w_0}{r}e^{-\frac{r^2}{w_0^2}} J_1(2
\frac{r b}{w_0^2})
\end{equation}

\begin{figure}[ht]
\begin{center}
\includegraphics[width=0.9\textwidth]{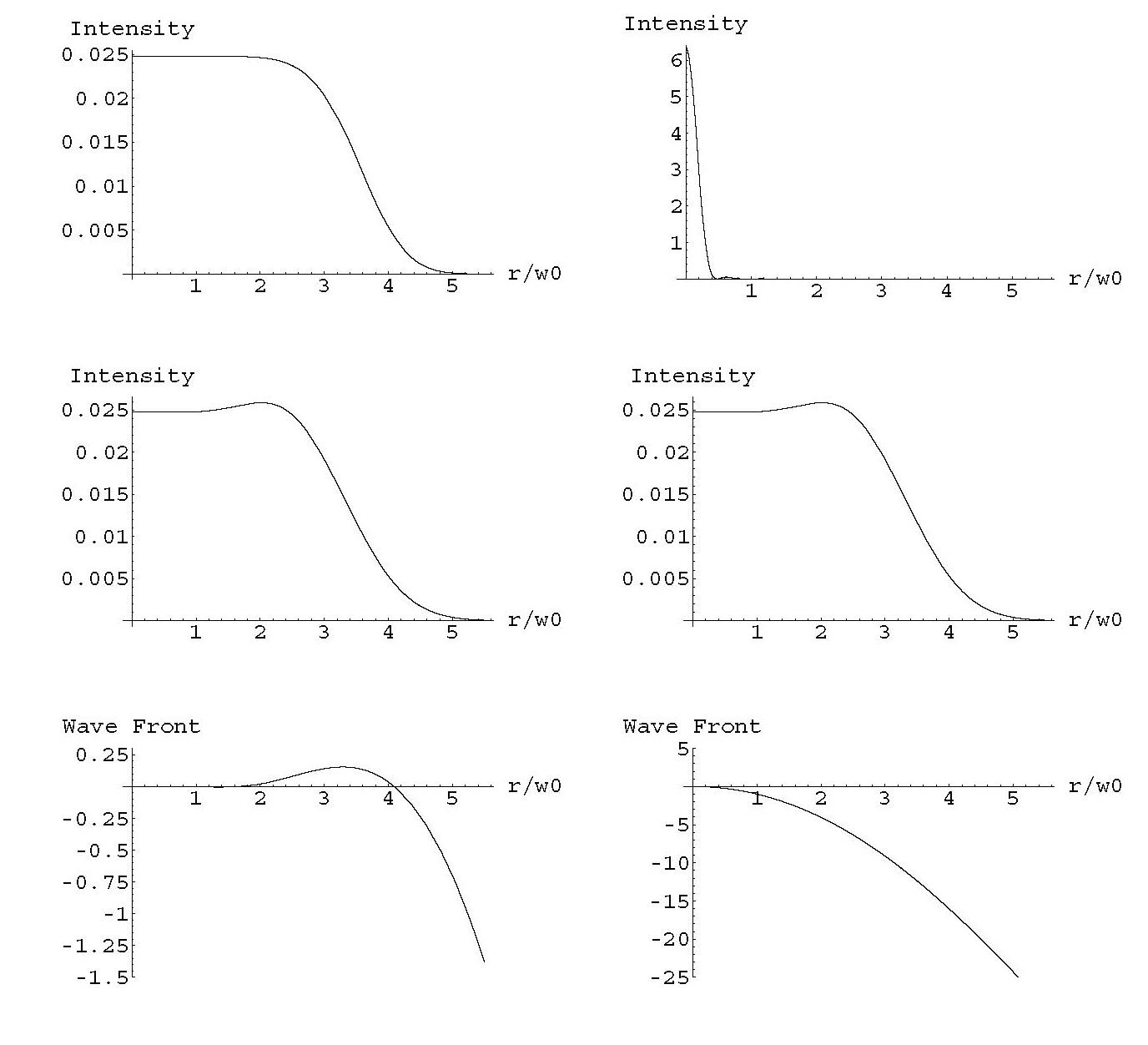}
\end{center}
\caption{Comparison between nearly flat (left panels) and nearly
concentric (right panels) Mesa beams. Upper panels: normalized
intensity profiles at the center of the cavity. Middle panels:
normalized intensity profiles at mirror surfaces Lower panels:
phase fronts at the position of the mirrors. \label{fig:aa}}
\end{figure}

This field can then be propagated along the optical axis using the
Huygens propagator \eqref{Huyg-ABCD-Cyl} adapted to a simple free
space propagation.  Examples of normalized power distributions of
nearly flat and nearly concentric mesa beams are plotted in the
upper panels of Fig.~\ref{fig:aa}. In these plots, we take
$b=4w_0$, which corresponds to the configuration proposed for
Advanced LIGO .

 \begin{figure}[htb]
\begin{center}
\label{fig:subfig:FMpropag}
     \includegraphics[width=0.8\textwidth]{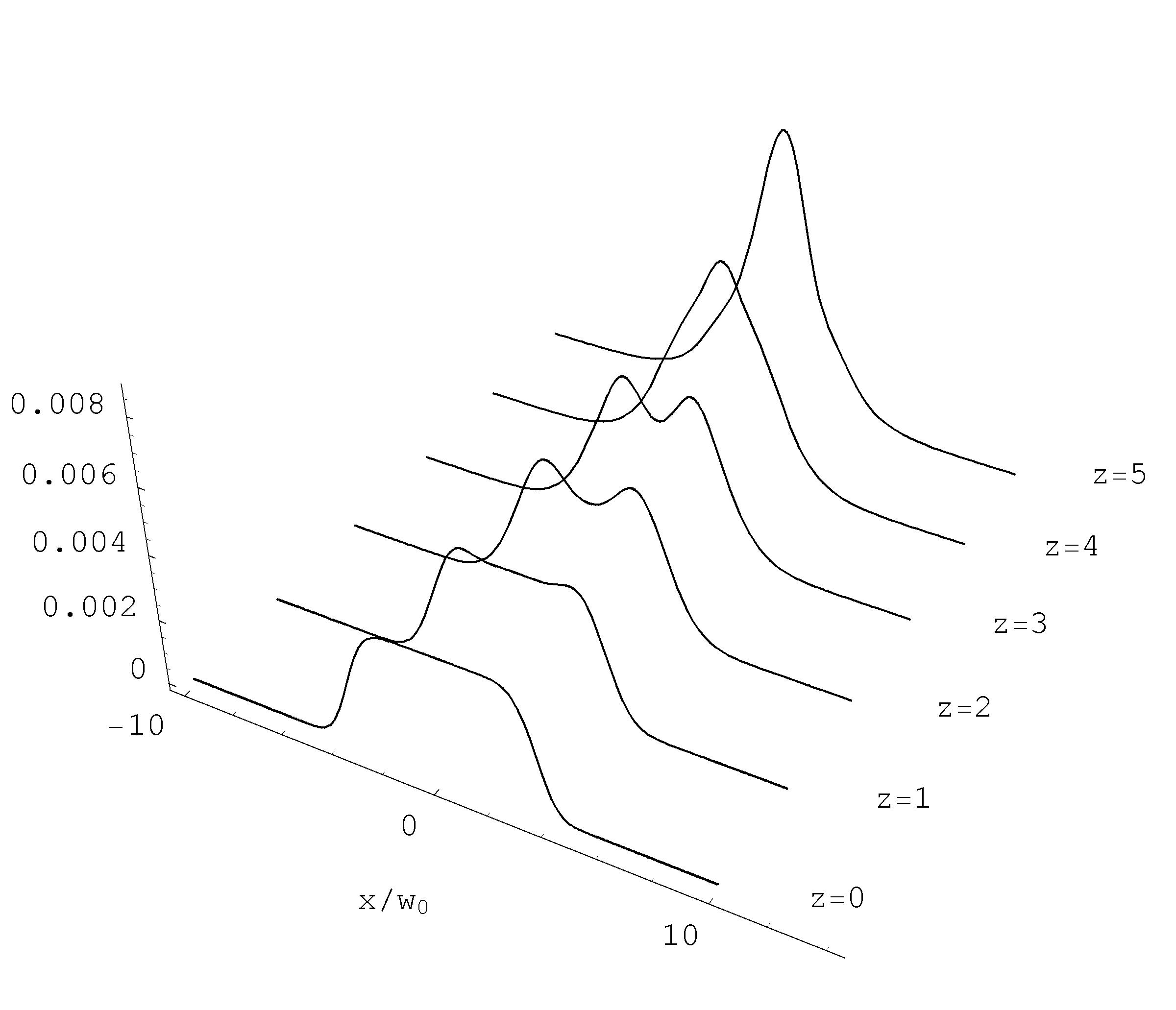}
\end{center}
\caption{FM beam profiles along propagation on free space. $z$ is
in unit of $z_0$}
\end{figure}

\begin{figure}[htb]
\begin{center}
 \label{fig:subfig:CMpropag}
     \includegraphics[width=0.8\textwidth]{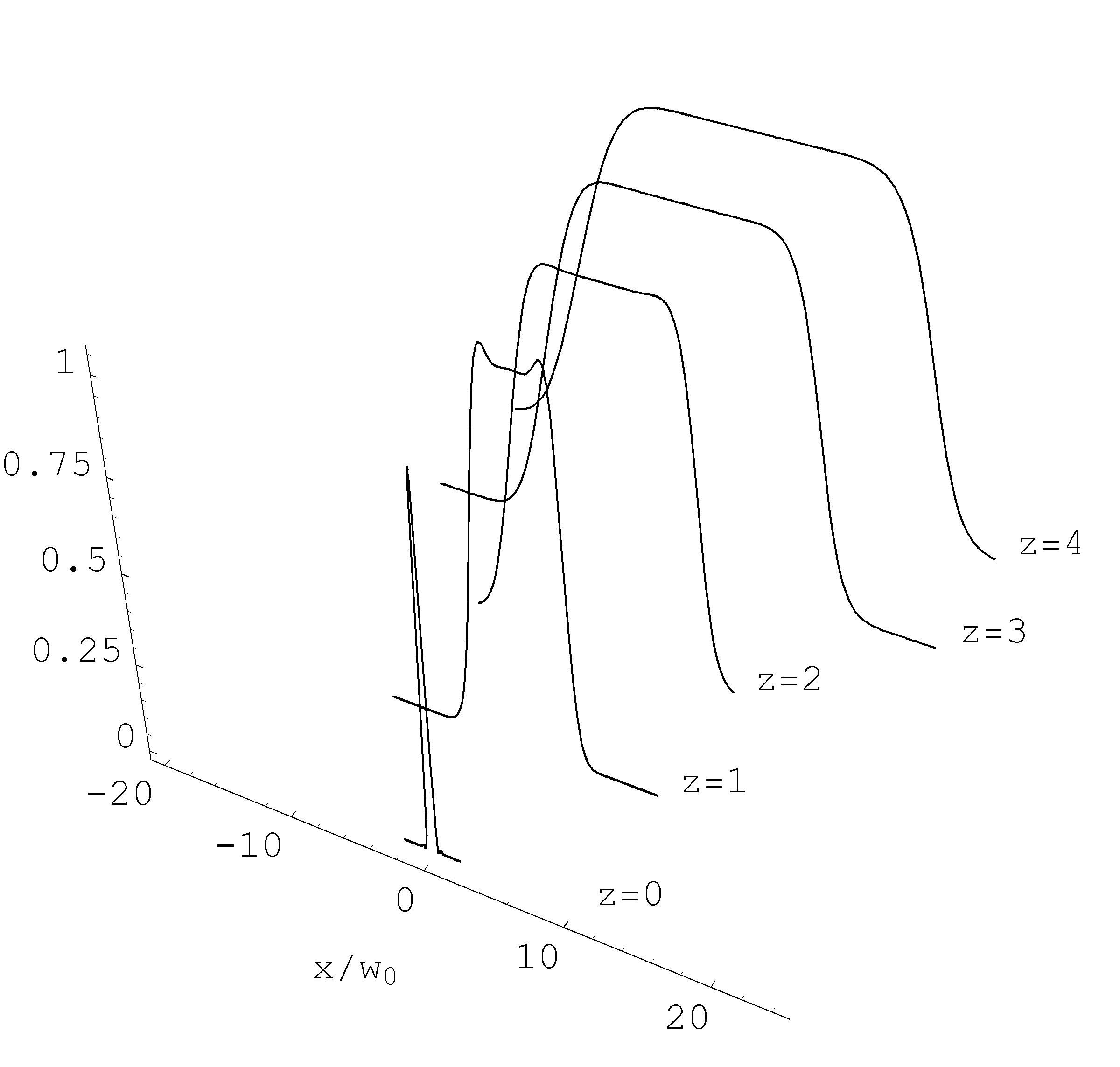}
\end{center}
\caption{CM beam profiles along propagation in free space.$z$ is
in unit of $z_0$. The intensity distribution are scaled to the
on-axis value}
\end{figure}

Even if the beam profile doesn't conserve the same shape along
propagation (see Fig.\ref{fig:subfig:FMpropag} and
\ref{fig:subfig:CMpropag}), which is the case of Gaussian beam, we
will show in the next section that is possible to characterize the
beam trough the moments definitions of Sec.~\ref{sec:GenParBeam}.

\subsection{Analytical investigations of mesa
beams}\label{sec:MBan}

Mesa beam have been studied mostly numerically~\cite{erika} and
Sec.~\ref{sub:MH modes}, perturbatively~\cite{oshaug, reducing}
and Sec. or using an expansion on Gauss Laguerre
functions~\cite{Hyperboloidal}. It is however of great utility to
have some analytical formula for studying the beam characteristics
and beam propagation trough an optical system without recurring to
numerical analysis. In this section we follow the guidelines given
in Sec.~\ref{sec:GenParBeam} for the definition of some parameters
which are the usually measurable quantities~\eqref{WRD-cyl} in an
optical beam and which behave in a simple way under a generic
$ABCD$ transformation. The expression of the mesa field as given
by equation \eqref{eq:MBint} seems quite intractable from an
analytical point of view, in particular the integral cannot
explicitly computed. However it is a convolution product of a
gaussian  with the disc function of radius $b$ and in the Fourier
domain this become a simple product of the transformed functions,
which are well known. The first important step is to express the
formula for the generalized beam width and radius of curvature in
the Fourier domain. Since we are dealing with cylindrically
symmetric problems it is useful to work with these definitions of
Fourier transform and Hankel transform

\begin{align}\label{eq:FTpol}
    & \widetilde{f}(k_x,k_y)=\int_{\mathbf{R}^2}dx\, dy\, e^{- i(k_x x +
    k_y
    y)}f(x,y)=\widetilde{f}(s) =2 \pi \int_0^\infty \, r dr\, f(r) J_0(r
    s)= H(f) \\
    \nonumber \\
    &f(x,y)= \frac{1}{4 \pi^2} \int_{\mathbf{R}^2} dk_x\, dk_y\,e^{ i(k_x x +
    k_y y)} \widetilde{f}(k_x,k_y) = f(r)=\frac{1}{2 \pi} \int_0^\infty \, s ds\, \widetilde{f}(s) J_0(r
    s)  = H^{-1}(\widetilde{f}) \nonumber
\end{align}

We recall here the formula of the generalized beam parameter
assuming a normalized field $\psi$ \footnote{Note that the
coefficient of the divergence contains different $\pi$ factors due
to the different definition of the Fourier transform.}

\begin{align}\label{eq:WRD}
    w&= 2 \sqrt{\pi\int_0^\infty |\psi(r)|^2 r^3
    dr}\\
    \frac{1}{R}&=\frac{i \lambda}{
    w^2(\psi)}\int_{0}^\infty \left (\frac{\partial \psi(r)}{\partial
    r}\psi^*(r) - \psi(r)\frac{\partial \psi^*(r)}{\partial
    r}\right ) r^2  dr \\
    \theta_0&= \frac{\lambda}{2} \sqrt{\int_0^\infty |\widetilde{\psi}(s)|^2 s^3  ds}
\end{align}

Using the definition \eqref{eq:FTpol} and the properties of Bessel
functions is easy to prove the following equalities

\begin{align}
    & 2 \pi \int_0^{\infty} r^2 \psi(r) J_0(r s) r dr = - \left ( \frac{\partial ^2 \widetilde{\psi}(s)}{\partial s^2}
    + \frac{1}{s} \frac{\partial \widetilde{\psi}(s)}{\partial
    s}\right)
    \nonumber \\
    & 2 \pi \int_0^{\infty} r \frac{\partial \psi(r)}{\partial r} J_0(r s) r
    dr = - \frac{1}{s}\frac{\partial}{\partial s}(s^2
    \widetilde{\psi}(s))
\end{align}

It is then possible to express the equations \eqref{eq:WRD} in a
different way

\begin{align}\label{eq:WRD-FT}
    w&= \sqrt{- \frac{1}{\pi}\left [\int_0^\infty s \, ds\, \widetilde{\psi}^*\left ( \frac{\partial ^2 \widetilde{\psi}}{\partial s^2}
    + \frac{1}{s} \frac{\partial \widetilde{\psi}}{\partial
    s}\right)  \right]}\\
    \frac{1}{R}&=\frac{i \lambda}{4 \pi^2
    w^2}\int_{0}^\infty \left [\widetilde{\psi} \frac{\partial}{\partial
    s}(s^2 \widetilde{\psi}^*) - \widetilde{\psi}^* \frac{\partial}{\partial
    s}(s^2 \widetilde{\psi})\right ] \, ds \\
    \theta_0&= \lambda \sqrt{-\frac{1}{\pi}\left [ \int_0^{\infty}\,r\, dr\,\psi^* \frac{\partial^2 \psi}{\partial r^2} +
    \int_0^{\infty}\, dr \psi^* \frac{\partial \psi}{\partial r}   \right]}
\end{align}

First of all we need to find a normalization for our flat mesa
(FM) beam \eqref{eq:MBint} otherwise the above formulas are not
analytically computable. We use the Parseval relation\footnote{The
same technique has been used in~\cite{Virgo-Book}. Our results
confirms their calculation of the normalization factor and correct
some errors for the Gaussian beam coupling efficiency calculated
there.} applied to the transformation \eqref{eq:FTpol} in polar
coordinates

\begin{equation}\label{Parseval}
    \int_{\mathbf{R}^2}dx\, dy\,|f(x,y)|^2= 2 \pi
    \int_0^{\infty} r \, dr |f(r)|^2= \frac{1}{2 \pi}
    \int_0^{\infty} s\, ds |\widetilde{f}(s)|^2
\end{equation}

Lets consider the mesa beam \eqref{eq:MBint} at the beam waist

\begin{equation}\label{eq:MBF-waist}
    U_0\propto \int_{D(b)}
    e^{-\frac{(x-x_0)^2+(y-y_0)^2}{w_0^2}}\, dx_0 dy_0 \qquad
    \mbox{where}\quad D(b)= \left\{ \begin{array}{ll}
0 & \textrm{if $r>b$}\\
1 & \textrm{if $r\leq b$}
\end{array} \right.
\end{equation}

Now, the  Hankel transformation of the gaussian and of the disc
function are well known

\begin{equation}\label{Htras-GeD}
    H(e^{-\frac{r^2}{w_0^2}})= \pi w_0^2 e^{-\frac{w_0^2 s^2}{4}}
    \qquad \mbox{and}\qquad H(D(b))= 2 \pi b \frac{J_1(b s)}{s}
\end{equation}

The integral of the intensity in the Fourier domain can be
performed analytically

\begin{equation}\label{F00}
   2  \int_0^{\infty}\,  e^{-\frac{w_0^2 s^2}{2}} \frac{J_1^2(b
    s)}{s^2}\, s \, ds = 1- e^{-\frac{b^2}{w_0^2}}\left(I_0\!\!\left(\frac{b^2}{w_0^2}\right) + I_1\!\!\left(\frac{b^2}{w_0^2}\right)
    \right) \equiv \Upsilon
\end{equation}

We have therefore the normalized field at the beam waist
$\psi_{0,\mathbf{FM}}$

\begin{align}\label{FMN-waist}
 \psi_{0,\mathbf{FM}}(r)&= \frac{1}{w_0^2 b \sqrt{\Upsilon}
\pi^{\frac{3}{2}}} \int_{D(b)}
    e^{-\frac{(x-x_0)^2+(y-y_0)^2}{w_0^2}}\, dx_0 dy_0 \nonumber
    \\
     &= \frac{2}{b
    \sqrt{\Upsilon}\sqrt{\pi}}\int_0^{\frac{b}{w_0}}
    e^{-(\frac{r^2}{w_0^2}- x^2)} I_0\!\!\left(\frac{2 r x}{w_0}\right)\, x\, dx
\end{align}

The expression of the Fourier transformed field is
\begin{equation}\label{FMN-waist-FT}
 \widetilde{\psi}_{0,\mathbf{FM}}(s)= \frac{2 \pi}{\sqrt{\pi}
 \sqrt{\Upsilon}} e^{- \frac{w_0^2 s^2}{4}} \frac{J_1(b s)}{s}
\end{equation}

The expression of the FM beam in other plane along $z$ is just a
straightforward application of the propagation rule for each
Gaussian composing the beam. The calculation led the following
result\footnote{Omitting the phase factor common to all the
minimal Gaussian given by the Gouy phase shift $e^{-i
\arctan\frac{z}{z_0}}$.}

\begin{align}
     \psi_{z,\mathbf{FM}}(r,z)&= \frac{1}{w_0 b  \pi \sqrt{\Upsilon} \sqrt{2}}\sqrt{\frac{2}{\pi w_z^2}} \int_{D(b)}
    e^{-\frac{(x-x_0)^2+(y-y_0)^2}{w_z^2}\left(1 + i \frac{z}{z_0}\right)}\, dx_0 dy_0 \nonumber
    \\
    &=\frac{2 w_z}{b w_0
    \sqrt{\Upsilon}\sqrt{\pi}}\int_0^{\frac{b}{w_z}}
    e^{-\left(\frac{r^2}{w_z^2}- x^2\right)\left(1 + i \frac{z}{z_0}\right)} I_0\!\!\left(\frac{2 r x\left(1 + i \frac{z}{z_0}\right)}{w_z}\right)\, x\, dx
\end{align}

where $z_0$ is the Rayleigh range and $w_z$ is the beam radius at
the $z$ plane of the minimal Gaussian field
\begin{equation}\label{ZR e wz}
    z_0= \frac{\pi w_0^2}{\lambda},\qquad w_z=w_0 \sqrt{1+\left(\frac{z}{z_0}\right)^2}
\end{equation}

The Fourier transformed field is

\begin{equation}\label{FMN-z-FT}
 \widetilde{\psi}_{z,\mathbf{FM}}(s,z)= \frac{2 \sqrt{\pi} w_0}{\sqrt{\Upsilon} w_z }\left(1 + i
 \frac{z}{z_0}\right) \frac{J_1(b s)}{s} \, e^{-\frac{w_0^2 s^2}{4}\left(1 + i \frac{z}{z_0}\right)}
\end{equation}

Using the expression \eqref{FMN-waist-FT} in equation
\eqref{eq:WRD-FT} we can calculate the FM beam waist
$W_{0,\mathbf{FM}}$ and the divergence $\Theta_{0,\mathbf{FM}}$.

\begin{align}\label{eq:W0FM}
W_{0,\mathbf{FM}}&=\sqrt{\frac{b^2 +2 w_0^2 -
e^{-\frac{b^2}{w_0^2}}\left[2(b^2 +w_0^2)
I_0\!\!\left(\frac{b^2}{w_0^2}\right)+(2 b^2 +w_0^2)
I_1\!\!\left(\frac{b^2}{w_0^2}\right)\right]}{\Upsilon}}\\\label{eq:D0FM}
\Theta_{0,\mathbf{FM}}&=\frac{\lambda}{\pi}
\sqrt{\frac{e^{-\frac{b^2}{w_0^2}}I_1\!\!\left(\frac{b^2}{w_0^2}\right)}{\Upsilon
w_0^2}}
\end{align}

Using \eqref{FMN-z-FT} with \eqref{eq:WRD-FT} we can calculate, at
any plane $z$, the beam width and the radius of curvature

\begin{align}
    W_{z,\mathbf{FM}}&=\sqrt{\frac{b^2 +2 w_0^2 -
e^{-\frac{b^2}{w_0^2}}\left[2(b^2 +w_0^2)
I_0\!\!\left(\frac{b^2}{w_0^2}\right)+(2 b^2 +w_0^2
-\frac{z^2}{z_0^2})
I_1\!\!\left(\frac{b^2}{w_0^2}\right)\right]}{\Upsilon}}\\
\frac{1}{R_{z,\mathbf{FM}}}&= \frac{\lambda
I_1\!\!\left(\frac{b^2}{w_0^2}\right) e^{-\frac{b^2}{w_0^2}}
}{\Upsilon \pi W_{z,\mathbf{FM}}^2}\frac{z}{z_0}
\end{align}

It is possible to verify that these quite complicated propagation
rules can be put in the standard form

\begin{align}
    W_{z,\mathbf{FM}}&= W_{0,\mathbf{FM}}\sqrt{\left(1+ z^2
    \frac{\Theta_{0,\mathbf{FM}}^2}{W_{0,\mathbf{FM}}^2}\right)}
    \label{WzFM}
    \\
R_{z,\mathbf{FM}}&= z \left(1+
    \frac{W_{0,\mathbf{FM}}^2}{z^2 \,
    \Theta_{0,\mathbf{FM}}^2}\right)\label{RzFM}
\end{align}

\subsection{Concentric Mesa beam}

In the Concentric Mesa (CM) beam case, \eqref{Uconc}, we have the
great advantage of having an explicit form of the field at the
waist plane. The normalization is thus immediately performed using
the same integral~\eqref{F00}.

\begin{equation}\label{CMN-waist}
    \psi_{0,\mathbf{CM}}(r)= \frac{1}{\sqrt{\pi \, \Upsilon}}
    \frac{e^{-\frac{r^2}{w_0^2}}}{r} J_1 \!\!\left(\frac{2 r
    b}{w_0^2}\right)
\end{equation}

and the propagation is obtained using the Huygens integral
\eqref{Huyg-ABCD-Cyl} with $L=B=z,\,A=D=1,\, C=0$.

\begin{equation}\label{CMN-z}
    \psi_{z,\mathbf{CM}}(r,z)= \frac{1}{\sqrt{\pi \,
    \Upsilon}}\frac{i\,e^{-i k z}
    k}{z}\int_0^{\infty}
    \frac{e^{-\frac{r'^2}{w_0^2}}}{r'} J_1 \!\!\left(\frac{2 r'
    b}{w_0^2}\right) e^{-\frac{i k}{2 \, z}(r'^2+r^2)}\,J_0\!\! \left(\frac{k r' r}{z}\right)
    r'dr'
\end{equation}

It interesting to observe that the field $
\psi_{0,\mathbf{CM}}(r)$ has exactly the same for of the Fourier
transformed $ \widetilde{\psi}_{0,\mathbf{FM}}(s)$. This is a
manifestation of the duality relation between these two kind of
beam; one beam is related to the other by a transformation which
has the form of an Hankel transformation \footnote{Further details
about this aspect of the duality relation are given
in~\cite{Duality}}. As seen for the FM case with equations
\eqref{WzFM} and \eqref{RzFM} it is sufficient to calculate the
beam waist and divergence to obtain the propagation formulas. For
the CM it is more useful to work with the field \eqref{CMN-waist}
and use the first of \eqref{eq:WRD} and the last of
\eqref{eq:WRD-FT} for this computation. The results are

\begin{align}\label{eq:W0CM}
    W_{0,\mathbf{CM}}&=\sqrt{\frac{w_0^2
    e^{-\frac{b^2}{w_0^2}}\,I_1\!\!\left(\frac{b^2}{w_0^2}\right)}{\Upsilon}}\\
\Theta_{0,\mathbf{CM}}&=\frac{\lambda}{\pi}\sqrt{\frac{b^2 +2
w_0^2 - e^{-\frac{b^2}{w_0^2}}\left[2(b^2 +w_0^2)
I_0\!\!\left(\frac{b^2}{w_0^2}\right)+(2 b^2 +w_0^2)
I_1\!\!\left(\frac{b^2}{w_0^2}\right)\right]}{\Upsilon \,
w_0^4}}\label{eq:D0CM}
\end{align}

\begin{figure}[htb]
\begin{center}
 \label{fig:subfig:MBwidthpropag}
      \includegraphics[width=0.7\textwidth]{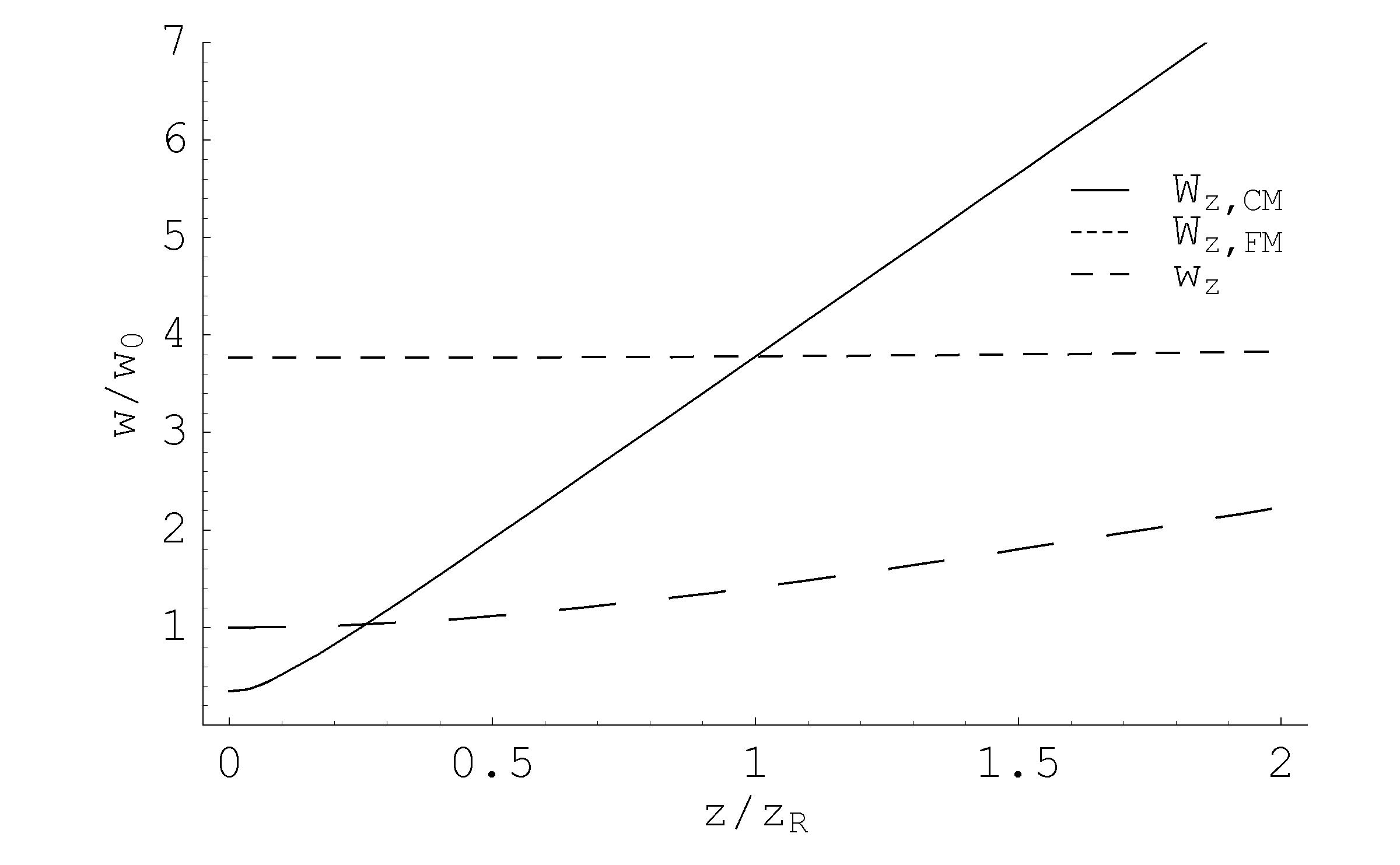}
\end{center}
\caption{Beam width evolution along propagation in free space.}
\end{figure}

\subsection{$M^2$ parameters and misalignment sensitivity}\label{sec:MBmis}

The parameter $M^2$, introduced in \eqref{M2-gen}, is an invariant
of the beam along the propagation and is a very useful tool in
beam propagation analysis. At the beam waist is given simply
by\footnote{For a pure Gaussian beam $M^2=1$.}
\begin{equation}\label{M2-waist}
    M^2= \frac{\pi}{\lambda} W_0 \Theta_0
\end{equation}

Combining \eqref{eq:W0FM} and \eqref{eq:D0FM} we have

\begin{equation}\label{M2-FM}
    M^2_{\mathbf{FM}}=\frac{1}{\Upsilon w_0}\sqrt{
    e^{-\frac{b^2}{w_0^2}}\,I_1\!\!\left(\frac{b^2}{w_0^2}\right)\left\{ b^2 +2
w_0^2 - e^{-\frac{b^2}{w_0^2}}\left[2(b^2 +w_0^2)
I_0\!\!\left(\frac{b^2}{w_0^2}\right)+(2 b^2 +w_0^2)
I_1\!\!\left(\frac{b^2}{w_0^2}\right)\right]  \right\}}
\end{equation}

\begin{figure}[htb]
\begin{center}
\includegraphics[width=0.7\textwidth]{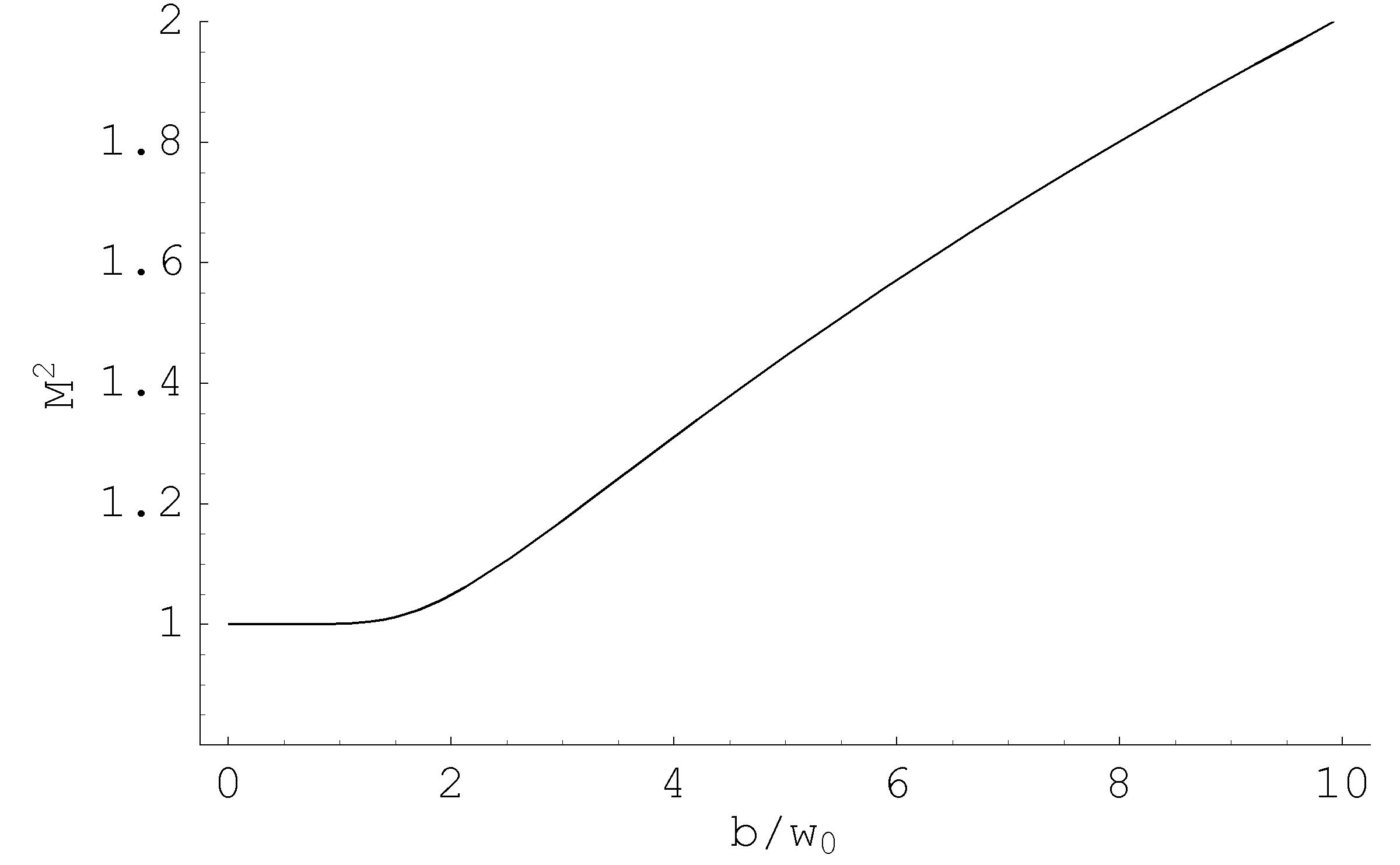}
\end{center}
\caption{$M^2$ parameter for the mesa beam as function of the
integration disc radius $b$ .} \label{fig:M2MB}
\end{figure}

In Fig.\ref{fig:M2MB} is shown $ M^2_{\mathbf{FM}}$ as a function
of the integration disc radius $b$.

 Looking at the formulas for the CM
beam \eqref{eq:W0CM} and \eqref{eq:W0CM} we obtain the important
result

\begin{align}
    W_{0,\mathbf{CM}}&= \frac{\pi
    w_0^2}{\lambda}\Theta_{0,\mathbf{FM}}, \quad
    \Theta_{0,\mathbf{CM}}=\frac{\lambda}{\pi
    w_0^2}W_{0,\mathbf{FM}}\quad \Rightarrow \, M^2_{\mathbf{CM}}= M^2_{\mathbf{FM}}
\end{align}

which is another manifestation of the duality relation.

The misalignment sensitivity of a generic paraxial beam is
quantified by the misalignment factor $|\eta_m|^2(\alpha,\delta)$
introduced in~\cite{Morin} and briefly recalled on
pg.~\pageref{misa-small} (where $\delta$ and $\alpha$ represent
the transverse and angular shifts, respectively, of the misaligned
beam in a given plane). The misalignment superposition integral of
Eq.~\ref{misa-gen} can not be analytically performed for FM and CM
beams but the approximated expression Eq.~\ref{misa-small} offers
a good approximation for small misalignments.

\begin{figure}[htbp]
  \centering
  \subfigure[Transverse shift.]{
     \label{fig:subfig:MBFtras}
     \includegraphics[width=0.7\textwidth]{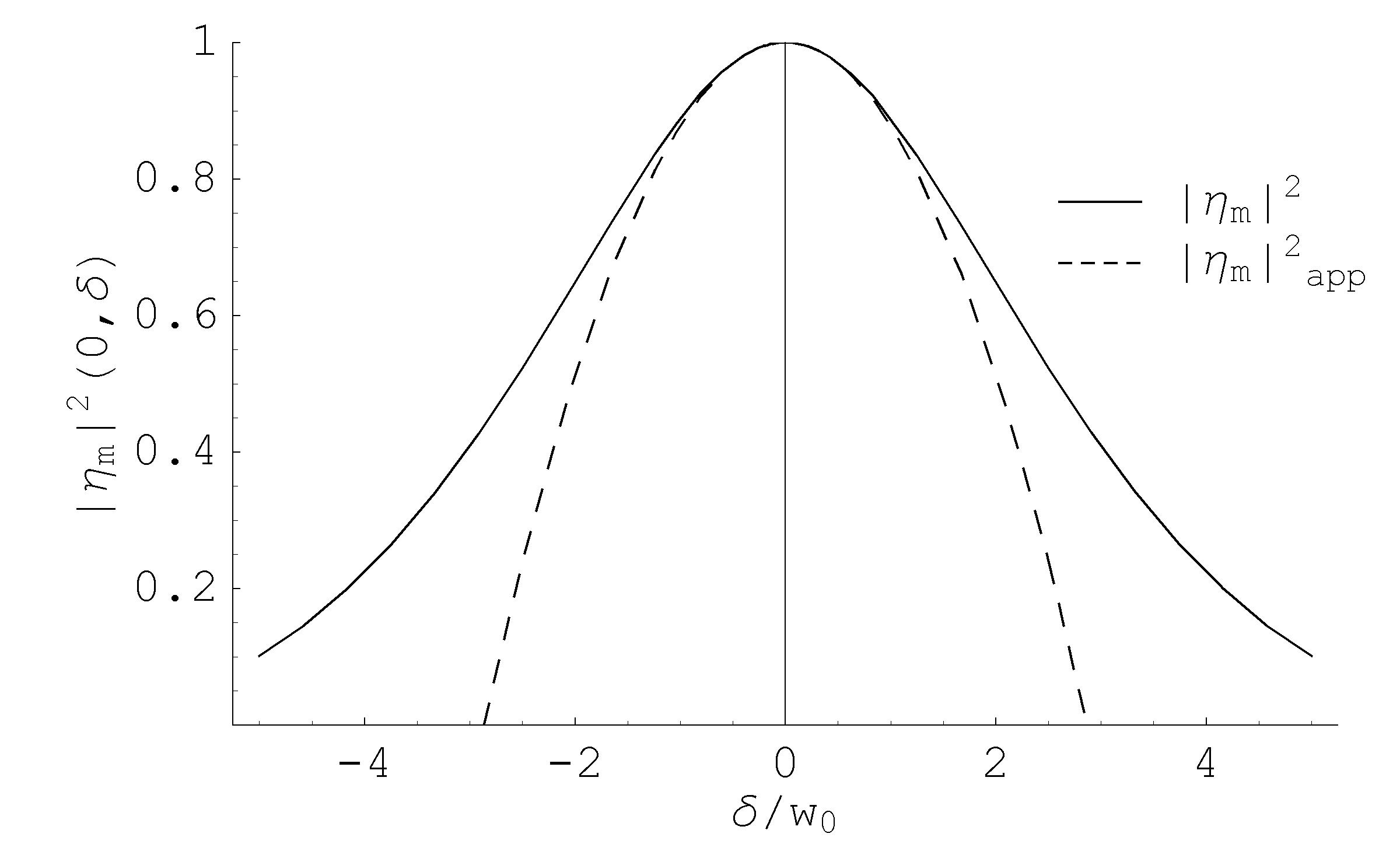}}
     \hspace{0.2in}
\subfigure[Angular shift.]{
      \label{fig:sudfig:MBFtilt}
      \includegraphics[width=0.7\textwidth]{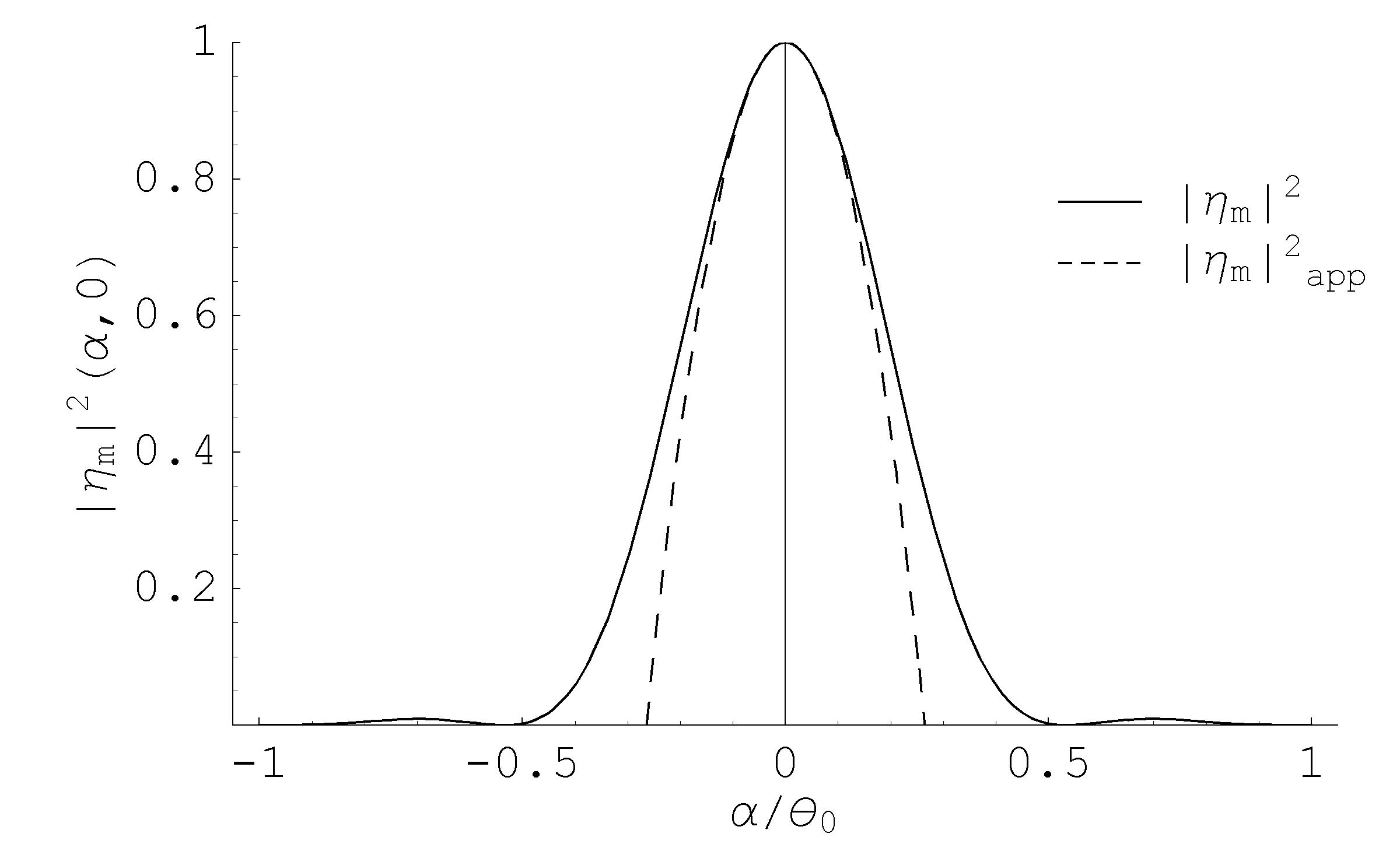}}
 \caption{Comparison between the exact and approximated expression of $|\eta_m|^2(\alpha,\delta)$ for FM beam with $b=4 w_0$.}
 \label{fig:eta}
 \end{figure}

\begin{figure}[htbp]
  \centering
  \subfigure[Divergence.]{
     \label{fig:subfig:MBdiver}
     \includegraphics[width=0.7\textwidth]{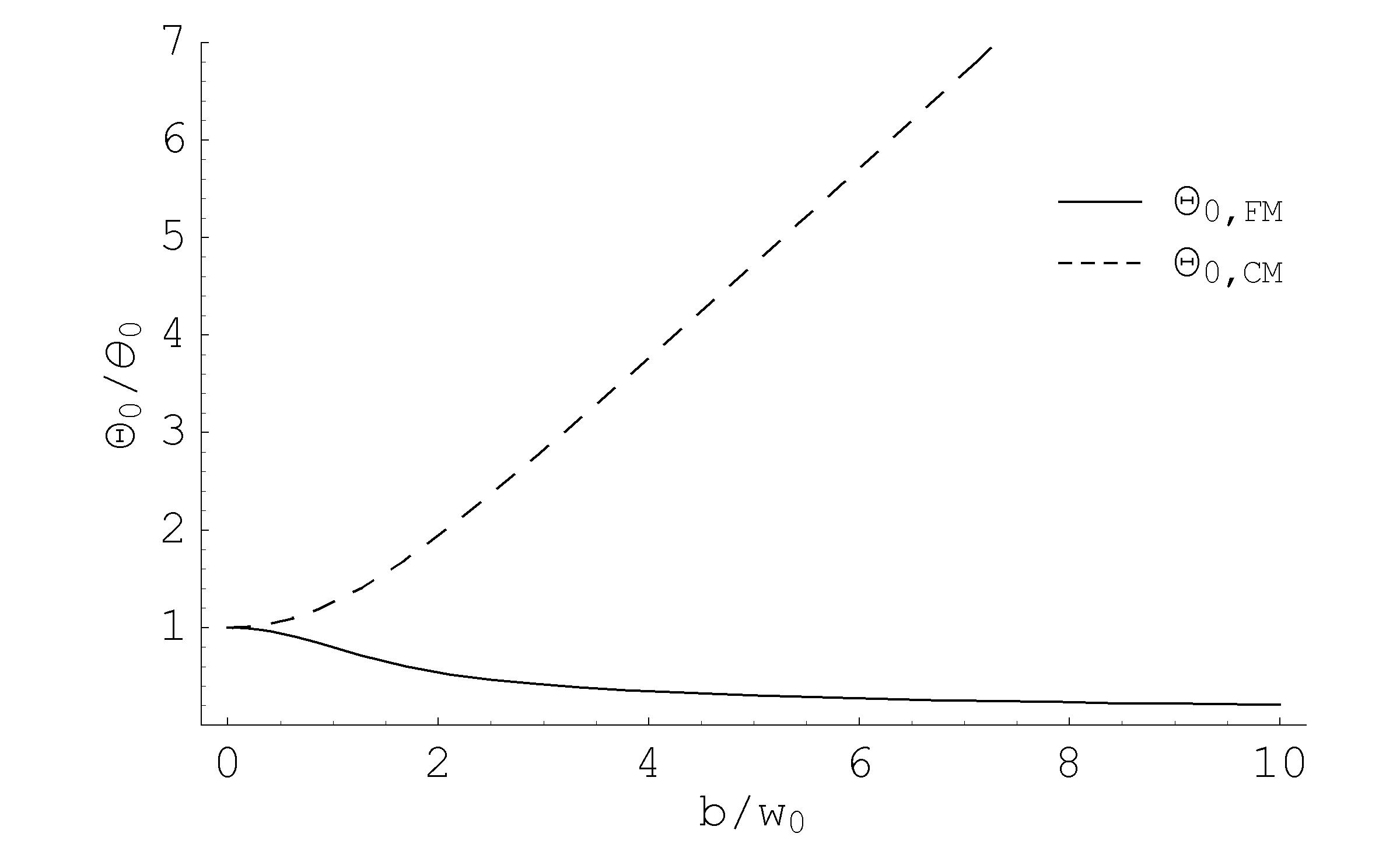}}
     \hspace{0.2in}
\subfigure[Waist width.]{
      \label{fig:sudfig:MBwaist}
      \includegraphics[width=0.7\textwidth]{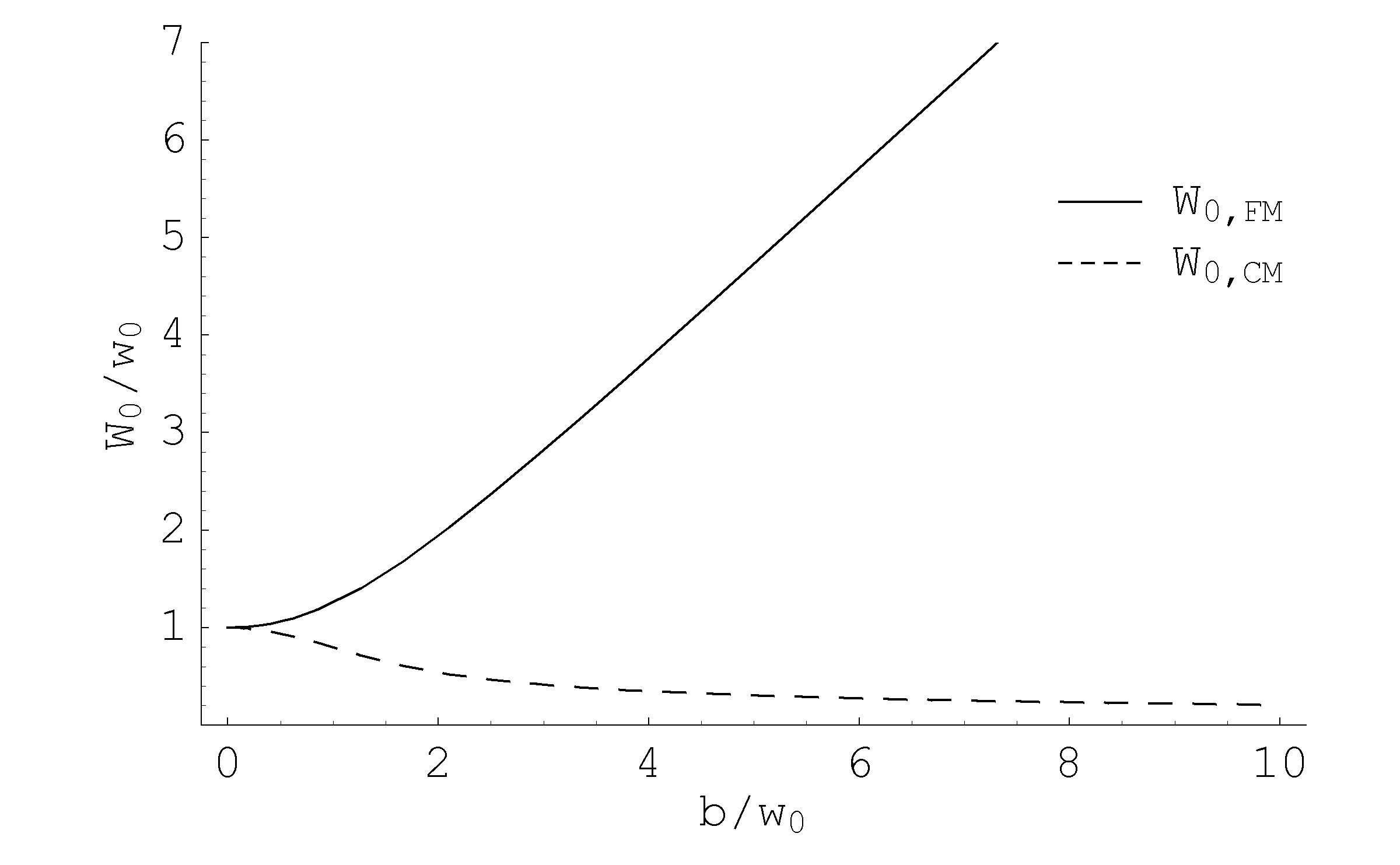}}
 \caption{The scaling to the divergence and waist of the minimal Gaussian reveals the duality relation between the FM and CM design. }
 \label{fig:MBdiver-waist}
 \end{figure}

We will therefore continue the misalignment analysis using the
approximated expression

\begin{equation}\label{mis-smallMB}
|\eta_m|^2= 1- (M^2)^2 \left (
\frac{\alpha^2}{\Theta_0^2}+\frac{\delta^2}{W_0^2}\right)
\end{equation}

where $\Theta_0$ and $W_0$ are the beam divergence and beam waist
respectively. This factor is trivially computable for Gaussian
beam.  To analyze the mesa beam misalignment sensitivity in both
configuration, FM and CM, we plot in Fig.~\ref{fig:MBdiver-waist}
the beam divergence and beam waist for different value of the
constituent parameter $b$.

\begin{figure}[htb]
\begin{center}
\includegraphics[width=0.8\textwidth]{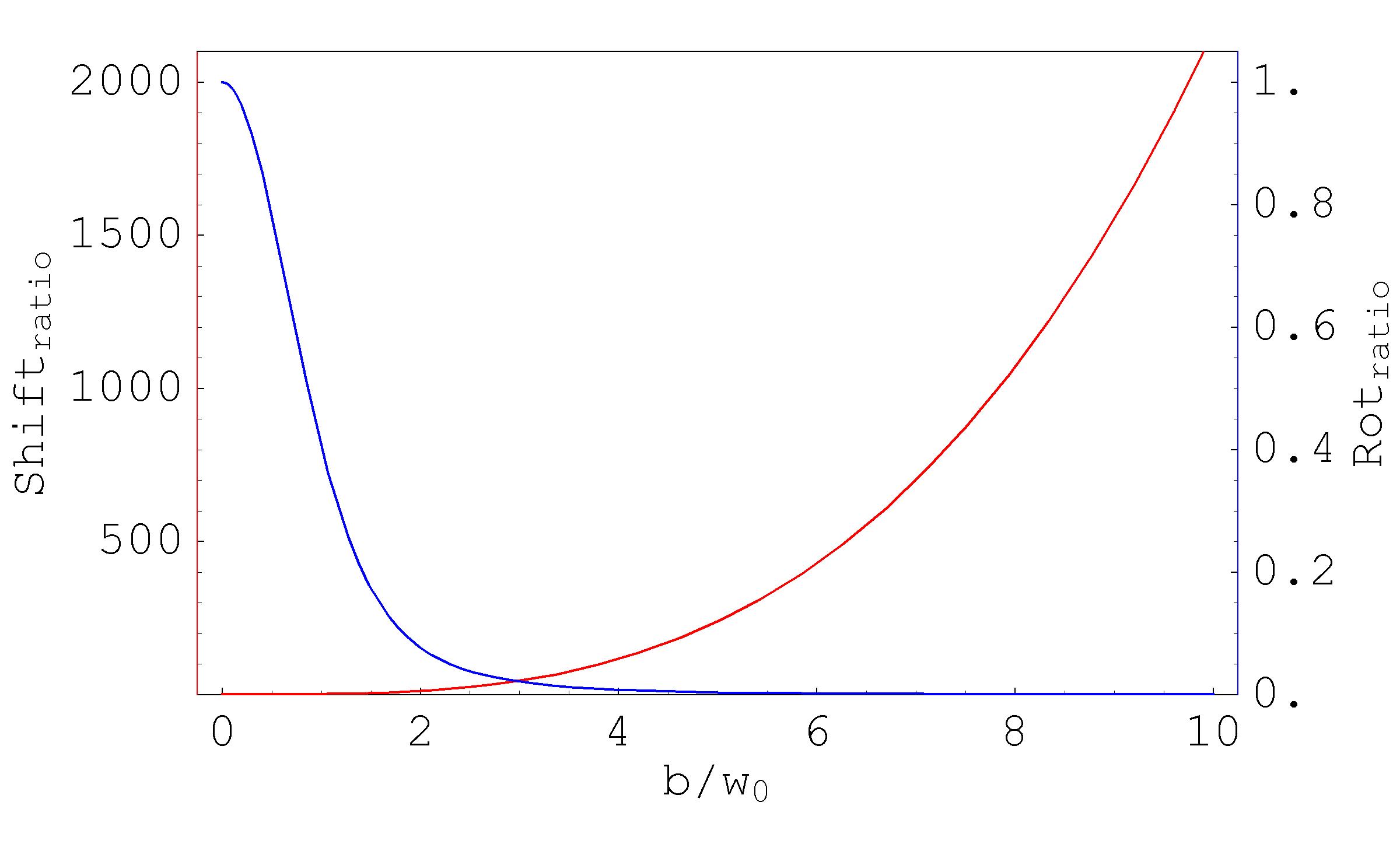}
\end{center}
\caption{Ratios between translation and rotation sensitivities for
FM and CM.} \label{fig:FMBCMBmisal}
\end{figure}

The misalignment sensitivity ratios between the FM and CM
configurations are represented in Fig.~\ref{fig:FMBCMBmisal}

\subsection{Gaussian beam coupling to FM and CM beams}

For the application to advanced GW interferometers FP cavities it
is of fundamental importance to find the best Gaussian injection
beam which has the maximum coupling with the Mesa beam resonating
in the Mexican hat cavities. This can be computed analytically
using the above formulas. Lets take the scalar product, at the
waist plane, between a gaussian field of waist $w_g$ and the FM
beam field \eqref{eq:MBF-waist}

\begin{equation}\label{CouplingFM}
    {\cal C}_{FM}=<\psi_G|\psi_{0,\mathbf{FM}}>=\sqrt{\frac{2}{\pi w_g^2}}\frac{4 \pi^2}{w_0^2 b \pi^{\frac{3}{2}} \sqrt{\Upsilon}}
    \int_0^{\infty}\,dr \int_0^b \, dr_0 e^{-\frac{r^2}{w_g^2}}\,
    e^{-\frac{r^2+r_0^2}{w_0^2}}\, I_0\!\!\left(\frac{2 r
    r_0}{w_0^2}\right)\, r \,r_0
\end{equation}

 After the integration  we have

\begin{equation}\label{CouplingFM-final}
 {\cal C}_{FM}=\frac{\sqrt{2}}{b \sqrt{\Upsilon}}w_g \left
 (1-e^{-\frac{b^2}{w_0^2+w_g^2}} \right)
\end{equation}

\begin{figure}[htb]
\begin{center}
\includegraphics[width=0.7\textwidth]{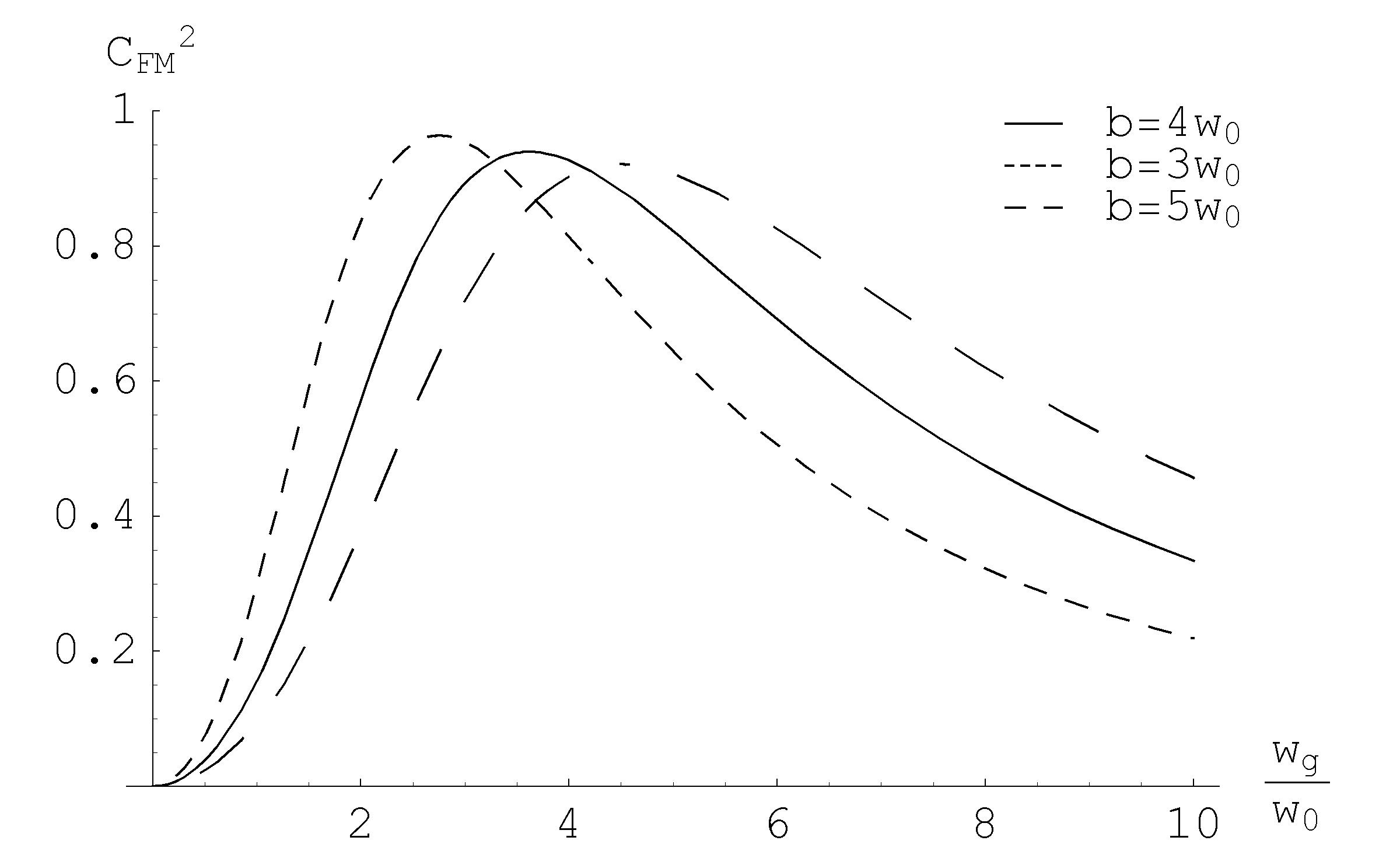}
\end{center}
\caption{Power coupling between a Gaussian beam and a FM beam.}
\label{fig:coupligFM}
\end{figure}

Fig.~\ref{fig:coupligFM} shows the power directly coupled, ${\cal
C}_{FM}^2$ from a gaussian beam into a FM beam. The maximum
coupling, which occurs for $w_g$ solution of

\begin{equation}\label{wgFM-max}
    1- e^{-\frac{b^2}{w_0^2+w_g^2}}\left(1+\frac{2 b^2
    w_g^2}{(w_0^2+w_g^2)^2}\right)=0
\end{equation}

in agreement with~\cite{erika}, can be quite large, i.e for $b=4
w_0$ we have a power coupling of $94\%$ with a gaussian beam of
$w_g\simeq 3.62 w_0$.

 We now proceed to the same calculation for
the CM beam, which seems the most promising design for advanced
interferometers

\begin{equation}\label{Coupling-CM}
     {\cal C}_{CM}=<\psi_G|\psi_{0,\mathbf{CM}}>=\sqrt{\frac{2}{\pi w_g^2}}\frac{2 \pi}{ \sqrt{\pi} \sqrt{\Upsilon}}
    \int_0^{\infty}\,dr  e^{-\frac{r^2}{w_g^2}}\,
    \frac{e^{-\frac{r^2}{w_0^2}}}{r}\, J_1\!\!\left(\frac{2 r
    b}{w_0^2}\right)\, r
\end{equation}

Carrying out the integration we have

\begin{equation}\label{Coupling-CMfinal}
{\cal C}_{CM}= \frac{\sqrt{2}}{\sqrt{\Upsilon}}\frac{w_0^2}{w_g
b}\left(1- e^{-\frac{b^2}{w_0^2+w_g^2}\frac{w_g^2}{w_0^2}}\right)
\end{equation}

The maximum coupling occurs for $wg$ solution of

\begin{equation}\label{maxCM}
      1- e^{-\frac{b^2 w_g^2}{(w_0^2+w_g^2})w_0^2}\left(1+\frac{2 b^2
    w_g^2}{(w_0^2+w_g^2)^2}\right)=0
\end{equation}

\begin{figure}[htb]
\begin{center}
\includegraphics[width=0.7\textwidth]{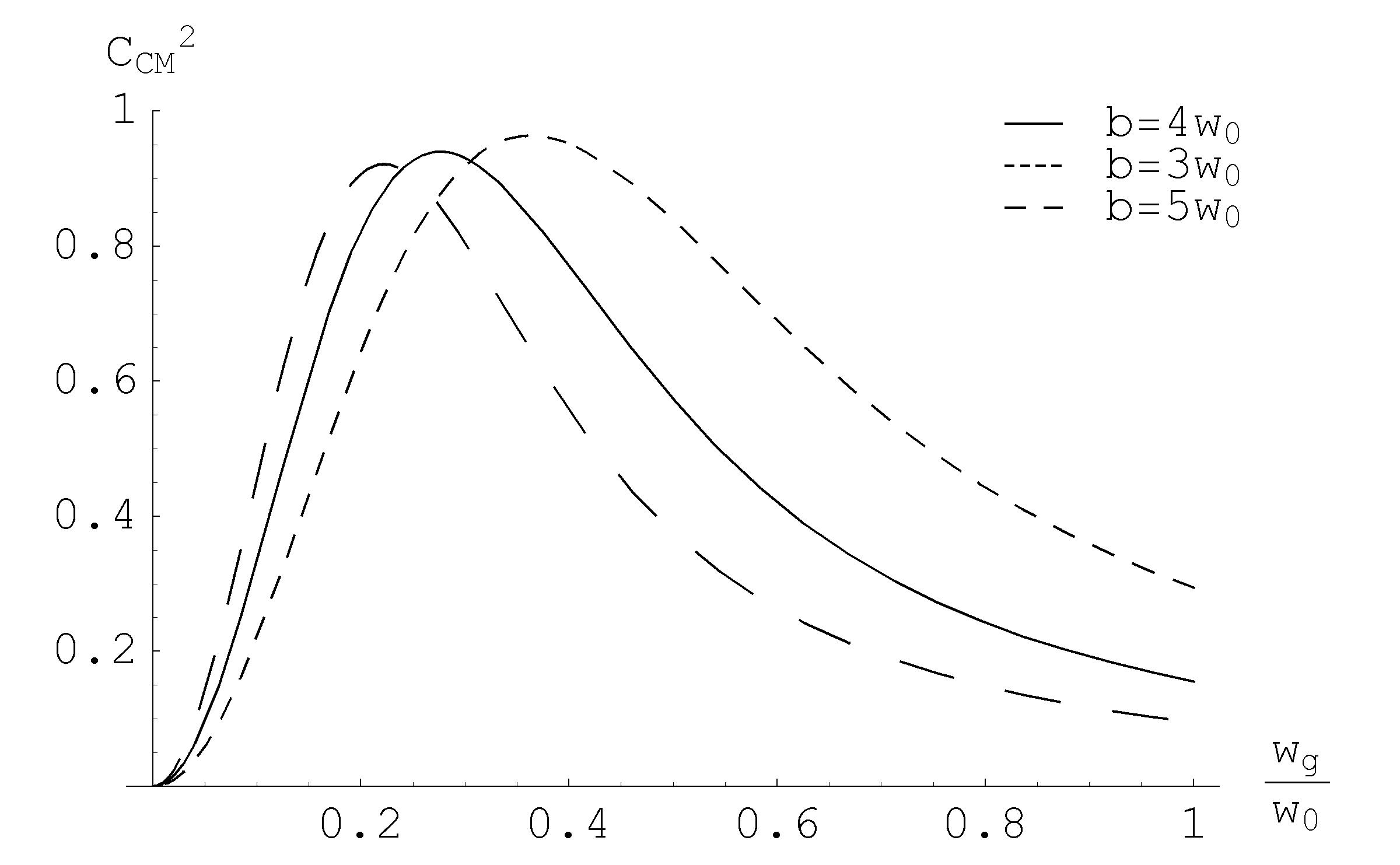}
\end{center}
\caption{Power coupling between a Gaussian beam and a CM beam.}
\label{fig:coupligCM}
\end{figure}

As shown in Fig.\ref{fig:coupligCM},for the proposed design $b=4
w_0$, a power coupling of $94\%$ occurs with $w_g\simeq 0.28 w_0$.

Therefore, direct coupling of a gaussian beam with a mesa beam can
be very efficient in both configurations.

\clearpage

\section{Application of the duality relation to real problems}

Spherical cavities are not optimal in terms of their thermal
noise: (the two types of) mesa beams, whose intensity profiles are
flatter given the same loss specification, turn out to provide
much lower thermal noises. For these beams, the larger the
parameter $b$, the lower the thermal noises, but the higher the
diffraction loss. The loss specification of Advanced LIGO
corresponds to $b=4 w_0$ which is the case we study in
Fig.~\ref{fig:comparison}. While having the same diffraction
losses and thermal noises, dual configurations do differ
significantly in a very important aspect: their eigenspectra are
different. Thus, any problem using modal analysis of optical
cavities will reveal these differences.

\begin{figure}[htb]
\begin{center}
\includegraphics[angle=90,width=0.9\textwidth]{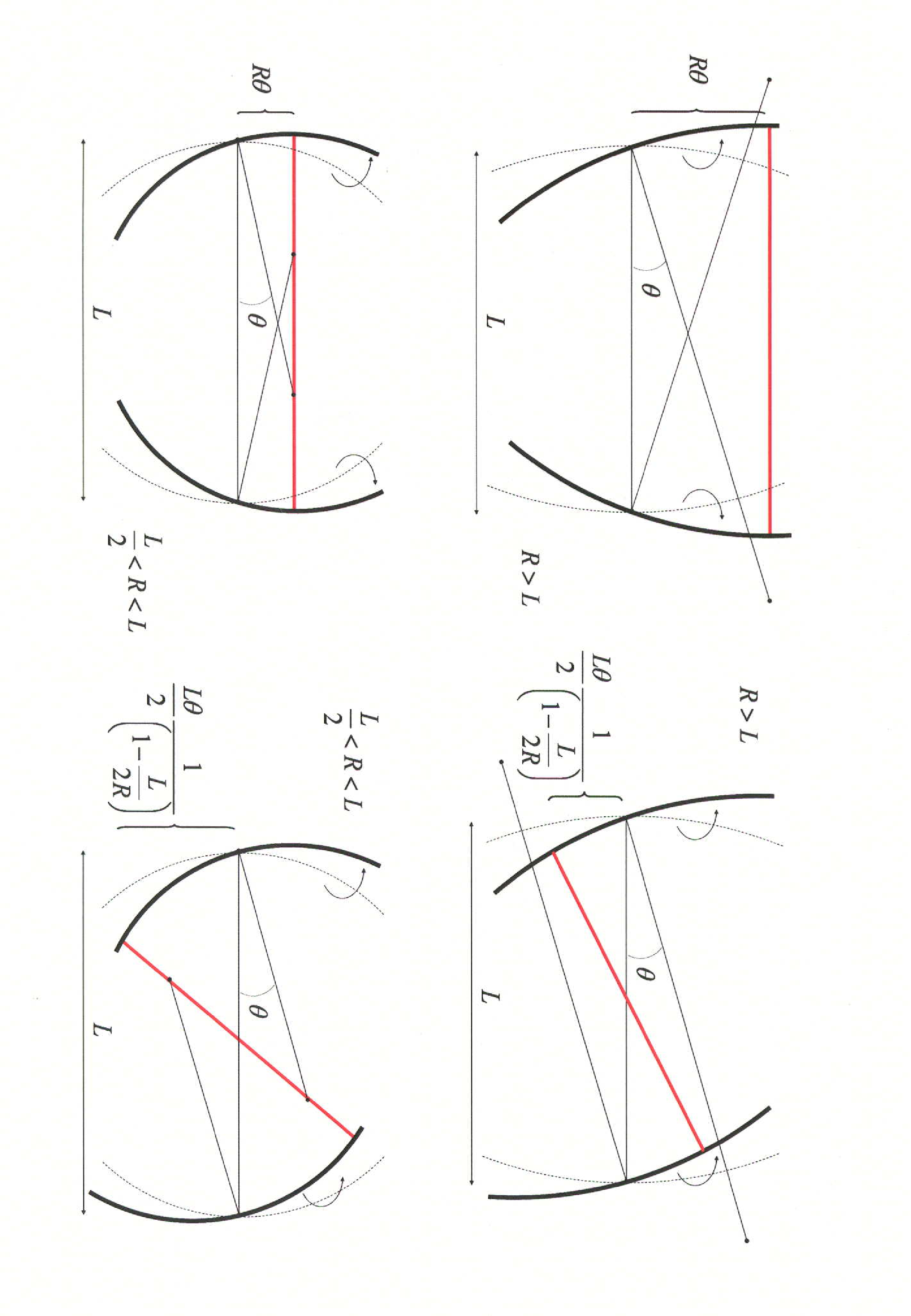}
\end{center}
\caption{Each of these optical resonators is formed by two
identical spherical mirrors, with radius of curvature much larger
(nearly flat cavity) and smaller (nearly concentric) then the
distance between the optics. When the mirrors are tilted in a
common mode (on the left) by an angle $\theta$ the optical axis
shifts by $R\theta$. The resulting torque is smaller in nearly
concentric cavities than in nearly flat resonators.  The torque
restores the cavity alignment when the mirrors are tilted in a
differential mode (on the right). The change in the optical axis
is a rotation $\theta/(1-\frac{L}{2 R})$ around the center of the
cavity. The resonating beam is displaced by
${\frac{L\theta}{2}}/(1-\frac{L}{2 R})$ on the optics. This
displacement occurs in opposite directions and tends to restore
the correct alignment. Since this rotation is larger in nearly
concentric cavities, these benefit more from the restoring torque.
Quantitative assessments can be found in literature and
substantiate these considerations.} \label{fig:tilt}
\end{figure}

It has been recently pointed out~\cite{sidles} that the force
induced by radiation pressure, when the end mirrors of an optical
resonator are misaligned in tilt, can induce a positive feedback
and further increase the tilt. This effect is obviously power
dependent and increases with power. In the case of anti-symmetric
tilt, the coupling between the laser radiation pressure and the
misalignment makes the mirrors tilt further, while in the
symmetric case the torque induced by the light beam impinging on
the mirrors, counteracts the tilt. Calculations show that the
antisymmetric tilt is much more critical, and is therefore
advantageous of reducing the effects of the antisymmetric case
even at the expense of worsening the symmetric case. Although
control systems are designed to keep the mirrors aligned, they
have limited authority, and minimized coupling between radiation
pressure and cavity misalignment is desirable, when there is
positive feedback. Quantitative assessments of the
problem~\cite{sidles} (that depends on the geometry of the
resonator) have shown that optical cavities with almost flat
mirrors are more prone to the antisymmetric instability with
respect to almost concentric cavities (where the comparison is
done between geometrical configurations that correspond to the
same transverse distribution of the fundamental mode on the
reflective surface of the mirrors). In Fig.~\ref{fig:tilt}  there
is a geometrical explanation of this phenomenon for the spherical
mirror case.

 Because of this issue, the current baseline design for Advanced LIGO has been
changed from nearly flat to nearly concentric~\cite{AdLIGORefDes}.

 It is important to
evaluate the radiation pressure problem  for the case of mesa
beams. We will use our proved duality-relation to show that, in
the quasi-concentric Mexican-Hat configuration, the cavity is less
prone to become unstable and therefore  easier to control
~\cite{poster}. For general, non-spherical cavities, a
perturbative approach  must be used to calculate the tilt
instability. The torque due to radiation pressure is expressed in
terms of numerically found eigenvalues and intensity profiles of
the cavities's spatial eigenmodes.
 Using the numerical work done for
the mesa beam in the nearly flat configuration, and using the
mapping of eigenvalues and eigenmodes between the nearly flat and
nearly concentric cavities (to obtain the estimation of the tilt
instability for nearly concentric Mexican-Hat cavities without
having to solve the eigenvalue problem again) we proved that, for
the antisymmetric case, the nearly concentric MH configuration is
much more stable than the nearly flat MH one, and even slightly
more stable than the corresponding nearly concentric spherical
configuration proposed for Advanced LIGO. The same results were
obtained by Savov and Viatchanin~\cite{Savov-Vyat} using a similar
modal analysis.

\subsection{Perturbation theory for finite size mirrors}

Since the propagation between finite mirrors is described by
non-unitary operator, Eq.\eqref{eq:eigen2} we can't apply the
standard perturbation theory, based on a set of power-orthogonal
modes.  We derived a rigorous procedure for perturbative analysis
in the finite mirror case,  which converges to the standard
approach in the limit of infinite mirrors. With this method we can
apply a perturbative approach even in those cases where the
diffraction losses are large and the optical modes are greatly
affected by the mirror finiteness. The drawback is that, since the
modes of an open resonator are the eigenmodes of a non-unitary
operator, it cannot be rigorously guaranteed that they form a
complete set. We simply make the assumption that these modes can
be used as a basis set and check this assumption numerically by
examining the convergence of the series expansion.

In general we have that for a non hermitian operators, a
biorthogonality relation holds between the eigenfunctions of the
operator $K=K(\vec r,\vec r_0)$, the transposed operator
$K^T=K(\vec r_0,\vec r)$ and the hermitian adjoint operator
$K^H=K^*(\vec r_0,\vec r)$. $K$,$K^T$ and $K^H$ will have separate
and different sets of eigensolutions fulfilling these relations

\bea && K \phi_n = \gamma_n \phi_n \qquad K^T \varphi_n =\kappa_n \varphi_n \qquad K^H \omega_n =\alpha _n \omega_n \nonumber \\
&& \alpha _n^* = \kappa_n = \gamma_n  \quad\mbox{and  }\varphi_n =
\omega_n^*\quad \nonumber \eea

but no simple relationship between eigenfunctions $\phi_n$ and
$\omega_n$ for non-hermitian operator ($K \neq K^H$)

Instead of being power-orthogonal ($\int \phi_n^*(\vec r)
\phi_m(\vec r) d\vec r \neq \delta_{nm}$) the eigenfunctions of
$K$ will be biorthogonal to the eigenfunctions of the
corresponding transpose (or hermitian adjoint) operator

\beq \int \phi_n(\vec r) \varphi_m(\vec r) d\vec r \equiv \int
\phi_n(\vec r) \omega_m^*(\vec r) d\vec r = \delta_{nm} \nonumber
\eeq

In the special case of simple optical cavities the kernel
\eqref{eq:eigen2} is always symmetric in $\vec r \leftrightarrow
\vec r '$ and this implies $\phi_n = \varphi_n$\footnote{This is
not true for more general types of resonator like ring resonator.
These cases are discusses in~\cite{siegman}. }.

In the following  we will use a compact formalism for hermitian
scalar product and usual integration and we will assume that the
eigenmodes are power-normalized even if not power-orthogonal.

\bea
\left\langle u |u \right\rangle = \int d \vec r u^*(\vec r) u(\vec r)= 1\\
\left[u |v \right] = \int d \vec r u(\vec r) v(\vec r)\\
\left[ u |G|v\right] = \int d \vec r d\vec r' u(\vec r')G(\vec
r,\vec r') v(\vec r) \eea

To keep light notation we will indicate simply with $n$ the set of
integers which characterize the eigenvalues and eigenvectors of
our problem (e.g. $p,m$ for unperturbed aligned cylindrically
symmetric cavity).

Assume that we solved the eigenvalues problem for the kernel
$K_0$, which corresponds to the unperturbed cavity

\beq K_0 \phi_n = \gamma_n^{(0)}\phi_n \eeq

and we want to study the effect of a small perturbation (e.g.
small mirror tilting) so that the resulting full kernel can be
written as \beq K = K_0 + K_1 \quad \mbox{ and the eigenvalues
equation as }\quad (K_0 + K_1)\psi_n = \delta_n \psi_n \eeq

 We'd like to solve this problem using a perturbation series and
to keep track of powers of the perturbation we will make the
substitution $K_1 \rightarrow \epsilon K_1$ where $\epsilon$ is
assumed to be a small parameter in which we are making the series
expansion of our eigenvalues and eigenvectors.

\bea &&\delta_n  = \gamma_n^{(0)} + \epsilon \gamma_n^{(1)} +\epsilon^2 \gamma_n^{(2)}+....\\
&& \psi_n = N(\epsilon) \Big( \phi_n + \sum_{k\neq n}c_{n k}(\epsilon) \phi_k \Big)\\
&& c_{n k}(\epsilon) = \epsilon c_{n k}^{(1)} + \epsilon^2 c_{n
k}^{(2)} +.... \eea

where the superscript $ (0),(1),(2) $ are the zeroth, first, and
second order terms in the series and $N$ is a normalization
factor.

Using these equations we can write

\beq  (K_0 + K_1)\Big( \phi_n + \sum_{k\neq n}c_{n k}(\epsilon)
\phi_k \Big)= (\gamma_n^{(0)} + \epsilon \gamma_n^{(1)}
+\epsilon^2 \gamma_n^{(2)}+....) \Big( \phi_n + \sum_{k\neq n}c_{n
k}(\epsilon) \phi_k \Big) \eeq For this equation to hold as we
vary $\epsilon$, it must hold for each power of $\epsilon$.
Looking at the first three terms
$(\epsilon^0,\epsilon^1,\epsilon^2)$

\bea &&K_0 \phi_n = \gamma_n^{(0)} \phi_n \\
&& K_1 \phi_n + K_0 \sum_{k\neq n}c_{n k}^{(1)}\phi_k = \gamma_n^{(1)} \phi_n + \gamma_n^{(0)}\sum_{k\neq n}c_{n k}^{(1)}\phi_k \nonumber \\
&& K_0 \sum_{k\neq n}c_{n k}^{(2)}\phi_k + K_1 \sum_{k\neq n}c_{n k}^{(1)}\phi_k = \gamma_n^{(0)}\sum_{k\neq n}c_{n k}^{(2)}\phi_k + \gamma _n^{(1)}\sum_{k\neq n}c_{n k}^{(1)}\phi_k +\gamma_n^{(2)}\phi_n \nonumber \\
\eea

The zero order term is the solution of the unperturbed problem and
so there is no new information there, but we can extract the first
and second order correction to eigenvalues and eigenvectors from
the other two terms using the orthonormality of the functions
$\phi_i$. Multiplying these equations by $\phi_n$ or $\phi_k$ and
integrating over the free coordinates we have after
straightforward calculations

\bea && \gamma_n^{(1)} = \frac{\left[ \phi_n | K_1 | \phi_n \right]}{\left[ \phi_n | \phi_n \right]} \nonumber \\
&&c_{nk}^{(1)} = \frac{\left[ \phi_k | K_1 | \phi_n \right]}{\gamma_n^{(0)}-\gamma_k^{(0)}}\frac{1}{\left[ \phi_k | \phi_k \right]} \nonumber \\
&& \gamma_n^{(2)} = \frac{1}{\left[ \phi_n | \phi_n \right]}\sum_{k\neq n}\frac{(\left[ \phi_k | K_1 | \phi_n \right])^2}{\gamma_n^{(0)}-\gamma_k^{(0)}} \nonumber\\
\eea Where the denominators take care of the normalization
properties for the eigenfunctions. The normalization factor
$N(\epsilon)$ played no role in the solutions to the eigenvalues
equation since that equation is independent of normalization. We
can easily derive the normalization factor for the first order
corrected eigenfunction .
\bea && \frac{1}{N(\epsilon)^2}=\left\langle  \phi_n + \sum_{k\neq n} \epsilon c_{n k}^{(1)} \phi_k | \phi_n + \sum_{k\neq n} \epsilon c_{n k}^{(1)} \phi_k  \right\rangle \\
&&= 1 + 2 Re \left(\sum_{k\neq n} \epsilon c_{n k}^{(1)} \left\langle  \phi_n|\phi_k  \right\rangle \right) + O(\epsilon^2) \nonumber \\
&& N(\epsilon)\approx 1 - Re\left(\sum_{k\neq n} \epsilon c_{n
k}^{(1)} \left\langle  \phi_n|\phi_k  \right\rangle \right)+
O(\epsilon^2) \nonumber \eea The correction is of order $\epsilon$
and can't  be neglected at this level of approximation.

These results are nearly formally the same as in quantum mechanics
but we have to keep in mind the different meaning of our notations
(the expression of the normalization constant is completely
different due to the fact that we don't have power-orthogonal
functions).

\subsection{Evaluating tilt instability}\label{sec:TiltInst}

The case of antisymmetric mirror tilt, as indicated above, is the
possible source of laser-power driven instability.

The mechanical torque exerted by the misaligned fundamental mode
on the mirror is given by

\begin{equation}\label{Torque}
    T=\frac{2 P}{c}\int_{S}r \cos(\varphi)
    |\psi_{00}(\myvec{r})|^2 d\myvec{r}^2
\end{equation}

For a symmetric spherical mirror cavity, the torque can be
computed, at the leading order, using very simple ray-beam
physics. The pure geometric contribution is given by $T= 2 P/c\,
\delta x$ where $\delta x$ is the displacement of the optical axis
(see Fig.~\ref{fig:tilt} given by $L \theta /(1-g)$. This very
quick result is completely in agreement
with~\cite{sidles,Savov-Vyat}.

 In the special
case of mirror's antisymmetrical tilt by a small angle $\theta$ we
have that the full kernel describing the propagation from one
mirror's surface to the other is

\beq K^{tilt}_{flat-flat}(\vec r,\vec r')=\frac{i k}{2\pi L}e^{
 ik\theta r \cos(\varphi) + i k h(r)-\frac{ik}{2L}|\vec r-\vec r'|^2+i k h(r')+ik\theta r'\cos(\varphi') }
\eeq

If $ \theta \ll \frac{1}{k a }=\frac{\lambda}{2\pi a }$ we can
write this kernel as \beq  K^{tilt}_{flat-flat}(\vec r,\vec r')=
K_0(\vec r,\vec r')\big( 1+ik\theta r'\cos(\varphi') +ik\theta r
\cos(\varphi) \big) + O(\theta^2) \eeq and we can identify \beq
K_1(\vec r,\vec r')= K_0(\vec r,\vec r')ik\theta \big(
r'\cos(\varphi') + r \cos(\varphi) \big) \eeq

We now calculate the lowest order correction to the fundamental
eigenmode and eigenvalue.

\bea && \gamma_0^{(1)} \propto \left[ \phi_0 | K_1 | \phi_0 \right] = \left[ \phi_0(\vec r) | K_0(\vec r,\vec r')ik\theta \big(r'\cos(\varphi') + r \cos(\varphi) \big)| \phi_0(\vec r') \right] \nonumber \\
&& =\left[ \phi_0(\vec r) | K_0(\vec r,\vec r')ik\theta r'\cos(\varphi')| \phi_0(\vec r') \right]  +\left[ \phi_0(\vec r) | K_0(\vec r,\vec r')ik\theta r cos(\varphi)| \phi_0(\vec r') \right]  \nonumber \\
&& = 2 \gamma_0^{(0)} i k \theta \left[ \phi_0(\vec r) | r
\cos(\varphi)| \phi_0(\vec r) \right] = 0 \quad \mbox{since}\quad
\phi_0(\vec r)= \phi_0(r)\nonumber \eea so the first order
correction to the lowest eigenvalue is zero and before calculating
the second order correction let us calculate the first order
coupling coefficient $c_{0k}^{(1)}$

\bea && c_{0k}^{(1)} = \frac{\left[ \phi_k | K_1 | \phi_0 \right]}{\gamma_0^{(0)}-\gamma_k^{(0)}}\frac{1}{\left[ \phi_k | \phi_k \right]} = \nonumber \\
&&\frac{\left[ \phi_k(\vec r) | K_0(\vec r,\vec r')ik\theta  r'cos(\varphi')| \phi_0(\vec r') \right]  +\left[ \phi_k(\vec r) | K_0(\vec r,\vec r')ik\theta r \cos(\varphi)| \phi_0(\vec r') \right] }{\gamma_0^{(0)}-\gamma_k^{(0)}\qquad\left[ \phi_k (\vec r)| \phi_k (\vec r) \right]} \nonumber \\
&&= i k \theta \frac{\gamma_k^{(0)}+
\gamma_0^{(0)}}{\gamma_0^{(0)}-\gamma_k^{(0)}} \frac{\left[
\phi_k(\vec r) | r \cos(\varphi)| \phi_0(\vec r) \right]}{\left[
\phi_k (\vec r)| \phi_k (\vec r) \right]} \nonumber \eea

Remembering the form of the unperturbed eigenfunctions $
\phi_{pm}(\vec r) = R_{pm}(r)\cos(m\varphi)$ we have that the
possible non zero terms involve only $\phi_{p 1}$(the radial index
may vary) and the result is

\beq  c_{0,p1}^{(1)}= i k \theta \frac{\gamma_{p1}^{(0)}+
\gamma_{00}^{(0)}}{\gamma_{00}^{(0)}-\gamma_{p1}^{(0)}}
\frac{\left[ \phi_{p1}(\vec r)| r \cos(\varphi)| \phi_{00}(\vec r)
\right]}{\left[ \phi_{p1} (\vec r)| \phi_{p1} (\vec r) \right]}
\eeq

Since the same kind of integral enters in the definition of
$\gamma_0^{(2)}$ we can easily see that the sum over $k\neq 0$
reduces to just one contribution

\beq \gamma_{00}^{(2)}= - k^2 \theta^2\frac{(\gamma_{p1}^{(0)}+
\gamma_{00}^{(0)})^2}{\gamma_{00}^{(0)}-\gamma_{p1}^{(0)}}
\frac{(\left[ \phi_{p1}(\vec r)| r \cos(\varphi)| \phi_{00}(\vec
r) \right] )^2}{\left[ \phi_{00} (\vec r)| \phi_{00} (\vec r)
\right]} \eeq

We now calculate the first order corrections in the $p=0$ case
(lowest radial eigenfunctions) so that the eigenfunctions entering
in the integrations are of the form

\beq  \phi_{00}(\vec r) = R_{00}(r) \qquad \phi_{01}(\vec r) =
R_{01}(r)\cos(\varphi)\nonumber \eeq

From Eq.~\ref{Torque} and the above results we have the dominant
contribution is given by

\begin{equation}\label{Torque2}
    T \simeq \frac{2 P}{c} 2 \Re\left( c_{0,01}^{(1)} \langle \phi_{00}| r \cos(\varphi) |\phi_{01} \rangle \right)
\end{equation}

This can be written as explicit function of the eigenvalues and
radial eigenfunctions

\begin{equation}\label{Torque3}
    T \simeq \frac{2 P}{c} 2 \Re\left(i k \theta \frac{\gamma_{01}+\gamma_{00}}{\gamma_{00}-\gamma_{01}}\frac{\int_0^a
dr \,r\, R_{01}(r) r R_{00}(r) }{\int_0^a dr \,r\, R_{01}(r)
R_{01}(r)}\, \pi\,\int_0^a dr \,r\, R^*_{01}(r) r R_{00}(r)\right)
\equiv \frac{2 P}{c} \alpha \, \theta
\end{equation}

\begin{figure}[htb]
\begin{center}
\includegraphics[width= 0.9 \textwidth]{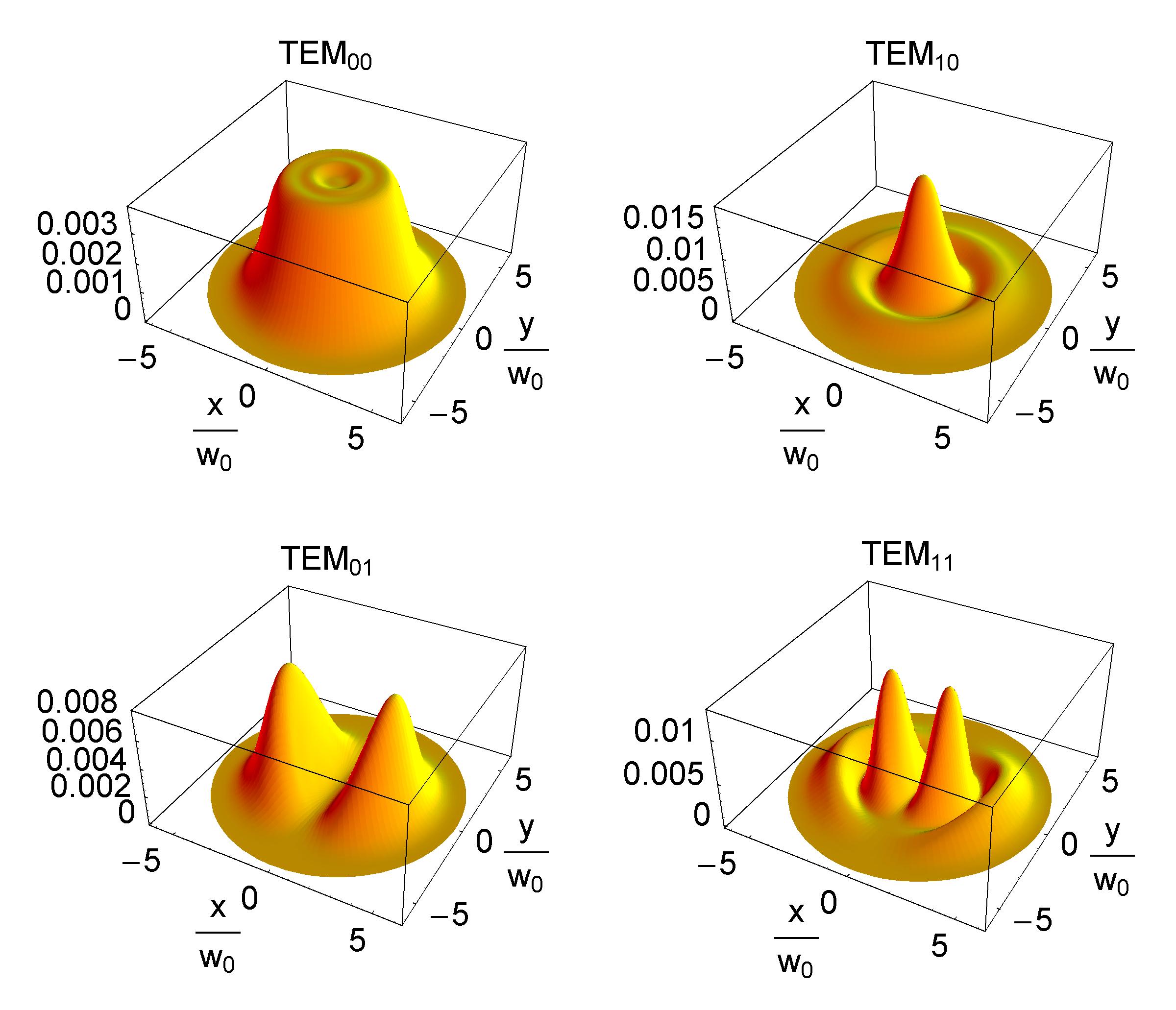}
\end{center}
\caption{First Mexican Hat cavity modes (power distributions in
a.u.): they replace the Gauss-Laguerre modes of the spherical
cavities.} \label{fig:modi2}
\end{figure}

 From our numerical calculation for the solution of the
eigenvalues equation using the algorithm explained in
Sec.~\ref{sec:FEM} follows these results: Fig.~\ref{fig:modi2}
shows the intensity profiles of the lowest order modes of a
symmetric Mexican hat symmetric cavity (both FM and CM).
Tab.~\ref{tab:eig-FMH} shows an example of eigenvalue calculation
for a Mexican hat mirrors cavity like Ad-LIGO proposal ($a=15.7
cm, b= 4\cdot w_0$)

\begin{table}[htb]
\begin{center}
\begin{tabular}{||p{2cm}||*{2}{c|}|}
\hline
  $\gamma_{pm}$\, FMH    & $ m=0$      & $  m=1$\\
\hline \hline
 $p=0$   &$ 0.759684-0.650267 i $ & $0.835831-0.548805 i$ \\
\hline
 $p=1$   &$ 0.979863-0.199425 i  $& $0.984052+0.175851 i$ \\
\hline
\end{tabular}
\end{center}
\caption{First eigenvalues for a nearly flat Mexican Hat cavity
with $L=4 Km$, $a=15.7 cm$, $\lambda=1.064 \mu m$ and $b=4 w_0$.}
\label{tab:eig-FMH}
\end{table}

Using our equivalence relation we can immediately write the
eigenvalues and eigenvectors in the nearly concentric case

\beq \gamma_{p m}^{CMH}=(-1)^{m+1}(\gamma_{p m}^{FMH})^* e^{-2ikL}
\nonumber \eeq

\begin{equation}\label{R-CMH}
    R^{CMH}_{pm}(r) = \left[ R^{FMH}_{pm}(r)   \right]^*
\end{equation}

and the numerical calculations for the overlapping integrals are
performed using the eigenfunctions of the FMH configuration
(Fig~\ref{fig:modi2}).

 We
calculate the ratio between the coupling coefficients of the
nearly-flat and nearly-concentric configurations

\beq  \frac{\alpha_{CMH}}{\alpha_{FMH}}= \approx \frac{1}{247}
\nonumber \eeq

which means that the MH-concentric configuration is about 250
times less sensitive to antisymmetrical-tilt induced torque  than
the MH-flat configuration.

\begin{figure}[htb]
\begin{center}
\includegraphics[width= 0.9 \textwidth]{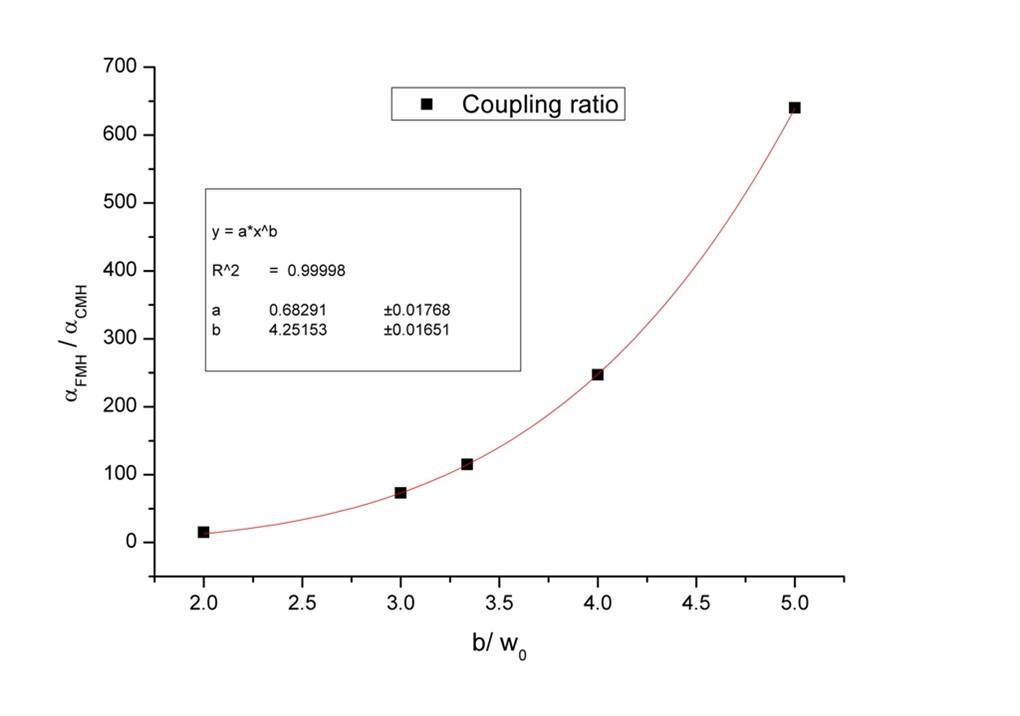}
\end{center}
\caption{Radiation pressure torque sensitivity ratios of flat
Mexican hat (FMH) to concentric Mexican hat (CMH) mirrors cavity.}
\label{fig:FM-CMcoupling}
\end{figure}

In Fig.~\ref{fig:FM-CMcoupling} we investigate the sensitivity of
different mesa beam design with respect to the baseline Advanced
LIGO configuration.

As we will see in Sec.~\ref{sec:TNMBG}, increasing the mesa beam
radius has the advantage of reducing the mirror thermal noise. In
particular the configuration with $b \approx 4.65 w_0$ should be
the optimal choice for Ad-LIGO in terms of thermal noise, and this
correspond to a factor of $470$ between the torques in the FM and
CM configuration.

\clearpage

\section{Mexican Hat cavity experiment}\label{sec:MHprot}

This work is a collaboration with  M. Tarallo , J. Miller, R.
DeSalvo, E. D'Ambrosio, P. Willems, B. Simoni, J.M. Mackowski and
A. Remillieux, colleagues at University of Pisa, the LIGO
Laboratory at Caltech and LMA (Laboratoire des Mat\'{e}riaux
Avanc\'{e}s) in Lyon. The main motivation of this
project~\cite{SPIEmio,mesabeam} is to demonstrate the feasibility
of building  a Fabry-Perot cavity with Mexican Hat mirrors, which
can support a Mesa beam.

 The aim of this experiment is to explore the main properties of
a single optical cavity before any eventual use in a second
generation gravitational-wave interferometer. In particular, we
are interested in studying the experimental mesa field achievable
with realistically imperfect mirrors and how its behavior differs
from that of a gaussian field with respect to perturbations such
as cavity misalignments. Other groups have demonstrated mesa beam
cavities using deformable mirrors~\cite{beyersdorf, derosa}, but
such mirrors are not obviously usable in low-noise gravitational
wave interferometry.

The  MH mirror production technique sets the main constraint to
the prototype geometry.

The production of the Mexican hat mirror has been undertaken in
the Laboratoire des Mat\'{e}riaux Avanc\'{e}s in Lyone (LMA). They
use a three steps deposition process over a micro-polished flat
substrate: the general shape coating, the corrective coating and
the multi-layer coating. In the first step a ``rough'' Mexican hat
shape is deposited with a precision of about $60$ nm using a
profiled mask and rotating the substrate to generate the
cylindrical symmetry. The mask, calculated from the thickness
profile of the ideal Mexican hat, is placed between the sputtered
flow of silica and the rotating substrate.

The second step is a more precise correction of the ``general
shape'' previously obtained. This method controls the deposited
profile with a precision of about $10$ nm Peak-to-Valley (PV).
Nevertheless, it is not possible to coat more than $100$ nm with
this technique, because it would require a deposition time which
is too long.
  The measurement of
  the achieved mirror shape after the first deposition is performed using an interferometric technique.
   The comparison between the achieved and the desired mirror shape generates
   a data file which is used to move the robot arm that positions the mirror
    in front of the corrective silica beam. The main limitations of the
     corrective technique come from the measurement of the wavefront,
     the precision of the robot arm movement and  the size and resolution
     of the ${\rm SiO}_2$ corrective beam. The maximum achievable slope
      is $500$ nm/mm, thus setting a limit on the smallest feasible Mexican hat mirror.
       Finally, a high reflectivity ${\rm SiO}_2/{\rm Ta}_2{\rm O}_5$ multi-layer
        coating is deposited on the corrected substrate.
The slope of the Mexican Hat mirror profile ~\ref{fig:subfig:mirr}
sets the radius of the smallest feasible mesa beam made using our
technique to about $6$ mm.

Our smallest practical mirror size sets our cavity length to $\sim
16$ meters. We further reduced the physical length of the
structure to $\sim 8$ m by building a half-symmetric cavity (a
single MH mirror paired with a flat mirror at what would be the
midpoint of a full length cavity) and then to $\sim 4$ m by
folding. In this way it was possible to build a rigid suspended
cavity.

\begin{figure}[htbp]
\subfigure[Picture of the apparatus.]{
     \label{fig:cavity}
     \begin{minipage}[b]{0.7\textwidth}
     \centering
     \includegraphics[width=4in]{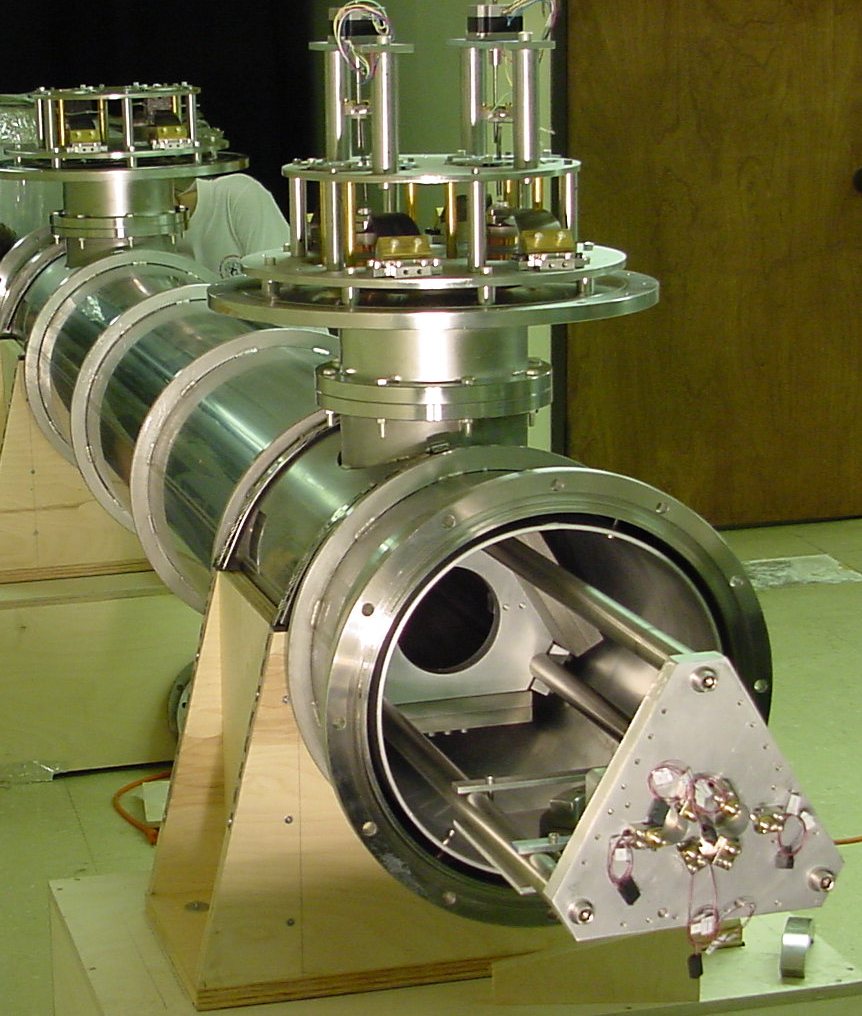}
     \end{minipage}}
  \subfigure[Schematic diagram of the cavity prototype]{
    \label{fig:diagprot}
    \begin{minipage}[b]{0.8\textwidth}
      \includegraphics[width=5in]{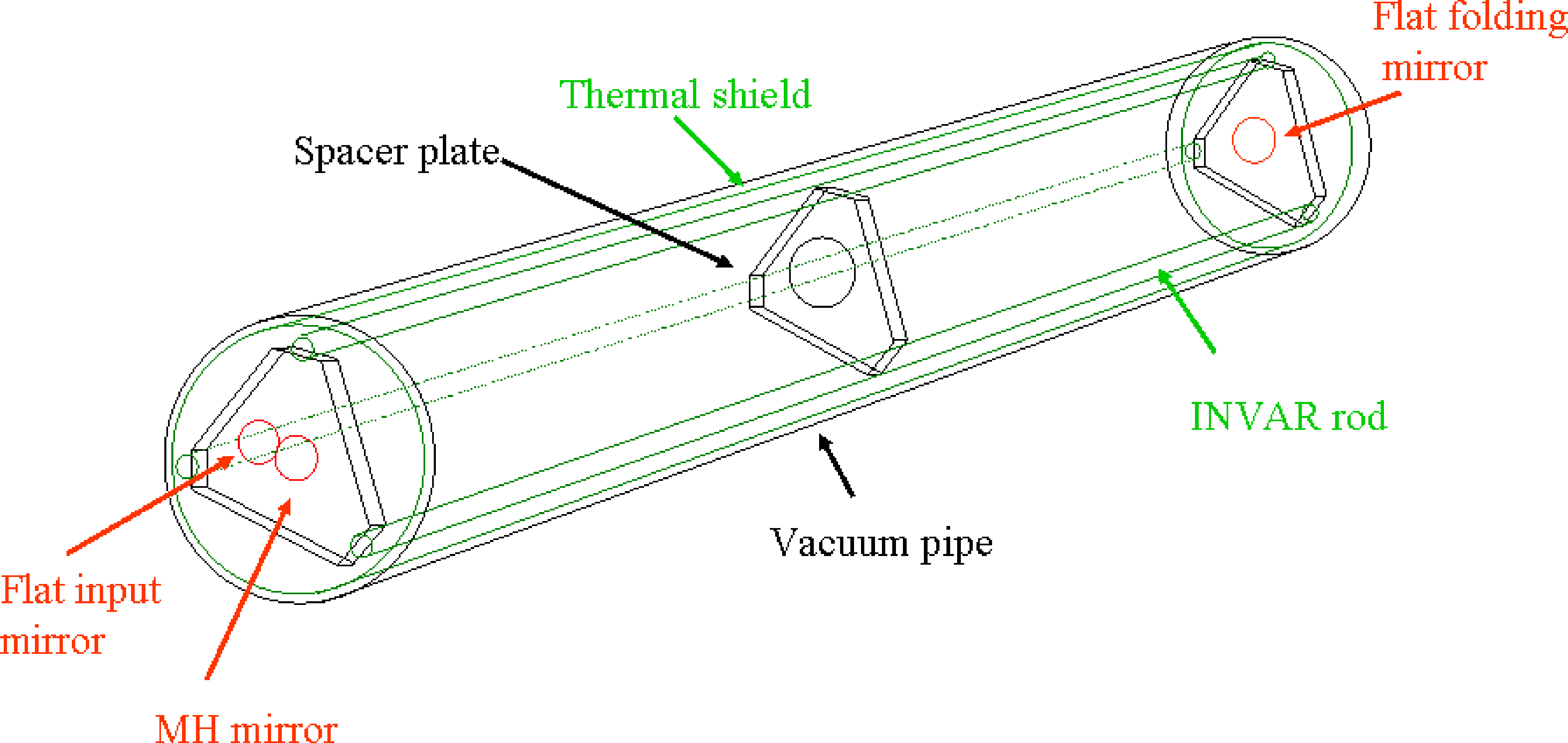}
      \end{minipage}}
 \caption{Mesa beam cavity prototype.}
 \label{fig:prot}
 \end{figure}

Fig.~\ref{fig:cavity} shows the suspended cavity. Three Invar rods
fix the cavity length to $L_{prototype}$= 2$\times $3.657 m = 7.32
m, with a folding mirror at one end of the structure and the input
and end mirrors on the other end. Five triangular spacers maintain
structural rigidity, with the outer two spacers bolted at the ends
of the structure and containing the mirror mounts. The cavity is
suspended by two pairs of maraging steel wires from GAS
(Geometric-Anti-Spring)~\cite{cella} blades, providing both
horizontal and vertical isolation. The whole is suspended in an
aluminum chamber for thermal stability and protection from air
currents; this chamber is not evacuated.

The test MH mirror was designed using the waist size of the
minimal gaussian with $L=2L_{prototype}$ as a reference length, so
that the resulting mesa beam had a radius $r_{mesa}=6.30$ mm. We
required that our test mirror have similar diffraction loss around
its aperture as an Advanced LIGO test mass; in order to have 1 ppm
diffraction loss the mirror radius was set to $a=13$ mm.

\begin{figure}[htb]\label{fig:mir5008central}
\begin{center}
\includegraphics[width=0.8\textwidth]{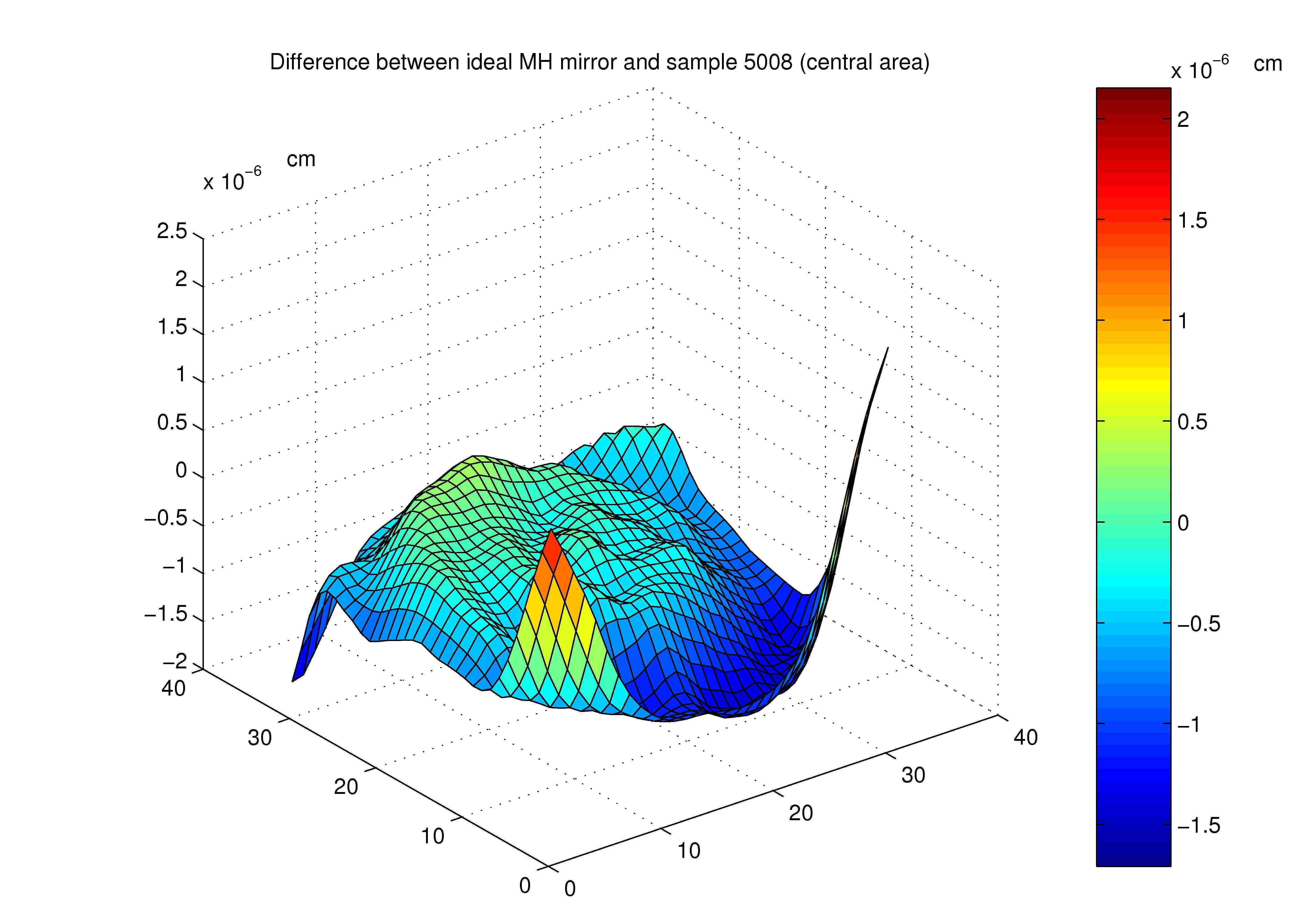}
\end{center}
\caption{Deviation from idea MH shape in the central mirror area.
x-y pixel size is $0.35$ mm.}
\end{figure}

Due to the technical difficulties of the MH figure deposition on
the flat substrate, the MH mirror had non-negligible figure error
[see Fig.~\ref{fig:mir5008central}], mostly in the central bump
where the height is just $27$ nm. In particular, the figure error
reaches a maximum of $5$ nm at the edge of the central bump.

Further details on the experimental set-up and the development of
the project are given in the MS thesis of Simoni~\cite{SimoniTesi}
and Tarallo~\cite{TaralloTesi}.

\subsection{Comparison between experiment and simulations}

In order to simulate our FP cavity we used a dedicated version of
the FFT implementation described in Sec.~\ref{sec:FFT}. Its aim is
to investigate the impact of misalignments and mode matching with
the input Gaussian beam driving the FP cavity and to study the
cavity behavior with imperfect mirrors using real mirror maps for
setting requirements and tolerance for Mexican hat mirror
manufacture and control constraints. From the mirror's maps we can
make theoretical prediction of the sensitivity of this new type of
optical cavity to mirror imperfections and alignment. Since the
geometry of the Mexican hat mirror of this FP cavity is completely
different from the standard spherical mirrors supporting Gaussian
beams, the effects of misalignment for mesa beams are more
important. In the Mexican hat case, the change in the surface of
the mirror due to orientation has a more general effect: not only
does the cavity have a new optical axis but the phase profile
sensed by the beam is quite different. For spherical mirrors this
is not the case since an incident beam experiences the same
curvature at all points on the mirror surface.

\begin{figure}[htbp]
  \centering
  \subfigure[Before intrinsic tilt correction.]{
     \label{fig:5008}
     \includegraphics[width=0.45\textwidth]{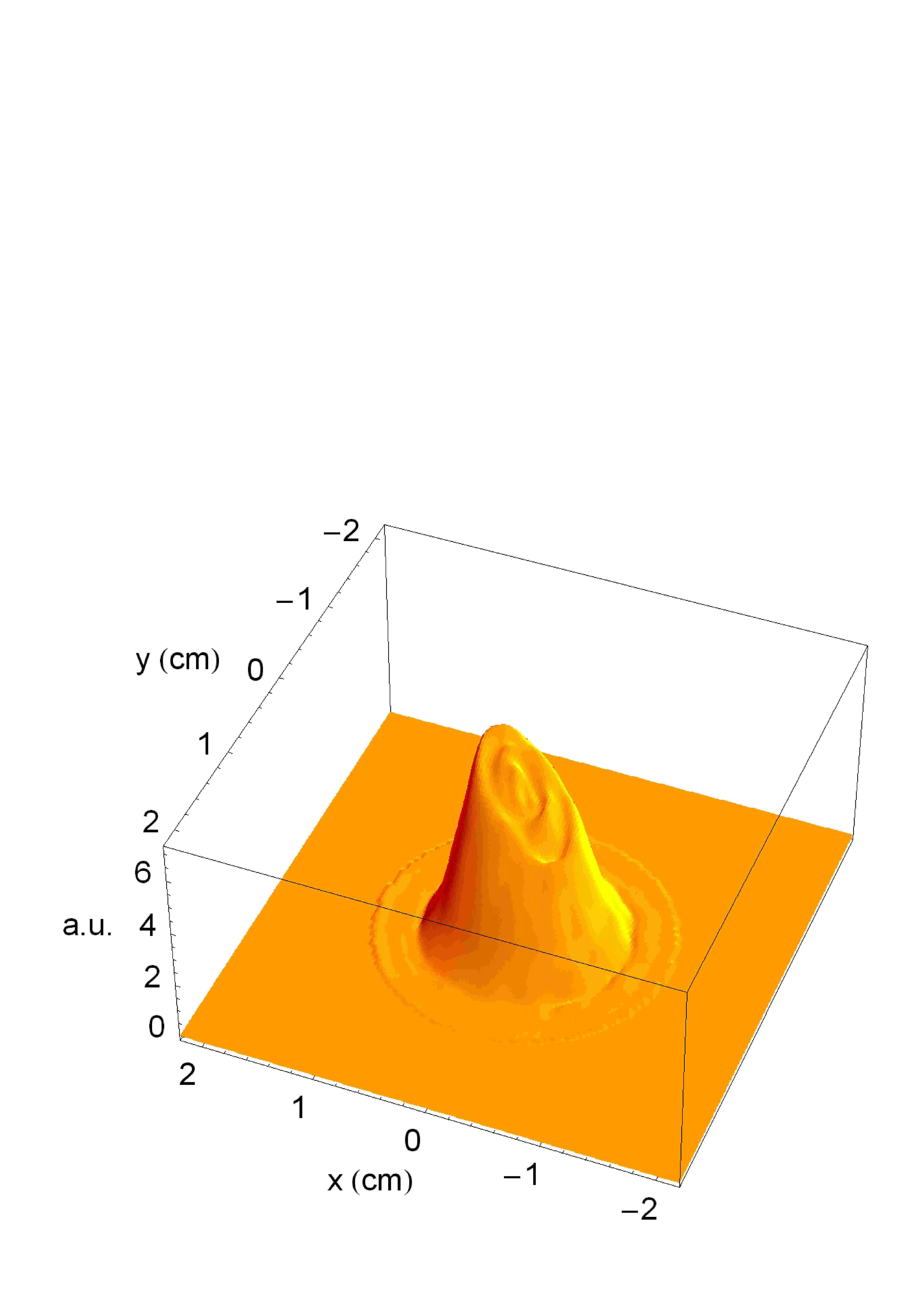}}
     \hspace{0.1in}
\subfigure[After mirror tilt of $1\,\mathrm{\mu rad}$.]{
      \label{fig:5008tilted}
      \includegraphics[width=0.45\textwidth]{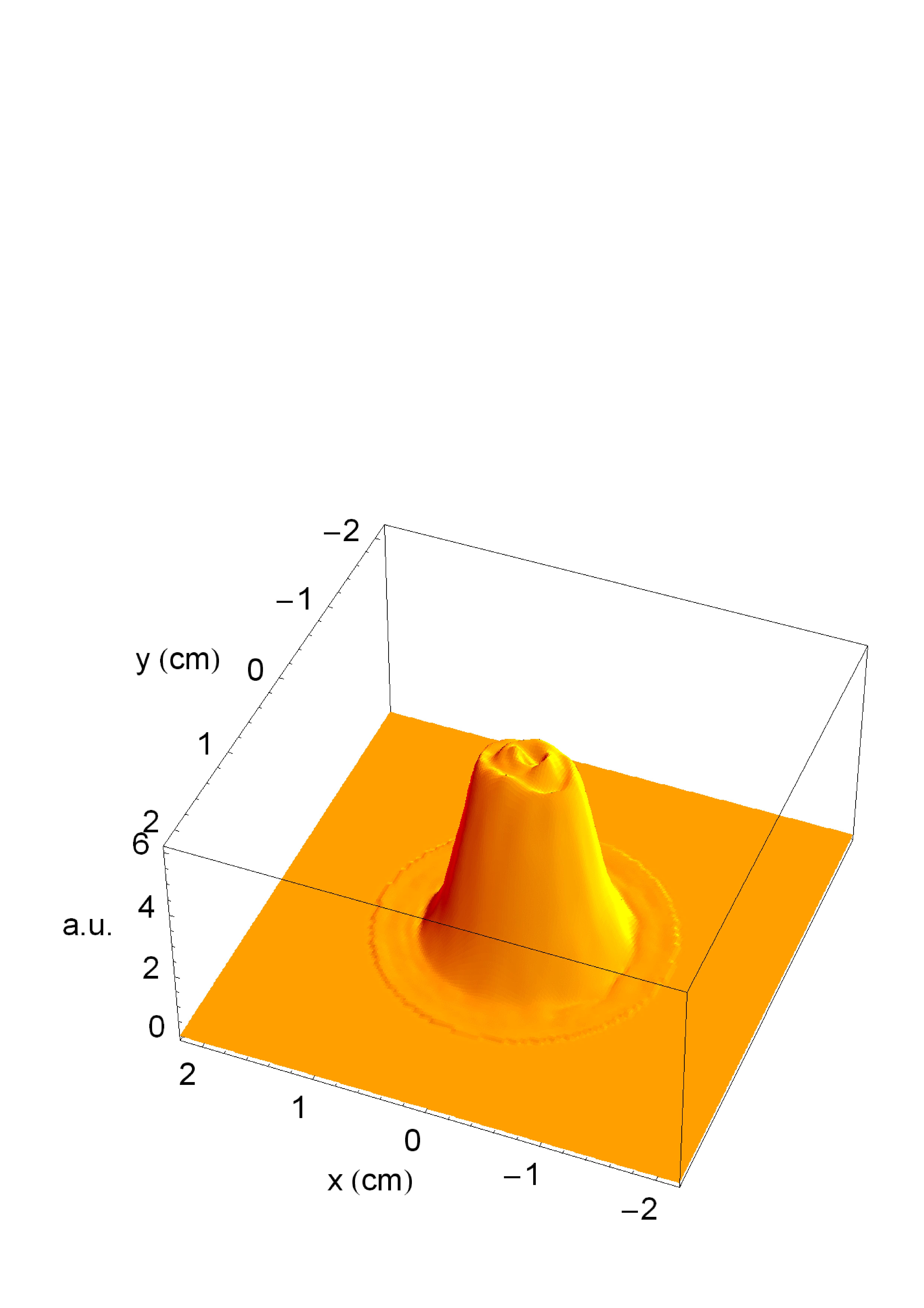}}
 \caption{FFT simulations with the real mirror map of the sample 5008.}
 \label{fig:mir5008}
 \end{figure}

We have mounted the sample $5008$, one of the three made in the
LMA laboratory. As shown in Fig.~\ref{fig:mir5008central} it
presents significant deviations from the design mirror shape. In
particular, there is an evident slope in the central part of the
mirror, which manifests itself by modifying the expected beam
shape, as shown in the FFT simulation of Fig.~\ref{fig:5008}. We
used the map of this sample mirror in our simulation program to
characterize the properties of the resonating light. Our
simulations have shown that the resulting beam shape could fit the
expected flat top behavior: by slightly changing  ($\approx 1 \mu$
rad) the alignment of the mirror we can partially correct for this
intrinsic tilt. The best beam we could obtain with this imperfect
mirror is shown in Fig.~\ref{fig:5008tilted}(assuming perfect flat
input and folding mirrors). The ripples in the central area are
inevitable due to the roughness of the mirror surface (limited by
the accuracy of the corrective coating deposition) and set a
limitation of about $10\%$ PV on the flatness of the power
distribution on the top of the beam.

\subsubsection{Alignment and intrinsic defect correction }

The greatest experimental difficulty was found in obtaining a
sufficiently precise alignment to achieve a flat top power
distribution in the cavity. In a cavity made with spherical
mirrors, a tilted mirror presents the same spherical profile to
the opposite mirror, but shifted sideways, and the cavity simply
resonates the same transverse mode spectrum centered upon a
shifted optical axis. In contrast, the MH mirror has a
non-spherical shape, and any misalignment destroys the cylindrical
symmetry of the cavity. In such a situation the resonant beam
senses a mirror with a suboptimal profile and the cavity mode will
thus have a radically different intensity distribution and phase
front. When our cavity was in such a misaligned state, higher
order and distorted modes were found easily. We were able to
reproduce these patterns with our simulation program by changing
the alignment of the MH mirror and/or shifting the input Gaussian
beam as shown in Sec.~$6.3$ of ~\cite{TaralloTesi}.

 \begin{figure}[htb]
\begin{center}
\includegraphics[width=\textwidth]{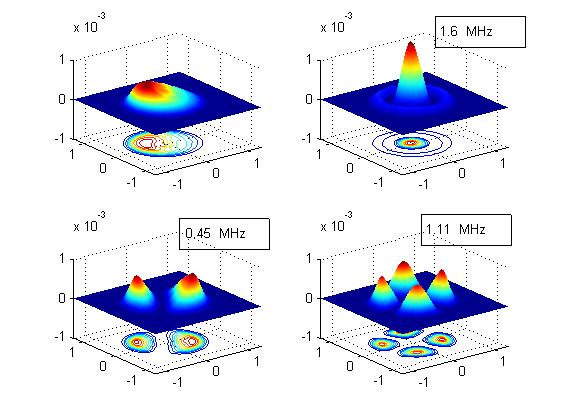}
\end{center}
\caption{Effect of 1 $\mu$rad tilt on the simulated cavity mode
spectrum.} \label{fig:modiTILT}
\end{figure}

Fig.~\ref{fig:modiTILT} shows the simulated cavity spectrum when a
tilt of $1$ $\mu$rad is applied to the MH mirror. The higher order
modes are less effected by the MH mirror tilt with respect to the
fundamental mode; this is traduced in the experiment in a greater
difficulty in aligning the mirrors to obtain a fundamental flat
mode. On the other side higher order (misaligned) modes are
obtained very easily.

\subsubsection{Cavity modes  and spectral distribution}\label{sub:MH modes}

\begin{figure}[htbp]
    \begin{center}
        \includegraphics[width=0.7\textwidth]{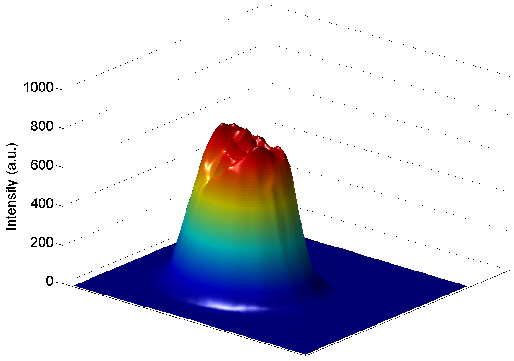}
        \end{center}
    \caption{Three-dimensional profile of the mesa fundamental mode. Experimental data.}
    \label{fig:Best3D}

\end{figure}

\begin{figure}[htbp]
    \begin{center}
        \includegraphics[width=0.8\textwidth]{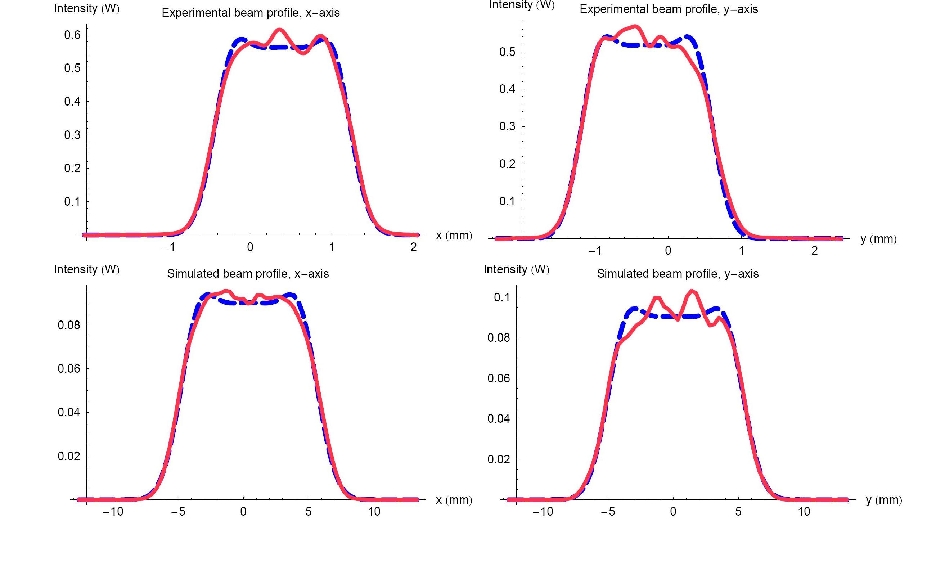}
        \end{center}
    \caption{ One-dimensional profiles of fits to the mesa beam profiles. The top row shows normalized experimental data as measured at the CCD camera. The dashed line is the best fit mesa profile. The bottom row shows profiles extracted from the FFT simulation with best corrective tilt applied. In this case, the transverse scale is taken
at the MH mirror. }
    \label{fig:XYcomparison}
\end{figure}

The extreme sensitivity to misalignment proved to be caused by our
nominally flat folding and input mirrors, which had surface
deviations of order $60 \div 100$ nm from flatness before we fixed
aluminium compensating/strengthening rings to them. Having reduced
the flat mirrors' flexure, it was possible to lock the cavity to a
stable fundamental mode with nearly uniform distribution. In fact,
taking into account the residual warping of the flat mirrors and
MH mirror imperfections, this mesa beam is consistent with the
best achievable using our current prototype MH mirror [see
Fig.~\ref{fig:Best3D}]. Figure \ref{fig:XYcomparison} shows four
beam profiles. Two are experimental data, smoothed with a $0.1$ mm
($5$ pixels) gaussian kernel to clean them of digitization noise
and dust diffraction rings. The other two are profiles simulated
using our FFT code. The simulated profiles represent the leakage
field at the output bench ($\sim 5$ meters from the input mirror)
achieved applying the ideal corrective tilt at the MH mirror.

\newpage

 Analyzing the profile in
Fig.~\ref{fig:XYcomparison}, the normalized absolute power in the
simulated profile not fitted by the mesa TEM$_{00}$ is $3.4\%$. By
comparison, the normalized absolute power in the experimental
profile not fitted by the mesa TEM$_{00}$ is $3.8\%$, after the
normalization of the power on the CCD, with a peak to valley
deviation from the flat profile of about $9.4\%$. These numbers
suggest that the resulting mesa beam is very close to the
experimental limit due to the imperfect MH test mirror.

\begin{figure}[htbp]
    \begin{center}
        \includegraphics[width=0.8\textwidth]{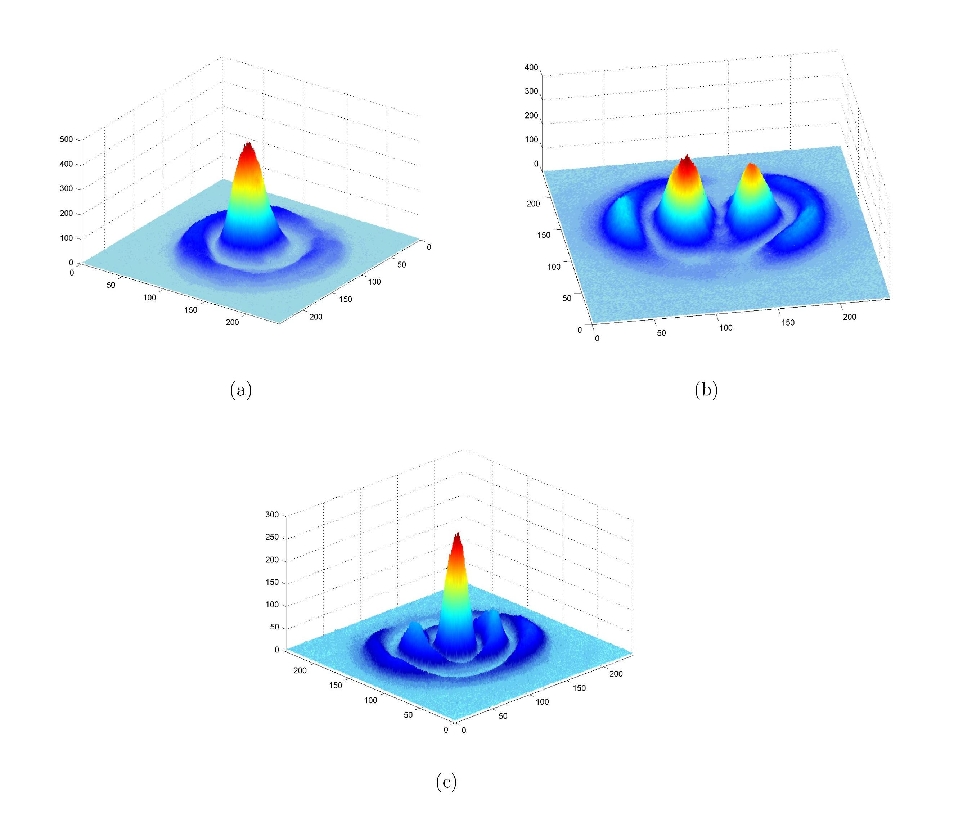}
        \end{center}
    \caption{ High order mesa beam transverse modes: (a) TEM$_{10}$, (b) TEM$_{11}$, (c) TEM$_{20}$ . Experimental data. }
    \label{fig:tem10 data}
\end{figure}

\begin{figure}
\begin{center}
    \includegraphics[width=0.7\textwidth]{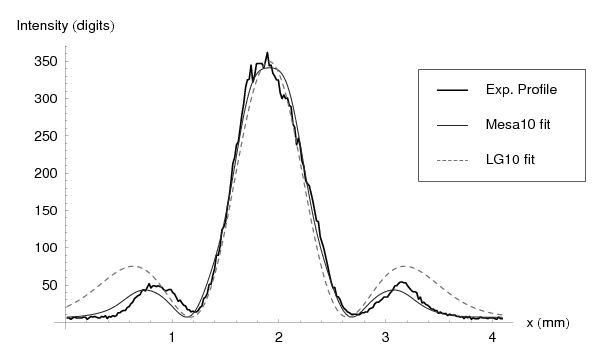}
    \end{center}
  \caption{The mesa TEM$_{10}$ profile (thick black line).
The light gray line show the theoretical mesa TEM$_{10}$, which better fits the data than a Laguerre-Gauss TEM$_{10}$ mode (dashed line). }% caption for the whole figure
\label{fig:fittem10mesa}
\end{figure}

In Fig.~\ref{fig:tem10 data} we show some higher-order transverse
mesa beam modes observed. These modes are superficially quite
similar to the Laguerre-gaussian modes for a spherical mirror
Fabry-Perot cavity. However, in the precise power distribution
there are differences as shown in Fig.~\ref{fig:fittem10mesa}. The
experimental data is in good agreement with the expected mesa
TEM$_{10}$ profile.  As for the fundamental mode, there is some
asymmetry due to the mirror imperfections.

A simple study has been done to compare the numerical predictions
for our resonator and cavity length sweeps. Results are shown in
Tab.~\ref{tab:exp spectrum2}.The second column is calculated
assuming perfect MH and flat Input mirrors. Although the estimate
has large uncertainty ($0.2$ MHz due to frequency spacing and
systematic effects) we can conclude that the numerical predictions
are well respected.

\begin{table}
\begin{center}
       \begin{tabular}{c|c c c}
            \hline
            \hline
             Peak & $\Delta f_{exp}$(MHz) & TEM$_{pl}$ (expected) & $\Delta f(TEM_{pl})$\\
            \hline
             1 & 0 & 00 & 0\\
           2 & 0.4413 $\pm$0.2    & 01  & 0.4141\\
           3 & 1.198  $\pm$0.2    & 02  & 1.0945\\
           4 & 1.574  $\pm$0.2    & 10  & 1.6542\\
           5 & 2.144  $\pm$0.2    & 03  & 1.9905\\
           6 & 2.900   $\pm$0.2    & 11  & 2.8789\\
           7 & 4.161   $\pm$0.2     & 12  & 4.1754\\
           8 & 4.414   $\pm$0.2   & 20  & 4.4050\\
           9 & 5.549   $\pm$0.2     & 13  & 5.5523\\
           10 & 5.801   $\pm$0.2   & 21  & 6.0031\\
            \hline
            \hline
        \end{tabular}
        \end{center}
        \caption{Frequency spacing between eigenmodes for the $7.32$m long
MH cavity prototype. The frequency values are expressed in MHz.
The error is $0.2$ MHz (sampling spacing and systematic effects).}
    \label{tab:exp spectrum2}
\end{table}

\begin{figure}[htb]
\begin{center}
\includegraphics[width=\textwidth]{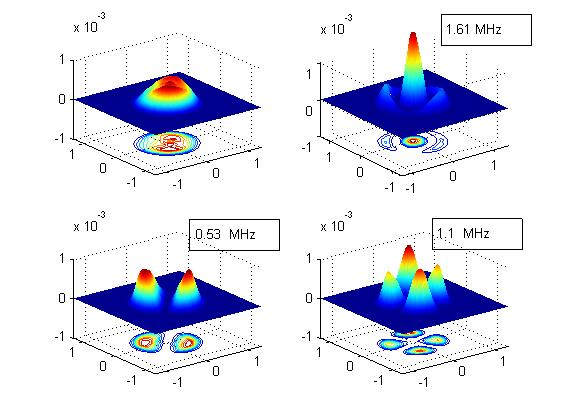}
\end{center}
\caption{Lowest order eigenmodes of the MH cavity with real
deformed mirrors(both MH and Input). FEM eigenmodes calculation.}
\label{fig:modiREAL}
\end{figure}

The mirrors imperfections and the quite large deformations of the
Input mirror cause a deformation in the mode shapes and some
shifting in the resonances spectrum. The simulation shown in
Fig.~\ref{fig:modiREAL} is performed using our algorithm described
in Sec.~\ref{sec:FEM-Gen}. The cylindrical symmetry is broken by
the mirror's imperfection/deformations and the power distribution
over the lobes of the higher order modes is no more symmetric (as
experimentally observed)

\clearpage

\subsubsection{Tilt sensitivity}

Since the mesa beams are intended for use in actual
interferometers it is important to study their ease of control.
Some theoretical and experimental investigations have been carried
out in this area \cite{reducing,oshaug,erika,beyersdorf}.  The
alignment tolerances for a mesa beam arm cavity are expected to be
$\sim 3$ times more stringent than those of the gaussian arm
cavity in an advanced gravitational wave detector.

In our $7.3$ m cavity (just as in a long baseline GWID), extremely
small tilts create significant modification of the mesa beam
profile.

The tilt precision of the optical lever is estimated to be $0.05$
$\mu$rad.

\begin{figure}[htbp]
  \centering
  \includegraphics[width=0.8\textwidth]{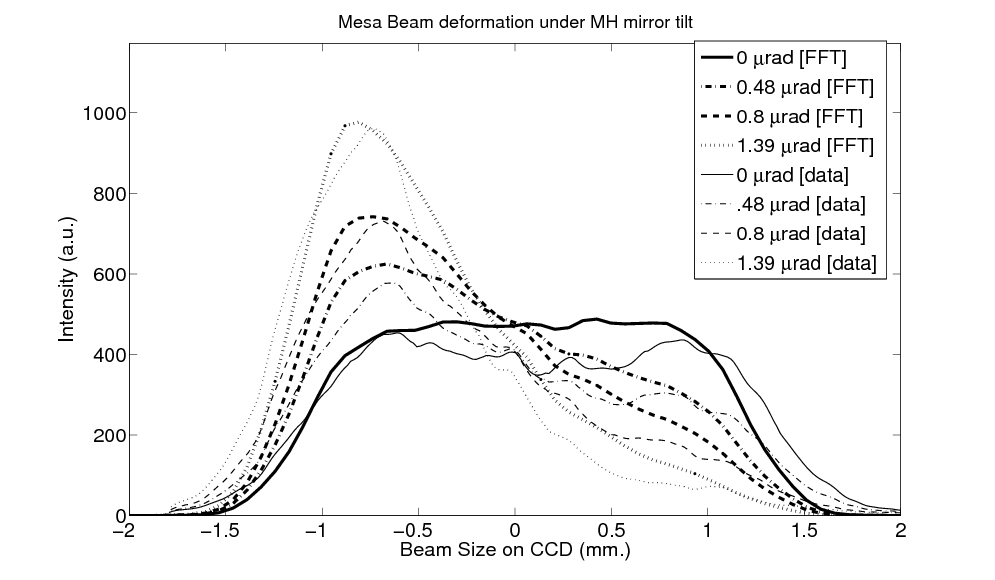}
  \caption{Comparison of FFT simulated (thick) and experimental profiles (thin) with same integrated power for various tilts. }
  \label{fig:TiltResults}
  \end{figure}

Figure \ref{fig:TiltResults} shows the good agreement between our
recorded profiles (thin lines) and FFT simulated data (thick
lines). Note that these FFT simulated profiles were constructed
using a two-mirror cavity (as opposed to the real three-mirror
cavity).

% ------------------------------------------------------------------------

%%% Local Variables:
%%% mode: latex
%%% TeX-master: "../thesis"
%%% End:

\chapter{Thermal noise in view of GW sensitivity improvements}
\ifpdf
    \graphicspath{{Chapter3Figs/PNG/}{Chapter3Figs/PDF/}{Chapter3Figs/}}
\else
    \graphicspath{{Chapter3Figs/EPS/}{Chapter3Figs/}}
\fi

\section{Introduction}

Thermal noise is one of fundamental noise sources in precision
measurement, such as gravitational
experiments~\cite{Saulson:1990jc}. It is expected that the
sensitivity of the next generation interferometric gravitational
wave detectors (see Fig.~\ref{fig:AdLIGOB}) will be limited by the
thermal noise of its optical components (i.e. mirrors substrate
and high-reflective dielectric coatings). Therefore, it is
important to study the thermal noise in view of sensitivity
improvement. In this chapter, the fundamental theorem and the
estimation method of the thermal noise are introduced. Moreover,
the evaluation of the thermal noise of the interferometric
gravitational wave detectors are discussed. We will analyze two
different method for thermal noise reduction: the first based on
beam shaping whereas the second on coating optimization. In this
thesis we calculate for the first time the coating noise reduction
using mesa beam instead of standard Gaussian beam and we  show
that the mesa beam is a good candidate\footnote{There are
currently different studies on other beam shapes~\cite{Galdi,
vinetLG} (and so other mirror shapes) to minimize the mirror
thermal noise. We will comment briefly on this aspect later.} for
application in advanced GW interferometers.

\section{Fluctuation Dissipation Theorem}\label{FDT}
The fluctuation-dissipation theorem (FDT) is one of the most
important theorems of non-equilibrium statistical mechanics. The
FDT predicts the relationship between the spectrum of the thermal
noise and the dissipation of linear\footnote{The linearity states
that the amplitude of the response is proportional to the
amplitude of the driving force.} dissipative\footnote{A system is
considered dissipative if, under an external solicitation, it
absorbs energy through irreversible processes.} systems. Since the
thermal fluctuations and the dissipation are governed by the same
interaction between a system and the heat bath, there must be a
relation between both phenomena. This theorem was established by
Callen \textit{et al.}~\cite{PhysRev.83.34}. In their fundamental
work, they established a generally valid connection between the
response function and the associated equilibrium quantum
fluctuations, i.e., the quantum fluctuation-dissipation theorem.
Another key development must be credited to Lars Onsager: via his
regression hypothesis, he linked the relaxation of an observable
in the presence of weak external perturbations to the decay of
correlations between associated microscopic
variables~\cite{Onsager1,Onsager2}. This all culminated in the
relations commonly known as the Green--Kubo
relations~\cite{Callen-Green,Kubo}. This notion of ``linear
response'' related to the fluctuation properties of the
corresponding variables constitutes the (response)
\textit{fluctuation-dissipation theorem}.

For a macroscopic system, the fluctuation-dissipation theorem
describes the behavior of any generalized coordinate $X$ of any
system that is weakly coupled to a thermalized bath with many
degrees of freedom. The interaction between the system and
externals is represented by a fluctuating generalized force,
$F(t)$. The information of the dissipation of this system is
included in the response of $X$ to $F$. This response is derived
from the equation of motion of the system.

The information of the dissipation of the system is embedded in a
generalized impedance function $Z(\omega)$, and the FDT
establishes a relation between the spectral density of the
spontaneous fluctuations exhibited by the system in equilibrium
and this generalized admittance (the inverse of the impedance) of
the system. The impedance $Z$ is defined as

\begin{equation}\label{Z_FDT}
    Z(\omega)=\frac{\widetilde{F}(\omega)}{\widetilde{\dot{X}}(\omega)}
\end{equation}

where $\widetilde{F}$ and $\widetilde{\dot{X}}$ are the Fourier
components of the generalized force and velocity, respectively. If
the system were completely conservative, then the impedance would
be perfectly imaginary. The bath prevents the system from being
conservative: energy can be fed back and forth between the
generalized coordinate $X$ and the bath's many degrees of freedom.
This energy coupling changes the impedance $Z(\omega)$ from pure
imaginary to complex, and correspondingly, the resulting motions
of $X$ feed energy into the bath and vice versa. The FDT states
that the spectral density (power spectrum) of the thermal
fluctuation for the quantity $X$ is
\begin{equation}\label{FDT-X}
    S_{X}(\omega)=\frac{4 k_B T}{\omega^2}\Re
    \left[Y(\omega)\right]
\end{equation}
and that the spectral density of the generalized force
$S_{F}(\omega)$ \footnote{This is also known as generalized
Nyquist noise formula} is given by

\begin{equation}\label{FDT-F}
    S_{F}(\omega)= 4 k_b T \Re
    \left[Z(\omega)\right]
\end{equation}

 The fluctuation dissipation theorem is derived from the
assumption that the response of a system in thermodynamic
equilibrium to a small external perturbation is the same as its
response to a spontaneous fluctuation. Therefore it seems natural
to look for a direct relation between the fluctuation properties
of the thermodynamic system and its linear response properties to
an external perturbation, which is coupled exactly to the
generalized coordinate we are interested in. Levin~\cite{Levin}
proposed a direct and very powerful method for the application of
the FDT and derived a formula which requires no specific
information about the dissipation mechanism, apart from the fact
that it should be dissipation of mechanical energy  and that the
localization and the amount of dissipation are known. As a
consequence, the Levin formula can be used to study a broad range
of thermal noise induced fluctuations associated with all specific
dissipations. We imagine to apply to the system an external
oscillatory force $F(t)= F_0 \exp(i \omega_0 t)$ that derives the
generalized momentum conjugate to $X$ but does not derive the
other generalized momenta of the system. In order to introduce
this force into this equation of the motion, the new term,

\begin{equation}\label{H-int}
    H_{int}= - F(t) X
\end{equation}

is added to the Hamiltonian of the system. From the definition of
the generalized impedance $Z$ we have that
$\dot{X}(t))=\frac{1}{Z(\omega_0)}F(t)$ and the average dissipated
power within the system as a function of the driving frequency is
given by

\begin{equation}\label{Wdiss}
    \langle W_{diss}(\omega_0)\rangle =\left\langle \dot{X}F \right\rangle =
    \frac{1}{2}F_0^2 \Re\left[Y(\omega_0)\right]
\end{equation}

Substituting Eq. \eqref{Wdiss} into the expression \eqref{FDT-X}
of the FDT, we get

\begin{equation}\label{FDT-X-levin}
    S_{X}(\omega)=\frac{8 k_B T}{\omega^2} \frac{\langle W_{diss}(\omega)\rangle}{F_0^2}
\end{equation}
where we have done the substitution $\omega_0\rightarrow \omega$
since the same reasoning can be repeated for any driving frequency
(in the range coupled to the bath) and the spectral density $S_X$
has the claimed form anywhere in this range. This formula is
particulary useful because for many systems, the calculation of
the dissipated energy is simpler than that of the real part of the
generalized admittance function.

\subsection{FDT: superposition principle}

In many practical situations it is necessary to evaluate the
behavior of system in which the generalized coordinate we are
looking at is a superposition of other generalized coordinates of
the system. This can be the case of the surface of a mirror as
seen by a laser beam profile, since the displacement sensed by the
laser beam can be expressed as a spatially weighted superposition
of the internal modes of the mirror interacting with the beam
itself. Since the FDT  is based on the linear response theory, it
is obvious that for any linear combination of generalized
coordinates we can repeat the reasonings of the previous section
and obtain analogous formulas for the FDT. Lets define a new
coordinate as a weighted superposition of certain system's
generalized coordinates
\begin{equation}\label{Xnew}
    X_{new}=\int P(\mathbf{r}) X(\mathbf{r}) d^3r
\end{equation}

The generalized force $F_{new}$ that drives the momentum
conjugated to $X_{new}$ is introduced in the  Hamiltonian of the
system, trough the following interaction term

\begin{equation}\label{Hint}
    H_{int}= - \int F_{new}(t)  P(\mathbf{r}) X(\mathbf{r}) d^3r =
    - F_{new}(t) X_{new}
\end{equation}

and the generalized admittance of the system becomes

\begin{equation}\label{Z_FDT}
    Y_{new}(\omega)=\frac{\widetilde{\dot{X}}_{new}(\omega)}{\widetilde{F}(\omega)}
\end{equation}

where $\widetilde{F}$ and $\widetilde{\dot{X}}_{new}$ are the
Fourier components of the generalized force and of the velocity of
the new generalized coordinate respectively. The spectral density
of the thermal fluctuation for the coordinate $X_{new}$ is given
by

\begin{equation}\label{FDT-Xnew}
    S_{X_{new}}(\omega)=\frac{4 k_B T}{\omega^2}\Re
    \left[Y_{new}(\omega)\right]
\end{equation}

\section{Mirror thermal noise}
The standard thermal noise is the phase noise caused by random
motions of the reflecting faces of mirrors in a GW interferometer.
A reflecting face can move either because it is displaced by its
suspension systems or because it undergoes internal stresses. At
finite temperatures both effects are possible. We address here the
internal stresses. Consider a massive body at temperature $T$. If
$T
>0$ the atoms constituting the body are excited and have random
motions around their equilibrium position. The fact that they are
strongly coupled to neighboring atoms makes possible propagation
of elastic waves of various types, reflecting on the faces and the
onset of stationary waves. One can show that, for a finite body
(like for instance a cylinder of silica) there is a discrete
infinity of such stationary waves, each corresponding to a
particular elastic normal mode. At thermal equilibrium, the state
of the body can be represented by a linear superposition of all
the modes, with random relative phases, and, due to the energy
equipartition theorem, the same energy $k_B\,T$ ($k_B$ is the
Boltzmann constant). The motion of atoms near a limiting surface
of the body will slightly modify its shape, and if we consider the
reflecting face of a mirror, a surface distortion is cause of
phase change in the reflected beam; in other words, of a noise.
Estimation of the resulting spectral density of phase noise in the
probe beam is the internal thermal noise problem in massive
mirrors.

Modal expansion is a traditional method to calculate thermal
noise. Although now superseded by the Levin approach, we find it
useful to first discuss the general concept of this method and
then discuss its application to the mirror. In a frequency domain,
the equation of motion of a one-dimensional harmonic oscillator is
written as

\begin{equation}\label{oscill-loss}
    - m \omega^2 X(\omega) + m \omega_0^2 [1+ i \phi(\omega)]
    X(\omega)= F(\omega)
\end{equation}
where m is the mass of the oscillator, and $\omega_0$ is the
angular resonant frequency of the oscillator. We introduced the
mechanical loss as the loss angle $\phi(\omega)$, and $m
\omega_0^2 [1+ i \phi(\omega)]$ is called the complex spring
constant. For a mechanical system with a spring constant, we
describe the dissipation as the imaginary part of a complex spring
constant. In the standard theory of inelasticity, Hooke's law is
extended to represent the internal friction process, leading to a
description of the dissipation of energy in the oscillator in
terms of its complex stiffness spring. As a matter of fact,
Eq.~\eqref{oscill-loss} has merely phenomenological relevance to
express the experimental appearance of a lag angle between the
stress and the strain and the behavior with respect to the
frequency. The mechanical impedance of the system is given by

\begin{equation}\label{Z-oscill}
    Z(\omega)=\frac{m}{\omega}\left[\phi(\omega) \omega_0^2 + i
    (\omega^2- \omega_0^2)\right]
\end{equation}

and the application of the FDT \eqref{FDT-X} allows the
calculation of the spectral density of the thermal fluctuations of
the oscillator's position

\begin{equation}\label{SX-oscill}
    S_{X}(\omega)=\frac{4 k_b T}{m \omega}\frac{\phi(\omega) \omega_0^2}{(\omega^2-\omega_0^2)^2+\omega_0^4 \phi^2(\omega)}
\end{equation}

Before giving a short overview of the modal expansion approach to
the calculation of the FDT, we have to summarize and fix the
notation for the elastic theory of continuum media which will be
used in the following Sections.

\section{Basic elasticity theory}

We label the position of a point (a tiny bit of solid) in an
unstressed body, relative to some convenient origin, by its
position vector $\mathbf{x}$. Let a force be applied so the body
deforms and the point moves from $\mathbf{x}$ to
$\mathbf{x}+\mathbf{u}(\mathbf{x})$; we call $\mathbf{u}$ the
point's displacement vector. The strain tensor is defined as a
second-rank tensor field given by the covariant derivative of the
displacement vector which can be written in arbitrary system of
coordinates using index notation and the symbol `$;$' for
covariant derivative

\begin{equation}\label{strain-cov}
    S_{ij}=u_{i ;\, j}
\end{equation}

In a Cartesian coordinate system the components of the gradient
are always just partial derivatives.

The strain tensor $\mathbf{S}$  is a general, second-rank tensor.
Therefore, its irreducible tensorial parts are its trace
$\Theta=Tr(\mathbf{S}) = S_{i;\,i}$, which is called the deformed
body's expansion; its symmetric, trace-free part
$\mathbf{\Sigma}$, which is called the body's shear; and its
antisymmetric part $\mathbf{R}$, which is called the body's
rotation:

The strain tensor can be reconstructed from these irreducible
tensorial parts in the following manner
\begin{equation}\label{strain-decom}
    S_{i j}=\frac{1}{3} \Theta g_{i j} + \Sigma_{i j} + R_{i j}
\end{equation}

Since the third term represent a rigid rotation which does not
deform the solid and we are interested in the deformation itself,
it is useful to rewrite the strain tensor as

\begin{equation}\label{strain-split}
    S_{i j}= \varepsilon_{i j} + R_{i j}=\frac{1}{2}(u_{i ; j}+u_{j ;
    i}) + \frac{1}{2}(u_{i ; j}-u_{j ;
    i})
\end{equation}

The first part $\mathbf{\varepsilon}$ is a symmetric tensor and
represents elongation, compression, and shear. In many textbooks
this part alone is called strain tensor. The displacement due to
rotation will be not considered in the following since it is not
related to the deformations of the elastic body.

The forces acting within an elastic solid are measured by a second
rank tensor, the stress tensor. Consider two small, contiguous
regions in a solid. If we take a small element of oriented area
$d\mathbf{\Sigma}$ in the contact surface with its positive sense
 pointing from the first region toward the second, then the first
region exerts a force $d\mathbf{F}$ (not necessarily normal to the
surface) on the second through this area. The force the second
region exerts on the first (through the area $d\mathbf{\Sigma}$)
will, by Newton's third law, be equal and opposite to that force.
The force and the area of contact are both vectors and there is a
linear relationship between them.  The two vectors therefore will
be related by a second rank tensor, the stress tensor
$\mathbf{\sigma}$:

\begin{equation}\label{stress-force1}
    dF^i= \sigma^{i j} d\Sigma_j
\end{equation}

and the elastic force per unit volume acting on a solid can be
expressed as the divergence of the stress tensor
\begin{equation}\label{stress-force}
    f^i = \sigma^{i j}_{ ;j}
\end{equation}

The diagonal terms of the stress tensor define the normal stresses
in the coordinate system in which $\sigma^{i j}$ is expressed. The
off-diagonal terms define the shear stresses. Normal stresses act
parallel to the unit vector orthogonal to the surface while shear
stresses act perpendicular to it. The conservation of angular
momentum requires that the stress tensor be symmetric.
Consequently, there are only 6 independent stress components to be
determined.

To understand the relationship between stress and strain in
elastic media, we must generalize Hooke's law for continuous,
linear, elastic media.

The most general linear equation relating two second rank tensors
will involve a fourth rank tensor known as the elastic modulus
tensor or elastic stiffness tensor, $\mathbf{C}$. In tensor
notation we have

\begin{equation}\label{stiffness-gen}
    \sigma^{i j}= C^{i j k m} \varepsilon_{k m}
\end{equation}
The inverse relation defines the fourth rank compliance tensor
$\mathbf{G}$

\begin{equation}\label{compliance-gen}
    \varepsilon_{i j}= G_{i j k m} \sigma^{k m}
\end{equation}

 Now, a general fourth rank tensor in three dimensions
has $3^4 = 81$ independent components. However there are several
symmetries that we can exploit. As the stress tensor is symmetric,
and only the symmetric part of the strain tensor creates stress,
$\mathbf{C}$ is symmetric in its first pair of indices and also in
its second pair. Moreover, from the expansion of the elastic
internal energy in terms of components of strain, it can be shown
that the elastic tensor must be symmetric under the interchange of
the first two indices with the second ones. There are therefore $
21$ independent components in $\mathbf{C}$.

Many substances, notably crystals, exhibit additional symmetries
and this can reduce the number of independent components
considerably. The simplest, and in fact most common, case arises
when the medium is isotropic. In other words, there are no
preferred directions in the material. This occurs when the solid
is polycrystalline or amorphous and completely disordered on a
scale large compared with the atomic spacing, but small compared
with the solid's inhomogeneity scale.

If a body is isotropic, then its elastic properties must be
describable by scalars. Now, the stress tensor $\mathbf{\sigma}$,
being symmetric, must have just two irreducible tensorial parts, a
scalar part and a trace-free symmetric part; and the parts of the
strain that can produce this behavior are the scalar expansion
$\Theta$ and the trace-free, symmetric shear $\mathbf{\Sigma}$,
but not the rotation. The only linear, coordinate-independent
relationship between these tensorial quantities involving solely
scalars is

\begin{equation}\label{isotro1}
    \sigma_{i j}= K \Theta g_{i j} + 2 \mu \Sigma_{i j}
\end{equation}

 Here $K$ is called the bulk modulus and $\mu$ the shear modulus, and
the factor $2$ is included for purely historical reasons.
Sometimes it is convenient to introduce the Lam\'{e} coefficients
$\lambda, \mu$ which are related to $K$ by the relation
\begin{equation}\label{K-Lame}
    K= \lambda + \frac{2}{3}\mu
\end{equation}
 and write the elastic tensor in a generic coordinate system as
\begin{equation}\label{isotro2}
    C^{i j r s}= \lambda  g^{i j}g^{r s}  + \mu
    ( g^{i r}g^{j s} + g^{j s}g^{i r})
\end{equation}
and the relation between stress and strain tensor as

\begin{equation}\label{isotroL}
    \sigma^{i j} = 2 \mu \varepsilon^{i j} + \lambda
    \varepsilon^k_k g^{i j}
\end{equation}

 In terms of
the Young modulus $Y$ and the Poisson coefficient $\nu$ the linear
elasticity relation \eqref{stiffness-gen} for an isotropic and
homogeneous body can be written as

\begin{equation}\label{isotro3}
    \sigma^{r s}=  \frac{Y}{1+\nu} \left (
     g^{m r}g^{n s} + g^{m s}g^{n r}+ \frac{2 \nu}{1- 2 \nu}g^{r s} g^{m n}\right) \varepsilon_{m n}
\end{equation}

If one has coordinate components, found from generalized
coordinate tensor analysis, for some quantity, such as stress or
strain, one needs to be able to translate those into the values
measured in the experiment.  The method for doing this is reviewed
in Appendix~\ref{App:A1}. The measured value for a given vector
component, unlike the coordinate component, is unique within a
given reference frame. In differential geometry (tensor analysis),
that measured value is called the ``physical
component''(App.~\ref{App:A1}). It is important to recognize that
physical components do not transform as true tensor components and
one can not simply use physical components in tensor analysis as
if they were. Typically, one starts with physical components as
input to a problem. These are converted to coordinate components,
and the appropriate tensor analysis carried out to get an answer
in terms of coordinate components. One then converts these
coordinate components into physical components as a last step, in
order to compare with values measured with instruments in the real
world. Since we will deal with only orthogonal coordinate
(Cartesian and Cylindrical) we can restrict our discussion to
orthogonal coordinate systems in which the metric tensor is
diagonal and

\begin{equation}\label{metric}
    ds^2= g_{i j} dx^i dx^j  \qquad g_{i i}= h_i^2, i \,\,\mbox{not summed} \quad g_{i
    j}=0 \; i\neq j
\end{equation}

\textit{The physical components of a tensor are given by the
components of the tensor with respect to unit basis vectors.} In
the following sections we will deal with orthonormal cylindrical
basis and the components denoted by $r,\varphi,z$ are referred to
this basis. More details about the relation between physical
components and tensorial components is given in App.~\ref{App:A1}.

\subsection{Thermal noise with mode expansion}

When the thermal noise of the internal modes of a mirror in
interferometric gravitational wave detectors is calculated, the
mirror is treated as a elastic cylinder.  Cylindrical coordinates
are employed with the origin at the center of the mirror and the
$z$-axis along the cylindrical axis. The observable coordinate,
$X$, which is the surface displacement averaged by the beam power
distribution, is expressed as

\begin{equation}\label{X-mirr}
    X=\int_S u_z(\mathbf{r})P(\mathbf{r}) dS
\end{equation}

where $u_z$ is the $z$ component of the displacement vector
$\mathbf{u}$, and the weighting function $P$ is the normalized
distribution of the beam power over the mirror surface. The
external generalized force that drives the momentum conjugated to
$X$ is introduced in the equation of motion without dissipation
for the elastic body in the following form

\begin{equation}\label{EQ-mot-elas}
   \rho \frac{\partial^2 \mathbf{u}}{\partial t^2} - \frac{Y}{2 ( 1 +
   \nu)}\triangle \mathbf{u} - \frac{Y}{2 ( 1 + \nu) (1 + 2 \nu
   )}\nabla (\nabla \cdot \mathbf{u})= F(t)P(\mathbf{r})\hat{z}
\end{equation}

where $\rho$ is the density and we have used the equation of
motion of a linear and homogeneous elastic body
(Eq.~\eqref{equi-displa-gen} in App.~\ref{App:A2}). We can look
for a solution of \eqref{EQ-mot-elas} in the form of a normal
modes expansion, that is, we write the following superposition of
basis functions $\mathbf{w}_n$

\begin{equation}\label{u-decompo}
    \mathbf{u}(\mathbf{r},t)=\sum_n \mathbf{w}_n(\mathbf{r})
    q_n(t)
\end{equation}

The basis functions, $\mathbf{w}_n$, are the solution of the
eigenvalue problem, for the free elastic body, written as

\begin{equation}\label{EQ-modi-elas}
  - \rho \omega^2_n \mathbf{w}_n = \frac{Y}{2 ( 1 +
   \nu)}\triangle \mathbf{w}_n + \frac{Y}{2 ( 1 + \nu) (1 + 2 \nu
   )}\nabla (\nabla \cdot \mathbf{w}_n)
\end{equation}

where $\omega_n$ and $\mathbf{w}_n(\mathbf{r})$ correspond to the
angular resonant frequency and the displacement of the n-th
resonant mode of the system, respectively.

The function, $q_n(t)$ in Eq.\eqref{u-decompo}, represents the
time development of the n-th mode. The equation of motion of $q_n$
is derived substituting the equation Eq. \eqref{u-decompo} for
$\mathbf{u}$ in Eq. \eqref{EQ-mot-elas} and using the
orthogonality property of the complete set of functions
$\mathbf{w}_n$. The result is written in the form

\begin{equation}\label{modi-q}
    m_n \ddot{q}_n(t) + m_n \omega^2_n q_n(t)=F(t)
\end{equation}
where the parameter $m_n$, called effective mass of the n-th mode,
is defined by the normalization choice for the functions
$\mathbf{w}_n$

\begin{align}\label{nor-modes}
  & \int w_{n,z}(\mathbf{r}) P(\mathbf{r}) dS = 1 \nonumber \\
   & m_n = \int \rho(\mathbf{r})|\mathbf{w_n(\mathbf{r})}|^2 dV
\end{align}

where $w_{n, z}$ is the $z$ component of $\mathbf{w}_n$.
Therefore, the time evolution of the n-th mode is the same as that
of a harmonic oscillator of mass $m_n$ and angular resonant
frequency $\omega_n$ with an external force $F(t)$ acting on it.
Using Eq. \eqref{u-decompo} and Eq. \eqref{nor-modes}in Eq.
\eqref{X-mirr} it follows

\begin{equation}\label{X-modi}
    X(t) = \sum_n  q_n(t)
\end{equation}

showing that the observable coordinate $X$ can be simply described
as a superposition of the motions of the harmonic oscillators
$q_n$. Up to now we  neglected dissipation in the system. To
describe dissipation, the loss angles,$ \phi_n(\omega)$, are
introduced in the frequency domain equation for the normal modes
evolution. Equation \eqref{modi-q} is rewritten as

\begin{equation}\label{modes-diss}
    - m_n \omega^2 \tilde{q}_n + m_n \omega_n^2 \left [1 + i
    \phi_n(\omega) \right ] \tilde{q}= \tilde{F} ,
\end{equation}

From Eq. \eqref{X-modi} and the equations \eqref{modes-diss}, the
generalized admittance of the system is

\begin{equation}\label{Y-modes}
    Y(\omega)=\frac{\tilde{\dot{X}}(\omega)}{\tilde{F}(\omega)}= i
    \omega \frac{\sum_n \tilde{q}_n(\omega)}{\tilde{F}(\omega)}= i
    \omega \sum_n \frac{1}{- m_n \omega^2 + m_n \omega^2_n \left [1 + i
    \phi_n(\omega) \right ]}
\end{equation}

The power spectral density of $X$, $S_X$ can now be derived
applying the FDT, Eq. \eqref{FDT-Xnew} and Eq. \eqref{Y-modes}:

\begin{equation}\label{SX-modes}
    S_X(\omega) = \frac{4 k_B T}{\omega}\sum_n \frac{\omega_n^2 \phi_n(\omega)}
    {m_n \left[ \left (\omega^2- \omega_n^2 \right )^2 + \omega_n^4 \phi_n^2(\omega) \right]}
\end{equation}

Therefore the thermal motion of the system is expressed as the sum
of the contributions given by the harmonic oscillators of the
normal-mode expansion. Equation~\eqref{SX-modes} allows to
calculate the thermal noise from the angular resonant frequency
$\omega_n$, the effective mass $m_n$ and the loss angle
$\phi_n(\omega)$ of each mode. The angular resonant frequency and
the displacement of the mode $\mathbf{w}_n$ are obtained from the
eigenvector problem, Eq.~\eqref{EQ-modi-elas} . The effective mass
$m_n$ is calculated from the $\mathbf{w}_n$  found and equation
\eqref{nor-modes}. The loss angle is derived from the experiments
but its measurement in a wide frequency range is commonly
difficult; thus, it is usual to estimate it from the Q-value on
the resonant frequencies for different modes, according to the
following relation:

\begin{equation}\label{Q-n}
    Q_n = \frac{1}{\phi_n(\omega_n)}
\end{equation}

There are two methods to solve the eigenvalue problem. The first
one is the method proposed by Hutchinson~\cite{Hutch}. This is a
very accurate semi-analytical algorithm to simulate resonances of
an isotropic elastic cylinder. The second method is the finite
element method which is a numerical method. The thermal noise can
be calculated using~\eqref{SX-modes}, where the summation has to
be done over a number of modes that assures a good convergence of
the series.
 Using the mode expansion formalism described above, the
mirror thermal noise of interferometric gravitational waves
detector has been calculated by Gillespie--Raab~\cite{Gillespie},
Bondu--Vinet~\cite{Bondu:1995ur} under the assumption of
homogeneously distributed loss inside the mirror bulk.

 Recently, a problem of the modal
expansion method was clarified in~\cite{yamamoto}. The
introduction of the mechanical loss after the system is divided
into basis functions, is not equivalent to the solution that is
directly obtained from the equation of motion with the loss
included since the beginning. The modal expansion fails,
especially when the loss is inhomogeneously distributed in the
system, $\phi(\mathbf{r}, \omega)$, because of a coupling between
the internal modes. The LIGO mirrors, composed of several parts,
mirror substrate, coating, magnet, each with its own loss
mechanism, correspond to this case.

\section{Thermal noise: direct calculation}

At present the technique recognized to be the most appropriate to
calculate the thermal noise is the ``direct'' method. It has the
advantage that leaves aside the modal decomposition. There are
many different ways of implementing this computation, Levin's
approach~\cite{Levin}, Nakagawa's approach~\cite{Nakagawa} and
Numata's numerical dynamic approach~\cite{Numata}. In the
following we limit the discussion to Levin's method explained in
Sec.~\ref{FDT}, since this is the approach used in this thesis
work.

Consider the surface of the mirror invested by the laser beam
which has a profile given by the weighting function
$P(\mathbf{r})$. The read-out variable will be the generalized
coordinate $X$ expressed as a continuous combination of system
coordinates in the form of Eq.~\eqref{X-mirr}. Using the FDT,
Eq.~\eqref{FDT-Xnew}, the calculus of the thermal noise reduces to
the calculus of the real part of the admittance. Levin's method
consist in the calculation of $Re[Y_{new}]$ from the average
mechanical energy dissipated by a particular driving force.
Suppose to apply on the mirror surface an oscillatory continuous
force $F(\mathbf{r}, t)$ that mimics the profile $P(\mathbf{r})$
of the laser beam:

\begin{equation}\label{F-lev}
F(\mathbf{r}, t)= F_0 \cos(\omega t) P(\mathbf{r})
\end{equation}

This force corresponds to a generalized force

\begin{equation}\label{F-gen-lev}
    F(t) = F_0 \cos(\omega t)
\end{equation}

that drives only the momentum conjugated to $X$ as is expressed in
Eq.~\eqref{H-int}; then we can apply the
equations~\eqref{FDT-X-levin} for the spectral density of the
fluctuations of our observable coordinate $X$.

\section{Mirror thermal noise contributions}

Mirror thermal noise (in the bulk and in the coating),  can be
divided into:

\begin{itemize}
    \item  Brownian thermal noise,
    \item thermoelastic noise,
    \item thermorefractive noise.
\end{itemize}

In this thesis, we will focus our attention on these  kinds of
mirror thermal noises and we will show the possible sensitivity
improvements in using a mesa beam instead of a standard Gaussian
beam.

\subsection{Brownian thermal noise}

Internal friction in solids was identified by Kimball and
Lovell~\cite{PhysRev.30.948}, who described it as a phase shift
between stress and strain. Brownian thermal noise due to mirrors,
can be interpreted as a fluctuation of the mirror surface position
coming from the mirror modes thermal excitation or, using the FDT,
as fluctuations induced by a structural damping. In the real
lattice, impurities, dislocations and imperfections play the role
of dissipation sources in the phonon dynamics. Irreversible
processes can also be attributed to relaxation mechanism in
asymmetric potential-well models~\cite{Two-level,Diss-asi}.
Thermal noise due to homogeneously distributed damping processes,
such as Brownian bulk noise in GW detector mirrors, can be
estimated using complex valued elastic coefficients whose
imaginary part is related to the dissipation mechanism. For
example, the structural damping can be parameterized by a complex
material's Young modulus, were the imaginary part is related to
the energy dissipation of the system

\begin{equation}\label{Y-imag}
    Y \longmapsto Y (1 + i \phi )
\end{equation}

Measurements~\cite{Freq-silica} have shown that $\phi$ is
dependent on the frequency $f$, but its dependence is sufficiently
slow that it can be neglected in the frequency region of interest.
To calculate the power spectral density associated to Brownian
bulk motions, Levin expressed $\langle W_{diss} \rangle$ in
Eq.~\eqref{FDT-X-levin} as

\begin{equation}\label{W-diss-BS}
    \langle W_{diss} \rangle = 2 \omega \phi_s \langle U_s\rangle
\end{equation}

where $\pi_s$ is the loss angle of the substrate, $\langle ..
\rangle$ denotes the time average over the period of  the
oscillatory pressure previously defined and $U_s$ is the elastic
energy stored in the substrate of the test mass. This energy is
derived by using elasticity theory, and will depend on which
monitoring beam profile you use as well as the mirror's
dimensions. Levin supposed that the mirror is an infinite
half-space and that the elastic deformation follows
quasistatically the oscillatory pressure (a reasonable hypothesis
because the first resonance mode of the mirror is far higher than
the region of interest). In the case when the beam has a gaussian
profile and the mirror is approximated by a semi infinite
half-space, the displacement spectral density will be~\cite{Levin}

\begin{equation}\label{SX-B-SUB-INF-GB}
    S_X^{B , s}= \frac{4 k_B T}{\sqrt{\pi}}\frac{1- \nu^2}{\omega Y
    w} \phi_s
\end{equation}

The calculation was also made by Bondu \textit{et
al.}~\cite{Bondu:1998sm} , and reviewed by Liu and
Thorne~\cite{LT} in the case of a finite size mirror.

The correction to the noise spectral density is an infinite series
of Bessel function with coefficients depending on the material
properties and on mirror and beam-spot size dimensions. If
compared with the result in the approximation of infinite mirror
mass, this calculation gives a correction to the thermal noise
that, e.g. in the case of the new Advanced LIGO mirrors, starts to
be greater than $30 \%$ for beam spot sizes greater than $6$ cm.

Considering that there is the project to use flat beams, this
correction becomes crucial in estimating the Brownian thermal
noise correctly.

\subsection{Thermoelastic noise}\label{sub:TEnoise}

Thermoelastic noise is intended as a noise that comes from the
coupling of thermal fluctuations with displacement fluctuations
thanks to a non-null coefficient of thermal expansion. The
oscillating squeeze and stretch of the substrate material causes
an oscillating, inhomogeneous temperature distribution: heat flows
down the temperature gradient in such a way that it converts
oscillation energy into additional heat.

The inhomogeneity of the deformations causes the temperature
perturbation $\delta T$ to be inhomogeneous, and that
inhomogeneity produces a heat flux $\mathbf{q} = - \kappa \nabla
\delta T$. Whenever an amount of heat flows from a region of
temperature $T$ to one of slightly lower temperature $T - dT$ ,
there is an increase of entropy and the resulting rate of entropy
increase per unit volume is

\begin{equation}\label{dS-TE}
    \frac{dS}{dV dt}=\frac{- \mathbf{q}\cdot \nabla \delta T}{T^2}
    = \frac{\kappa \left (\nabla \delta T \right )^2}{T^2}
\end{equation}

This entropy increase entails a creation of new thermal energy at
a rate per unit volume $dE_{th} / (dV dt) = T dS /(dV dt)$. Since,
for our thought experiment with temporally oscillating applied
stress, this new thermal energy must come from the oscillating
elastic energy, the rate of dissipation of elastic energy must be

 \begin{equation}\label{W_diss-TE}
    W_{diss} = \int \frac{\kappa \left (\nabla \delta T \right
    )^2}{T}dV
\end{equation}

 To calculate this noise
for the mirror it is necessary to solve a system of two coupled
equations, the first one is the elasto-dynamic equation for the
displacement $\mathbf{u}(\mathbf{r},t)$ including a stress term
coming from the temperature inhomogeneity, and the second one is
the thermal conductivity equation for the temperature perturbation
$\delta T(\mathbf{r},t)$ including an heat-source term coming from
the non uniform expansion. As the time required for sound to
travel across the mirror is usually smaller than the oscillatory
period, it is often used a quasistatic approximation in the
equation of elasticity  for a field of deformation $\mathbf{u}$

\begin{equation}\label{stress-TE}
    \nabla ( \nabla \cdot \mathbf{u}) + (1- 2 \nu) \triangle
    \mathbf{u}= 2 \alpha (1 + \nu) \nabla \delta T
\end{equation}

The thermal conductivity equation for the temperature
inhomogeneity $\delta T$

\begin{equation}\label{cond-TE}
    \frac{\partial \; \delta T}{\partial t} - \frac{\kappa}{
    C} \triangle \delta T = - \frac{\alpha Y T}{ C (1 - 2 \nu)}\frac{\partial (\nabla \cdot \mathbf{u})}{\partial t}
\end{equation}

where $Y,\nu , \alpha, \kappa$ and $C$ are Young modulus, Poisson
ratio, the coefficient of linear thermal expansion, the thermal
conductivity and the specific heat per unit volume at constant
volume of the mirror. Braginsky \textit{et
al.}~\cite{Braginsky:1999rp} solved the problem using the
fluctuation-dissipation theorem in the form~\eqref{FDT-Xnew},
approximating the mirror with an infinite half-space and
considering that the oscillatory period is far higher with respect
the typical time scale of the diffusive heat flow. This is the so
called adiabatic limit in which the second term on the left hand
side of~\eqref{cond-TE} is neglected. In this approximation,
$\delta T$ is simply proportional to the elastic expansion
$\Theta$.

 Liu
and Thorne~\cite{LT} solved the same problem with the same
approximations applying the FDT in Levin's form,
Eq.~\eqref{FDT-X-levin}, and using \eqref{W_diss-TE}. They found
that the displacement power spectrum for half infinite mirror is

\begin{equation}\label{SX-TE-SUB-INF-GB}
    S_X^{TE , s}= \frac{16}{\sqrt{ \pi}} \alpha^2 (1 + \nu)^2 \frac{k_B T^2 \kappa }{C^2 \omega^2}\frac{1}{w^3}
\end{equation}

They also made the calculations for a finite cylindrical test mass
finding the correction factor w.r.t. the infinite test mass
solution. Depending on the mirror/beam geometry and the mirror's
material this correction can be several $10 \%$.

Cerdonio \textit{et al.}~\cite{Cerdonio} extended the previous
analysis releasing the assumption of the adiabatic limit, i.e.
performing the analysis that is valid also at low temperatures
and/or small beam spot size. Equation~\eqref{SX-TE-SUB-INF-GB} has
to be multiplied by a correction factor, which depend on the ratio
between the beam radius and the thermal diffusivity length (that
is frequency dependent).

\section{Coating thermal noise}

It has been pointed out~\cite{Levin} that it is critically
important how the losses are distributed inside the test masses.
Losses far from the beam spot contribute less to the total thermal
noise, whereas losses near the spot, for example in the dielectric
coating directly reflecting the beam, contribute more.

The coatings on the mirrors play the crucial role in reflecting
the laser used for position sensing off of the test mass. They
must be highly reflective and able to handle high optical power,
to reduce shot noise, and have low levels of intrinsic noise.
Primary among intrinsic noise are thermally driven motions of the
coating face and the optical path length.

 To obtain the required high reflectivity, multi-layers,
dielectric coatings are used. Such coatings consists of
alternating ion-beam deposited layers  of optical thickness equal
to $\lambda/4 $ (at $1.064 \mu$ m) of two dielectric materials
with differing refractive indexes (i.e., $Ta_2O_5/SiO_2$ for
initial LIGO).  It was found~\cite{Penn-loss} that the mechanical
loss comes from internal dissipation of the coating materials, and
it is the high index Tantalum-pentoxide that is primarily
responsible for the thermal noise in first generation
interferometers.

Another source of thermal noise is thermally driven fluctuations
in the optical path length of the coating. This can occur from
thermal fluctuations in length, i.e. thermoelastic
noise~\cite{BVcoat, Fejer}, or from fluctuations in the refraction
index, i.e thermorefractive noise~\cite{Braginsky:2000wc}. These
two mechanisms are driven by the same thermal fluctuations,
therefore will add coherently. This collective noise source is
referred to as \textit{thermo-optic} noise. The level of
\textit{thermo-optic} noise in next generation gravitational-wave
interferometers depends on the thermal, elastic, and optic
properties of the coating materials, primarily on $\alpha (=
dL/dT)$ and $\beta(= dn/dT )$. Values in
literature~\cite{Fejer,BVcoat}for ion beam deposited coatings
indicate that thermo-optic noise could contribute to Advanced LIGO
sensitivity and that the high index coating material may be the
primary problem and lots of effort by the optics research groups
is dedicated to measure these parameters.

\subsubsection{Brownian coating thermal noise}

 Coating thermal
noise due to internal losses is expected to be the dominant
contribution to the thermal noise for mirrors with a
$SiO_2/Ta_2O_5$ coating on a fused silica substrate.  Nakagawa
\textit{et al.}~\cite{Nakagawa} calculated the dissipation induced
by an inhomogeneous loss distribution on a half-infinite mirror
due to the fact that loss angle in the bulk material $\phi_s$ is
different from loss angle in the coating $\phi_c$ but neglecting
the differences in the elastic properties of the substrate and the
coating

\begin{equation}\label{SX-COAT-Na}
 S_X^{B , c}= \frac{4 k_B T}{\sqrt{\pi}}\frac{1- \nu_s^2}{\omega
 Y_s w}\frac{1}{\sqrt{\pi}}\frac{(1- 2 \nu)}{(1-
 \nu)}\frac{d}{w} \phi_c
     \frac{1}{\sqrt{\pi}}\frac{d}{w}
\end{equation}

Harry \textit{et al.}~\cite{HarryCQG}, using Levin's method,
generalized this result for a non isotropic coating loss, assuming
that the loss angle $\phi_{\parallel}$ associated with energy
stored in strains parallel to the plane of the coating, is
different from the loss angle $\phi_{\perp}$ associated with the
energy stored in strains perpendicular to the surface. For a thin
coating, provided that $\nu_c \simeq \nu_b$, $Y_c =
Y_{\parallel}\simeq Y_{\perp}$ and $\nu_c=\nu_{\parallel}\simeq
\nu_{\perp}$, they found

\begin{align}\label{SX-B-COAT-INF-GB}
    S_X^{B , c}= \frac{4 k_B T}{\sqrt{\pi}}\frac{1- \nu^2}{\omega Y
    w} \frac{1}{\sqrt{\pi}}\frac{d}{w} & \Bigg ( \frac{Y_c (1 + \nu_s) (1- 2 \nu_s)^2 +
    Y_s \nu_c (1+ \nu_c) (1- 2 \nu_s)}{Y_s (1- \nu_c^2)(1-
    \nu_s)}\, \phi_{\parallel}+ \nonumber \\
    & + \frac{Y_s (1+ \nu_c) (1- 2 \nu_c)-Y_c \nu_c (1+ \nu_s)(1-2 \nu_s)}{Y_c (1- \nu_c) (1-
    \nu_s^2)} \, \phi_{\perp}\Bigg )
\end{align}

Equation \eqref{SX-B-COAT-INF-GB} is valid provided that the
coating loss occurs in the coating materials themselves and not
for rubbing in the bulk/coating or layer/layer and this has been
experimentally investigated by Penn \textit{et
al.}~\cite{Penn-loss,Penn-loss2}.

\subsubsection{Coating thermoelastic noise}

Temperature fluctuations (that couple with displacement thanks to
$\alpha$ can be originated either by intrinsic thermodynamical
fluctuation, or by laser-photon absorption. We will briefly
describe the first mechanism and refer to Rao~\cite{Rao} for the
description of the latter one since that is a negligible effect in
LIGO and Ad-LIGO mirrors.

The diffusive heat characteristic length $r_T$ of the substrate
and coating (of the order of mm) is far larger than the coating
thickness (a few $\mu$ m). Because diffusive heat flow in the
longitudinal direction is not negligible, heat flow in the
direction normal to the coating cannot be treated adiabatically.
The problem was extensively treated by Braginsky and
Vyatchanin~\cite{BVcoat} and by Fejer et al.~\cite{Fejer} in the
case of Gaussian beam. In the frequency band $f < \tau_c^{-1}$ ,
where $\tau_c$ is the thermal diffusion time across the coating
($\tau_c^{-1}\sim 6$ kHz for $Ta_2O_5/SiO_2$ multi-layer coating),
coating thermoelastic noise can be estimated by the equation

\begin{equation}\label{SX-TE-COAT-INF-GB}
    S_X^{TE , c}= \frac{8 \sqrt{2 k_B T^2}}{\pi
    \sqrt{\omega}}\frac{d^2}{w^2}(1+ \nu_s)^2 \frac{C_c^2}{C_s^2}\frac{\alpha_s^2}{\sqrt{\kappa_s
    C_s}}\Delta^2
\end{equation}

where $\Delta^2$ is a dimensionless combination of material
constants that vanishes when the film and substrate are identical

\begin{equation}\label{Delta2}
    \Delta^2= \left ( \frac{C_s}{2 \alpha_s C_c} \frac{\alpha_c}{(1 - \nu_c)}
    \left [\frac{1+ \nu_c}{1+ \nu_s} + (1 - 2 \nu_s) \frac{Y_c}{Y_s}\right ] -1
    \right)^2
\end{equation}

\subsubsection{Coating thermorefractive noise}

The thermo-refractive noise is generated by temperature
fluctuations that couple with phase fluctuations of the reflected
laser beam (and therefore with measured displacement), thanks to
the non-vanishing coefficient $\beta = \frac{d n}{dT}$, where $n$
is the refractive index. Considering the frequency range (that is
typical for the ground interferometer) around $100$ Hz and an half
infinite mirror with a Gaussian beam, Braginsky \textit{et
al.}~\cite{Braginsky:2000wc} calculated the equivalent
displacement noise spectral density

\begin{equation}\label{SX-TR-COAT-INF-GB}
    S_X^{TR , c}= \frac{2 \sqrt{2}}{\pi}\frac{k_B}{\sqrt{C \kappa \omega}}\left (\frac{\beta_{eff}\lambda
    T}{w}\right )^2
\end{equation}
where $\beta_{eff}=\frac{ n_2^2\beta_1 + n_1^2 \beta_2}{4
(n_1^2-n_2^2)} $ for a  multi-layer coating which consists of
alternating sequences of quarter-wavelength dielectric layers
having refractive indices $n_1$ and $n_2$\footnote{This formula
has been revisited many times during the last few years.
Originally was proposed in this form in ~\cite{Braginsky:2000wc};
subsequently was modified in ~\cite{BVcoat}, and recently an
independent analysis~\cite{Castaldi} confirmed the original
formula.}.

\section{Thermal noise: Gaussian VS mesa beam}\label{sec:TNMBG}

Coating thermal noise \eqref{SX-B-COAT-INF-GB} due to internal
losses is expected to be the dominant contribution to the thermal
noise for mirrors with a $SiO_2/Ta_2O_5$ coating on a fused silica
substrate, whereas the thermoelastic noise of the substrate is the
dominant contribution for sapphire mirrors at room temperature.
The local surface fluctuations produced by thermal noise are
averaged by the intensity distribution of the laser beam spot over
the mirror surface. Reading the entire mirror surface with uniform
sensitivity would minimize the thermal noise. The standard design
of interferometers uses light beams with a Gaussian distribution
of power, which are eigenfunctions of cavities with spherical
mirrors, a well-developed and understood technology.
Table~\ref{tab:GBnoises} summarizes the geometrical dependence of
each kind of mirror thermal noise on the Gaussian beam radius $w$
for half infinite test mass.
\begin{table}
\begin{center}
\begin{tabular}{|c|c|}
  \hline
  % after \\: \hline
   Thermal noise & $S_X \propto w^{-n}$ \\
  \hline \hline
  Substrate Brownian & $n= 1$ \\
  \hline
  Substrate Thermo-elastic & $n=3$ \\
  \hline
  Coating (all kinds) & $n=2$ \\
  \hline
\end{tabular}
\end{center}
\caption{Dependence of the thermal noise spectral densities on the
Gaussian beam radius.} \label{tab:GBnoises}
\end{table}

The larger is the beam radius, the better is the averaging of the
fluctuations and thus lower will be the noise. However the beam
size is constrained by the allowable diffraction losses
requirements, which cannot exceed a few ppm. Taking into account
the diffraction loss constraints, a Gaussian beam, effectively
averages out the thermal fluctuations only over a few percent of
the mirror surface. A significant reduction in all kinds of
mirror's thermal noises can be achieved by using modified optics
that reshape the beam from a conventional Gaussian profile into
the ``mesa-beam'' profile described in Sec.~\ref{sec:MesaBeam}. A
large-radius, flat-topped beam with steeply dropping edges
(necessary to satisfy the diffraction loss constraint) will lead
to a better sampling of the fluctuating surface, lower noise in
the determination of the mirror surface position and better
sensitivity for GW detectors. The calculation of substrate
thermo-elastic noise reduction using Mesa beam has been done
in~\cite{oshaug} for sapphire test mass. More recently
Vinet~\cite{VinetFB} calculated the substrate Brownian thermal
noise reduction using Mesa beam for Virgo mirror size.

Calculation of the coating thermal noise, which is expected to be
the most significant contribution to the thermal noise budget for
the test masses of the next generation of GW interferometers, has
never been published for non Gaussian beams and finite cylindrical
test masses. In this thesis we present a comparative study of the
various sources of thermal noise in different mirror and beam
configurations, considering both Gaussian and Mesa beam profiles,
addressing the problem of thermal noise reduction, through mirror
size aspect-ratio and beam size optimization. (Some of these
results have been already presented in~\cite{Aspen} and
~\cite{TNmeeting}).

\subsection{Elastic solution for a cylindrical test mass}

We used the Levin approach to the Fluctuation Dissipation
Theorem~\eqref{FDT-X-levin} in order to calculate the power
spectral density of the test mass displacement. The generalized
coordinate $X$ is given by the average of the normal displacement
of the test mass surface, weighted by the beam spot power
distribution as introduced in Eq.~\eqref{X-mirr}. In this way the
thermal noise evaluation reduces to the calculation of $\langle
W_{diss}\rangle$ in accordance to the specific mechanism and
localization of dissipation, i.e. Brownian or thermoelastic and
substrate or coating.

The main ingredient for our calculations is the model proposed by
Bondu, Hello and Vinet \cite{Bondu:1998sm} and corrected by Liu
and Thorne \cite{LT} of the approximate solution of the elasticity
equations for a cylindrical test mass subjected to the oscillatory
pressure with the same spatial profile as the beam power
distribution (assumed cylindrically symmetric). Consider a finite
sized, cylindrical test mass with radius $a$ and thickness $H$ and
with a cylindrically symmetric light spot centered on the
cylinder's circular face, which applies an oscillating pressure

\begin{equation}\label{pressure}
    P(r)= F_0 f(r) \cos(\omega t)
\end{equation}
where $F_0$ is a constant force amplitude and $f(r)$ beam power
distribution normalized over the mirror surface

\begin{equation}\label{f-norm}
    2 \pi \int_0^a f(r) r dr =1
\end{equation}

 The fundamental
approximation underneath these calculation is the quasi-static
approximation for the calculation of the displacement fields
according to the oscillatory pressure which is a good
approximation for oscillatory period larger than the time required
for sound to travel across the test mass and far away from the
resonances of the test mass.

The static stress balance equations in cylindrical coordinates
derived in Appendix~\ref{App:A2} have to be solved together with
the following boundary conditions:

\begin{itemize}
    \item No shear on the cylinder's surfaces, i.e.
    \begin{equation}\label{stress-surf}
    \sigma_{r z}(r=a , z)=0 , \quad \sigma_{r z}(r , z=0)=0, \quad
    \sigma_{r z}(r , z=H)=0
\end{equation}
    \item Pressure of the beam on the front face and no normal
    stress at the back surface

    \begin{equation}\label{stress-norm}
\sigma_{z z}(r , z=0)=- P(r) , \quad \sigma_{z z}(r , z=H)=0,
\end{equation}
    \item No radial stress on the cylindrical edge
    \begin{equation}\label{stress-radial}
     \sigma_{r r}(r=a , z)=0
\end{equation}
\end{itemize}

The displacement vector satisfying the mentioned boundary
conditions \footnote{The boundary condition \eqref{stress-radial}
is not fulfilled exactly, in fact, as discussed in
\cite{Bondu:1998sm}, the $c_0$ and $c_1$ terms in the displacement
are a correction to the leading-order displacement, designed to
improve the satisfaction of the $\sigma_{r r}(r=a , z)=0$ boundary
condition as far as $\sigma_{r r}(r=a , z)$ calculated without
these corrections can be approximated by a linear function.} is
given by

\begin{align}\label{u-r}
    u_r(r,z,t)=F_0 \cos(\omega t) \Bigg [ & \sum_{m=1}^\infty A_m(z)
    J_1(k_m r)+ \frac{\lambda + 2 \mu}{2 \mu (3 \lambda + 2
    \mu)}(c_0 r + c_1 r\, z) \nonumber \\
    &+ \frac{\lambda p_0 r}{2 \mu (3 \lambda + 2
    \mu)}\left( 1- \frac{z}{H} \right)\Bigg]
\end{align}

\begin{align}\label{u-phi}
     u_{\theta}(r,z,t)=0 \qquad \qquad \qquad \qquad \qquad \qquad  \qquad \qquad \qquad \qquad &&
\end{align}

\begin{align}\label{u-z}
    u_z(r,z,t)= F_0 \cos(\omega t) \Bigg [&\sum_{m=1}^\infty B_m(z)
    J_0(k_m r)-\frac{\lambda}{\mu (3 \lambda + 2 \mu)}\left (c_0 z
    +\frac{c_1 z^2}{2}\right) \nonumber \\
    & -\frac{\lambda + 2 \mu}{4 \mu (3\lambda
    +2\mu)}c_1 r^2
 + \frac{\lambda p_0 r^2}{4 \mu H
(3\lambda+2\mu)} \nonumber \\
&-\frac{(\lambda+\mu)p_0}{\mu (3 \lambda +2\mu)}\left(z
-\frac{z^2}{2H}\right) \Bigg]
\end{align}

where $J_0$ and $J_1$ are the Bessel function of order zero and
order one respectively, $\zeta_m$ is the $m-$th zero of the Bessel
function $J_1(x)$, $k_m=\zeta_m/a$ and the other coefficients are

\begin{align}\label{LTcoeff}
     p_0 &= \frac{1}{\pi a^2}, \qquad p_m = \frac{2}{a^2
    J_0^2(\zeta_m)}\int_0^a f(r) J_0(k_m r) r dr \nonumber \\
     c_0 &=6 \frac{a^2}{H^2}\sum_{m=1}^\infty\frac{J_0(\zeta_m)
    p_m}{\zeta_m^2}, \qquad c_1= - \frac{2 c_0}{H}  \nonumber \\
     A_m(z)&= \gamma_m e^{- k_m z} + \delta_m e^{k_m z} + \frac{k_m z}{2}\frac{\lambda+\mu}{\lambda +2\mu}
    \left( \alpha_m e^{- k_m z} + \beta_m e^{k_m z} \right)
    \nonumber \\
     B_m(z)&= \frac{k_m z}{2}\frac{\lambda+\mu}{\lambda +2\mu}
    \left( \alpha_m e^{- k_m z} - \beta_m e^{k_m z} \right) +
    \left[\frac{\lambda + 3\mu}{2(\lambda+2\mu)}\beta_m -\delta_m
    \right] e^{k_m z}\nonumber \\
    &+\left[\frac{\lambda + 3\mu}{2(\lambda+2\mu)}\alpha_m +\gamma_m
    \right] e^{-k_m z}
\end{align}

and $\alpha_m, \beta_m, \gamma_m, \delta_m$ are constants given by

\begin{align}\label{LTcoeff2}
    & Q_m = e^{-2 k_m H} \\
    & \alpha_m= \frac{p_m (\lambda+2\mu)}{k_m \mu (\lambda+\mu)}
    \frac{1-Q_m + 2k_m H Q_m}{(1-Q_m)^2-4 k_m^2 H^2 Q_m}\nonumber \\
    & \beta_m=\frac{p_m (\lambda+2\mu)Q_m}{k_m \mu (\lambda+\mu)}
    \frac{1-Q_m + 2k_m H}{(1-Q_m)^2-4 k_m^2 H^2 Q_m} \nonumber \\
    & \gamma_m = -\frac{p_m}{2 k_m \mu (\lambda + \mu)}\frac{[2 k_m^2 H^2 (\lambda+\mu)+2 \mu k_m H]Q_m
    +\mu (1-Q_m)}{(1-Q_m)^2-4 k_m^2 H^2 Q_m} \nonumber \\
    & \delta_m=-\frac{p_m Q_m}{2 k_m \mu (\lambda+\mu)}\frac{2k_m^2 H^2 (\lambda+\mu) - 2 \mu k_m H-\mu(1-Q_m)}
    {(1-Q_m)^2-4 k_m^2 H^2 Q_m}
\end{align}

In the axi-symmetric case, the non-zero (physical) components of
the strain tensor in cylindrical coordinates are given by

\begin{equation}\label{strain-cyl}
    \varepsilon_{r r}= \frac{\partial u_r}{\partial r}, \quad
    \varepsilon_{\theta \theta}=\frac{u_r}{r}, \quad
    \varepsilon_{z z}=\frac{\partial u_z}{\partial z}, \quad
    \varepsilon_{r z}=\frac{1}{2}\left (\frac{\partial u_r}{\partial z}+\frac{\partial u_z}{\partial
    r} \right)
\end{equation}

and, for an homogeneous and isotropic body, the stress components
$\sigma_{i j}$ are related to the strain by the Lam\'{e}
coefficients $\lambda, \mu$

\begin{align}\label{stress-cyl}
    &\sigma_{r r}= \lambda \Theta + 2 \mu \varepsilon_{r r}
    \nonumber \\
    &\sigma_{\theta \theta}= \lambda \Theta + 2 \mu \varepsilon_{\theta
    \theta}\nonumber \\
    &\sigma_{z z}=\lambda \Theta + 2 \mu \varepsilon_{z z}
    \nonumber \\
    &\sigma_{r z} = 2 \mu \varepsilon_{r z}
\end{align}

where $\Theta = \varepsilon_{r r}+\varepsilon_{\theta
\theta}+\varepsilon_{z z}$ is the expansion.

\subsection{Thermal noise calculations for finite size test masses}

In order to manipulate these quite complicated expressions, we
developed a $Mathematica^\circledR$ Notebook called Thermal Noise
Notebook (TNN)\footnote{Available at
\url{http://www.ligo.caltech.edu/~jagresti/}} with the intent of
providing a user-friendly tool which can be used by GW
interferometer researchers to evaluate quickly the expected mirror
thermal noise. We chose to implement the calculation using this
program because of its great versatility for numerical and
analytical calculation and the possibility of changing very easily
all the parameters involved in the calculations, beam shape,
mirror's aspect ratio, mechanical and thermal properties of the
material of the substrate and coating. We now give an overview of
the calculation for the different types of thermal noise.

\subsubsection{Substrate Brownian thermal noise}

The conventional thermal noise of the substrate is given by Eq.
\eqref{FDT-X-levin} with the time averaged dissipation $\langle
W_{diss}\rangle$ given by

\begin{equation}\label{WBS}
    \langle W_{diss}\rangle= 2 \omega \phi_s \langle U_s\rangle
\end{equation}
where $\phi_s$ is the loss angle of the
substrate,$\langle..\rangle$ denotes the time average over the
oscillatory period and $U_s$ is the elastic energy stored in the
test mass that can be determined integrating, over the test mass
volume, the elastic energy density given by

\begin{equation}\label{uS}
    \varrho(r,z)=\frac{1}{2}\varepsilon_{i j} \sigma_{i j}=
    \frac{1}{2}\left( \lambda \Theta^2 + 2 \mu (\varepsilon_{r
    r}^2+\varepsilon_{\theta \theta}^2+ \varepsilon_{z z}^2+ 2
    \varepsilon_{r z}^2) \right )
\end{equation}

\begin{equation}\label{US}
    U_s = \int_V \varrho(r, z) dV
\end{equation}

$\varepsilon_{ij}$ and $\sigma_{ij}$ are the component of the
strain and stress tensor respectively calculated from the
displacement vector using equations~\eqref{strain-cyl} and from
the constitutive relations (elastic moduli tensor) for an
homogeneous and isotropic material, Eq.~\eqref{stress-cyl}. These
cumbersome calculations are performed very efficiently by
$Mathematica^\circledR$ and will not be reported here.

\subsubsection{Substrate thermo-elastic noise}

As reported in Sec.~\ref{sub:TEnoise}, the thermo-elastic
dissipation is given by

\begin{equation}\label{WTE}
    \langle W_{diss} \rangle=\int \frac{\kappa_s \left (\nabla \delta T_s \right
    )^2}{T_s}dV
\end{equation}
Where $\kappa_s$ is the substrate thermal conductivity and $\delta
T_s$ is the substrate temperature perturbation induced by the
elastic deformation due to the oscillatory pressure and is given
 by

\begin{equation}\label{deltaTs}
    \delta T_s = - \frac{\alpha_s Y_s T}{C_s (1-2\nu_s)}\Theta_s
\end{equation}

Where $\alpha_s$ is the linear thermal expansion coefficient,
$Y_s$ and $\nu_s$ are the Young modulus and Poisson ratio
respectively, $C_s$ is the specific heat per unit volume of the
substrate.

The equation \eqref{deltaTs} follows from the adiabatic
approximation of the general thermal conductivity
equation~\eqref{cond-TE}. The adiabatic approximation, already
discussed in Sec.~\ref{sub:TEnoise}, consist in the following: if
the time scale for diffusive heat flow is much longer than the
pressure oscillating period, we can approximate the oscillations
of stress, strain and temperature as adiabatic, neglecting the
heat flow term in the thermal conductivity equation. Using
equations \eqref{strain-cyl} for the calculation of the expansion
$\Theta$, substituting in \eqref{deltaTs} we can calculate the
thermo-elastic dissipated energy in \eqref{W_diss-TE} and then the
spectral density of the displacement noise given by
Eq.~\eqref{FDT-X-levin}.

\subsubsection{Coating thermal noise}

The geometry we consider for the reflective surface of the mirrors
consists of a thin film of thickness on a substrate whose
thermo-mechanical properties are different from those of the film.
To simplify the analysis, we assumed that the multi-layer coating
can be approximated as a uniform layer with appropriately averaged
properties.

\subsubsection{Coating Brownian thermal noise}

In this case the averaged energy dissipated by the intrinsic
losses in the coating is given by an analogous formula of
\eqref{WBS}

\begin{equation}\label{WBc}
 \langle W_{diss}\rangle= 2 \omega \phi_c \langle U_c\rangle
\end{equation}

 Where $\phi_c$ is the loss angle of the coating and $U_c$ is the portion of elastic energy stored in the coating
 (in this calculation we assume an isotropic and homogeneous coating with averaged elastic coefficient: Young modulus $Y_c$ and Poisson ratio $\nu_c$).
 In the thin film approximation $d\ll H$, we assume that the energy stored in the coating is given by $U_c= d \delta
 U_c$, where $d$ is the thickness of the coating and $\delta U_c$ is the energy density stored at the surface, integrated over the
 surface, which is expressed in terms of the coating stress and
 strain as

\begin{equation}\label{deltaUc}
    \delta U_c = \int_S \frac{1}{2}\varepsilon_{i j}^c \sigma_{i
    j}^c  \, dS
\end{equation}

In the thin film approximation, the coating is approximated as a
thin layer in which the stress and strain do not vary considerably
as a function of depth within the coating and we shall approximate
them as being constant. Following~\cite{HarryCQG}, the stresses
and strains in the coating can be calculated in terms of the
stresses and strains at the surface of the substrate because of
the boundary condition between coating and substrate: the coating
must have the same tangential strains as the surface of the
substrate and the coating experiences the same perpendicular
pressure as the surface of the substrate. Since the only exerted
force is normal to the plane we must have no shear stress on the
coating, $\sigma_{r z}^c=0$,($\sigma_{r z}^s(r, z=0)=0$ is a
boundary condition for the elastic problem of the substrate). The
coating stress and strain can thus be calculated by the following
equations

\begin{align}\label{bound-coat}
    &\varepsilon_{r r}^c(r)= \varepsilon_{r r}^s(r,z=0), \qquad \varepsilon_{\theta
    \theta}^c(r)=\varepsilon_{\theta \theta}^s(r, z=0), \qquad \varepsilon_{z
    z}^c(r)= \varepsilon_{z z}^s(r, z=0) \nonumber \\
    & \nonumber \\
    & \sigma_{k k}^c = \lambda_c \Theta^c + 2\mu_c \varepsilon_{k
    k}^c \quad \mbox{with}\quad k= r,\theta,z \quad
    \mbox{and}\quad \Theta^c=\varepsilon_{rr}^c +
    \varepsilon_{\theta \theta}^c + \varepsilon_{zz}^c
\end{align}

In this way we can calculate all the fields necessary for the
computation of the elastic energy stored in the coating, $U_c$,
using the expressions already found for the substrate.

\subsubsection{Coating thermo-elastic noise}

In the thermo-elastic problem of the coating is important to note
that the coating thickness, the diffusive heat transfer length and
the beam radius satisfy the following relation

\begin{equation}\label{TEC-con}
    d\ll r_T \ll w, \qquad \mbox{with}\quad r_T=\sqrt{\frac{\kappa}{C \omega}}
\end{equation}

This relation justify the approximation of the multi-layer film as
a uniform film with averaged properties and when computing the
oscillating temperature distribution we can consider the
temperature variation as adiabatic in the transversal direction
and that only the thermal diffusion orthogonal to the surface of
the mirror need to be considered. The two thermo-elastic coupled
equations \eqref{stress-TE} and \eqref{cond-TE} are solved
perturbatively at the first order in the linear thermal expansion
coefficient $\alpha$. We first consider the quasi-static
stress-balance equation at the zeroth order in $\alpha$ which has
been solved in the preceding section (Eqs.~\ref{bound-coat}).
Then, we solved the thermal conductivity equation
Eq.~\eqref{cond-TE} for a one-dimensional heat flow and using the
zeroth order elastic fields as the source term.

\begin{equation}\label{eqTECcond}
    \left (\frac{\partial}{\partial t}+ K_{\beta} \frac{\partial^2}{\partial
    z^2} \right ) \delta T_{\beta}= - \left (\frac{Y \alpha T}{(1- 2\nu) C}\frac{\partial \Theta (z=0)}{\partial
    t}\right)_{\beta}= - B_{\beta}
\end{equation}

where $\Theta (z=0)$ is the expansion at the mirror surface
associated with the zeroth-order elastic fields calculated for the
previous sections and $\beta = s, c$ indicates quantities
evaluated in the substrate and the coating respectively. For a
multi-layer coating this equation determines an averaged
temperature field and the coating quantities are averaged
following~\cite{Fejer}. If $d_1$ and $d_2$ are the thickness of
the two materials composing the coating ($d_1+d_2=d$), we have

\begin{align}\label{averageTE}
   &\left (X\right)_{avg}\equiv \frac{d_1}{d_1+d_2}X_1 +
   \frac{d_2}{d_1+d_2}X_2 \nonumber \\
    & K_c = \frac{\kappa_c}{C_{avg}}, \quad \kappa_c^{-1}=
    (\kappa^{-1})_{avg}, \quad B_c=\frac{(C B)_{avg}}{C_{avg}}
\end{align}

Assuming a time dependence of the form $e^{i \omega t}$ for the
oscillatory thermal and elastic fields, equations
\eqref{eqTECcond} can be cast in this form

\begin{equation}\label{eqTECcond-rid}
    \left (i \omega - K_{\beta}\frac{\partial^2}{\partial
    z^2} \right) \delta T_{\beta}= - i \omega B_{\beta} \qquad
    \beta= s,c
\end{equation}

with the boundary conditions of zero heat flux at the surfaces of
the test mass and continuity of temperature and heat flux at the
boundary between coating and substrate

\begin{equation}\label{bound-condTE}
    \frac{\partial \delta T_c}{\partial z}\Bigg |_{z=0}=0, \quad \frac{\partial \delta T_s}{\partial z}\Bigg |_{z=H}=0,
    \quad \delta T_c = \delta T_s \Bigg |_{z=d}\quad \kappa_c \frac{\partial \delta T_c}{\partial
    z}=\kappa_s \frac{\partial \delta T_s}{\partial
    z}\Bigg |_{z=d}
\end{equation}

The general solution of \eqref{eqTECcond-rid} is given by

\begin{equation}\label{gensol-deltaT}
    \delta T_{\beta}= - B_{\beta} + C1_{\beta}\, e^{\gamma_{\beta}
    z} + C2_{\beta} \,e^{- \,\gamma_{\beta}z}, \qquad \gamma_{\beta}=
    (1+i)\sqrt{\frac{\omega}{2 K_{\beta}}}
\end{equation}

The boundary conditions \eqref{bound-condTE} determine the four
arbitrary constants $C1_{\beta}, C2_{\beta}$ and the solutions for
the temperature variation induced in the coating and in the
substrate are

\begin{align}\label{deltaTcTE}
    \delta T_c = - B_c +\, &e^{\gamma_s d} (e^{2 \gamma_s d} - e^{2 \gamma_s H})\kappa_s
    (B_c-B_s)\gamma_s  \left[e^{\gamma_c z}+ e^{-\,\gamma_c z} \right] \cdot  \\
    &\Bigg( e^{2 \gamma_s H} (\kappa_c \gamma_c - \kappa_s \gamma_s)
    + e^{2 d (\gamma_c+\gamma_s)}(-\kappa_c
    \gamma_c+\kappa_s+\gamma_s)+\nonumber \\
&+ e^{2 \gamma_s d} (\kappa_c \gamma_c + \kappa_s \gamma_s)
    - e^{2 d \gamma_c+ 2 H\gamma_s}(\kappa_c
    \gamma_c+\kappa_s+\gamma_s)\Bigg)^{-1} \nonumber
\end{align}

\begin{align}\label{deltaTsTE}
    \delta T_s= - B_s + \, & e^{\gamma_s d} (e^{2 \gamma_c d}-1)
    \kappa_c (B_c-B_s) \gamma_c  \left[e^{\gamma_c z}+ e^{2 H \gamma_s}\, e^{-\,\gamma_c z} \right] \cdot  \\
&\Bigg( e^{2 \gamma_s H} (\kappa_c \gamma_c - \kappa_s \gamma_s)
    + e^{2 d (\gamma_c+\gamma_s)}(-\kappa_c
    \gamma_c+\kappa_s+\gamma_s)+\nonumber \\
&+ e^{2 \gamma_s d} (\kappa_c \gamma_c + \kappa_s \gamma_s)
    - e^{2 d \gamma_c+ 2 H\gamma_s}(\kappa_c
    \gamma_c+\kappa_s+\gamma_s)\Bigg)^{-1} \nonumber
\end{align}

The  dissipated power for the coating thermo-elastic noise is
given by

\begin{equation}\label{WdissTEC}
    W_{diss}\simeq \int_{V_s} \frac{\kappa_s}{T}\left (\frac{\partial \delta T_s}{\partial
    z}\right)^2 \, dV_s\, +\,\int_{V_c} \frac{\kappa_c}{T}\left (\frac{\partial \delta T_c}{\partial
    z}\right)^2 \, dV_c
\end{equation}

where we are neglecting the dissipation due to the radial heat
flow \footnote{The correction is completely negligible for the
coating contribution, and is around $3\%$ for the substrate
dissipated power.} and the integrals are extended over the
substrate and coating volume respectively. Averaging
Eq.~\eqref{WdissTEC} over the oscillatory period, and inserting
the result in Eq.~\eqref{FDT-X-levin} we have the spectral density
of displacement noise due to coating thermo-elastic fluctuations.

\subsubsection{Coating thermo-refractive noise}\label{sec:TRFB}

Thermodynamical fluctuations of temperature in mirrors of
gravitational wave antennae may be transformed into additional
noise not only through thermal expansion coefficient but also
through temperature dependence of refraction index. In this case
we cannot use the original Levin's method\footnote{Recently Levin
proposed a variation of its original direct method, to calculate
the thermo-refractive noise [arXiv:$0710.2710$]. } to calculate
this noise because it is not associated with mechanical energy
dissipation. Following~\cite{Braginsky:2000wc}, the spectral
density of the equivalent displacement noise can be written as

\begin{equation}\label{TRC}
    S_X (\omega) = \lambda^2 \beta_{eff}^2 S_T(\omega)
\end{equation}

where $\lambda$ is the laser wavelength, $S_T$ is the spectral
density of temperature fluctuation in the coating and
$\beta_{eff}$ for a multi-layer quarter wavelength optical coating
is given by

\begin{equation}\label{beta-eff}
    \beta_{eff}=\frac{ n_2^2\beta_1 + n_1^2 \beta_2}{4
(n_1^2-n_2^2)}
\end{equation}

The spectral density of the temperature fluctuation can be
calculated using the Langevin approach explained
in~\cite{Braginsky:2000wc}. For semi infinite mirror and for
cylindrically symmetric beam profile $f(r)$, it is easy to derive
the following equation

\begin{align}\label{SpecDensT}
    S_{T}(\omega)&=\frac{4 k_B T^2 K}{C}\int_{-\infty}^{\infty}
    dq_{z} \int_{0}^{\infty} \frac{q_{\perp}dq_{\perp}}{( 2 \pi)^2}\frac{2 q^2}{K^2 q^4 +\omega^2}
    \frac{1}{1+ q_{\perp}^2 d^2}|\tilde{f}(q_{\perp})|^2 \nonumber \\
    q^2 & = q_{\perp}^2 + q_z^2 = q_x^2 + q_y^2 + q_z^2
\end{align}

where $\tilde{f}(q_{\perp})= 2 \pi \int_0^{\infty} r dr f(r)
J_0(q_{\perp} r)$ is the Hankel transform of the normalized power
distribution over the mirror surface.

In this section we want to give just an estimation of the coating
thermo-refractive noise reduction using a flat beam instead of a
Gaussian beam and for this purpose we will approximate the real
mesa beam as a perfect Flat Top beam

\begin{equation}\label{FlatTop}
    f_{FT}(r)=\left\{
\begin{array}{ll}
    \frac{1}{\pi b^2} & \quad \hbox{$r \leq b$} \\
    \\
    0 & \quad \hbox{$r > b$} \\
\end{array}
\right.
\end{equation}

For Gaussian and Flat Top beam the Hankel transform can be
analytically performed but the integral in \eqref{SpecDensT} must
be numerically evaluated (unless other approximations are
accepted, as done in REF). For this comparison we chose a value of
$b=4 w_0, \,(w_0 = 2.6 cm)$ which correspond to the ``standard''
radius of the integration disc for Mesa Beam and compare this
ideally flat beam with the Advanced LIGO Gaussian baseline design
$w \simeq 6 cm$. The displacement noise is reduced by a factor
$1.7$ in the case of a Flat Top beam.

\begin{equation}\label{TRref-red}
    \sqrt{\frac{S_X^{GB}}{S_X^{FT}}} \Bigg |_{f= 100 Hz}\simeq 1.7
\end{equation}

\subsection{Results and discussion}

\begin{figure}[htb]
\begin{center}
\includegraphics[width=0.7\textwidth]{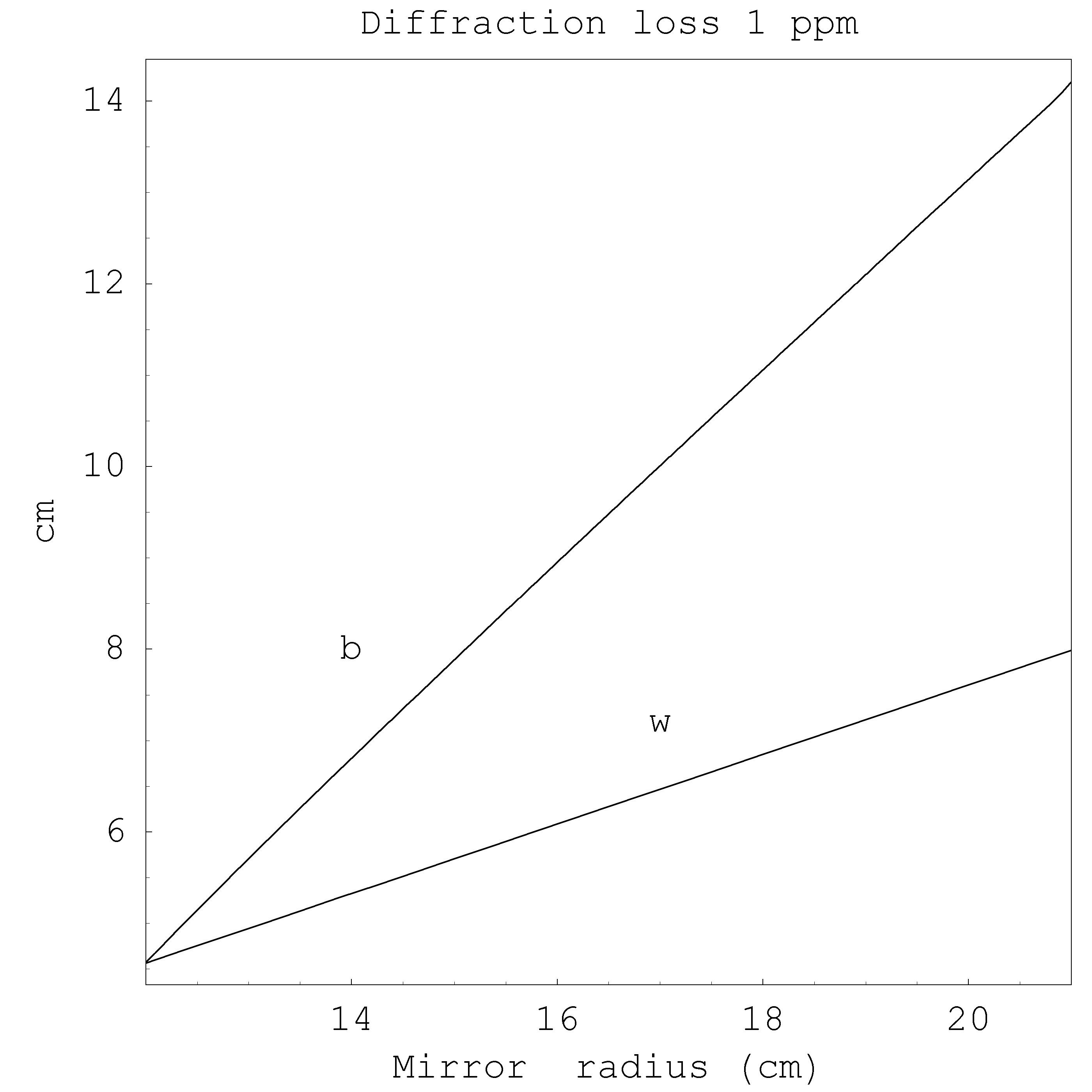}
\end{center}
\caption{Mesa beam and Gaussian parameters for Ad-LIGO like
cavity.} \label{fig:MB-G-radius}
\end{figure}

 This is a scheme of the calculations performed with TNN
for the analysis of the expected mirror thermal noise in Ad-LIGO
interferometer:
\begin{itemize}
    \item we fixed the mirror mass at $40 \,Kg$, constrained by the Advanced
LIGO suspension system design~\cite{AdLIGORefDes}.
    \item We fixed the diffraction loss constraint at $1\, ppm\, (10^{-6})$ for both the
    Gaussian and the Mesa beam, calculating the diffraction losses with the so
     called clipping approximation; in this approximation the losses are computed
      by the amount of light that falls outside the mirror and the beam profile is
       assumed to retain its shape even though the diffraction from the edge of the mirror.
    \item We changed the mirror radius from $12\, cm$ to $21 \,cm$ (with $1\, cm$ per step) and at the
     same time we increased the Gaussian beam radius $w$ and the Mesa beam integration disc
      radius $b$ to satisfy the $1\, ppm$ constraint for diffraction losses. The corresponding values are shown in Fig.~\ref{fig:MB-G-radius}. The thickness $H$ of the
       mirror is reduced correspondingly to satisfy the total mass constraint. In this way
        all the geometric parameters in the problem are functions of the mirror radius $a$.
\end{itemize}

\subsubsection{Fused silica substrate}

\begin{figure}[htbp]
  \centering
  \subfigure[Fused Silica (FS) substrate.]{
     \label{fig:subfig:SBFS}
     \includegraphics[width=0.8\textwidth]{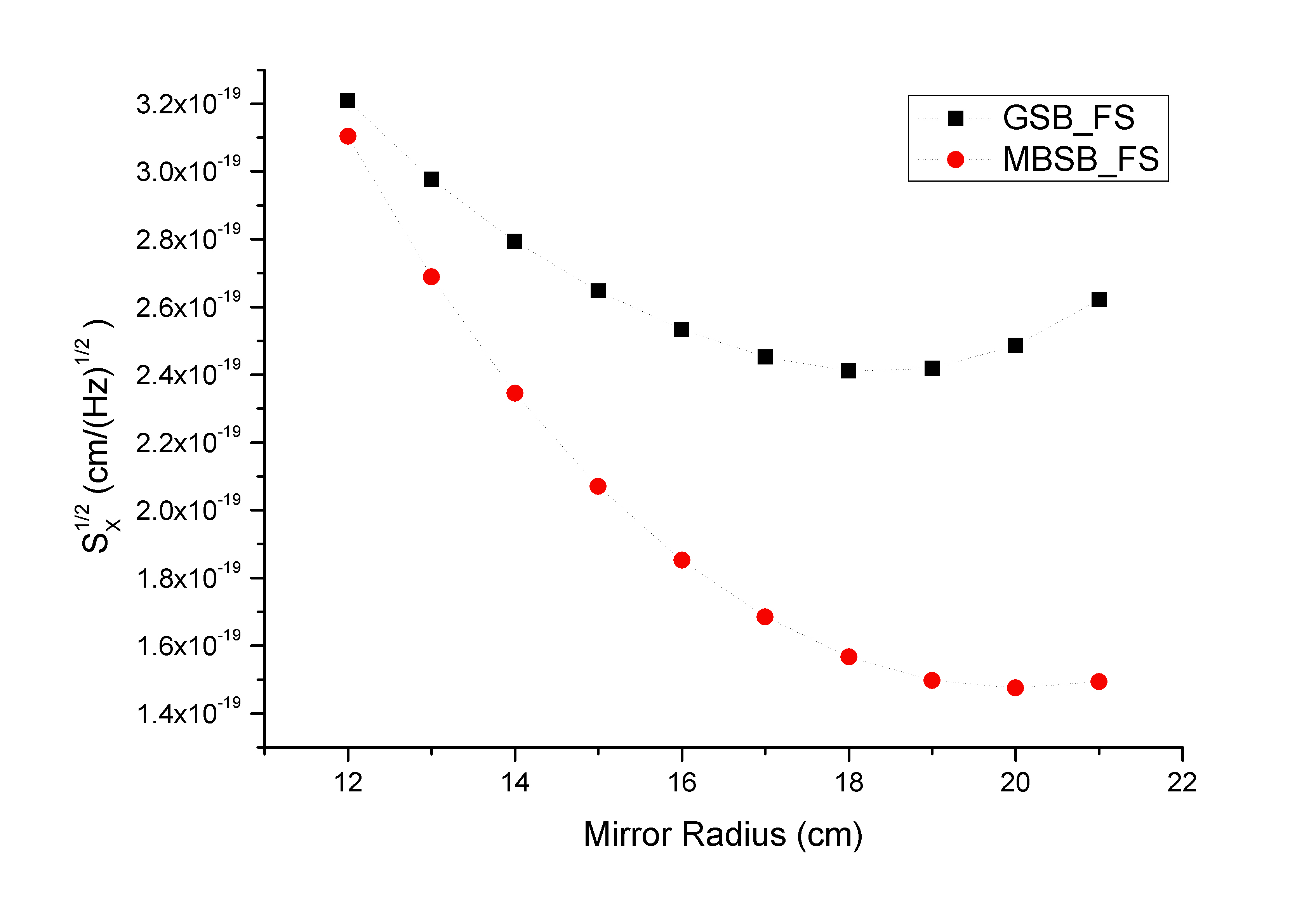}}
     \hspace{0.1in}
\subfigure[Sapphire (S) substrate.]{
      \label{fig:subfig:SBS}
      \includegraphics[width=0.8\textwidth]{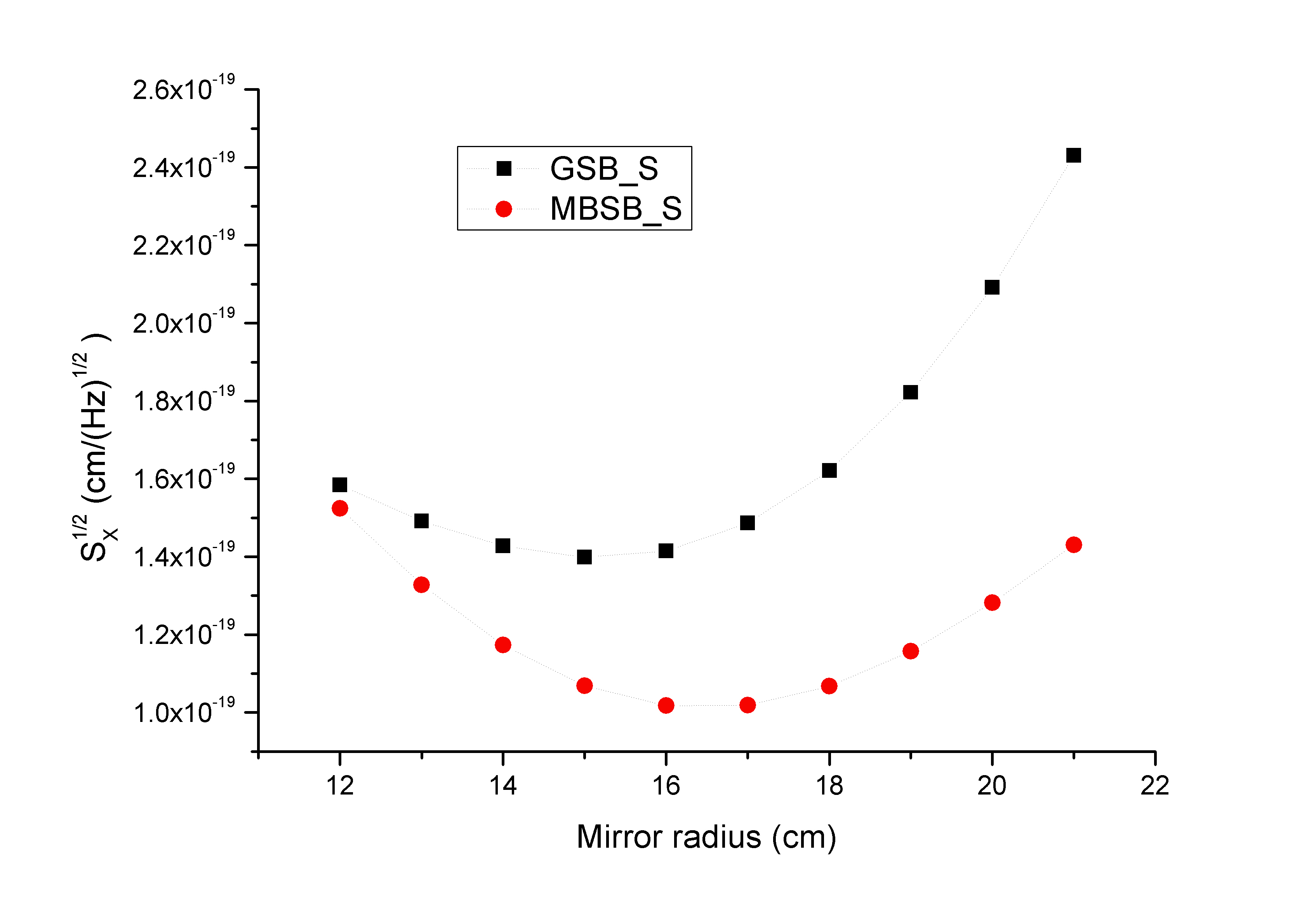}}
 \caption{Substrate Brownian (SB) thermal noise.}
 \label{fig:SB}
 \end{figure}

\begin{figure}[htbp]
  \centering
  \subfigure[Fused Silica (FS) substrate.]{
     \label{fig:subfig:STFS}
     \includegraphics[width=0.8\textwidth]{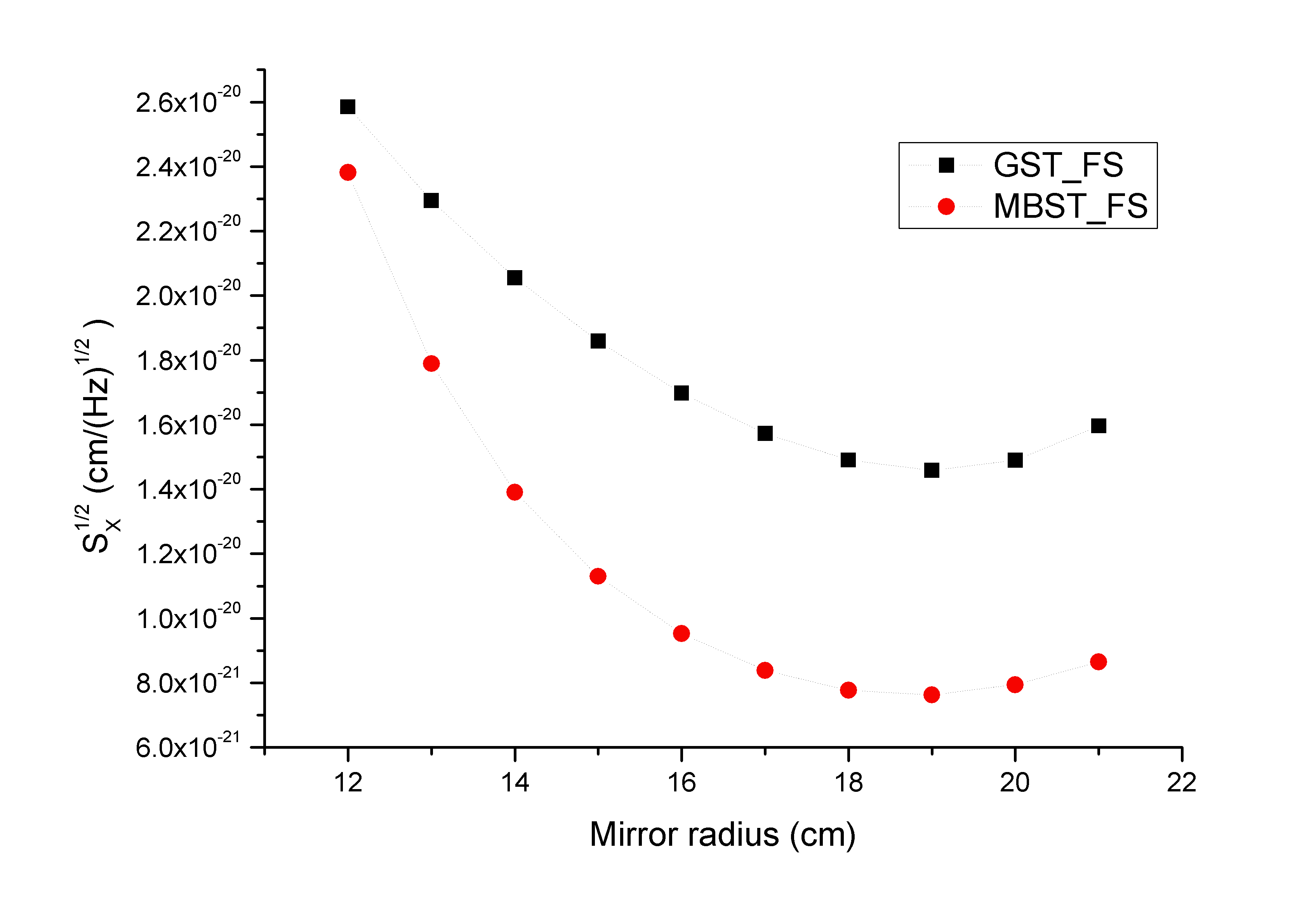}}
     \hspace{0.1in}
\subfigure[Sapphire (S) substrate.]{
      \label{fig:subfig:STS}
      \includegraphics[width=0.8\textwidth]{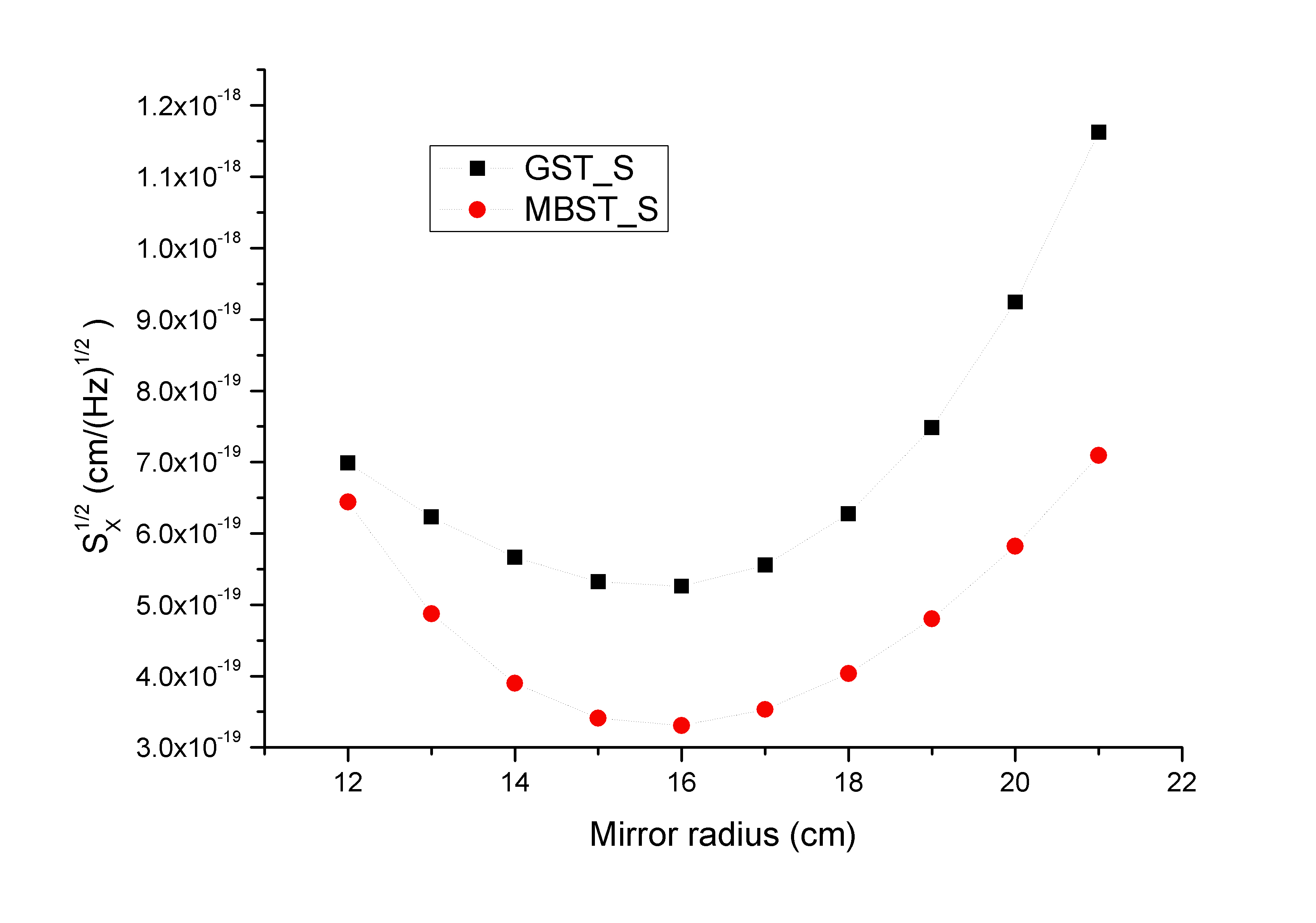}}
 \caption{Substrate thermoelastic (ST) thermal noise.}
 \label{fig:ST}
 \end{figure}

\begin{figure}[htbp]
  \centering
  \subfigure[Fused Silica (FS) substrate.]{
     \label{fig:subfig:CBFS}
     \includegraphics[width=0.8\textwidth]{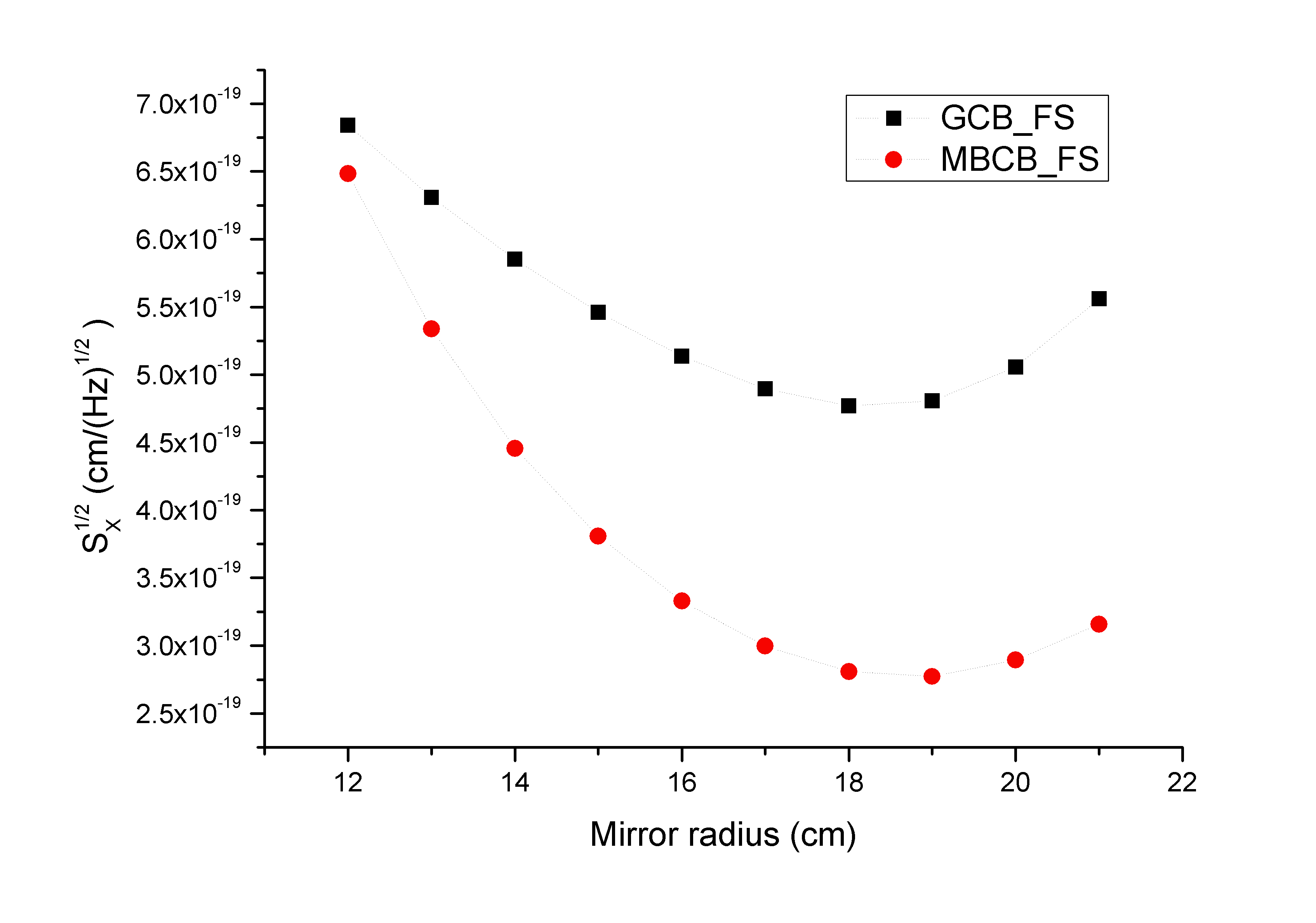}}
     \hspace{0.1in}
\subfigure[Sapphire (S) substrate.]{
      \label{fig:subfig:CBS}
      \includegraphics[width=0.8\textwidth]{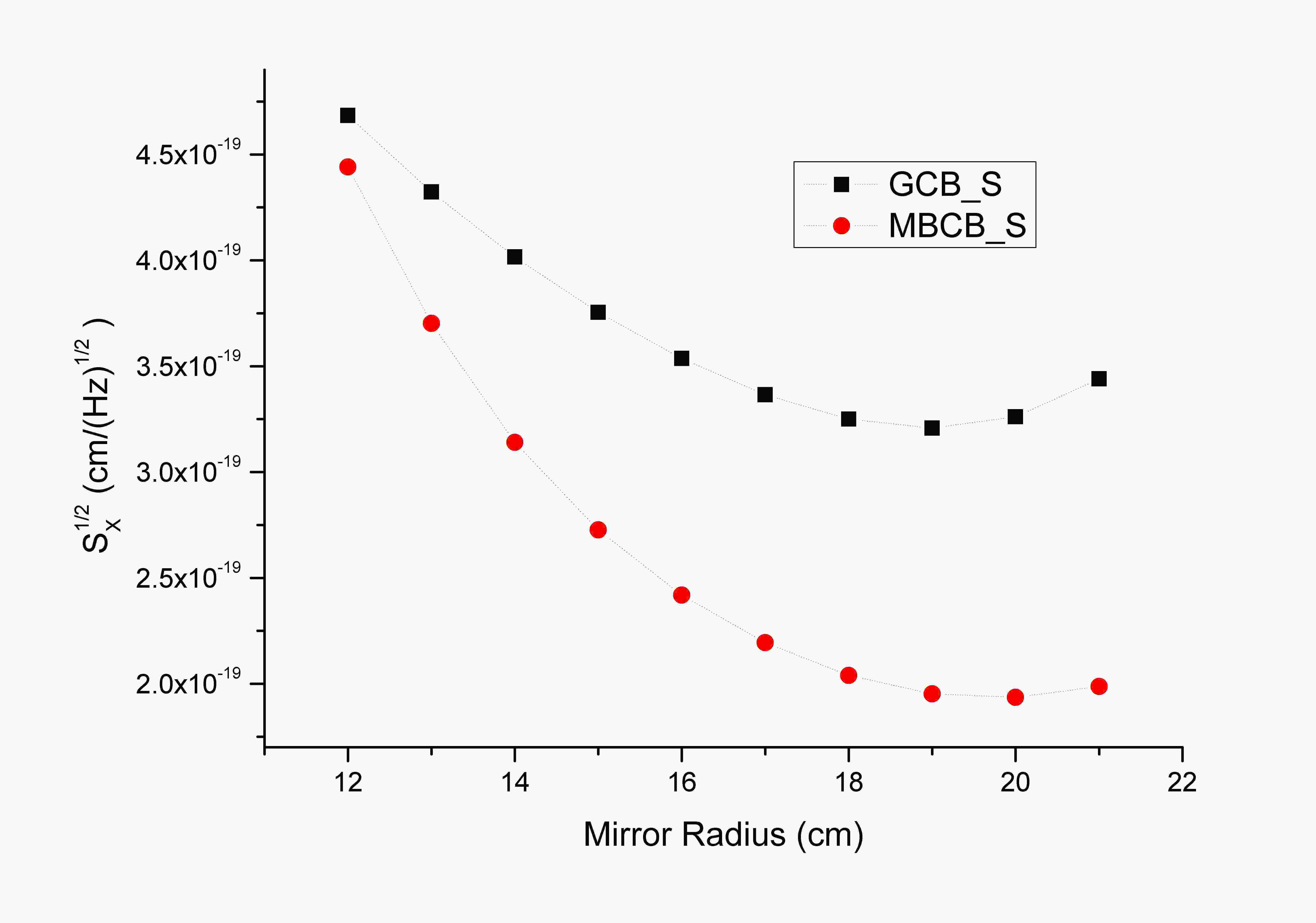}}
 \caption{Coating Brownian (CB) thermal noise.}
 \label{fig:CB}
 \end{figure}

\begin{figure}[htbp]
  \centering
  \subfigure[Fused Silica (FS) substrate.]{
     \label{fig:subfig:CTFS}
     \includegraphics[width=0.8\textwidth]{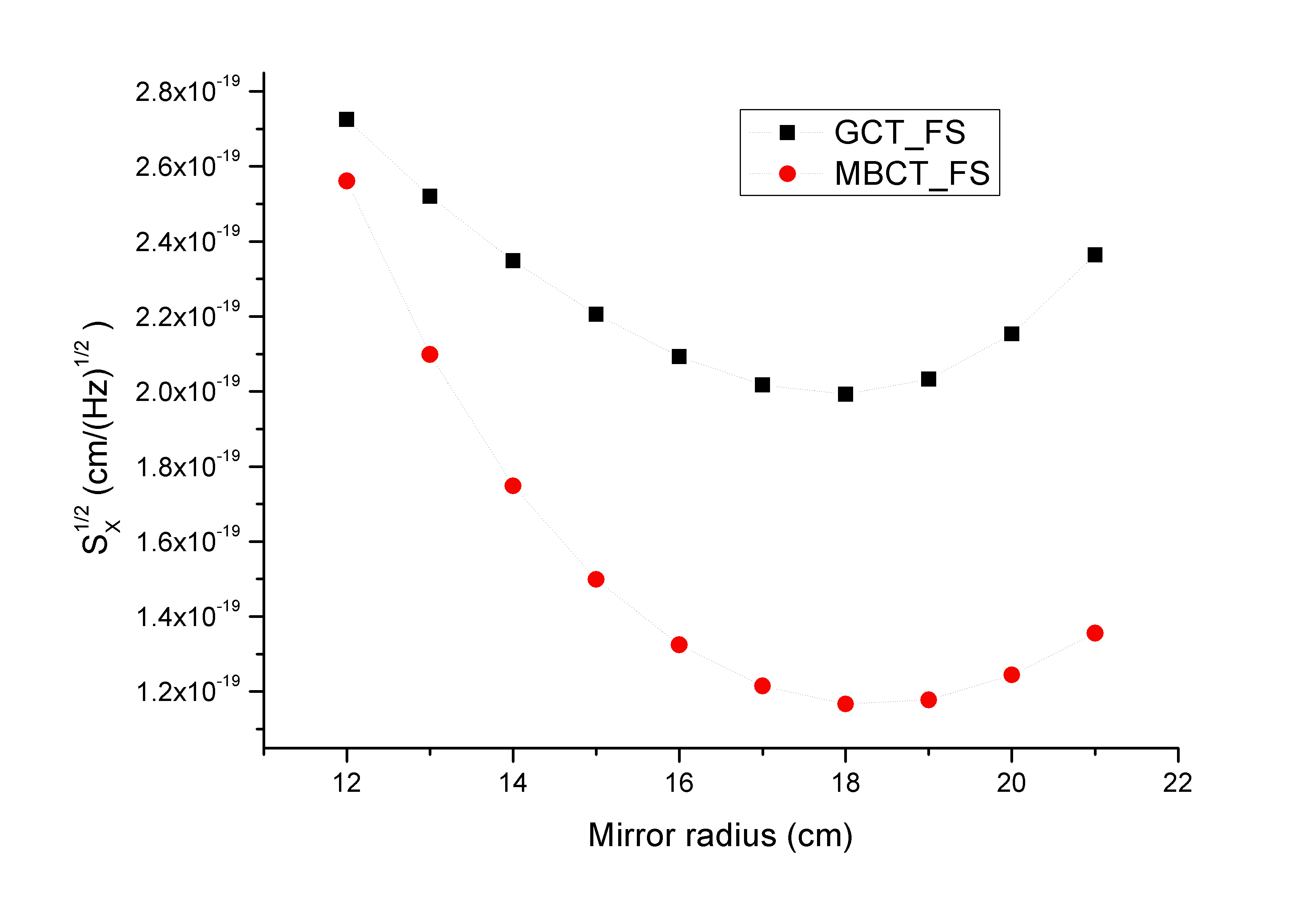}}
     \hspace{0.1in}
\subfigure[Sapphire (S) substrate.]{
      \label{fig:subfig:CTS}
      \includegraphics[width=0.8\textwidth]{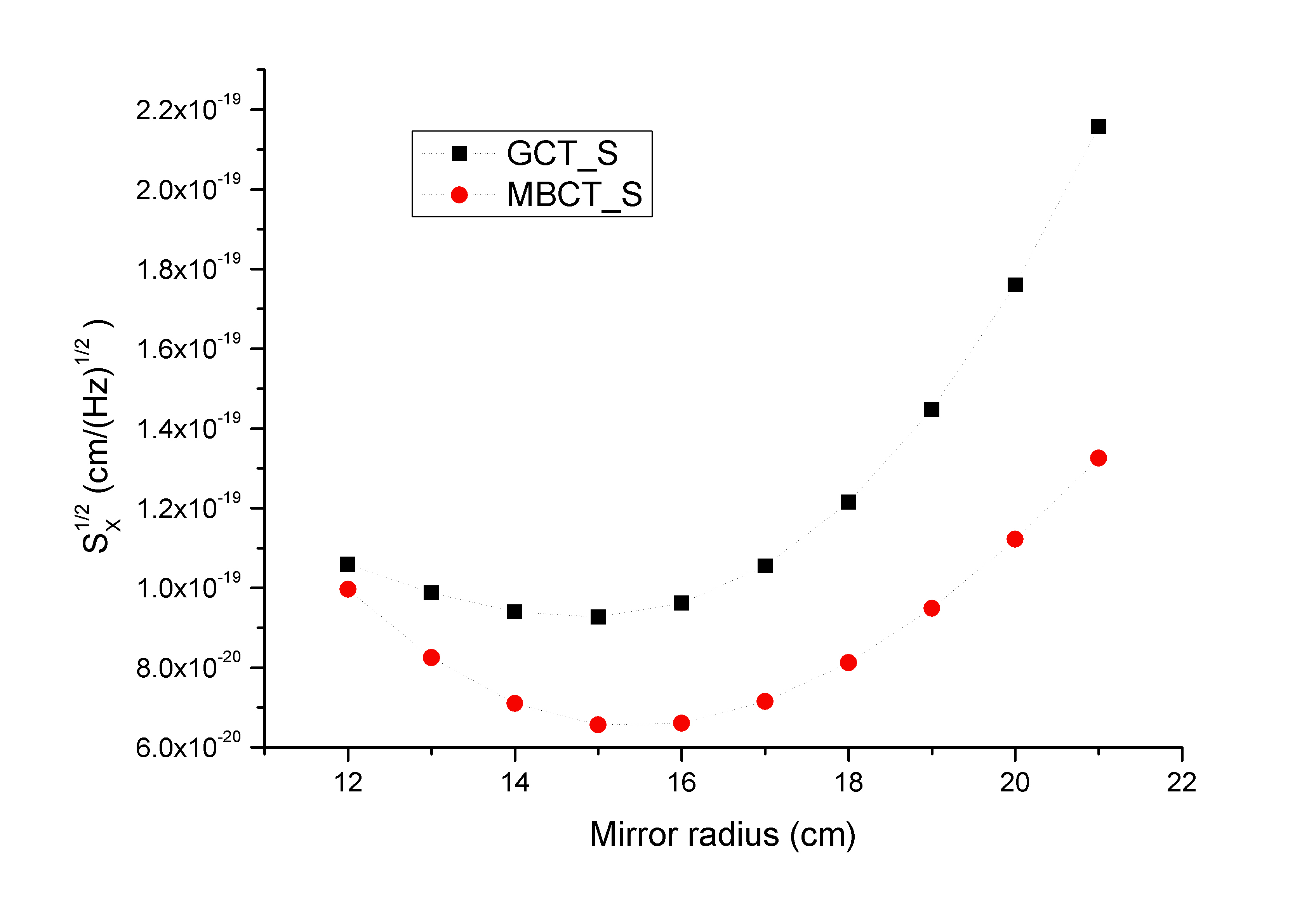}}
 \caption{Coating thermoelastic (CT) thermal noise.}
 \label{fig:CT}
 \end{figure}

Figures~\ref{fig:subfig:SBFS}, \ref{fig:subfig:STFS},
\ref{fig:subfig:CBFS}, \ref{fig:subfig:CTFS} show the displacement
noise of all the analyzed thermal noise contributions, for
Gaussian and Mesa beam in the case of Fused Silica substrate. The
dominant noise, the coating Brownian thermal noise, undergoes a
reduction of a factor $1.7$   for a mirror radius of $18$ cm. The
substrate thermal noise is reduced by a factor $1.55$  , whereas
the coating thermo-elastic and the substrate thermo-elastic are
reduced by factor $1.7$ and $1.9$ respectively.

\clearpage

It is interesting to note that all the thermal noise contributions
present a minimum in the finite cylindrical model. The minimum of
the sum of all contributions represents the best choice for the
mirror and beam dimensions. The descent part of the noise curves
reflects the basic idea that increasing the beam radius the noise
will get lower and the rising part of the curves can be explained
heuristically by the fact that all the noises contributions are
related somehow to the amplitude of the elastic deformation of the
test mass under a surface pressure and this effect is larger in
gong-shaped mirrors than in bar-shaped ones.

\begin{figure}[htb]
\begin{center}
\includegraphics[width=0.7\textwidth]{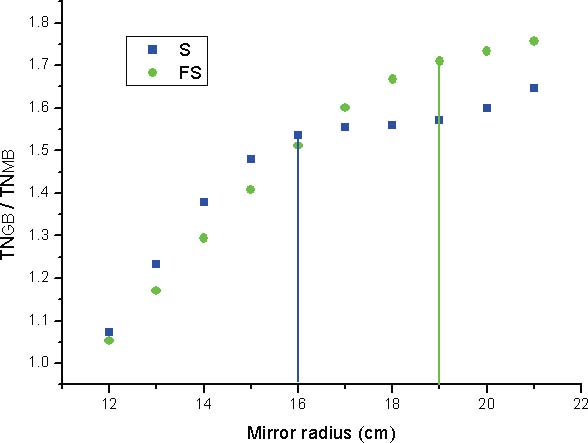}
\end{center}
\caption{Ratios between the total thermal noise with Gaussian and
mesa beams. The vertical lines correspond to the minima of the
noise curves.} \label{fig:TNGBMB}
\end{figure}

 From the incoherent sum of the four contributions we have that minimum thermal
noise occurs for a mirror radius of about 18 cm for Gaussian beam
and for about 19 cm for Mesa beam. The corresponding mirror aspect
ratios ($2\,a/H$ ) are $2$ for Gaussian beam and $2.4$ for Mesa
beam.  The aspect ratio chosen for Advanced LIGO baseline is
$1.7$. For this geometry the gain in sensitivity is about a factor
1.7 switching from Gaussian to Mesa beam at the minima of thermal
noise (see Fig.~\ref{fig:TNGBMB}).

\begin{figure}[htb]
\begin{center}
\includegraphics[width=0.9\textwidth]{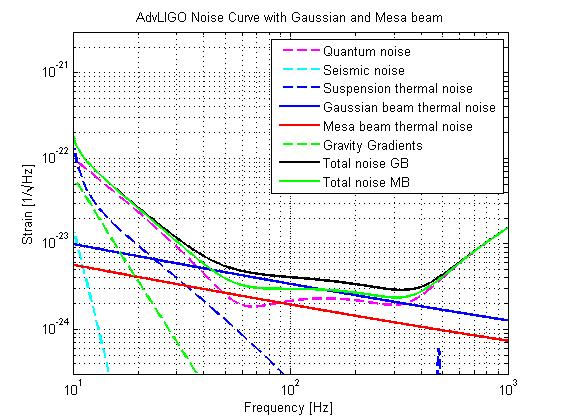}
\end{center}
\caption{Comparison between the Ad-LIGO expected sensitivity with
Gaussian or mesa beams.} \label{fig:AdLGMB}
\end{figure}

   Fig.\ref{fig:AdLGMB} shows the expected sensitivity of Advanced LIGO interferometer
    in the Gaussian or Mesa beam configuration calculated with \textit{Bench}\footnote{Bench is a simulation program available at \url{http://emvogil-3.mit.edu/bench/}}. The estimated range
     for NS-NS binary systems increases from $177 Mpc$ for Gaussian beam to $228 Mpc$ for Mesa beam. This is a
remarkable factor if we consider that we didn't optimize the other
interferometer's parameters to take full advantage of the reduced
mirror thermal noise floor. In this evaluation the
thermo-refractive noise of the coating was taken into account as
discussed in Sec.~\ref{sec:TRFB}.

\subsubsection{Sapphire substrate}

We have conducted the same kind of analysis for mirrors with
Sapphire substrate. This study was made mainly for historical
reasons as the idea of using sapphire substrates for the mirrors
is now abandoned. In this case the dominant noise is the substrate
thermo-elastic contribution. Some of the advantages of using Mesa
bean have been already analyzed in~\cite{oshaug} for this
particular noise source. Here we compute the various thermal noise
contribution for finite size test mass and show the relative gain
for each thermal noise employing a Mesa beam instead of a standard
Gaussian. For $40$ Kg sapphire substrates the minimum of thermal
noise occurs for a mirror radius of about 16 cm. The corresponding
mirror aspect ratio ($2\,a/H$) is about $2.6$. The total thermal
noise reduction for Mesa beam is about a factor $1.55$ around the
minimum.

\subsection{Further consideration }

\subsubsection{Finite size effect}

\begin{figure}[htbp]
  \centering
  \subfigure[Fused Silica (FS) substrate.]{
     \label{fig:subfig:FTMcorr}
     \includegraphics[width=0.8\textwidth]{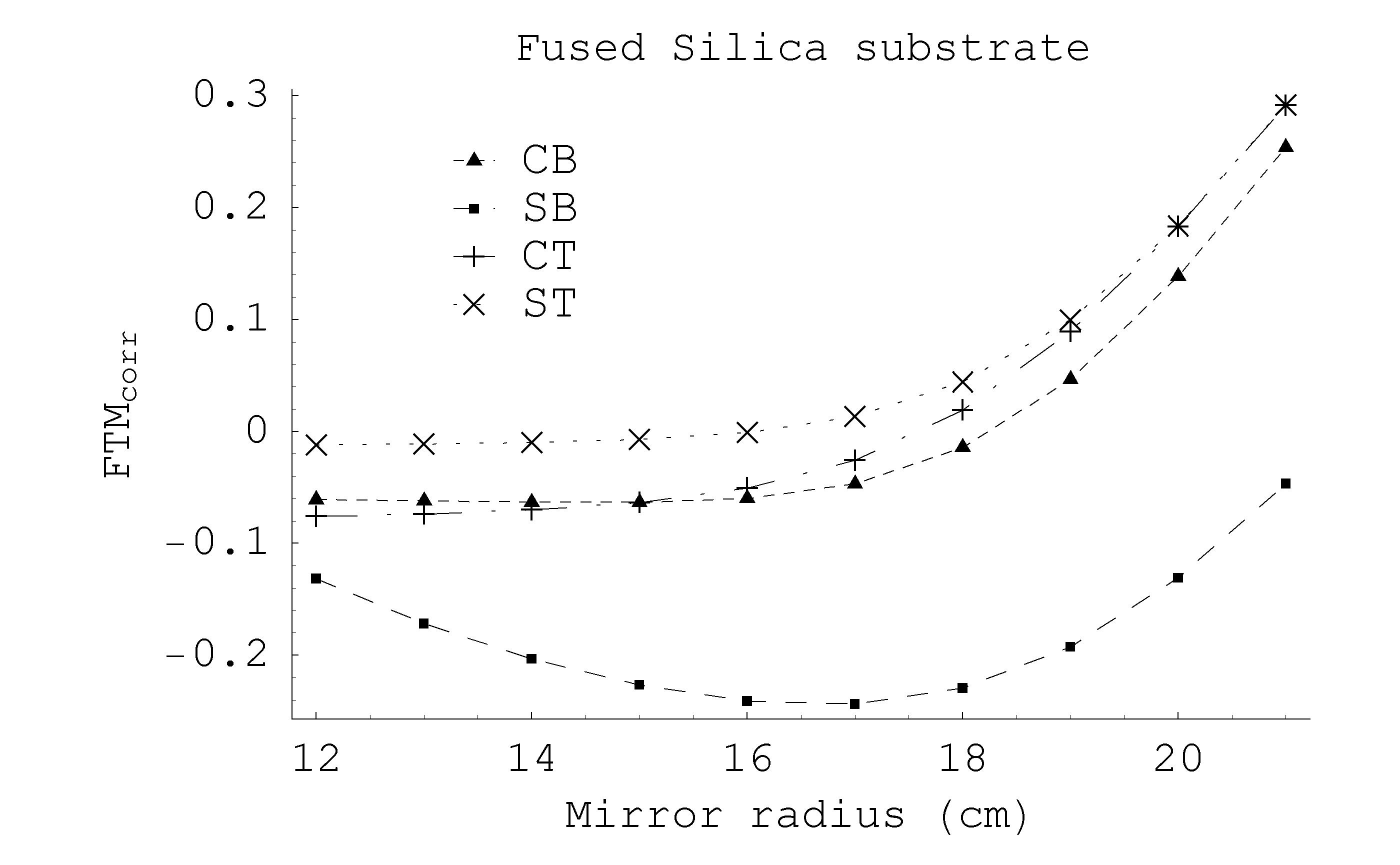}}
     \hspace{0.1in}
\subfigure[Sapphire (S) substrate.]{
      \label{fig:subfig:SFTMcorr}
      \includegraphics[width=0.8\textwidth]{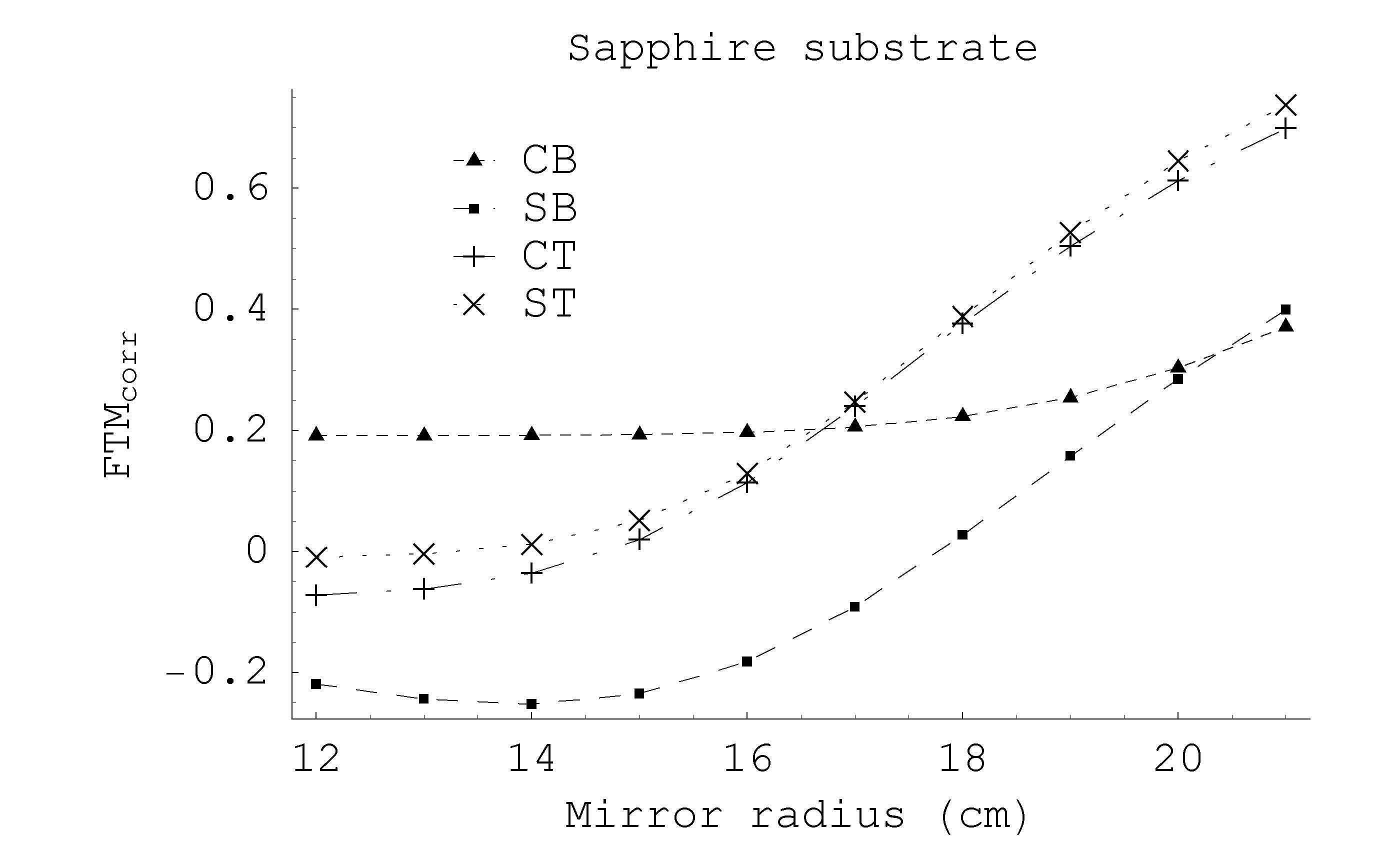}}
 \caption{FTM effects for Gaussian beam ($1$ ppm diff. loss).}
 \label{fig:FTMcorr}
 \end{figure}

 We first want to give a quantitative estimation of the finite
test mass (FTM) effect on the different kinds of thermal noise. In
Fig.~\ref{fig:FTMcorr} we plot the fractional correction provided
by our analysis with respect to the formulas found in literature
for Gaussian beams in the semi-infinite mirror case. For mirror
radius of $18$ cm, which provides the minimum of thermal noise,
the correction is around $20\%$ for substrate Brownian and less
than $10 \%$ for the other contributions.

Lovelace~\cite{Lovelace} performed the computation of simple
scaling laws for thermal noise with arbitrary beam shapes and
semi-infinite mirrors. He also analyzed the finite test mass
correction provided by our results with respect to his derivation
in the case of mesa beam and the conclusions are similar to the
FTM effect for Gaussian beam.

\clearpage

\subsubsection{High order LG modes}

It has been recently proposed~\cite{vinetLG} the use of high order
LG beams and excited modes of meas beams to reduce the thermal
noise in interferometric GW detectors. The problem is that
although these kind of beam provide a comparable reduction of the
substrate thermal noises with respect to the mesa beam, they are
sensibly worse than the mesa beam for the dominant coating thermal
noises.  The radial nodes of the higher order LG beam, cause
rapidly variation of the elastic field near the coating surface.
This produces an higher elastic energy density stored in the
coating, which boosts the coating Brownian thermal noise, and
higher thermal gradients, which are responsible for the
thermoelastic noise. To substantiate these considerations we shown
a comparison in the elastic energy distribution inside the mirror
between a standard Gaussian beam, a mesa beam and a $LG_{50}$
mode, all with $1$ ppm of diffraction losses.

 \begin{figure}[htbp]
  \centering
  \subfigure[Elastic energy distribution.]{
     \label{fig:subfig:GELEn}
     \includegraphics[width=0.45\textwidth]{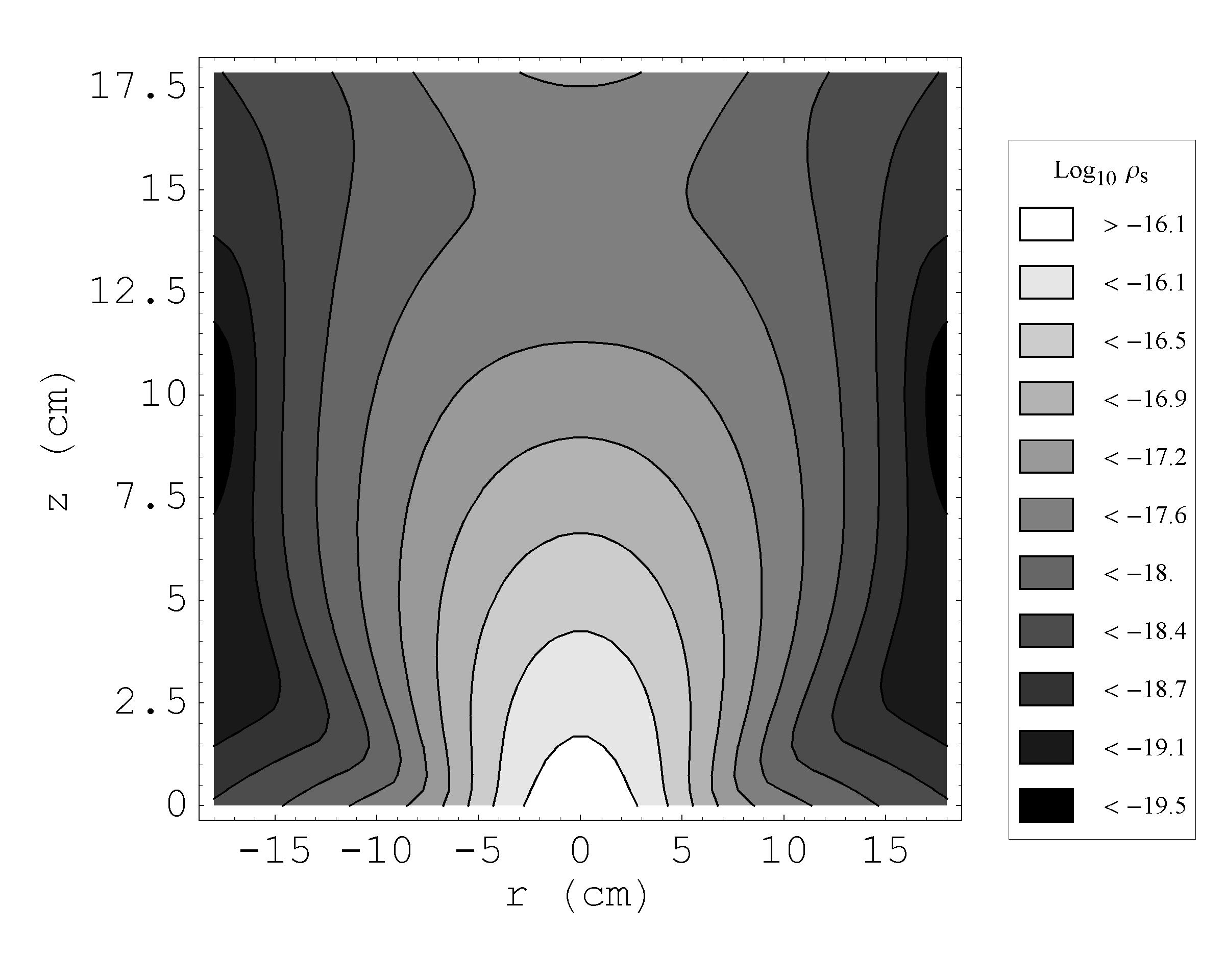}}
     \hspace{0.1in}
\subfigure[Coating elastic energy density.]{
      \label{fig:subfig:GdensC}
      \includegraphics[width=0.45\textwidth]{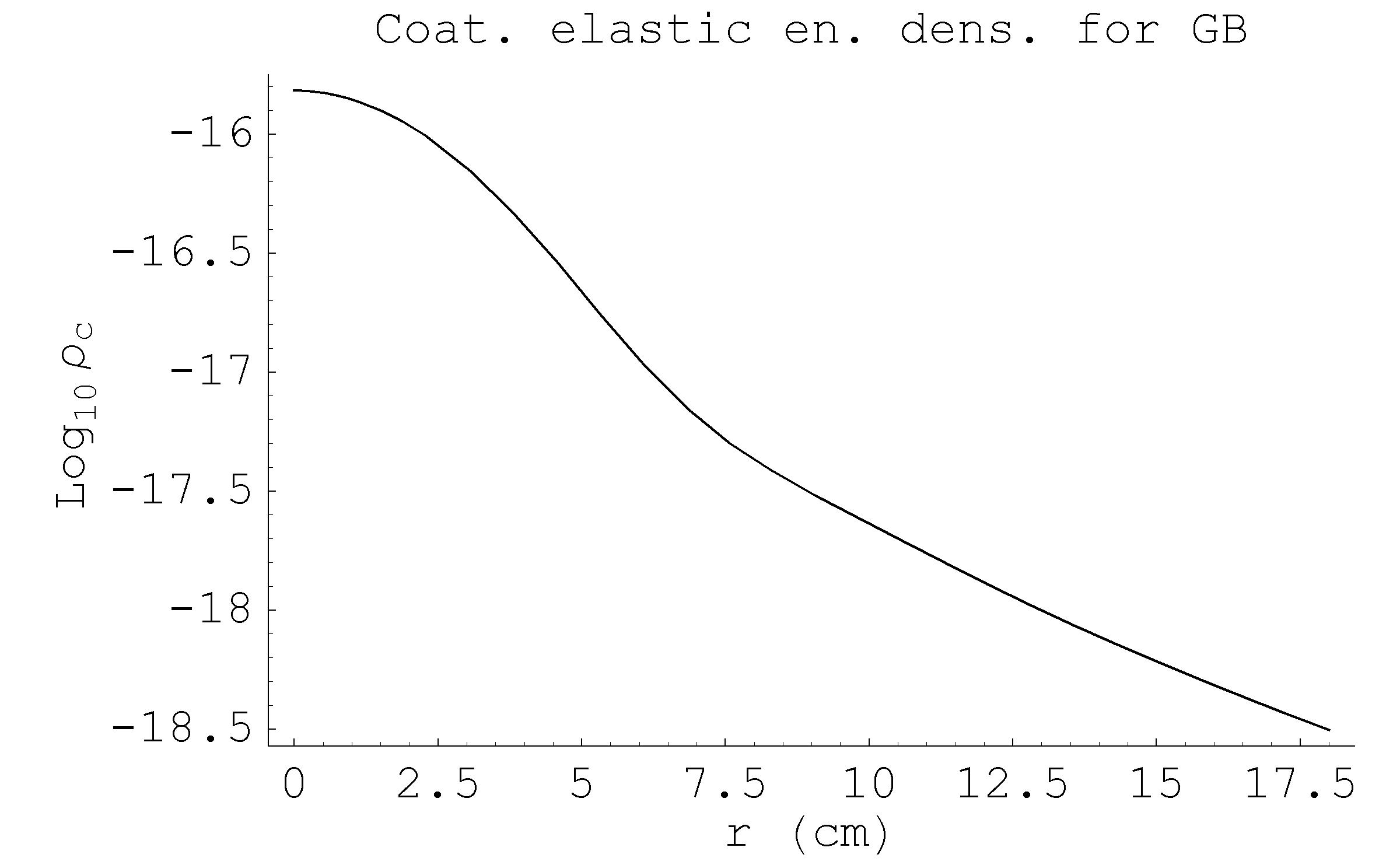}}
 \caption{Elastic energy for a Gaussian beam with $w=6.85$ cm.}
 \label{fig:Gelen}
 \end{figure}

\begin{figure}[htbp]
  \centering
  \subfigure[Elastic energy distribution.]{
     \label{fig:subfig:GMBELEn}
     \includegraphics[width=0.45\textwidth]{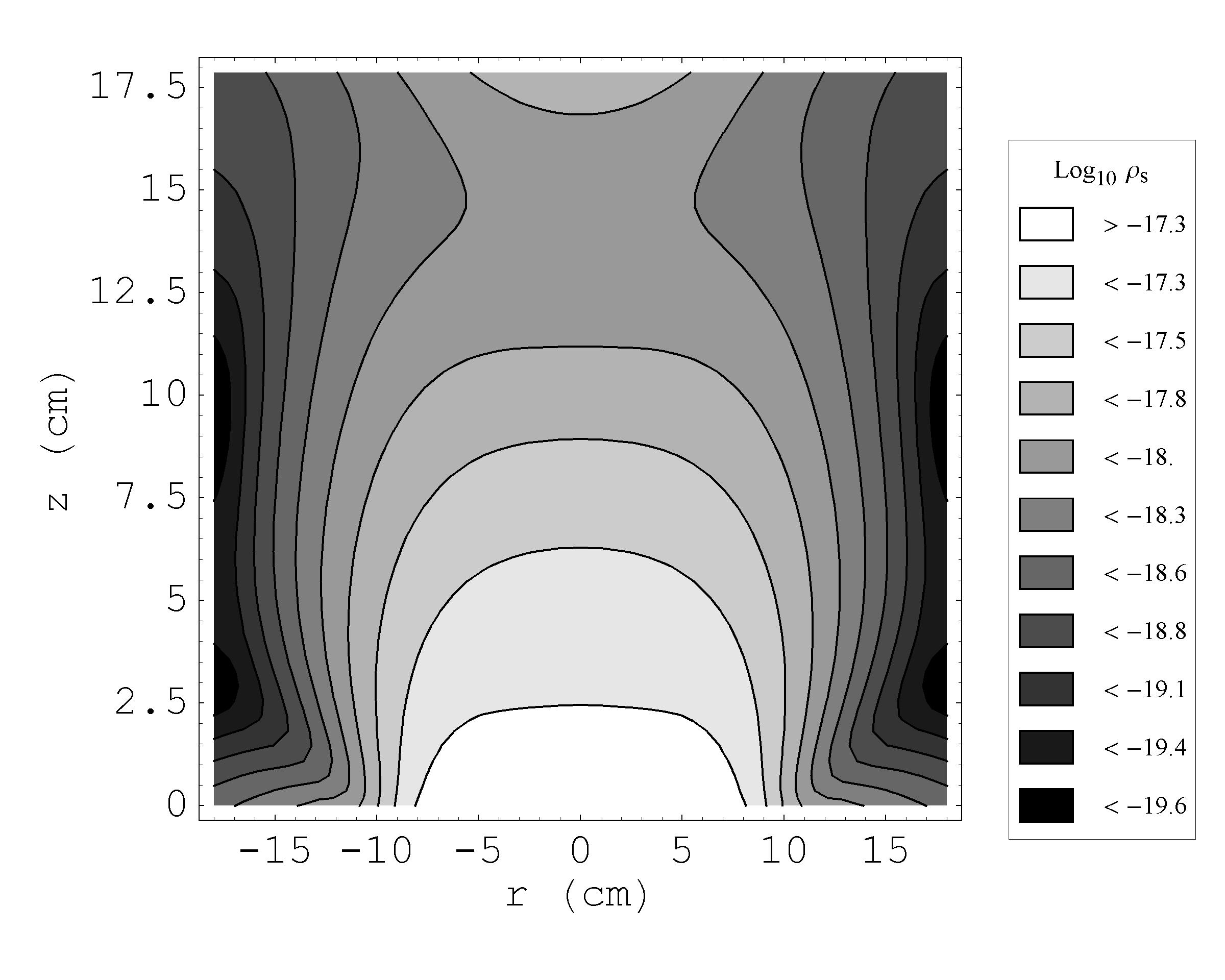}}
     \hspace{0.1in}
\subfigure[Coating elastic energy density.]{
      \label{fig:subfig:MBdensC}
      \includegraphics[width=0.45\textwidth]{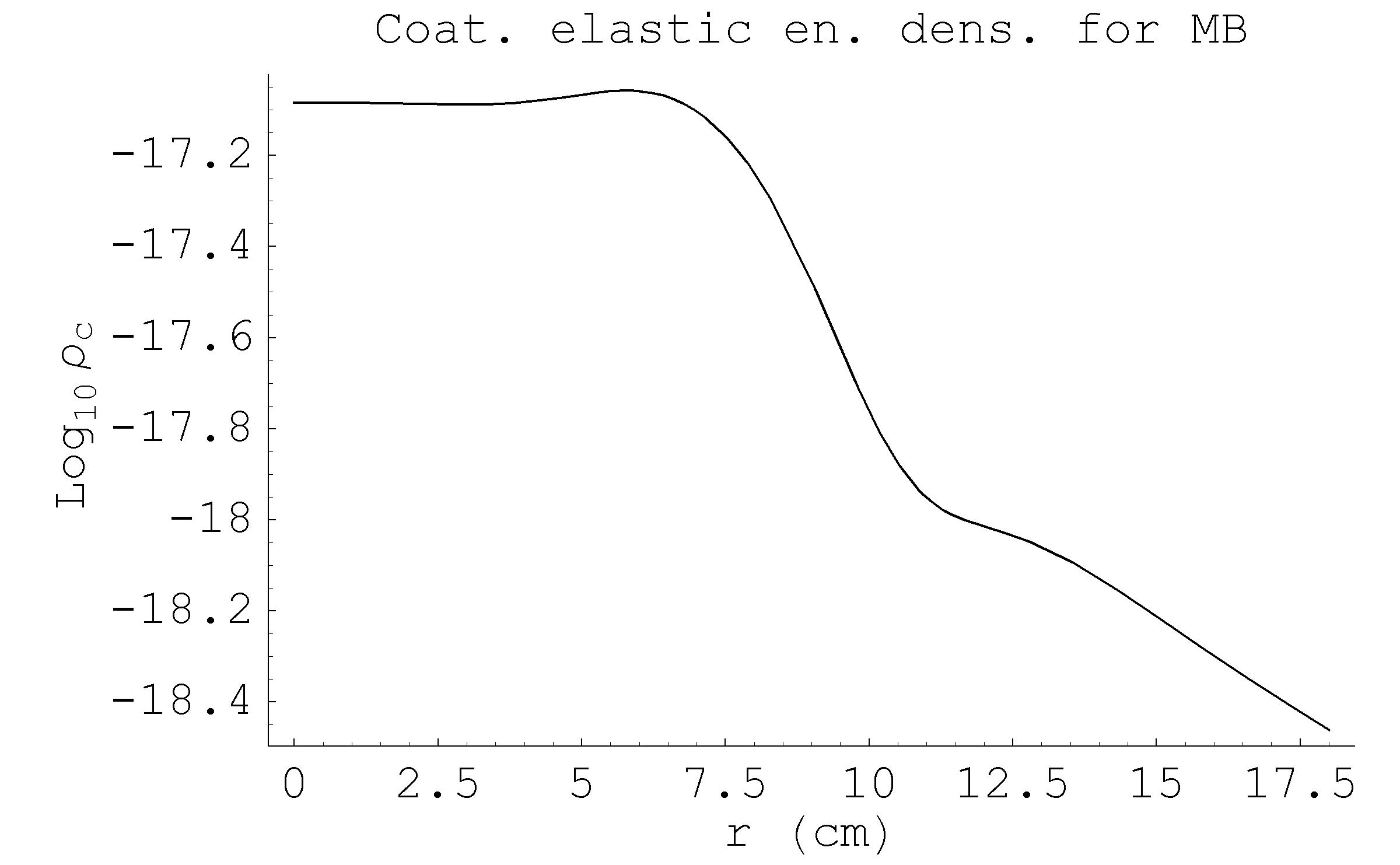}}
 \caption{Elastic energy for a mesa beam with $b=11$ cm.}
 \label{fig:MBelen}
 \end{figure}

\begin{figure}[htbp]
  \centering
  \subfigure[Elastic energy distribution.]{
     \label{fig:subfig:LGELEn}
     \includegraphics[width=0.45\textwidth]{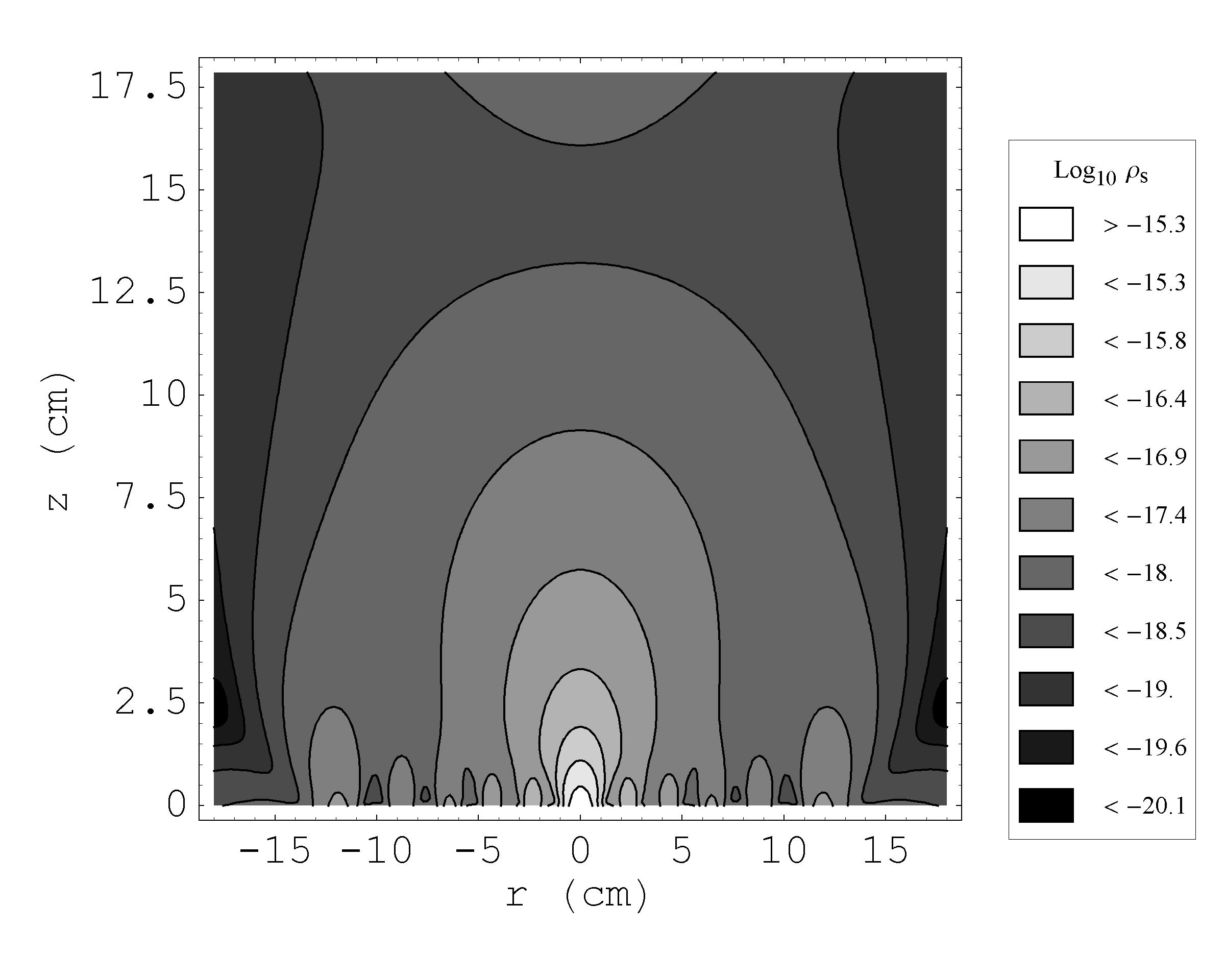}}
     \hspace{0.1in}
\subfigure[Coating elastic energy density.]{
      \label{fig:subfig:LGdensC}
      \includegraphics[width=0.45\textwidth]{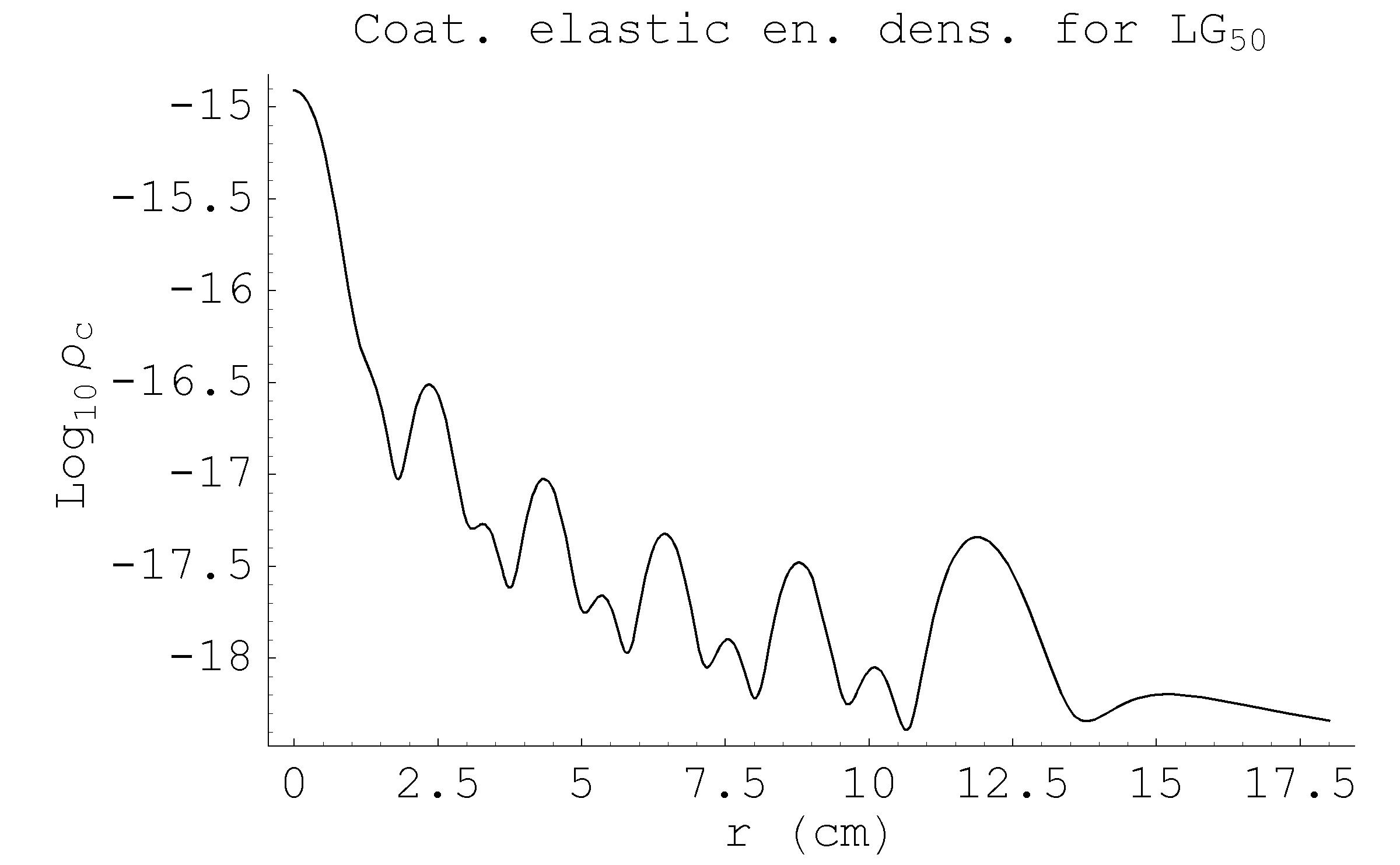}}
 \caption{Elastic energy for $LG_{50}$ with $w=4$ cm.}
 \label{fig:LGelen}
 \end{figure}

\begin{table}\label{tab:LGnoise}
\centering
\begin{tabular}{|c|c|c|c|}
  \hline
  Noise & GB & MB & LG$_{50}$ \\
  \hline
  CB & $4.72\cdot 10^{-19}$ & $2.8\cdot 10^{-19}$ & $3.2\cdot 10^{-19}$ \\
  CT & $1.99\cdot 10^{-19} $& $1.17\cdot 10^{-19}$ & $1.95\cdot 10^{-19}$ \\
  \hline
\end{tabular}
\caption{Noise calculation for GW sensitivity at $100$ Hz. Units
$cm/\sqrt{Hz}$}
\end{table}

\begin{figure}[htb]
\begin{center}
\includegraphics[width=0.7\textwidth]{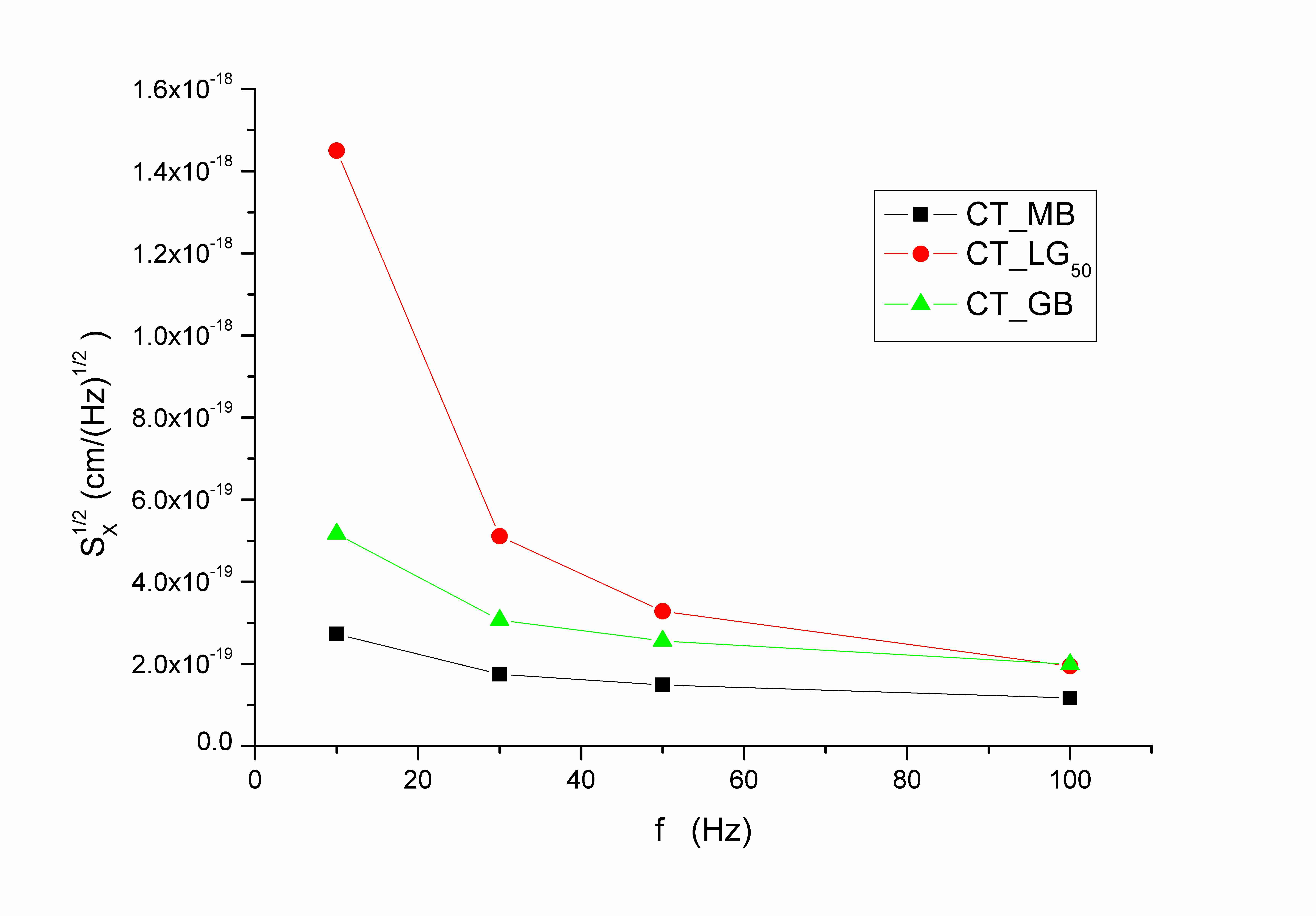}
\end{center}
\caption{Comparison between the coating thermo-elastic noise
trends with frequency.} \label{fig:LGthermoelastic}
\end{figure}

Because of the ripples in the elastic energy distribution, we
expect the coating thermoelatic noise for the higher order
Laguerre Gauss beams to become large in the low GW signal
frequency regime.  The characteristic thermal diffusivity length
$r_T= \sqrt{\frac{\kappa}{C_V \omega}}$ increases as the
sensitivity frequency decreases, in such a way that the
temperature fluctuations involves a larger volume of the test
mass. In Fig.~\ref{fig:LGthermoelastic} we show the comparison
between the Gaussian, mesa, and $LG_{50}$ beams with respect to
the coating thermoelatic noise. At low frequency ($\sim 10$ Hz)
the $LG_{50}$ noise is almost three times larger than for Gaussian
beam, which remains a factor of two above the mesa beam noise.

\clearpage

\section{Optimized coating}

The limit sensitivity of interferometric gravitational wave
antennas is set by the thermal noise generated in the dielectric
mirror coatings. These coatings are currently made of alternating
low-loss quarter-wavelength high/low index material layers. This
design yields the maximum reflectivity for a fixed number of
layers, but not the lowest noise for a prescribed reflectivity.

Realization that much of the thermal noise in the coating comes
from the material of the high index layer\cite{Penn-loss2} has
lead to investigation of optimizing the individual layer
thicknesses while preserving the needed reflectivity. This
motivated our recent investigation\footnote{ In collaboration with
the Waves Group, Department of Engineering, University of Sannio,
Benevento, Italy. } of optimal coating configurations, to
guarantee the lowest thermal noise for a targeted reflectivity.
This communication provides a compact overview of our results,
involving truncated periodically-layered
configurations\footnote{More general analysis involving
nonperiodic genetically-engineered coatings has been presented
in~\cite{SPIEcoat}}. Possible implications for the advanced Laser
Interferometer Gravitational wave Observatory (LIGO) are
discussed.

All interferometers presently in operation use quarter-wavelength
(QWL) designs, which are known (Bragg theorem) to be optimal, in
the sense that they yield the largest reflectivity for any fixed
number of layers (or, equivalently, the smallest number of layers
for any prescribed reflectivity). However, QWL coatings do not
yield the minimum TN for a prescribed reflectivity, and hence are
not optimal for GW interferometers, where the quantity that should
be maximized is the visibility distance of the instrument, which
in its turn is limited by the coating thermal noise.

This section addresses the above optimization problem, and is
organized as follows. We outline our comprehensive working model
for the multilayer mirror reflectivity and Thermal noise. Than we
present an examples of optimized coatings, focusing on periodic
stacked-doublet multilayers, and compare them with the standard
QWL syntheses. Conclusions and recommendations follow in the end
of this section.

\subsection{Multilayer Coating Reflectivity}

\begin{figure}[htb]
\begin{center}
\includegraphics[width=0.7\textwidth]{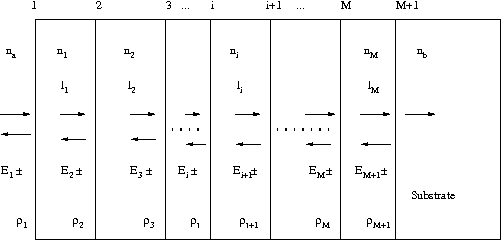}
\end{center}
\caption{Multilayer dielectric structure.} \label{fig:MultiL}
\end{figure}

We consider a planar multilayer dielectric coating composed of
alternating homogeneous layers (with variable thickness) of silica
($SiO_2$) and tantala ($Ta_2O_5$), illuminated by a
normally-incident plane wave, with implied time-harmonic
dependence. Referring to Fig.~\ref{fig:MultiL}, we consider $M$
layers, $M+ 1$ interfaces, and four dielectric media, including
the left and right semi-infinite media $a$ and $b$ (vacuum and
substrate).

 The incident and reflected
fields are considered at the left of each interface. The overall
reflection response, $\Gamma_1 = E_{1-}/E_{1+}$, can be obtained
recursively in a variety of ways, such as the propagation
matrices, the propagation of the impedances at the interfaces, or
the propagation of the reflection responses~\cite{Orfanidis}. The
elementary reflection coefficients $\rho_i$ from the left of each
interface are defined in terms of the refractive indices as
follows:

\begin{equation}\label{elem_refl}
    \rho_j=\frac{n_{j-1} - n_j}{n_{j-1}+ n_j}, \quad j= 1,\ldots ,
    M+1
\end{equation}

where $\rho_1=(n_a -n_1)/(n_a+n_1)$ and
$\rho_{M+1}=(n_M-n_b)/(n_M+n_b)$. The forward/backward fields at
the left of interface i are related to those at the left of
interface $j + 1$ by:

\[
\begin{bmatrix}
  E_{j +} \\
  E_{j -} \\
\end{bmatrix}
=\frac{1}{\rho_i +1}
\begin{bmatrix}
  e^{i k_j l_j} & \rho_j e^{- i k_j l_j} \\
  \rho_j e^{ i k_j l_j} &  e^{-i k_j l_j} \\
\end{bmatrix}
\begin{bmatrix}
  E_{j+1,+} \\
  E_{j+1,-} \\
\end{bmatrix} ,\quad j= M, M-1,\ldots,1
\]

where $k_j l_j$ is the phase thickness of the $j-th$ layer, which
can be expressed in terms of its optical thickness $n_j l_j$ and
the operating free-space wavelength by $k_j l_j = 2\pi(n_j
l_j)/\lambda$. The reflection responses $\Gamma_j = E_{j-}/E_{j+}$
will satisfy the recursions:

\begin{equation}\label{refl-i}
    \Gamma_j=\frac{\rho_j + \Gamma_{j+1} e^{-4\pi \frac{n_j l_j}{\lambda} }}{1+ \rho_j\Gamma_{j+1} e^{-4\pi \frac{n_j l_j}{\lambda}
    }},\quad j= M, M-1,\ldots,1
\end{equation}
and initialized by $\Gamma_{M+1} = \rho_{M+1}$. The mirror
reflectivity is given by $\Gamma\equiv \Gamma_1$.
It is useful for
the following analysis to introduce the scaled optical length
$z_j= n_j l_j/\lambda$.

The above formalism can be easily generalized~\cite{Orfanidis} to
the case of oblique incidence, via the introduction of
(polarization dependent) ``transverse'' wave-impedances and
refractive indexes.

\subsection{Multilayer  Coating Thermal Noise}

Since the function we want to minimize is the coating thermal
noise, we need the most accurate description of the thermal noise
for a multilayer structure. Eq.~\eqref{SX-B-COAT-INF-GB} has been
generalized by Harry~\cite{Harry-anisot} in order to incorporate
coating anisotropy into the thermal noise. A multi layer structure
composed of alternating layers of homogeneous materials exhibits a
mechanical behavior proper of a transversely isotropic material in
which the independent elastic coefficient in the stiffness
matrix~\eqref{stiffness-gen} are reduced to five by the rotational
symmetry around the axis orthogonal to the layers. $Y_{\parallel}$
and $Y_{\perp}$ are the Young modulus for stress causing strains
entirely within the plane parallel to the coating layers or
perpendicular to this one. There are also two Poisson ratios,
$\nu_{\parallel}$ for stresses and strains both with the plane
parallel to the coating layers, and $\nu_{\perp}$ for when either
the stress or the strain is perpendicular to the coating layers.
The displacement induced noise is given by

\begin{equation}
S_X(\omega)= \frac{4 K_B T}{\sqrt{\pi} \omega} \frac{1- \nu_s^2}{w
Y_s} \phi_{eff}^{coat}
\end {equation}

where the loss angle $\phi_{eff}^{coat}$ depends on the coating
thickness $d$, on the beam radius $w$, and on the loss angles in
the parallel and perpendicular direction with some coefficients
dependent on the elastic coefficient of the coating and substrate

\begin{eqnarray}
\phi_{eff}^{coat}&=& \frac{d}{\sqrt{\pi}w}\frac{1}{Y_{\perp}}
\Bigg( \Big( \frac{Y_s}{1-\nu_s^2}-\frac{2 \nu_{\perp}^2 Y_s
Y_{\parallel} }
{Y_{\perp}(1-\nu_s^2)(1-\nu_{\parallel})}\Big) \phi_{\perp} \nonumber\\
&+&\frac{Y_{\parallel}\nu_{\perp}(1-2\nu_s)}{(1-\nu_{\parallel})(1-\nu_s)}
(\phi_{\parallel}-\phi_{\perp})\nonumber \\
&+& \frac{Y_{\parallel} Y_{\perp}(1+\nu_s)(1-2\nu_s)^2}{Y_s
(1-\nu_{\parallel}^2)(1-\nu_s)} \phi_{\parallel}\Bigg)
\end{eqnarray}

In the paper~\cite{Harry-anisot}, $Y_{\perp}, Y_{\parallel},
\nu_{\perp}, \nu_{\parallel},\phi_{\perp}, \phi_{\parallel}$, are
calculated from the values of the isotropic materials that makes
up the layers of the coating in the following way

{\small
\begin{eqnarray}
Y_{\perp}&=& (d_1+d_2)/(d_1/Y_1 +d_2/Y_2) \nonumber \\
Y_{\parallel} &=& (Y_1 d_1 + Y_2 d_2)/(d_1 + d_2) \nonumber \\
\phi_{\perp}&=&Y_{\perp} (\phi_1 d_1/Y_1 + \phi_2 d_2/Y_2)/(d_1+d_2)\nonumber \\
\phi_{\parallel} &=& (Y_1 \phi_1 d_1 + Y_2 \phi_2 d_2)/[Y_{\parallel}(d_1+d_2)] \nonumber \\
\nu_{\perp}&=&(\nu_1 Y_1d_1 +\nu_2 Y_2 d_2)/(Y_1 d_1 + Y_2 d_2) \nonumber \\
\nu_{\parallel}&=&F(Y_1,Y_2,\nu_1,\nu_2,d_1,d_2) \nonumber
\end{eqnarray}
}

 $\nu_{\parallel}$  solution of
{\small
\begin{equation}
\frac{\nu_1 Y_1 d_1}{(1+\nu_1)(1-2\nu_1)}+\frac{\nu_2 Y_2
d_2}{(1+\nu_2)(1-2\nu_2)} = - \frac{Y_{\parallel}(\nu_{\perp}^2
Y_{\parallel} +
\nu_{\parallel}Y_{\perp})(d_1+d_2)}{(\nu_{\parallel}+1)(2\nu_{\perp}^2
Y_{\parallel}-(1-\nu_{\parallel})Y_{\perp})}\nonumber
\end{equation}
}

 When we first looked at the problem of coating optimization, we
were not completely satisfied with the derivation of the averaged
mechanical properties of the multi-layer coating quoted above.
Therefore we put some effort in the rigorous analysis of a
multi-layer structure and the results are presented in the
following section.

\section{Averaged elastic coefficients for multilayered optical coating}

Heterogeneity of a material or structure can be caused by two main
reasons.

\begin{enumerate}
    \item Non-uniformity of certain physical characteristic
(density, elastic modulus, conductivity, etc. ). Two or
multi-phase composites are typical examples of this type of
material behavior.
    \item  An additional source of heterogeneity is a geometrical one.
A multi-layer structure like the dielectric optical coating used
in gravitational waves interferometers is an example of this
effect.
\end{enumerate}

The replacement of heterogeneous media by an homogeneous
continuum, which is characterized by certain effective
constitutive equations, is the basic instrument for the effective
media theory (EMT), which is a well developed subject in composite
mechanics ~\cite{heterogeneous}. The purpose of this section is to
give a mathematical introduction to the problem of the calculation
of mechanical properties of heterogeneous solids with periodic
microstructure. The averaging procedure for mechanical fields and
the definitions for effective constitutive parameters (effective
elastic moduli, for instance) are described in the following for a
bilayered composite medium. We provide a rigorous derivation of
the coating effective mechanical parameters which is very
important in the study of coating optimization for GW
interferometers.

 Let us consider an elastic solid which consists
of a periodic array of two layers with thicknesses $d_1$ and
$d_2$, respectively in welded contact and stacked in the $x_3=z$
direction. It means that the components, $C_{ijkl}$, of the
elastic moduli tensor, which link strains $\varepsilon_{kl}$, and
stresses, $\sigma_{ij}$ are periodic functions,
$C_{ijkl}(z+d_1+d_2)=C_{ijkl}(z)$. Let us define the average over
the unit cell, composed dy a bilayer structure, as

\begin{equation*}
    \langle X \rangle = \frac{d_1 X + d_2 X}{d_1 + d_2}= \delta_1 X
    + \delta_2 X
\end{equation*}

One can introduce the definition of the effective moduli tensor
$C^\ast_{ijkl}$ as the coefficients which link the components of
effective fields variables defined by averaging the strain and
stress tensor and taking into account the continuity conditions at
the interface between the two layers. We skip all the mathematical
technicalities, referring the reader to~\cite{heterogeneous,
Wavepropag} and we report only the results of the homogenization
process

\begin{eqnarray}
    C^\ast_{1111}&=& \langle C_{1111}\rangle - \frac{\delta_1 \delta_2 \left(C_{3311}^{(2)}-C_{3311}^{(1)}\right)^2}{\delta_1 C_{3333}^{(2)} + \delta_2 C_{3333}^{(1)}} ,
    \nonumber \\
    C^\ast_{2211}&=&\langle C_{2211}\rangle + \frac{\delta_1 \delta_2 \left(C_{3311}^{(2)}-C_{3311}^{(1)}\right)
    \left(C_{2233}^{(1)}-C_{2233}^{(2)}\right)}{\delta_1 C_{3333}^{(2)} + \delta_2
    C_{3333}^{(1)}} \quad , \nonumber \\
    C^\ast_{3311}&=&\frac{\delta_1 C_{3311}^{(1)}C_{3333}^{(2)}+ \delta_2 C_{3311}^{(2)}C_{3333}^{(1)}}{\delta_1 C_{3333}^{(2)} + \delta_2
    C_{3333}^{(1)}}\quad , \nonumber \\
    C^\ast_{2222}&=& \langle C_{2222}\rangle-\frac{\delta_1 \delta_2 \left(C_{3322}^{(2)}-C_{3322}^{(1)}\right)^2}{\delta_1 C_{3333}^{(2)} + \delta_2
    C_{3333}^{(1)}}\quad , \nonumber \\
    C^\ast_{3322}&=& \frac{\delta_1 C_{3322}^{(1)}C_{3333}^{(2)}+ \delta_2 C_{3322}^{(2)}C_{3333}^{(1)}}{\delta_1 C_{3333}^{(2)} + \delta_2
    C_{3333}^{(1)}}\quad , \nonumber \\
    C^\ast_{1212}&=&\langle C_{1212}\rangle \quad , \nonumber \\
    \frac{1}{C^\ast_{p3p3}}&=&
    \left\langle\frac{1}{C_{p3p3}}\right\rangle \quad , p=1,2,3
    \quad.
\end{eqnarray}

It is interesting to point to the direct averaging law for the
composite planar shear modulus $C^\ast_{1212}$ and the inverse
averaging law for the transverse ones: $C^\ast_{1313},
C^\ast_{2323}$.

The constitutive relations for each of the two assumed isotropic
materials are given by \eqref{isotro2} as functions of the two
Lam\`{e} coefficients. In a cartesian coordinate system, the
components of the stiffness tensor are

\begin{equation}\label{C-isotro-cart}
    C_{ijkl}= \lambda \delta_{ij} \delta_{kl} + \mu (\delta_{ik}
    \delta_{jl} + \delta_{il}\delta_{jl})
\end{equation}

Let us remind in Tab.\ref{tab:YPisot} the relation between the
Lam\'{e} coefficients and the Young and Poisson coefficient for an
isotropic material

\begin{table}[htb]\label{tab:YPisot}
\begin{center}
\begin{tabular}{|c|c|}
\hline
$\mu = \frac{Y}{2(1+\nu)}$ & $\lambda = \frac{Y \nu}{(1+\nu) (1-2 \nu)}$ \\
\hline
$Y = \frac{\mu (3 \lambda + 2 \mu)}{\lambda + \mu}$ & $\nu = \frac{\lambda}{2 (\lambda + \mu)}$ \\
\hline
\end{tabular}
\end{center}
\caption{Elastic coefficients}
\end{table}

Since both materials are assumed isotropic, the resulting layered
medium is expected to posses transverse isotropy with isotropy
confined in the $x_1-x_2$ plane. Even in an anisotropic medium,
the axial components of the stiffness tensor are frequently
characterized by the parameters of uniaxial tension, namely Young
modulus $Y_i$, and the Poisson ratio $\nu_{ji}$, for the tension
in the $x_i$ direction:
\begin{equation}\label{Young-gen}
    Y_i=\frac{\sigma_{ii}}{\varepsilon_{ii}} \quad \mbox{and}\quad \nu_{ji}=
    -\frac{\varepsilon_{jj}}{\varepsilon_{ii}},
    \sigma_{jj}=0,j\neq i
\end{equation}

These parameters can be easily calculated for the homogenized
compound material by means of the components of the effective
stiffness tensor $C^\ast_{ijkl}$

\begin{align}
     & \nu^{\ast}_{ji}=\frac{C^{\ast}_{iijj} C^{\ast}_{kkkk}- C^{\ast}_{iikk} C^{\ast}_{jjkk}}{C^{\ast}_{jjjj}
     C^{\ast}_{kkkk}-(C^{\ast}_{jjkk})^2}, \\
    & Y^{\ast}_{i}= C^{\ast}_{iiii}- C^{\ast}_{iijj} \nu^{\ast}_{ji}
    - C^{\ast}_{iikk} \nu^{\ast}_{ki}, \nonumber \\
     & i\neq j, i\neq k, j\neq k
\end{align}

Planar, $Y^{\ast}_1=Y^{\ast}_2\doteq Y_{\parallel}$ and transverse
$Y^{\ast}_3\doteq Y_{\perp}$, Young moduli, share moduli and
Poisson ratios of binary multilayered coating composed of $SiO_2$
and $Ta_2O_5$ are shown in fig. REF as functions of the $Ta_2O_5$
layer thickness fraction. The explicit expressions of the elastic
coefficient of the multilayer structure as function of the elastic
coefficients of the two material are

{\small
\begin{align}
     Y^{\ast}_1&=Y^{\ast}_2= \frac{Y_1^2 (1-\nu_2^2) \delta_1^2 + 2 Y_1 Y_2 (1- \nu_1 \nu_2)\delta_1 \delta_2 + Y_2^2 (1-\nu_1^2) \delta_2^2}
    {Y_1 (1-\nu_2^2)\delta_1 + Y_2 (1- \nu_1^2) \delta_2}
    \label{Young-avg} \\
    Y^{\ast}_3 &= \frac{Y_1 Y_2 \left[ Y_1(1-\nu_2)\delta_1 + Y_2 (1-\nu_1)\delta_2 \right]}
    {Y_2^2 (1-\nu_1-2\nu_1^2)\delta_1 \delta_2 + Y_2^2(1-\nu_2 -2 \nu_2^2)\delta_1 \delta_2 + Y_1 Y_2
    \left[(1-\nu_2) \delta_1^2 +4\nu_1 \nu_2 \delta_1 \delta_2 + (1-\nu_1)
    \delta_2^2\right]}\nonumber \\
    \nu^{\ast}_{12} &= \frac{Y_1 \nu_1 (1-\nu_2^2)\delta_1 + Y_2 \nu_2 (1-\nu_1^2)\delta_2 }
    {Y_1(1-\nu_2^2)\delta_1 + Y_2 (1-\nu_1^2)\delta_2 }
    \label{Poiss-avg}
    \\
    \nu^{\ast}_{13} &= \frac{Y_1 Y_2 \left[(1-\nu_2) \nu_1 \delta_1 + (1-\nu_1) \nu_2 \delta_2\right]}
    {Y_1^2 (1-\nu_2 -2 \nu_2^2) \delta_1 \delta_2 +Y_2^2 (1-\nu_1 -2 \nu_1^2) \delta_1 \delta_2
    +Y_1 Y_2\left[(1-\nu_2)\delta_1^2 +4 \nu_1 \nu_2 \delta_1 \delta_2 + (1-\nu_1) \delta_2^2  \right]
    }\nonumber \\
    G^{\ast}_{\perp} & =\frac{Y_1 Y_2}{2 \left[ Y_2(1+\nu_1)\delta_1 + Y_2 (1+\nu_2) \delta_2\right]
    }\nonumber \\
    G^{\ast}_{\parallel} & = \frac{Y_1 \delta_1}{2 (1+\nu_1)}+\frac{Y_2 \delta_2}{2
    (1+\nu_2)} \nonumber
\end{align}}

\begin{figure}[htbp]
  \centering
  \subfigure[Effective Young moduli.]{
     \label{fig:subfig:YoungC}
     \includegraphics[width=0.8\textwidth]{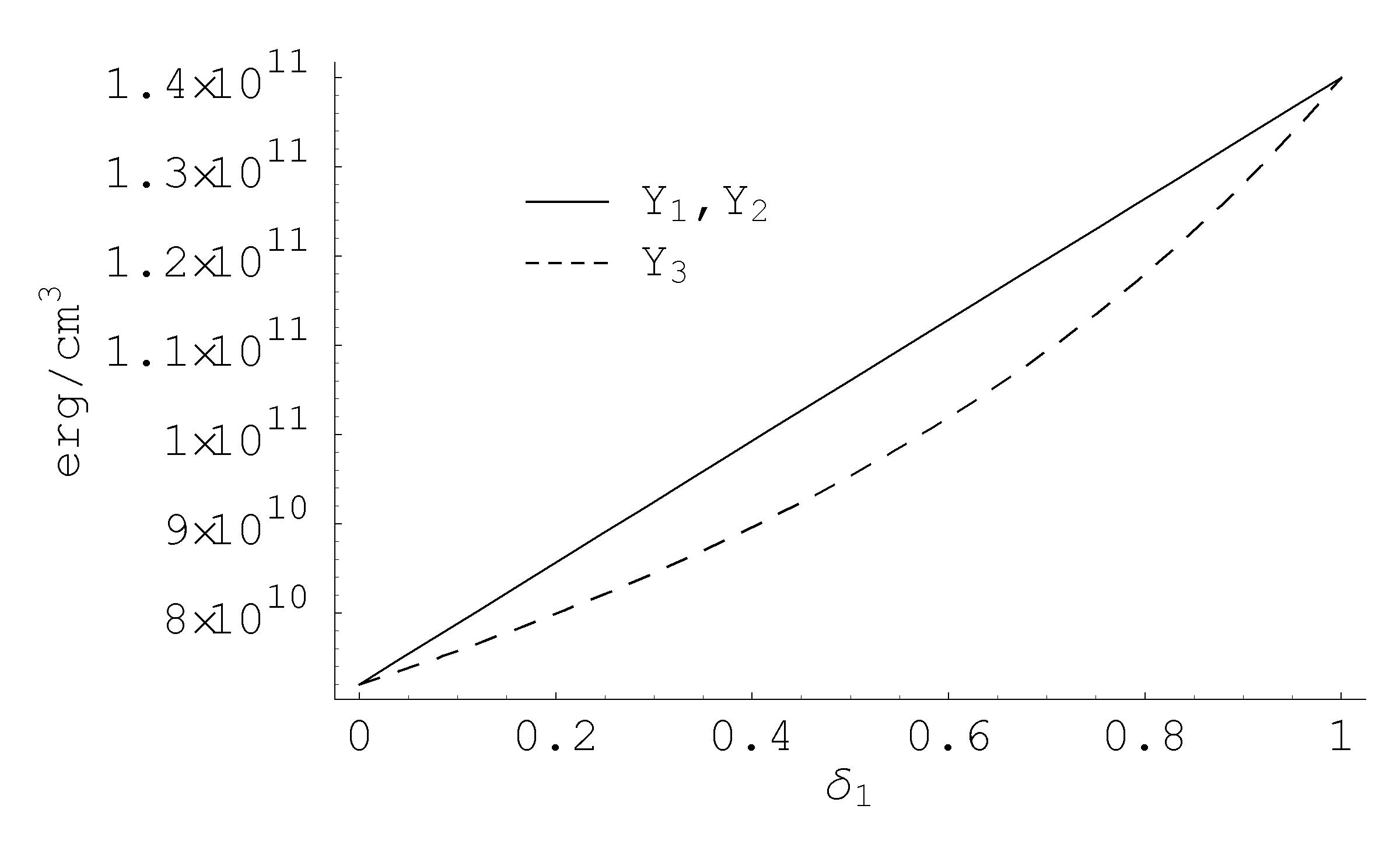}}
\subfigure[Effective Poisson ratios.]{
      \label{fig:subfig:PoissC}
      \includegraphics[width=0.8\textwidth]{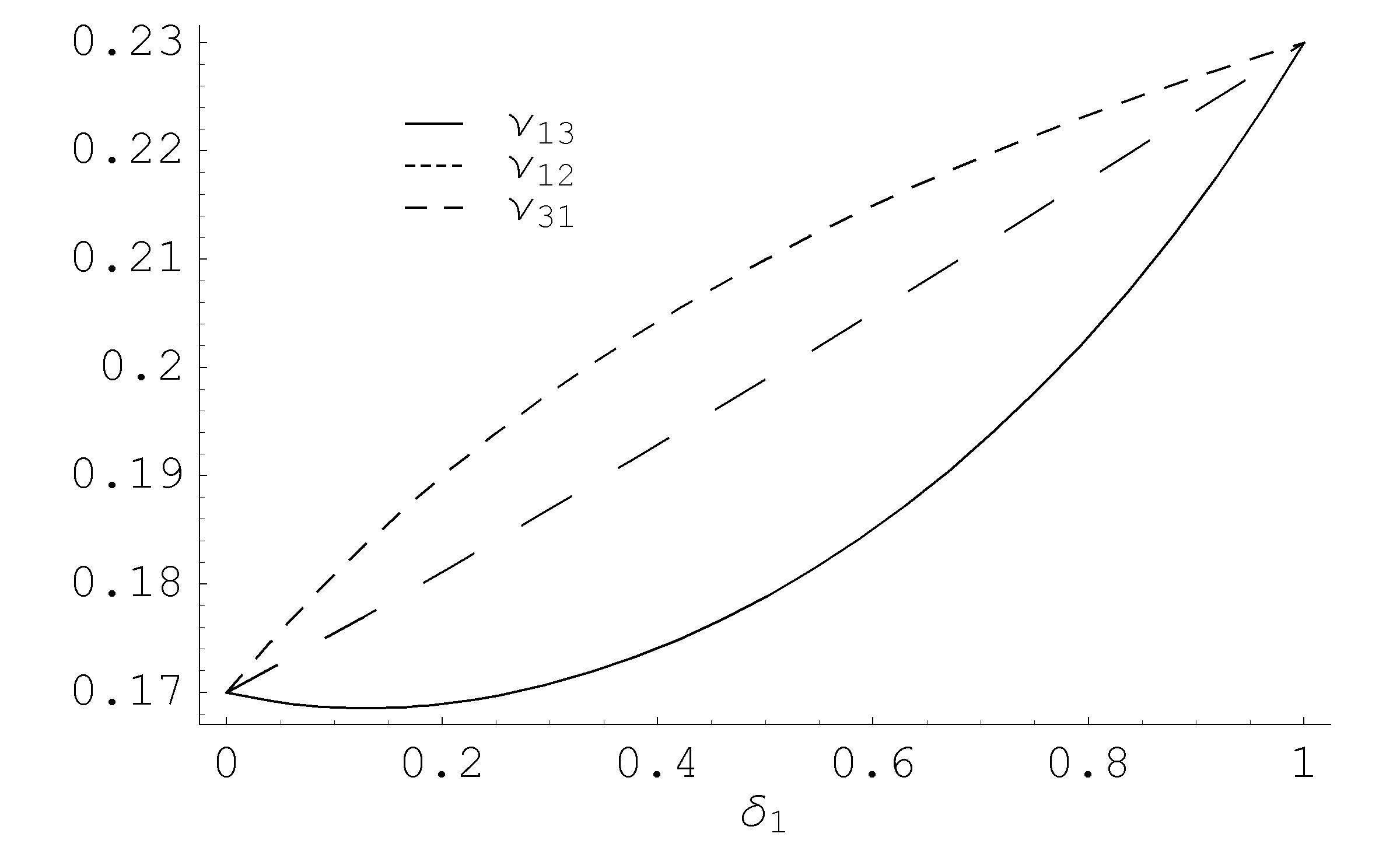}}
 \caption{Averaged elastic coefficients using material parameters of App.~\ref{App:A3} .}
 \label{fig:CB}
 \end{figure}

By convention, the $5$ elastic constants in transverse isotropic
constitutive equations are the Young's modulus and Poisson ratio
in the x-y symmetry plane, $Y_{\parallel}$ and $\nu_{\parallel}$,
the Young's modulus and Poisson ratio in the z-direction,
$Y_{\perp}$ and $\nu_{\perp}$, and the shear modulus in the
z-direction $G_{\perp}$. The compliance matrix takes the form

\[
\begin{pmatrix}
\varepsilon_{x x}\\\varepsilon_{y y}\\\varepsilon_{z
z}\\\varepsilon_{y z}\\\varepsilon_{x z}\\\varepsilon_{x y}
\end{pmatrix}
=
\begin{pmatrix}
\dfrac{1}{Y_{\parallel}} & -\dfrac{\nu_{\parallel}}{Y_{\parallel}} & - \dfrac{\nu_{\perp}}{Y_{\perp}} & 0 & 0 & 0\\
-\dfrac{\nu_{\parallel}}{Y_{\parallel}} & \dfrac{1}{Y_{\parallel}} & - \dfrac{\nu_{\perp}}{Y_{\perp}} & 0 & 0 & 0 \\
- \dfrac{\nu_{\perp}}{Y_{\perp}} & - \dfrac{\nu_{\perp}}{Y_{\perp}} & \dfrac{1}{Y_{\perp}} & 0 & 0 & 0 \\
0 & 0  & 0 & \dfrac{1}{2 G_{\perp}} & 0 & 0 \\
0 & 0 & 0 & 0 & \dfrac{1}{2 G_{\perp}} & 0 \\
0 & 0 & 0 & 0 & 0 & \dfrac{1+ \nu_{\parallel}}{Y_{\parallel}}
\end{pmatrix}
\begin{pmatrix}
\sigma_{x x}\\\sigma_{y y}\\\sigma_{z z}\\\sigma_{y z}\\\sigma_{x
z}\\\sigma_{x y}
\end{pmatrix}
\]

Let us now expand the exact expressions of the Young and Poisson
averaged coefficient, Eqs.~\eqref{Young-avg},\eqref{Poiss-avg}, in
power series of the two material's Poisson coefficient $\nu_1$ and
$\nu_2$

\begin{align}\label{elst-expans}
    Y_{\parallel} & = Y_1 \delta_1 + Y_2 \delta_2 + O(\nu^2) \\
    Y_{\perp} & = \dfrac{Y_1 Y_2}{Y_2 \delta_1 + Y_1 \delta_2} +
    O(\nu^2) \nonumber \\
    \nu_{\parallel} & = \dfrac{Y_1 \nu_1 \delta_1 + Y_2 \nu_2 \delta_2}{Y_1 \delta_1 + Y_2
    \delta_2} + O(\nu^2) \nonumber \\
    \nu_{\perp} & = \dfrac{Y_1 Y_2 \nu_1 \delta_1 + Y_1 Y_2 \nu_2 \delta_2}
    {(Y_1 \delta_1 + Y_2 \delta_2)(Y_2 \delta_1 + Y_1 \delta_2)}
  + O(\nu^2)   \nonumber
\end{align}

These more manageable expressions\footnote{The firsts two
expression reproduce the results of ~\cite{Harry-anisot}, whereas
the other two are different.} offer excellent approximation for
typical LIGO mirrors coating: for the standard quarter wavelength
design, the difference between the approximated values and the
exact values is within $0.5\%$.

\section{Thermal noise for an anisotropic coating}

The thermal noise for a semi-infinite mirror with an anisotropic
coating is given by \footnote{This is a corrected form of the
equation $(8)$ in~\cite{Harry-anisot}. }

\begin{equation}
S_X(f)= \frac{2 K_B T}{\pi^{\frac{3}{2}} f} \frac{1- \nu_s^2}{w
Y_s} \phi_{eff}^{coat}
\end {equation}

\begin{align}
\phi_{eff}^{coat}=& \frac{d}{\sqrt{\pi}w}\frac{1}{Y_{\perp}}\Bigg( \Big( \frac{Y_s}{1-\nu_s^2}-\frac{2 \nu_{\perp}^2 Y_s Y_{\parallel} }{Y_{\perp}(1-\nu_s^2)(1-\nu_{\parallel})}\Big) \phi^c_{\perp} \nonumber \\
 & +\frac{Y_{\parallel}\nu_{\perp}(1-2\nu_s)}{(1-\nu_{\parallel})(1-\nu_s)}(\phi^c_{\parallel}-\phi^c_{\perp})\nonumber \\
+& \frac{Y_{\parallel} Y_{\perp}(1+\nu_s)(1-2\nu_s)^2}{Y_s
(1-\nu_{\parallel}^2)(1-\nu_s)} \phi^c_{\parallel}\Bigg)
\end{align}

In the limit of isotropic coating this formula reduces to
\eqref{SX-B-COAT-INF-GB}. Since the loss angle is introduced in
the theory of elasticity as an imaginary part of the elastic
coefficient of the material, it is natural to obtain the values of
$\phi^c_{\parallel}$ and $\phi^c_{\perp}$ from the isotropic
values $\phi_1 , \phi_2$, using the same averaging rules as the
Young moduli. Therefore we have

\begin{equation}\label{aver-phi}
    \phi^c_{\parallel}= \frac{(Y
    \phi)_{\parallel}}{Y_{\parallel}}\quad \mbox{and} \quad
    \phi^c_{\perp}= \frac{(Y \phi)_{\perp}}{Y_{\perp}}
\end{equation}

and the expansion in the Poisson coefficients gives

\begin{equation}\label{aver-phi-appr}
     \phi^c_{\parallel}=\frac{Y_1 \phi_1 \delta_1 + Y_2 \phi_2 \delta_2}{Y_1 \delta_1 + Y_2
     \delta_2}+ O(\nu^2)\quad \mbox{and} \quad \phi^c_{\perp}= \frac{(Y_2 \delta_1 + Y_1 \delta_2)\phi_1 \phi_2}{Y_1 \phi_1
     \delta_2+Y_2 \phi_2 \delta_1} + O(\nu^2)
\end{equation}

We now have all the ingredient to express the coating thermal
noise as function of the characteristic elastic coefficient of the
substrate and coating materials, which for the present analysis
will be treated as constants, and the thickness,
$\delta_1,\delta_2$, and the number of doublet,$N_d$ of the
layers. The function $S_X(\delta_1,\delta_2,N_d)$ results in a
complicated expression which is not particulary illuminating and
its analytical form will be omitted in this presentation. However
we can give a very good approximation of this function by a first
order Taylor expansion around the point which correspond to the
standard design, $z_1=z_2=1/4$, where $z$ is the scaled optical
path $z= n \delta /\lambda$.

\begin{equation}\label{coat-linear}
S_X(f)= \frac{2 K_B T}{\pi^{\frac{3}{2}} f} \frac{1- \nu_s^2}{w
Y_s} \frac{N_d}{w}(2.2 10^{-10} z_1 + 1.45 10^{-11} z_2)
\end{equation}

\begin{figure}[htb]
\begin{center}
\includegraphics[width=0.9\textwidth]{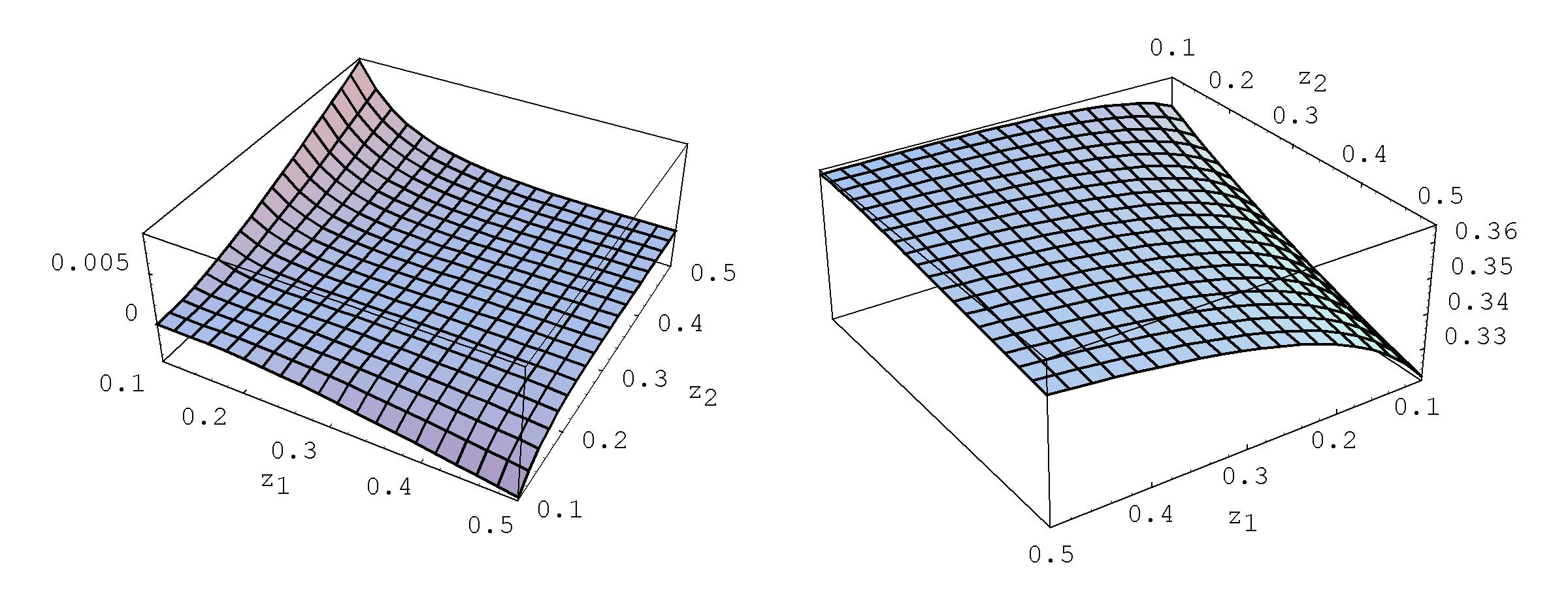}
\end{center}
\caption{Left: Fractional error using
expression~\ref{coat-linear}.
 Right: Fractional error for the expression ~\ref{coat-linear2}. }
\label{fig:coat-lin}
\end{figure}

 Fig.\ref{fig:coat-lin}(left panel) shows the accuracy level of this linear approximation. If
 we are looking for an analytical approximated expression we can
 take the limit of vanishing Poisson ratios in the general
 expression of the noise and then take the linear terms in the two
 optical thickness $z_1,z_2$

\begin{equation}\label{coat-linear2}
S_X(f)= \frac{2 K_B T}{\pi^{\frac{3}{2}} f} \frac{1- \nu_s^2}{w
Y_s} \frac{N_d \,\lambda}{w \sqrt{\pi}}\left
(\frac{\phi_1}{n_1}\left(\frac{Y_1}{Y_s}+\frac{Y_s}{Y_1}
\right)z_1 +
\frac{\phi_2}{n_2}\left(\frac{Y_2}{Y_s}+\frac{Y_s}{Y_2} \right)
z_2 \right) \qquad \nu_s,\nu_1,\nu_2 \ll1
\end{equation}

From Fig.\ref{fig:coat-lin}(right panel) we can easily see that
even if the expression above is very manageable, it provides a
rough approximation of the exact formula and it leads to error
above $30 \%$ in the interesting region.

\section{Coating thickness optimization}

Our collaborators from the University of Sannio, Benevento,
developed an analysis based on genetic algorithm (GA) to address
the problem of finding the best coating design for minimizing the
thermal noise. We briefly review their conclusions, in order to
validate our simple direct optimization. Genetic algorithms (GAs)

GAs have proven as effective tools for synthesizing general (e.g.,
multi-dielectric) reflective coatings, featuring several
heterogeneous (e.g., technological) design constraints, with
multiobjective (i.e., reflectivity and thermal noise) optimization
targets. The general analysis led to the result that, as the
number of generations is increased, the centermost part of the
GA-optimized coatings exhibits a neat tendency toward a cascade of
almost identical doublets, whose (total) optical lengths cluster
around $\lambda/2$. The above findings led us to investigate the
performance of coatings composed of cascaded identical multiplets
with the constraint $z_1+z_2= 1/2$. This constraint reduces the
number of independent variables to two, $z_1$ and $N_d$, the
optical thickness of the high index material and the number of
doublets.

\begin{figure}[htbp]
  \centering
  \subfigure[Thermal noise contour plot scaled to the Ad-LIGO baseline design.]{
     \label{fig:subfig:TNcontour}
     \includegraphics[width=0.7\textwidth]{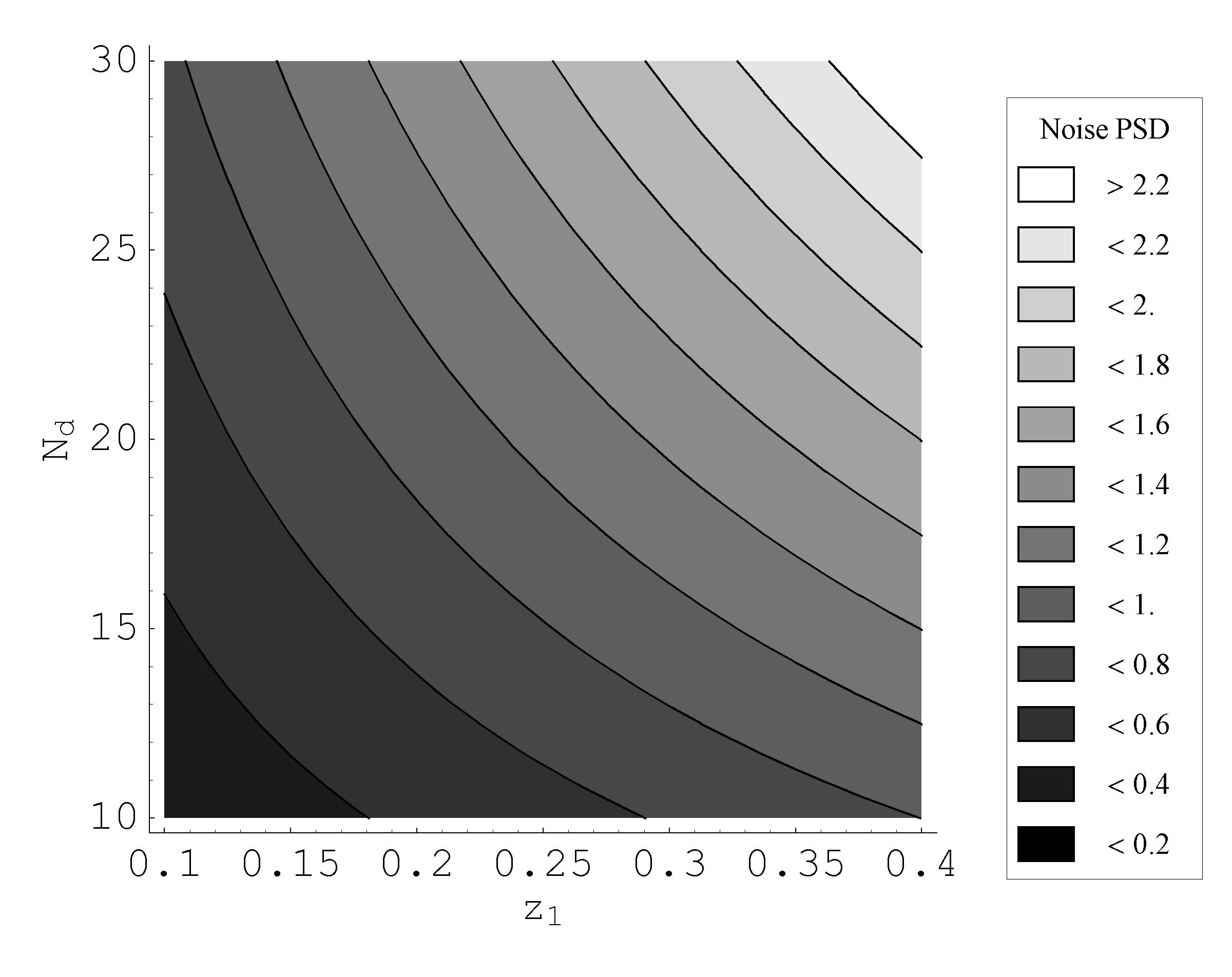}}
     \hspace{0.1in}
\subfigure[Iso-transmission curves.]{
      \label{fig:subfig:Transmiss}
      \includegraphics[width=0.7\textwidth]{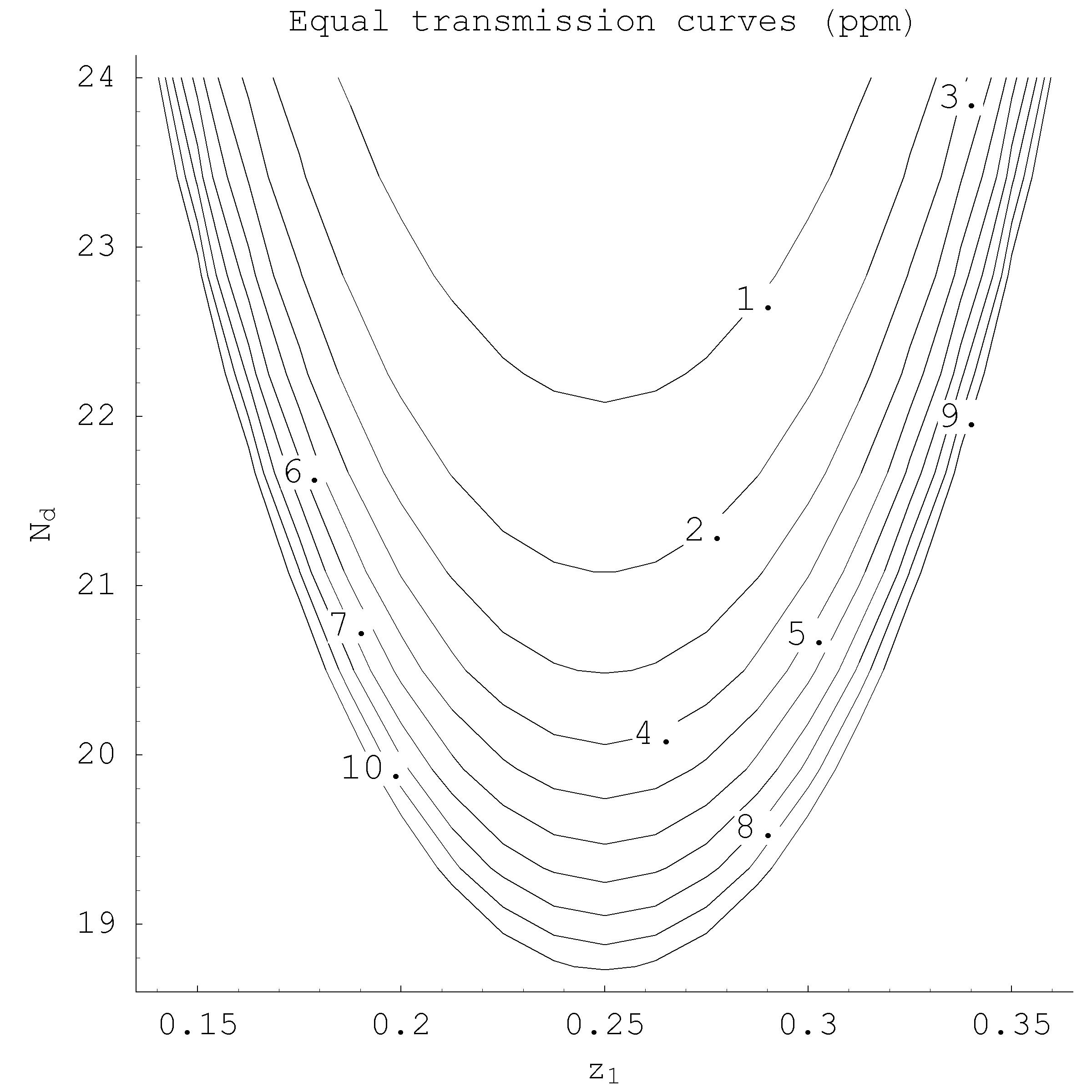}}
 \caption{The constraint $z_1+z_2=1/2$ allows a simple optimization by inspection of these plots .}
 \label{fig:TN-Trans}
 \end{figure}

Fig.~\ref{fig:TN-Trans} shows the noise function $S_X(z_1,N_d)$
scaled to the Ad-LIGO baseline value, and $1-|\Gamma|^2(z_1,N_d)$
in the range of interest.

 We followed the strategy:

\begin{itemize}
    \item Calculate the coating transmission \footnote{In the absence of (optical) losses. } for the baseline
    design, $z_1=z_2=1/4, \; N_d = 19$, using Eq.\eqref{refl-i}: $1-|\Gamma|^2 = 8.3 ppm$
    \item In the plane $(z_1,N_d)$,
     draw the curve of equal transmission constrained by the
    above value.
    \item Plot the noise function $S_X(z_1,N_d)$ in the same graph
    and find the point \emph{on} the above curve which correspond to the
    minimum noise. This correspond to the optimized design
    $(z_1,N_d)_{opt}$.
\end{itemize}

\begin{figure}[htb]
\begin{center}
\includegraphics[width=0.8\textwidth]{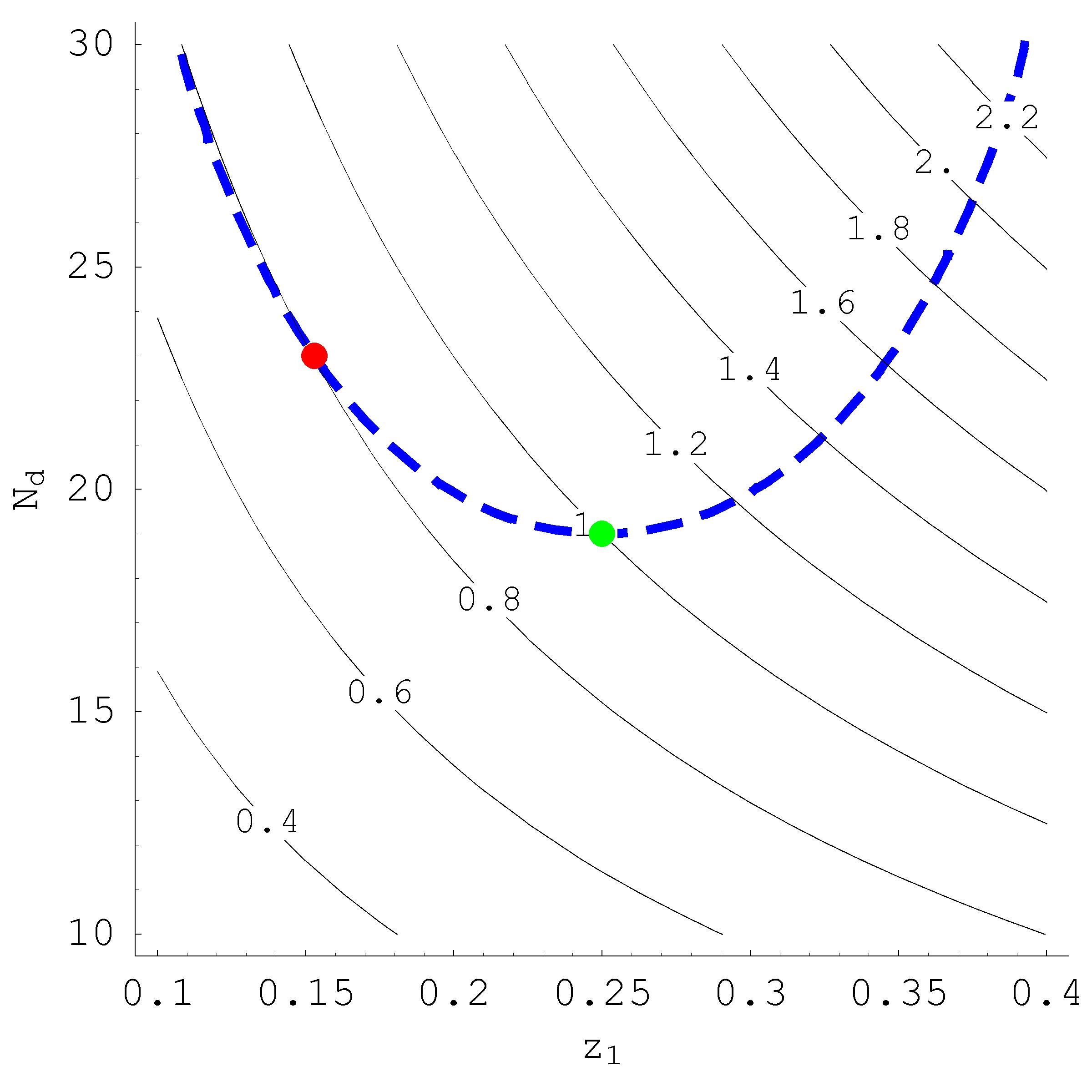}
\end{center}
\caption{The blue curve corresponds to a transmission of $8.3
ppm$, the green point corresponds to the Ad-LIGO coating standard
design, and the red point corresponds to the optimized design. }
\label{fig:optim}
\end{figure}

All the procedure is synthesized in Fig.\ref{fig:optim}. The
optimized design (for the known values of material losses, elastic
moduli, etc.) is given by $N_d=23$, $z_1=0.153$ and $z_2=0.347$;
for the same optical transmission of the $\lambda/4$-design, it
provides a $20 \%$ of reduction in noise spectral density. Under
the simplest assumption where the GW sources are distributed
homogeneously/isotropically throughout space, without
interferometer re-optimization, this may boost the event rate by
some $30 \%$.

The optimization of the coating structure to minimize the thermal
noise is therefore a very promising idea, which in principle could
stand beside the mesa beam proposal, increasing further the
detector sensitivity.
Coating thickness optimization should be considered as almost
mandatory to minimize coating noise, yielding in all cases\footnote{In this thesis
 we have focused the attention on the Brownian coating thermal noise, but on-going
 researches~\cite{Castaldi} have analyzed the total coating noise budget and provided the corresponding optimized design.} a substantial increase($> 30\%$ at $100$ Hz) in the expected event rate, as compared to the QWL design.
 A prototype of optimized coating
 has been designed by our collaboration and is
 scheduled for testing at Caltech in the TNI (Thermal Noise Interferometer)
 facility.
LMA is responsible for manufacturing the sample, using the ion
beam deposition techniques.

% ------------------------------------------------------------------------

%%% Local Variables:
%%% mode: latex
%%% TeX-master: "../thesis"
%%% End:

\def\baselinestretch{1}
\chapter{ Conclusions }
\ifpdf
    \graphicspath{{Conclusions/ConclusionsFigs/PNG/}{Conclusions/ConclusionsFigs/PDF/}{Conclusions/ConclusionsFigs/}}
\else
    \graphicspath{{Conclusions/ConclusionsFigs/EPS/}{Conclusions/ConclusionsFigs/}}
\fi

\def\baselinestretch{1.66}

This thesis presents a collection of different researches on
non-standard optics in view of enhancing the performances of  the
Advanced Gravitational waves interferometric detectors.

Although primarily based on simulations, calculations and
theoretical arguments, this work has been in continuous feedback
with the experimental side of the thermal noise minimization
effort. This work has provided the theoretical understanding and
guidance for the experimental side, while getting back continuous
and progressive validation of the predictions and prescriptions.
The end result was a very good example of synergy, and mutual
enrichment, for me and my introduction to laboratory work, and for
my colleagues that worked on my prescriptions and numbers.

 In the design of
the next generation Gravitational Waves interferometric detectors,
mirror's thermal noise plays a crucial role. The mirrors coating
thermal noise is expected to be the limiting factor in the GW
signal frequency region of highest sensitivity.

 The quest for increasing the event rate in the
observational band has motivated the exploration of various
techniques for reducing the mirror thermal noise toward the
development of second-generation detectors, such as
Advanced-LIGO~\cite{AdLIGO}. With specific reference to the
coating Brownian noise (dominant in the current baseline design
featuring fused-silica test masses), use of improved
(low-mechanical-loss) materials, geometric optimization of the
coating design, and flat-top (commonly referred to as ``mesa'')
beams seem the most promising. The latter option, intuitively
motivated by the potential capability of a mesa beam (MB) of
better averaging the thermally induced mirror surface fluctuations
as compared to a standard Gaussian beam (GB), has been numerically
proved in this thesis to yield significant reductions in the
overall thermal noise. For the first time we calculated the
coating types of thermal noise reduction in using mesa beam and
enlightened the issue of mirror geometric aspect ratio
optimization. We demonstrated that the coating induced
displacement noise can be reduced by a factor of $\simeq 1.8$
which increases the estimated binary neutron star inspiral range
\footnote{The effective range conventionally measures the
sensitivity of the interferometers to signals arising from binary
systems composed of $1.4$ solar mass neutron star pairs having a
signal-to-noise ratio of $8:1$ or better, and is computed by
averaging over both the antenna sensitivity pattern and over all
possible orientations of the binary systems.}, seen by Ad-LIGO,
from $175$ Mpc to $225$ Mpc.

During this work we implemented the different kinds of thermal
noise calculations for finite size test mass and arbitrary
cylindrically-shaped laser beams, in a simple
$Mathematica^\circledR$ notebook called Thermal Noise Notebook
(TNN). This package is available as a very simple tool for the
estimation of the thermal noise contributions taking into account
both the beam and the mirror geometry. We illustrated the
importance of uniform sampling of the mirror surface to reduce
thermal noise and the limitation brought by the use of excited
modes with nodes on the mirror surface, which cause non-negligible
gradients in the elastic and thermal fields in the test mass and
worsen the thermal noise performance.

In Sec.\ref{sec:MesaBeam} we further developed  the theory of mesa
beam and derived some unexplored analytical results for the dual
configuration, flat mesa beam and concentric mesa beam, which are
of interest in view of the implementation of this non-standard
optics in real systems. The expression of the beam width,
divergence and beam propagation parameter, $M^2$, are derived in
analytical form as functions of the characteristic parameters of
the mesa beams, thus allowing a straightforward  application of
the well known $ABCD$ formalism for the propagation of optical
beams trough paraxial systems.

 We also analytically
proved a new duality relation between optical cavities with
non-spherical mirrors. This derivation provides a unique mapping
between the eigenvalues and eigenvectors of two cavities  whose
mirrors shapes are related by a simple relation. This duality
allows the direct application of beam property calculations
performed in a case to geometries of the other configuration.

The growing interest in the GW community in the mesa beam idea,
has led to the development, by our group, of a Fabry-Perot cavity
prototype with non-spherical ``Mexican hat'' (MH) profile mirrors.
One of the main task in this thesis has been the development and
testing of optical simulation packages based on FFT routines or on
Huygens paraxial integral approach. The FFT based program can
simulate a Fabry-Perot optical cavity that includes non spherical
mirrors and/or ``realistic'' mirror deformations. Many different
cavity parameters or imperfections can be modelled as well as
mirror misalignment and input beam coupling sensitivity. These
programs have been successfully used to provide a theoretical
framework for the experimental work, i.e. the fundamental and
higher modes sensitivity to cavity misalignment and mirror
imperfections, for our Mexican hat mirror cavity, for the large
scale Ad-LIGO FP cavities or any interferometer to be designed.
Some examples of their application are given in this thesis as
well as a discussion on the accuracy and limitations they provide.

The research on the optical coatings for next generation detectors
requires the most possible accurate model of the coating elastic
and thermal parameters. In this thesis we present a rigorous,
never used in this problem (to our best knowledge), derivation of
the coating multi-layer effective elastic coefficients as function
of the constituents individual properties. Some simplified
expression are also derived, as leading term of the expansion of
the exact formulas, which are in agreement with published results.
This analysis is of fundamental importance when dealing with the
comparison of different coating design because the elastic
coefficients of the coating enter the thermal noise evaluation
formulas.

 The last part of the thesis has been dedicated to the
study of the coating thermal noise reduction by a modification of
the geometric structure of the multilayered coating. The scope of
the project has been to find an alternative coating design respect
to the standard quarter-wavelength one, which, for a prescribed
transmittance, gives the minimum thermal noise. We found that the
optimized design decreases the mirror thermal noise of about $10
\%$. Under the simplest assumption where the GW sources are
distributed homogeneously/isotropically throughout space, without
interferometer re-optimization, this may boost the event rate by
some $30 \%$.  A prototype of optimized coating
 has been designed by our collaboration and is
 scheduled for testing at Caltech in the TNI (Thermal Noise Interferometer)
 facility.

All these research topics are currently very actively investigated
by our collaborations ~\cite{Galdi}. The end result of our studies
will likely be the extension of the reach of GW observatories by a
factor of a few, and of the event rate by a factor of several,
maybe more than an order of magnitude.

%%% ----------------------------------------------------------------------

% ------------------------------------------------------------------------

%%% Local Variables:
%%% mode: latex
%%% TeX-master: "../thesis"
%%% End:

\appendix
\appendix
    \renewcommand{\thechapter}{A}
    \refstepcounter{chapter}
    \makeatletter
    \renewcommand{\theequation}{\thechapter.\@arabic\c@equation}
    \makeatother

    \chapter*{Appendix }
    \addcontentsline{toc}{chapter}{Appendix }

\section{Physical components of tensors in orthogonal coordinate
systems}\label{App:A1}

The physical components of a tensor are given by the components of
the tensor with respect to unit basis vectors. In an orthogonal
coordinate system, the metric tensor is diagonal and is
conventionally written as
\begin{equation}\label{orth-metric}
    g_{i i}= h_i^2, \quad g^{i i}=\frac{1}{h_i^2},\quad i \mbox{not
summed}, \mbox{and} \quad g_{i
    j}=0= g^{i j} \; i\neq j
\end{equation}

For example, the physical component of any vector in a given
direction is merely the projection of that vector onto that
direction, i.e., the inner product of the vector with a unit
vector in the given direction. This is simply the component of the
vector in a basis having a unit basis vector pointing in that
direction.

The unit contravariant vectors tangent to the three coordinate
lines are
\begin{equation}\label{unit-orthogonal}
    e^i_{(j)}= \frac{\delta^i_j}{h_j}
\end{equation}
The index enclosed in parentheses means a non tensorial index.
 In terms of the covariant/contravariant and metric
components, the physical components of an arbitrary  vector ,in an
orthogonal coordinate system, are given by

\begin{equation}\label{Phys-vec}
    T(i)= h_{i}  T^{i } = \frac{T_{i }}{h_{i}}
    \qquad \mbox {no summation over indices}
\end{equation}
and for a mixed tensor

\begin{equation}\label{Phys-tens}
    T(i..j \, k..l)= \frac{h_{i}..h_{j}}{h_{l}..h_{l}}  T^{i..j }_{k..l} \qquad \mbox {no summation over indices}
\end{equation}
The covariant derivative of a scalar function $\Phi$, which
reduces to the partial derivative, is a covariant vector
$\Phi_{,i}$, the gradient of $\Phi$. The physical components of
the gradient of a scalar function are given by

\begin{equation}\label{grad-phys}
    grad \Phi (i)=\frac{1}{h_i}\frac{\partial \Phi}{\partial x^i}
\end{equation}
The covariant derivative with respect to $x^i$ of a covariant
vector $A^i$ summed on $i$ is  the divergence of the vector
$\mathbf{A}$ and can be written in terms of the physical
components of the vector, $A(i)$

\begin{equation}\label{div-phys}
    div \mathbf{A}= \frac{1}{g^{1/2}}\frac{\partial}{\partial
    x^i}(g^{1/2} A^i) = \frac{1}{g^{1/2}}\frac{\partial}{\partial
    x^i} \left[ \left(\frac{g}{g_{ii}}\right)^{1/2} A(i) \right ]
\end{equation}

where $g$ denotes the determinant of the metric and in orthogonal
coordinates is expressed as $g^{1/2} = h_1 h_2 h_3$.

The Christoffel symbols in orthogonal coordinates are

\begin{equation}\label{Chris-ortho}
    \Gamma^i_{j k}= \sum_r \frac{1}{2}g^{i r} \left ( \frac{\partial g_{r j}}{\partial
    x^k} + \frac{\partial g_{r k}}{\partial x^j} - \frac{\partial g_{j k}}{\partial
    x^r}\right ) = \left\{%
\begin{array}{ll}
    0, & \mbox{when $i,j,k$ are all different;} \\
       \\
    \frac{1}{h_i}\frac{\partial h_i}{\partial x^j}, & \mbox{when $i=j=k,$}\\
    & \mbox{$i=j\neq k,i=k\neq j$;} \\
    -\frac{h_i}{h_j^2}\frac{\partial h_i}{\partial x^j}, & \mbox{when $j=k\neq i$.} \\
\end{array}%
\right.
\end{equation}

The physical components of the covariant derivative of a vector
field are

\begin{equation}\label{deriv-orthog}
    A(i;j)= \frac{1}{h_i h_j} A_{i ; j}=\frac{1}{h_i h_j}\left [\frac{\partial}{\partial
    x^j}{h_i A(i)}- \sum_k h_k A(k) \Gamma^k_{i,j} \right ]
\end{equation}

Using equation \eqref{deriv-orthog} and \eqref{Chris-ortho} the
physical components of the strain tensor can be written as
function of the physical components of the displacement vector

\begin{align}
    \varepsilon(i i) & =  \frac{\partial}{\partial x^i}\left (\frac{u(i)}{h_i} \right )
    +\frac{1}{2 h_i^2}\sum_m \frac{u(m)}{h_m}\frac{\partial}{\partial x^m}(h_i^2)\quad \mbox{no summation on}\; i   \\
     \varepsilon(i j) & = \frac{1}{2}\left [\frac{h_i}{h_j}\frac{\partial}{\partial x^j}\left (\frac{u(i)}{h_i} \right )
     + \frac{h_j}{h_i}\frac{\partial}{\partial x^i}\left (\frac{u(j)}{h_j}
     \right
     )\right] \quad \mbox{no summation on} \; i, j, \; i\neq j
\end{align}

The expansion $\Theta$ in terms of the physical components of the
displacement vector can be written using \eqref{div-phys}

\begin{equation}\label{Exp-phys}
    \Theta = \frac{1}{h_1 h_2 h_3} \left [\frac{\partial( h_2 h_3 u(1))}{\partial
    x^1} + \frac{\partial( h_1 h_3 u(2))}{\partial
    x^2} + \frac{\partial( h_1 h_2 u(3))}{\partial
    x^3} \right ]
\end{equation}

\section{Equilibrium equations}\label{App:A2}

The conditions for static equilibrium in a linear elastic material
are determined from the force balance equation in a generic
coordinate system

\begin{equation}\label{equi-gencoord}
    \sigma^{i j}_{; \,j} + \rho f^i=0
\end{equation}

where $\sigma_{i j}$ are the stress tensor components,$f^i$ are
the external body forces per unit mass and $\rho$ is the density
of the material. Using the Eq.\eqref{deriv-orthog} and
\eqref{Chris-ortho} we can express the equilibrium equation in
terms of physical components

\begin{equation}\label{equi-physical}
    \sum_j \frac{h_i}{\sqrt{g}}\frac{\partial}{\partial x^j}\left
    ( \frac{\sqrt{g}}{h_i h_j}\sigma(i j) \right ) + \sum_{j , k} \frac{h_i}{h_j
    h_k}\Gamma^i_{j k} \sigma(j k) + \rho f(i) =0 \quad \mbox{no
    summation on}\, i
\end{equation}

Using the constitutive equations for a linear isotropic material
Eq.\eqref{isotroL}, and the definition of the strain tensor in
terms of the displacement vector Eq.\eqref{strain-split} we have
that the equilibrium equations Eq. \eqref{equi-gencoord} becomes

\begin{equation}\label{equi-displa-gen}
    (\lambda + \mu ) u_{k ;\, k i } + \mu u_{i ;\, j j} + \rho f_i =0
\end{equation}

In an orthogonal coordinate system, the components of $\nabla
(\nabla\cdot \vec{u})$ can be expressed in terms of physical
components by the relation

\begin{equation}\label{grad-div-phys}
\left [\nabla (\nabla\cdot \vec{u})\right]_i =
\frac{1}{h_i}\frac{\partial}{\partial x^i}\left (\sum_k
\frac{1}{\sqrt{g}}\frac{\partial}{\partial x^k}\left
[\frac{\sqrt{g}}{h_k} u(k)\right ] \right)
\end{equation}

and the components of the second term becomes quite complicated in
terms of the physical components of the displacement vector

\begin{align}
 u_{i ;\, j j} & =  \sum_j \frac{1}{h_j^2} \Bigg [  \frac{\partial^2
(h_i u(i))}{\partial x^j \partial x^j}  - 2 \sum_m \Gamma^m_{i j}
\frac{\partial (h_m u(m))}{\partial x^j} - \sum_m \Gamma^m_{j j}
\frac{\partial (h_i u(i))}{\partial x^m} \nonumber \\
 & - \sum_m h_m u(m) \left ( \frac{\partial}{\partial
x^j}\Gamma^m_{i j} - \sum_p \left (\Gamma^m_{i p} \Gamma^p_{j j} +
\Gamma^m_{j p} \Gamma^p_{i j} \right ) \right ) \Bigg ]
\end{align}

\clearpage

\section{Mirror's elastic and thermal coefficients}\label{App:A3}

\begin{table}[htb]
\begin{center}
\begin{tabular}{|c|c|c|c|}
 \hline
  Parameters & Fused Silica & Sapphire & $Ta_2O_5$ \\
  \hline
  \hline
  Density $\rho$  ($g/cm^3$) & $2.2$ & $4$ & $6.85$ \\
  \hline
  Young modulus  $Y$ ($erg/cm^3$) & $7.2\cdot 10^{11}$ & $4 \cdot 10^{12}$ & $1.4\cdot 10^{12}$ \\
  \hline
  Poisson ratio  $\nu$ & $0.17$ & $0.29$ & $0.23$ \\
  \hline
   Loss angle $\phi$  & $5\cdot 10^{-9}$ & $3\cdot 10^{-9}$ & $4.5\cdot 10^{-4}$ \\
   \hline
  Linear thermal expansion  $\alpha$  ($K^{-1}$) & $5.5\cdot 10^{-7}$ & $5\cdot 10^{-6}$ & $3.6\cdot 10^{-6}$ \\
  \hline
  Specific heat per unit mass $C$  ($erg/(g\, K)$) & $6.7\cdot 10^6$ & $7.9\cdot 10^6$ & $3.06\cdot 10^6$ \\
  \hline
  Thermal conductivity  $\kappa$  ($erg/(cm \, s\, K)$) & $1.4\cdot 10^5$ & $4\cdot 10^6$ & $1.4\cdot 10^5$ \\
  \hline
  Refraction index (at $\lambda=1.064 \mu m$) $n$ & $1.46$ & $1.75$ & $2.06$ \\
  \hline
\end{tabular}
\end{center}
\caption{These parameters are referred to a temperature of
$300~K$. In reference~\cite{Fejer} the thermal conductivity of
$Ta_2O_5$ is chosen
  equal to that one of crystalline sapphire, but we think it would be more appropriate to take a
  value more similar to that of amorphous fused silica, being a ion-beam sputtered thin layer~\cite{cond-film}.
   Moreover the thermal conductivity of most thin films is
   found ~\cite{cond-film2} to be orders of magnitude lower than that of the material in bulk form.
    How this affects the thermal noise evaluation for the GW interferometers is under investigation. }
\end{table}

% ------------------------------------------------------------------------

%%% Local Variables:
%%% mode: latex
%%% TeX-master: "../thesis"
%%% End:

\bibliographystyle{unsrt}
\renewcommand{\bibname}{References} % changes default name Bibliography to References
\bibliography{references} % References file

\addcontentsline{toc}{chapter}{References} %adds References to contents page

\end{document}